\documentclass{ustthesis}
\usepackage{graphicx}
 \graphicspath{{Fig/}}
\usepackage{epsfig,graphics,subfigure,psfrag,amsmath,amssymb}
\usepackage{cite}
\usepackage{multirow}
\usepackage{epsf}
\usepackage{color}
\usepackage{algorithm}
\usepackage{algorithmic}
\usepackage{subfigure}
\usepackage{subfloat}
\usepackage{url}

\usepackage{etoolbox}

\newcommand{\tab}[1]{\hspace{3mm}}

\title{High Performance Network-on-Chips (NoCs) Design: Performance Modeling, Routing Algorithm and Architecture Optimization}  
\author{Zhiliang Qian}     
\degree{\PhD}              
\subject{Electronic and Computer Engineering}      
\department{Electronic and Computer Engineering}       
\advisor{Prof. Chi-Ying TSUI (ECE)}     
\member{Prof.  Albert C S CHUNG (CSE), Thesis Committee Chairman}
\member{Prof.  Chin-Tau LEA (ECE), Thesis Committee Member}
\member{Prof.  Jiang XU (ECE), Thesis Committee Member}
\member{Prof.  Chi-Keung TANG (CSE), Thesis Committee Member}
\depthead{Prof. Ross D. MURCH}    

\defencedate{2014}{03}{20}      

\begin{document}


\maketitle

%


\acknowledgments
\par This thesis summarizes my five and half years study and research experience in
VLSI Research Laboratory at Hong Kong University of
Science and Technology. I would like to take this opportunity to thank all people who have helped, accompanied and supported me during my PhD study in HKUST.
\par My foremost thanks belong to my supervisor, Prof. Chi-Ying Tsui, for his patient
guidance, encouragement, understanding, and support all over the time.
His brilliant insights on VLSI and highly motivation on NoC led me to the wonderful world of on-chip networks. He always inspires me and has provided me with valuable advice in my study. His brilliance, enthusiasm and hard working towards research,
and nice guidance not only have benefited my current research but also have
a lasting influence in my professional career and personal development.
\par I am deeply grateful to Prof. Chin Tau Lea, Prof. Jiang Xu, Prof.
Chi-Keung Tang and Prof. Albert C. S. Chung for serving on my Thesis Committee, and I am full of gratitude
to Prof. Oliver Chiu-Sing CHOY from Chinese University of Hong Kong for
being my Thesis External Examiner. They managed to take time to serve on my
committee in their tight schedule. I would also like to thank Prof. Radu Marculescu from Carnegie Mellon University for hosting my Fulbright visit. I benefit a lot through working together with Prof. Radu Marculescu, Prof. Diana Marculescu in CMU and Prof. Paul Bogdan from University of Southern California. Their valuable feedback helped improve my work in many ways.
\par I would like to thank Mr. Leo Fok and Mr Siu Fai Luk for providing me with great technical
and administrative support these years. I thank everyone in VLSI lab (especially Dr. Jie Jin, Dr. Liu Feng, Dr. Shao Hui, Mr Yunxiao ling, Mr Yingfei Teh, Mr Xing Li, Mr Youzhe Fan, Mr Jingyang Zhu and Mr Syed Abbas Mohsin) for creating
a family environment and the friends I met in HKUST for their friendship and
help. 
\par Last, but not the least, I would like to express my deepest gratitude to my parents
and girlfriend for their unconditional love and support through all these years during my Ph.D. study.
\endacknowledgments


\tableofcontents


\listoffigures


\listoftables


\abstract
With technology scaling down, hundreds and thousands processing elements (PEs) can be integrated on a single chip. Consequently, the embedded systems have led to the advent of multi-core System-on-Chip (MPSoC) design and the high performance computer architectures have evolved into Chip Multi-processor (CMP) platforms. A scalable and modular solution to the interconnecting problem becomes critically important. Network-on-chip (NoC) has been proposed as an efficient solution to handle this distinctive challenge by providing efficient and scalable communication infrastructures among the on-chip resources. In this thesis, we have explored the high performance NoC design for MPSoC and CMP structures from the performance modeling in the offline design phase to the routing algorithm and NoC architecture optimization. More specifically,  we first deal with the issue of how to estimate an NoC design fast and accurately in the synthesis inner loop. The simulation based evaluation method besides being slow, provides little insight to search the large design space in the NoC synthesis loop. Therefore, fast and accurate analytical models for NoC-based multicore performance evaluation are strongly desired to better explore the design space. For this purpose, we propose a machine learning based latency regression model to evaluate the NoC designs with respect to different configurations before the system is built or taped-out. Then, for high performance NoC designs, we tackle one of the most important problems, \textit{i.e.,} the routing algorithms design with different design constraints and objectives. For avoiding temperature hotspots, a thermal-aware routing algorithm is proposed to achieve an even temperature profile for application-specific Network-on-chips (NoCs). For improving the reliability, a routing algorithm to achieve maximum performance under fault is proposed.  Finally, in the architecture level, we propose two new NoC structures using bi-directional links for the performance optimization. In particular, we propose a flit-level speedup scheme to enhance the network-on-chip(NoC) performance utilizing bidirectional channels. In addition to the traditional efforts on allowing flits of different packets using the idling internal and external bandwidth of the bi-directional channel, our proposed flit-level speedup scheme also allows flits within the same packet to be transmitted simultaneously on the bi-directional channel. We also propose a flexible NoC architecture which takes advantage of a dynamic distributed routing algorithm and improves the NoC communication performance with minimal energy overhead. This proposed NoC architecture exploits the self-reconfigurable bidirectional channels to increase the effective bandwidth and uses express virtual paths, as well as localized hub routers, to bypass some intermediate nodes at run time in the network.  From the simulation results on both synthetic traffic and real workload traces, significantly performance improvement in terms of latency and throughput can be achieved. 
\endabstract




\chapter{Introduction}

\section{Challenges in computing platform design}
\begin{figure}
\includegraphics [width=1\textwidth]{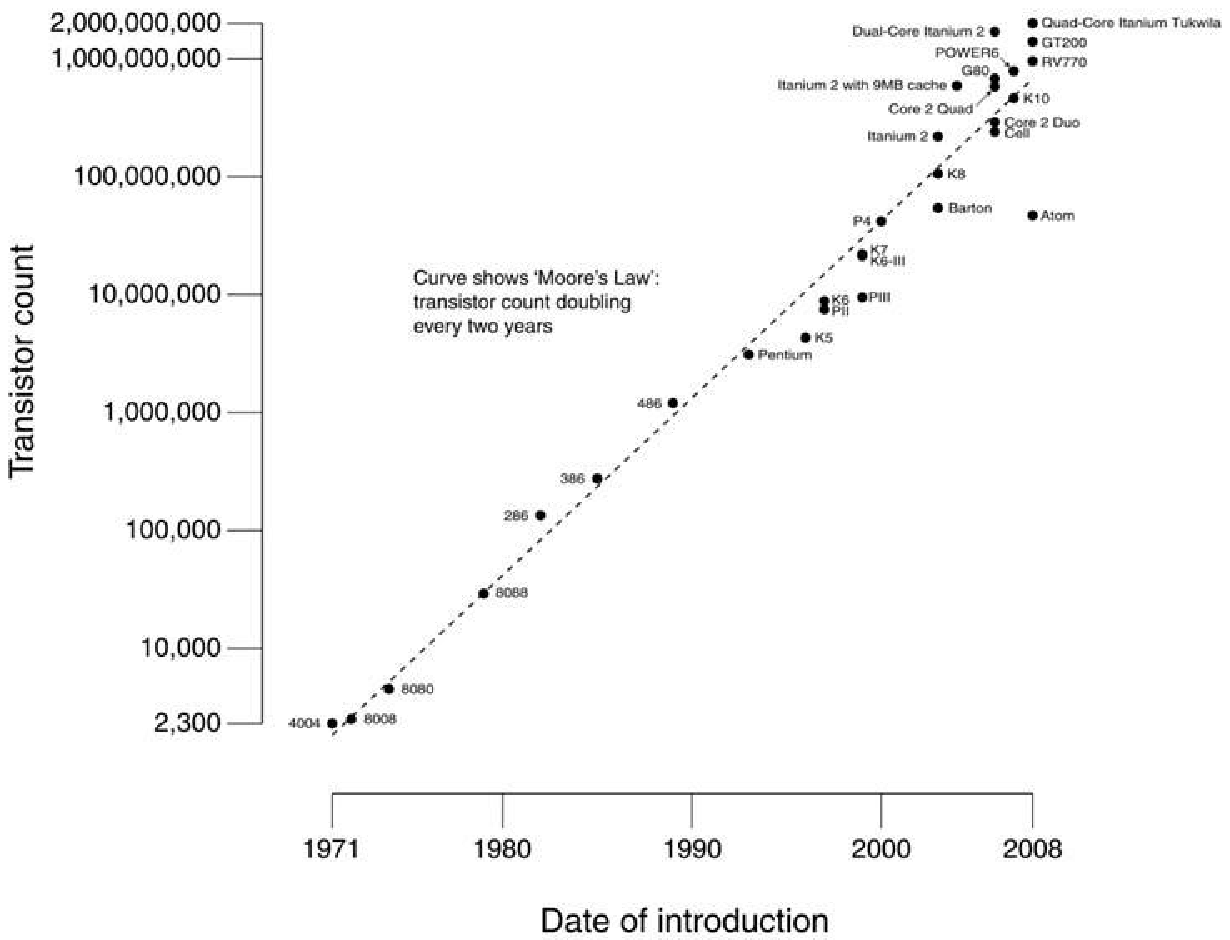}
\caption{\label{Moore_Law} The plots of transistor counts against dates and Moore's Law (courtesy of \cite{transistor_count})}
\end{figure}

\begin{figure}
\includegraphics [width=1\textwidth]{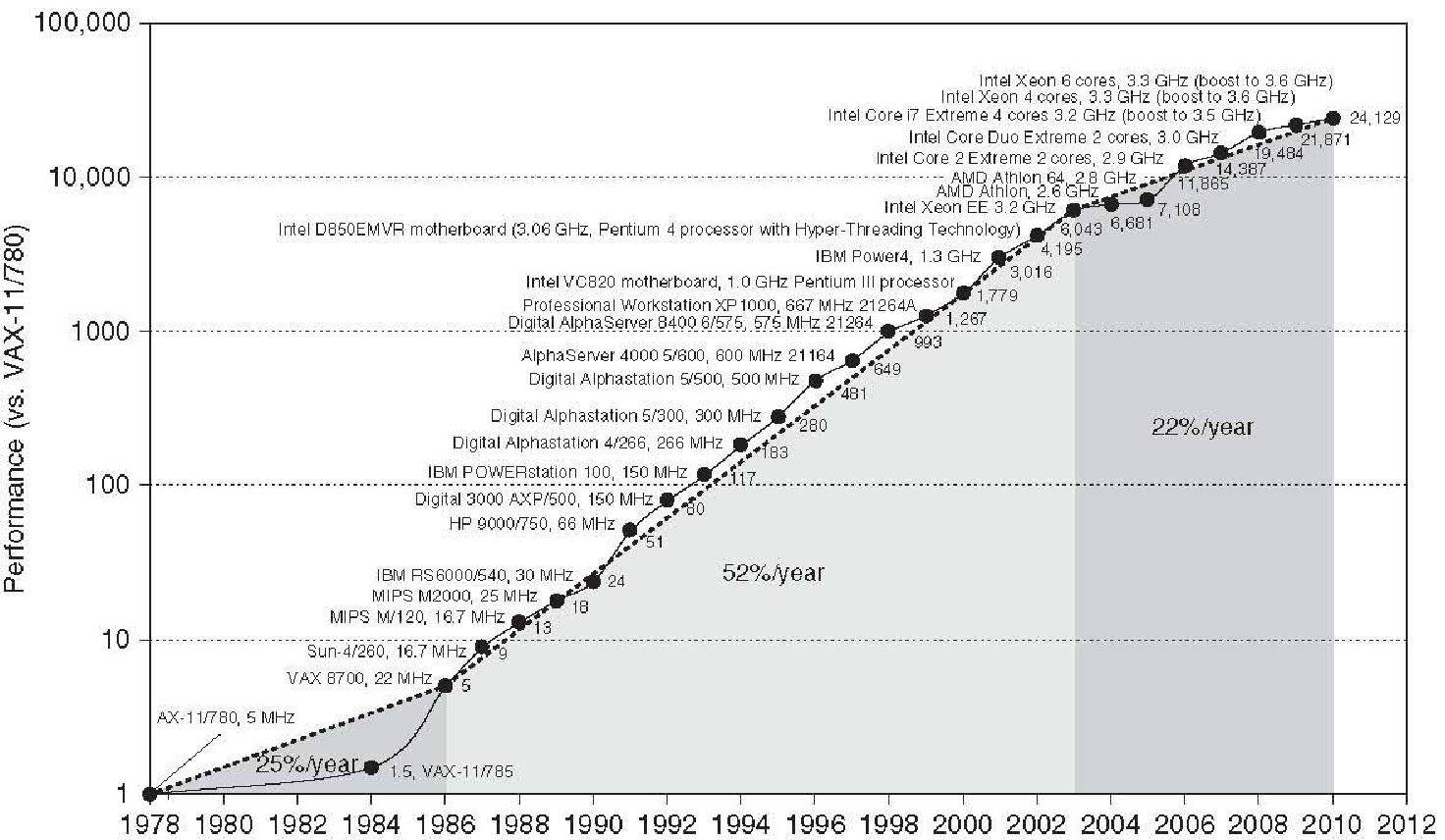}
\caption{\label{Growth_in_processor} The illustration of the processor performance growth since the late 1970s (from \cite{Computer_architecture_book})}
\end{figure}

\begin{figure}
\includegraphics [width=1\textwidth]{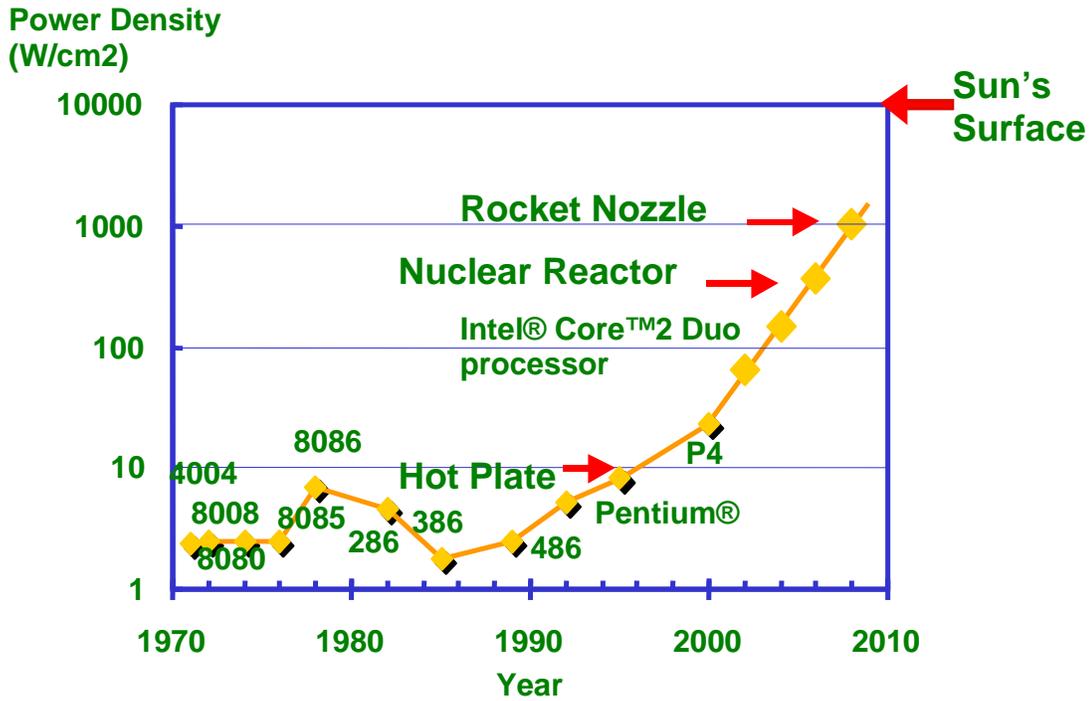}
\caption{\label{power_trends} Power density trends in Intel's CPU (from \cite{digital_system_design})}
\end{figure}

\begin{figure}
\includegraphics [width=1\textwidth]{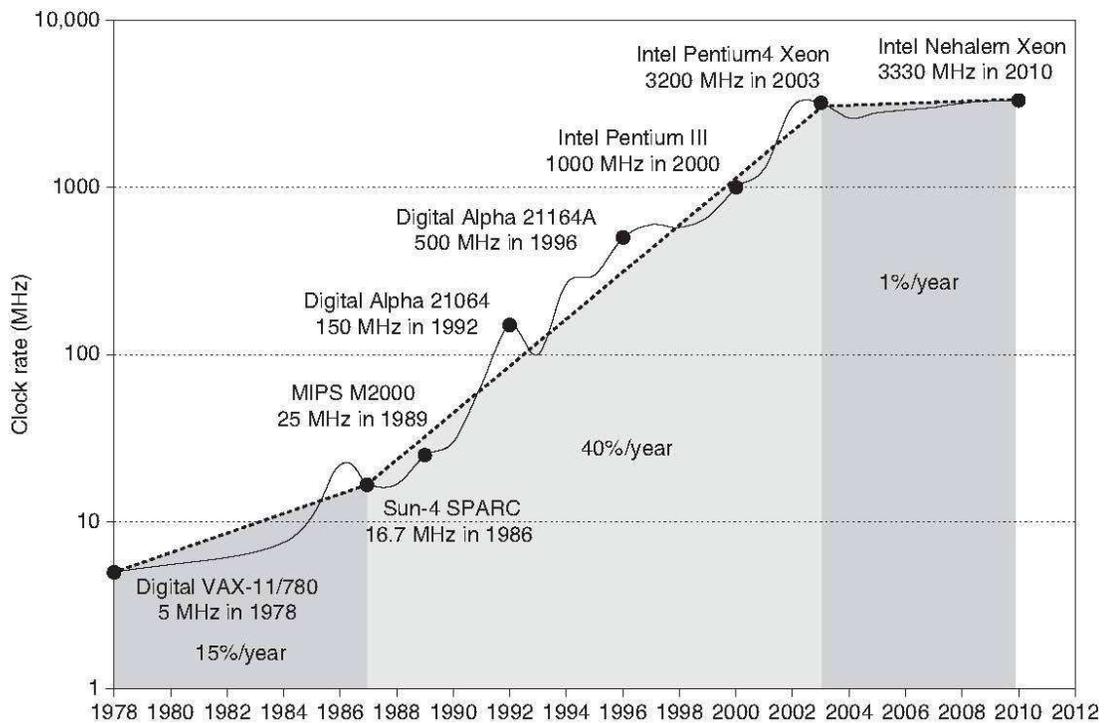}
\caption{\label{clock_rate_trends} Growth in clock rate of microprocessors (from \cite{Computer_architecture_book})}
\end{figure}

\begin{figure}
\includegraphics [width=1\textwidth]{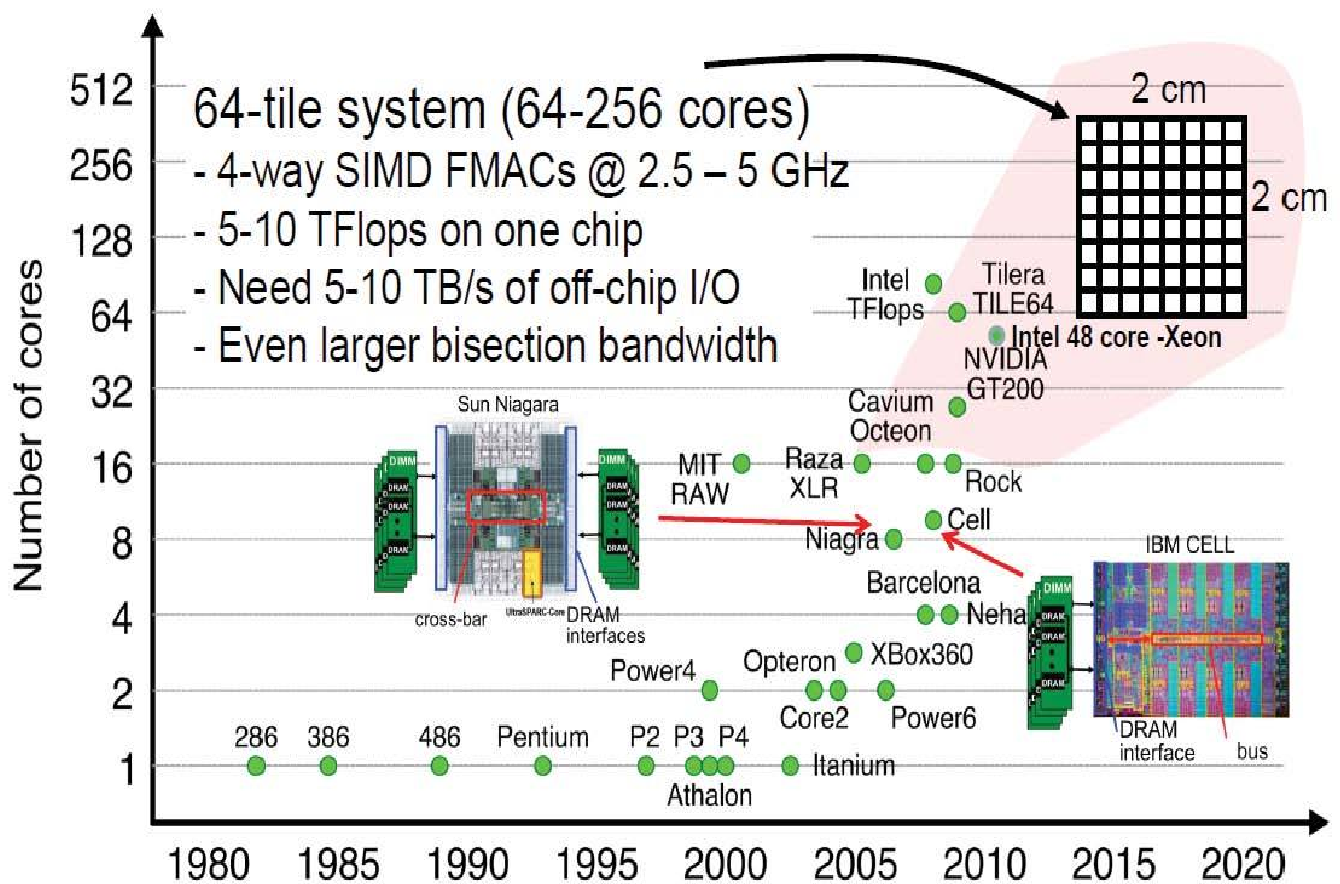}
\caption{\label{core_trends} Manycore system roadmap for improving computing platform performance (from \cite{isscc_noc_tutorial})}
\end{figure}

Computer and IC technology have made dramatic progress in the past few decades since the first generation of electronic computer was built \cite{Computer_architecture_book}. As indicated by the Moore's Law \cite{Moore_law} (shown in Fig. \ref{Moore_Law}), the transistor count on an integrated circuits(ICs) doubles every two years. Accordingly, by the year of 2012, the state-of-the-art processors already contain billions of transistors (such as the Intel's 10-core Xeon CPU with 2.5 billion transistors \cite{Xeon} and Nvidia's 7.08 billion transistors GPU \cite{Nvidia_gpu}). The steady technological improvements, together with the enhancement from better computer architectures, have contributed to a consistent performance improvement every year \cite{Computer_architecture_book}. Fig. \ref{Growth_in_processor} depicts the comparisons of computer processor performance relative to a VAX-11/780 processor which are measured using the standard SPEC benchmarks over the years \cite{Computer_architecture_book}. As shown in the figure, before 1980s, the growth in performance is largely driven by the technology advancement, which gives about $25\%$ performance increase per year \cite{Computer_architecture_book}. Then, this growth rate is improved to about $52\%$ due to the introduction of more advanced architectural and computer organizational concepts, such as the emergence of the RISC (Reduced Instruction Set Computer) based microprocessors \cite{Computer_architecture_book}. However, this trend begins to slow down again due to the constraints of power \cite{digital_system_design} \cite{Computer_architecture_book}. As shown in Fig. \ref{power_trends}, the power and power density on chip have been dramatically increased with the technology scaling down by the year of 2010. This is because as the technology process improves to a new generation, the increase in the number of transistors and the operating frequency overwhelms the decrease in load capacitance per transistor and the running voltage, which results in an overall growth in power density and energy \cite{Computer_architecture_book}. For example, the Intel processor Core2 Duo consumes as high as 130W power which is about 20 times power of the 10 years ago Pentium processor in market \cite{jounarth_thesis}. The high power density causes severe reliability issues and inevitably makes the die cost unaffordable due to the high cooling requirements and costs. In order to avoid reaching the power limit, the clock frequency growth with time has to be slowed down. This trend is shown in Fig. \ref{clock_rate_trends}, where the clock frequency remains around $2GHz$ after year 2003 \cite{Computer_architecture_book}. Instead of continuing the aggressive clock frequency scaling, the computer architects have proposed a new direction to achieve maximal performance under these tight constraints and budgets, \textit{i.e.,} by employing more processor cores on chip for the computation tasks \cite{isscc_noc_tutorial}. Consequently, the embedded systems have led to the multi-processor System-on-Chip (MPSoC) design and the high performance computer architectures have evolved into Chip Multi-processor (CMP) platforms, which involve tens or hundreds processor elements, memory blocks, ASIC acceleration engines to be inter-connected together on chip. Each processing element performs its tasks in a parallel way taking the advantage of the parallelism either in task, thread or system level. As shown in Fig. \ref{core_trends}, many recent chips have already switched to the paradigm of multi-core based platform for this purpose. For example, in \cite{CMP_ISSCC}, a $16$-core heterogeneous digital baseband IC for MIMO 4G Software Defined Radio (SDR) is proposed. Their proposed NoC-based prototype doubles the throughput and consumes only $39\%$ power over the previous MPSoC solutions. Another example is the Intel $80$-tile Teraflops processor \cite{Intel_teraflop} which is a homogeneous NoC-based CMP platform and delivers up to $1.28$ TFlops of performance. Recently, photonic on-chip network based multi-core systems have also been widely studied \cite{optical_noc,optical_noc_2}, which further attempts to optimize the traditional metal-based interconnect performance in terms of delay and power for future SoCs with thousands PEs.
  

\section{NoCs for multi-core communication}
\begin{figure}
\includegraphics [width=1\textwidth]{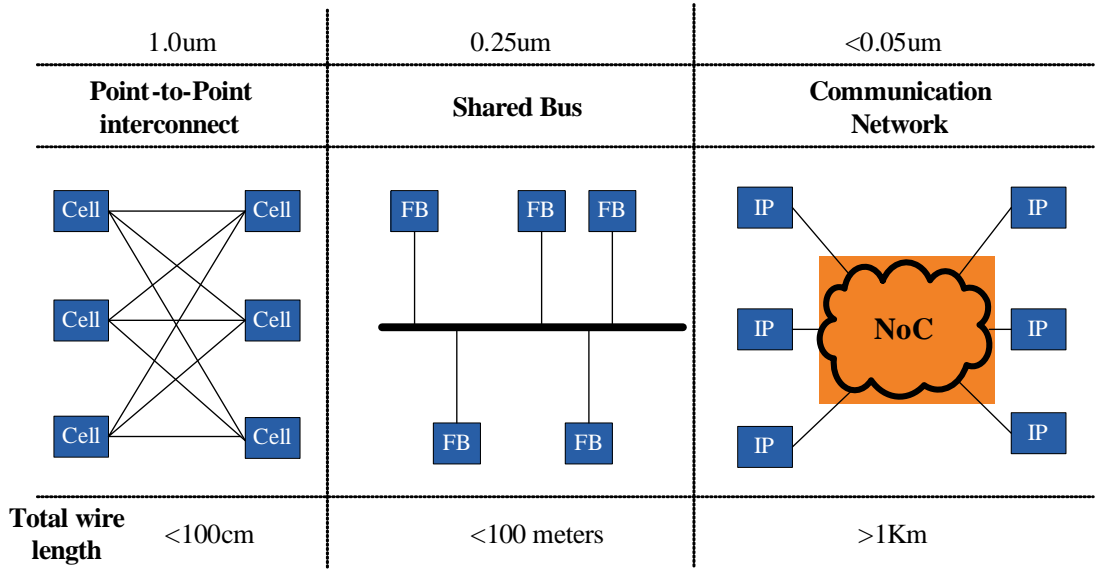}
\caption{\label{NoC_intro} The trend of on-chip interconnections (from \cite{Routing_algorithm_springer})}
\end{figure}

For multicore based computing platform, an efficient way to manage the communication among the on-chip resources become critically important. The peer-to-peer interconnection consumes large wire area, which leads to large area and fan-outs \cite{Routing_algorithm_springer}. The bus based architecture suffers from its limited bandwidth as well as scalability \cite{Routing_algorithm_springer}. The scalability issue of these two schemes brings significant overhead in power and transmission delay when the chip feature size reduces beyond 45nm. To satisfy the communication requirements with hundreds or thousands processor elements (PEs), Network-on-Chip has been proposed as an efficient and scalable solution. Borrowing the concepts in Internet and wireless network, NoC use routers to route packets instead of wires \cite{NoCbook_Peh}. The latency and throughput performance is improved due to the higher bandwidth offered by the network. Meanwhile, the power consumption can be significantly reduced by breaking long links between the processors and avoiding high fan-outs in the outputs. In summary, Fig. \ref{NoC_intro} shows the trend of on-chip interconnection and compares the total wire length under different technology nodes \cite{Routing_algorithm_springer}. As shown in the figure, for the technology nodes beyond 50nm, NoCs are more preferred over the other two paradigms in order to provide scalable communications for more than $1Km$ wire lengths. 
\\In Fig. \ref{NoC_VOPD_platform}, we show an example of a MPSoC design using the mesh topology NoC for the video object plane decoder (VOPD) application \cite{NoCbook_Peh}. The whole application is characterized by an application task graph \cite{NoCbook_Peh}, where the vertices in the graph represent certain computation tasks need to be performed and the edges indicate the communication bandwidth (MB/s) between two adjacent tasks. As shown in the figure, for the NoC-based VOPD platform, the whole system consists of twelve tiles organized in a rectangular mesh topology. Each tile is made up of a processor element (PE) and a router. The PE executes certain tasks in the application task graph while each router has five input/output ports that are connected to the four neighboring routers as well as the local PE. At run time, the packets are routed based on the routing algorithm which is designed to determine the order of the routers to be traversed for a specific communication flow. For the NoC-based multicore system, besides the latency and throughput improvement, it also brings the following advantages:\\
\\ \textbf{1) High reliability:} For Multicore systems, the complex system is highly susceptible to faults \cite{vicis}. Compared to point-to-point dedicated links and buses, NoC can achieve higher reliability by providing redundant paths among the cores. If some of the routers fell into permanent or temporary faults, the other routers can be utilized to re-route the packets to the destinations and hence packet acceptance rate will not drop dramatically.\\
\\ \textbf{2) Modular design and IP re-use:} NoC provides sufficient bandwidth for communication, while the processors can be designed without considering the network; therefore it supports modularity design and IP reuse \cite{Book}. Moreover, the global clock synchronization is not necessary in NoC which increase the overall system yield \cite{GALS,jounarth_thesis}.\\
\\ \textbf{3) Global asynchronous, locally synchronous (GALS) design:} For multicore systems, it is difficult to distribute a single clock over thousands processor cores. To deal with these issues, NoC offers a good platform for the GALS design style \cite{GALS} because each tile (processor elements and the router) can work separately within its own clock domain \cite{jounarth_thesis}. By employing GALS design, multiple Voltage-Frequency islands can be developed in different regions of NoC so as to achieve lower power consumption \cite{jounarth_thesis}.\\
\\ \textbf{4) Power and area efficiency:} Compared to the buses, the arbitration time for contention is much smaller as each router only needs to handle local contention scenario \cite{Comparison_NoC}. Therefore, large buffers to store the unserved packets are not needed in NoC routers, which result in a more compact router design and reduces the area/power overhead \cite{Comparison_NoC}. For power dissipation, because the buses are connected to all the PEs in the system, while the links in NoC only need to connect two neighboring routers (or a router and a PE) \cite{Comparison_NoC}. Therefore, with proper floorplanning, NoCs uses shorter wire length and occupies less load per transition \cite{Comparison_NoC}. Moreover, NoCs provide a variety of efficient power management strategies to further reduce power. This is because the NoC can be partitioned into sub-networks and each region can be powered-off or slowed down via dynamic voltage and frequency scaling (DVFS) individually \cite{Comparison_NoC}. High power efficiency can be achieved without significant degradation in the overall system performance. 
\section{Application-specific NoC design flow}
\begin{figure}
\includegraphics [width=1\textwidth]{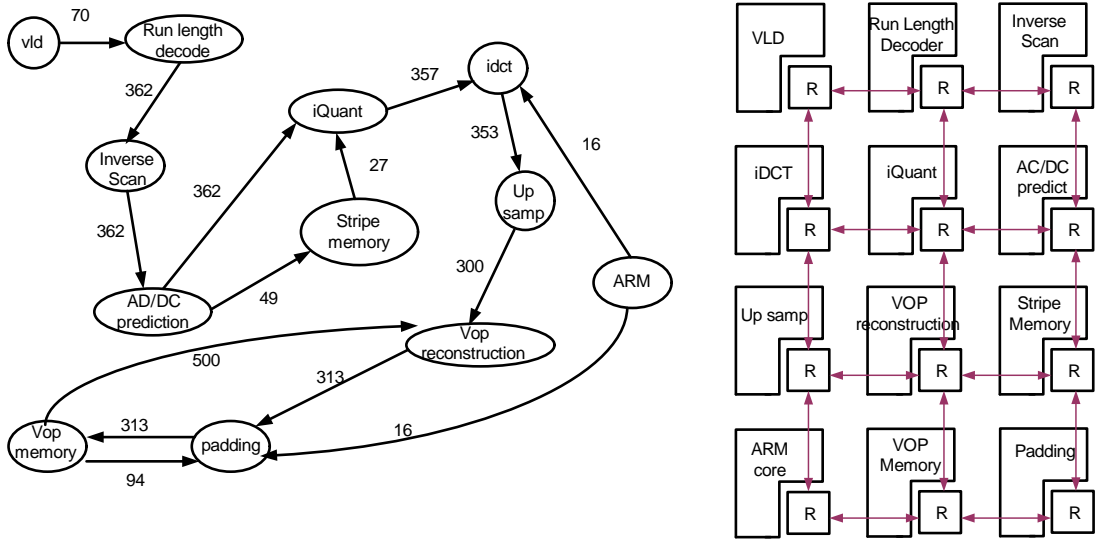}
\caption{\label{NoC_VOPD_platform} The NoC-based Video Object Plane Decoder (VOPD) platform in \cite{NoCbook_Peh}}
\end{figure}

\begin{figure}
\includegraphics [width=0.95\columnwidth]{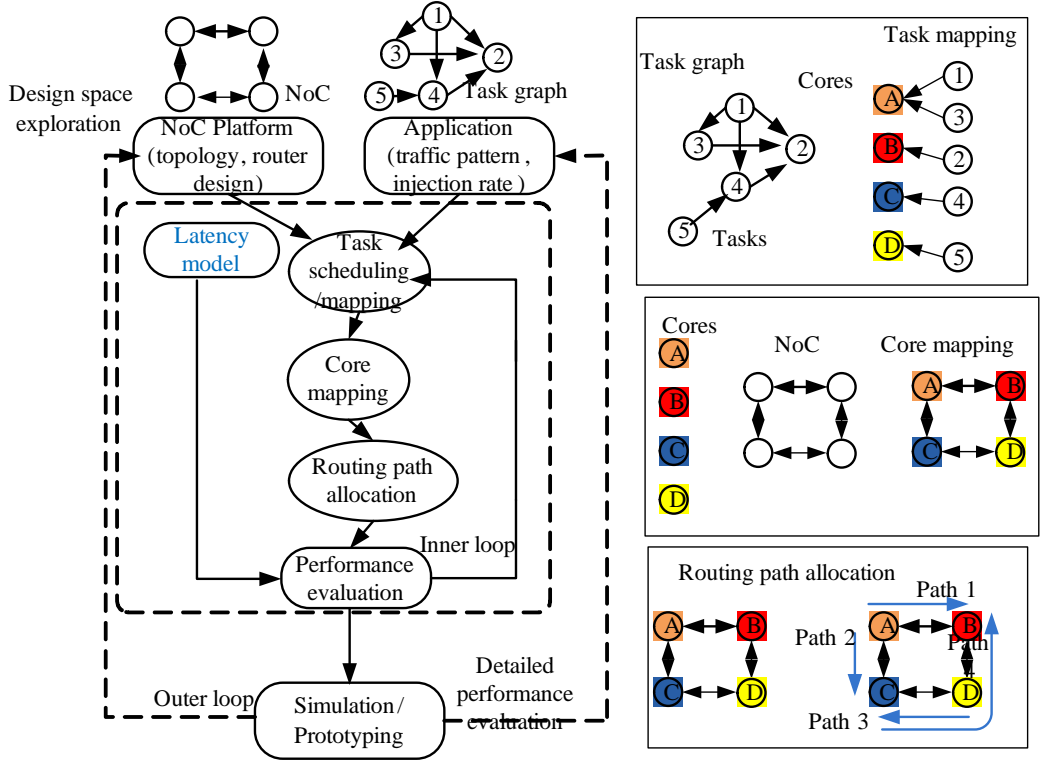}
\caption{\label{NoC_performance_flow} NoC synthesis flow for application specific multicore systems}
\end{figure}

In a typical NoC-based multicore system design, we begin with a specification of performance requirements combined with some cost constraints. These performance metrics, such as the latency/throughput, power consumption and hotspot temperature, drive the choice of NoC design parameters. A typical NoC synthesis flow works as follows (summarized from \cite{NoCbook_Peh}): \\
\\ \textbf{1) Task scheduling and mapping:} The first thing to determine is to allocate and schedule the tasks on the available processors. Usually a task graph is utilized to characterize the traffic patterns and the communication volumes of each traffic flow in the application (as shown in Fig. \ref{NoC_VOPD_platform}). Given the processors in the platform, task scheduling and mapping algorithms are developed to decide which processor that a specific task should be executed on as well as the order of the tasks to be executed on the same processor. In this step, bandwidth utilization, total delay and power consumption are major design objectives while physical bandwidth as well as hard or soft deadline of some particular tasks are the constraints that need to be considered.  \\
\\ \textbf{2) Core mapping:} After the tasks are scheduled and mapped onto processors, the next step is to place these processors onto the NoC architecture. A core mapping algorithm is developed for this purpose. The core communication graph derived in the previous step determines the placement of tiles in this step. A mapping solution with high throughput, low latency and low power is usually desired while it should not exceed the capacities of the physical link bandwidth.\\
\\ \textbf{3) Routing algorithm design:} After the task and processor mapping, routing algorithm is developed to decide the physical paths for sending the packets from the sources to the destinations. It will greatly affect the packet latency between the two cores as well as the overall chip power and thermal profile. For the routing algorithm design, one important issue is to avoid deadlock. The deadlock refers to the situation that the whole system stalls due to the circular dependencies \cite{application_specific_plaesi}. More specifically, for the deadlock scenario, it happens at run time, where flits from some packets occupies some resources in the router (such as the buffer). At the same time, they request to use other resources (such as the buffer in the downstream node), so the dependencies of the channels may have chances to form a cycle. In this case, all these packets are stalled in place and can not proceed to the destination anymore \cite{application_specific_plaesi}.
\\As there are a lot of possible design choices in each of the three steps above, NoC-based multicore system design produces a large space to be explored. Therefore, it is of utmost importance to provide an accurate performance evaluation with respect to the specific configurations in the synthesis inner loop. Both analytical models and simulations can be used in the NoC performance evaluations. To fully understand and model the details of the network situations occurred at run time, NoC simulators are developed and widely adopted with high fidelity. On the other hand, since NoC designs have many power, area and latency trade-offs in topology, task and core mapping algorithms \textit{etc.}, analytic models have also been deployed to allow fast design space explorations \cite{TVLSI12}. In general, it is more reasonable to work with simple analytical models first in the synthesis loops, while more detailed simulations become necessary to accurately characterize the exact performance of the network after only a few candidates being remained \cite{TVLSI12}. In Fig. \ref{NoC_performance_flow}, we summarized the synthesis flow for the NoC design. The usage of analytical models is highlighted within the inner loop in the figure.
\section{NoC characterization}
In order to better understand the terminologies, concepts and algorithms developed in this thesis, in the following, we briefly review the basic ideas and the models that characterizes an NoC platform. 
\subsection{NoC topology}
Network topology refers to the arrangement of various elements (links, nodes, \textit{etc.}) of the network \cite{Book}. Essentially, it is the topological structure of the multi-core platform and is dependent on the placement of the network's components, including the locations of the processors and the routers \cite{Book}. There are various NoC topologies, such as mesh, torus \cite{mesh_topo}, butterfly \cite{flattern_butterfly}, 3D-mesh \cite{3D-mesh} and fat trees \cite{fat_tree}. 
\\ In this thesis, we assumed the underlying NoC system is composed of a 2D mesh network (as shown in Fig. \ref{NoC_VOPD_platform}). The reason for using the 2D mesh network is due to its regularity and layout efficiency on silicon surface \cite{mesh_topo}. Moreover, the 2D mesh topology also matches well with the current IC manufacturing technology for the layout consideration, especifally for most IC components which have rectangular shape \cite{jounarth_thesis}. Therefore, this topology has attracted wide attention in most state-of-the-art NoC-based multicore prototypes (\textit{e.g.,} MIT’s 16-tile RAW chip \cite{raw_chip} and Intel’s 80-tile TFLOPS chip \cite{Intel_teraflop} ). 
\\ Another advantage of the mesh based topology is the high scalability to merge or combine building blocks which are developed with regular shapes \cite{jounarth_thesis}. When the complexity of the embedded systems is increasing, more PEs are trying to be put together on the chip. With regular shape, the additional PE blocks can be easily integrated on the original design \cite{jounarth_thesis} which eases the voltage/frequency island based control on NoC.
 
\subsection{NoC switching technique}
In general, based on the flow control granularity, the NoC routers can be classified into three types, namely the circuit switching, virtual cut-through switching and the wormhole switching \cite{NoCbook_Peh}. The reviews are done based on \cite{NoCbook_Peh,interconnection_network,jounarth_thesis}:
 \\ In the circuit switching paradigm, two PEs set up a specific communications channel (named as circuit path) in NoC first before they begin to transfer packets to each other. The circuit switching ensures the bandwidth for the channel settled and keeps connected during the whole communication period of the specific flow \cite{interconnection_network}. However, it is sometimes inefficient in using the channel bandwidth because the unused links reserved for one connection cannot be used by others when the circuit is set up \cite{NoCbook_Peh,interconnection_network}. 
 \\ In virtual cut-through switching \cite{virtual-cut-through}, the buffers are designed to be capable of storing the whole maximum packet. However, the whole packet is only stored into a router buffer if the downstream router buffer is already occupied by other packets \cite{jounarth_thesis}. Otherwise, the flits once arrived at the current buffer can be routed directly without the need to wait for the arriving of other flits in the same packet \cite{jounarth_thesis}. Hence, in virtual cut-through switching, if the packet stall \cite{Book} happens due to the failure of allocating a downsteam channel, the packet stays in the current node will not block any other packets \cite{jounarth_thesis}. Compared to the circuit switching, as the flits can be forwarded immediately, the network latency under no congestion is reduced; however, the virtual cut-through switching still requires the buffer size to be large enough in order to store the whole packet under congestion \cite{jounarth_thesis,NoCbook_Peh,interconnection_network}.
 \\ In order to overcome the limitations in the circuit and virtual cut through switching, the wormhole switching techniques have been proposed and widely used in the communication networks \cite{interconnection_network} as it requires fewer buffer resources than previous two techniques \cite{NoCbook_Peh,interconnection_network}. In particular, in wormhole NoC, each message consists of several packets. Furthermore, the packet is divided into several flits, which are the minimal flow control units in the routing. The header flit is utilized to settle the routing paths in the routers, while the body and tail flits simply follow the paths reserved by its header. When the tail flit leaves the router, it will release all the resources it reserved for the packet so that the consequent packets can use them again \cite{NoCbook_Peh,interconnection_network}. Therefore, one major advantage of wormhole routing is that it does not need a large enough buffer to hold the whole packet, which drastically reduces the overall latency \cite{jounarth_thesis}.   
 
\subsection{NoC router design} 
\begin{figure}
\includegraphics [width=0.95\textwidth]{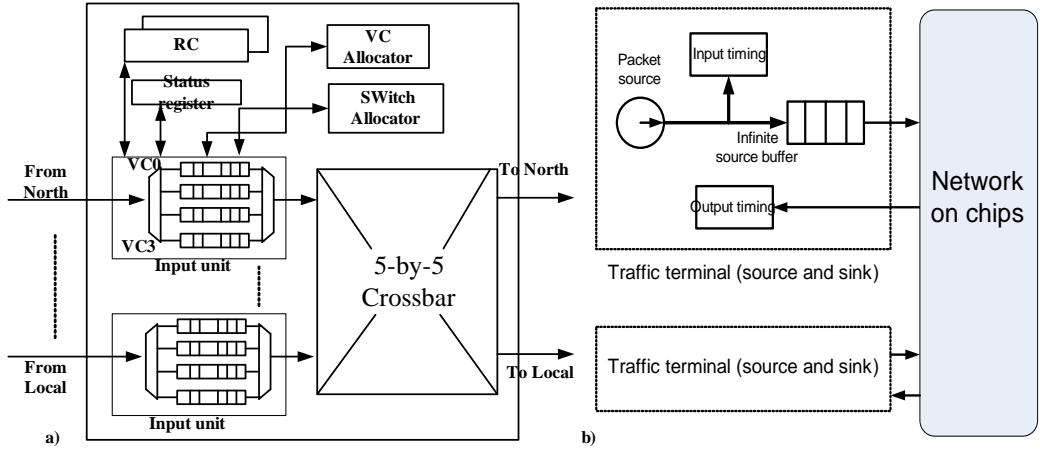}
\caption{\label{NoC_router_architecture} The NoC router architecture with virtual channels and the open-loop source queuing delay measurement}
\end{figure}

The typical structure of an on-chip router for a mesh NoC is shown in Fig. \ref{NoC_router_architecture}, where we show the basic control and data path with four virtual channels (VCs) in each input port \cite{Book, NoCbook_Peh}. For the wormhole router, it can be viewed as a special case of virtual channel routers where the number of VCs per input port equals to one. As shown in Fig. \ref{NoC_router_architecture}, the router control path usually consists of the routing computation (RC) module, the virtual channel allocation (VA) module and the switch allocation (SA) module. Its data path usually consists of the buffers (virtual channels or a single buffer), the switching crossbar and the output registers. Of note, the RC module is used to compute the output port according to the destination address recorded in the header flit in front of the buffer. The VA module is used to allocate the downstream virtual channels to the packet in the current virtual channel buffer. After a virtual channel is allocated to the packet, the switch allocation module works to arbitrate for the usage of the switch fabric (crossbar) entries among the packets. On top of this baseline architecture, many modifications on the router datapath and control path have been proposed to reduce the power consumption and latency (\textit{e.g.,} the speculative router \cite{Book} and the lookahead router with bypass architectures \cite{look_ahead_router}). 
\\The network interface (NI) is needed between the router and PE to convert and transfer messages. The standard measurement setup for interconnection networks is shown in Fig. \ref{NoC_router_architecture}-b \cite{Book}. To measure the performance of an interconnection network, we need to attach terminals (\textit{i.e.,} PEs)
to the local port of the adjacent router in the network. The NI is usually modeled with infinite buffer size to isolate the NoCs from the processors \cite{Book}. It is important that in this open-loop measurement, the monitors in NI is placed ahead of the source queue instead of after the queue (the monitor records the injection and ejection times of the packets to PE)\cite{Book}. In this way, the packets that have been generated and are still waiting to be injected into the network are considered \cite{Book}. Therefore, the overall packet latency under this set-up will not only include the NoC traverse time but also the source queuing time \cite{Book}. Various workload can be applied to the NI, including the traffics generated from real processor models, the trace-based workload as well as the synthetic workload without using any processor information \cite{Book}.

\section{Thesis overview}
In our work, we investigated the issues for a high performance NoC design, from analytical performance modeling to the routing algorithm design and NoC architecture optimization. In particular, we looked at the following areas: NoC performance modeling for the design space exploration (Chapter 2); routing algorithm design for the thermal-awareness and reliability objectives (Chapter 3 and 4); flexible NoC architecture design using self-reconfigurable bi-directional channels (Chapter 5 and Chapter 6). We aim to address several key problems in NoC design from both the algorithmic point of view as well as the hardware and architectural level optimization. To be more precise, the outline and
contributions of this thesis are summarized below:
\\ \textbf{Chapter 2: SVR-NoC: A learning based NoC latency model}
\\ In this chapter, instead of using conventional queuing-theory-based NoC latency model, which have several assumptions that comprise the overall prediction accuracy. We proposed a learning based NoC latency regression model, namely SVR-NoC, to accurately evaluate a candidate design in the inner loop for the design space exploration. Compared to the previous models, we showed that better accuracy over the queuing models and at the same time $70X-102X$ speedup over the detailed simulations can be achieved using the SVR-NoC model on both the synthetic traffic patterns and real application traces. Therefore, the SVR-NoC can benefit the exploration of numerous design configurations in the offline phase.
\\ \textbf{Chapter 3: A thermal-aware routing algorithm for application-specific Network-on-Chips (NoCs)}
\\ For NoC-based multi-core systems, the routing algorithm significantly affects the overall performance and needs to be tackled in the offline design phase to meet the certain design constraints. Among all the routing considerations, temperature is one of the most critical ones as the uneven temperature across the chip will introduce thermal hotspots and degrade the performance dramatically. Towards this end, in this chapter, we propose an offline thermal-aware routing algorithm to evenly distributed the traffic across the chip so to reduce the hotspot temperature. Specifically, we propose a deadlock free adaptive routing algorithm which provides maximal number of paths to route packets for the given application. Then, a linear programming based algorithm is used to find the optimal ratio to send packets among the paths. We show that as much as $10\%-20\%$ peak energy reduction can be achieved over a set of applications while the latency/throughput performance is maintained by using the proposed method.
\\ \textbf{Chapter 4: Fault-tolerant NoC routing algorithms design}
\\ In this chapter, we investigate and propose a highly resilient routing algorithm to tackle the router and link faults at run time. More specifically, we classify the permanent faults in NoC into two types (\textit{i.e.,} link faults and buffer faults). For the link faults, a highly resilient routing algorithm is used to re-route the packets from faulty links. While for the buffer faults, we propose two new schemes, namely dynamic buffer swapping and dynamic MUX swapping to handle the errors in the buffers and crossbar Muxes, respectively. We show that, higher packet acceptance rate as well as better latency and throughput performance can be achieved for a set of test traffics.
\\ \textbf{Chapter 5: FSNoC: A flit level speedup scheme for NoCs using self-reconfigurable bi-directional channels}
\\ Besides the optimization in the algorithm level, we also explored to improve NoC performance from the architectural level. In this chapter, we propose FSNoC, a new NoC router architecture that supports switching two flits from the same packet simultaneously by using the bi-directional channels. Compared to previous router architectures using bi-directional links, better link bandwidth utilization can be achieved and therefore FSNoC can lead to higher performance in latency and throughput. The channel direction control protocol as well as the router micro-architecture which supports flit-level parallel transmission have been proposed. We demonstrate the performance improvement of FSNoC using both synthetic traffic patterns as well as the traces from realistic applications. The hardware overhead of FSNoC in terms of area and power is also reported and analyzed in detail in this chapter.
\\ \textbf{Chapter 6: A traffic-aware adaptive routing algorithm on a highly flexible NoC architecture}
\\ In this chapter, we aims to add more flexible in the overall NoC architecture and propose a new platform which consists of i) self-reconfigurable bi-directional channels, ii) express virtual channels and iii) regional hub routers to improve the system performance. A fitness-based and traffic-aware adaptive routing algorithm is designed which is suitable for the proposed platform and chooses the routing path dynamically to adapt the traffic conditions at run time. Combining the routing algorithm and the platform, more than 80\% improvement in saturation throughput can be obtained, while involving less than 15\% overhead in power dissipation.
\\ \textbf {Chapter 7: Conclusion and future work}
\\ This chapter summarizes the works done in the whole thesis and discusses several future research directions.

\chapter{SVR-NoC: A Learning Based NoC Latency Performance Model}

\textit {In this Chapter, we propose SVR-NoC, a learning based Network-on-Chip
(NoC) latency model using support vector regression (SVR). Different
from the state-of-the-art NoC analytical models, which use queuing
models to compute the average channel waiting time and the source
queuing time, the proposed SVR-NoC model predicts the NoC latency
based on learning the typical training data. More specifically, given
the application communication graph, the NoC architecture and the
routing algorithm, we first analyze the links dependency and then
determines the ordering of latency analysis. The channel and source
queue waiting times are then estimated using a new generalized $GE/G/1/K$
queuing model, which can tackle bursty arrival times with general service time distributions. To improve the prediction
accuracy, the queuing theory based delays are included as one of the
features in the learning process. We propose a systematic learning
framework that uses the kernel-based support vector regression method
to collect training data and predict the traffic flow latency. The
proposed learning-based model can be used to analyze various traffic
scenarios for NoC platforms with arbitrary buffer and packet length.
Experimental results on both synthetic and real applications demonstrate
the accuracy and scalability of the proposed SVR-NoC model as well
as a $120$X speedup over simulation-based evaluation methods.} 

\section{Introduction}
\begin{figure}
\includegraphics[width=0.97\columnwidth]{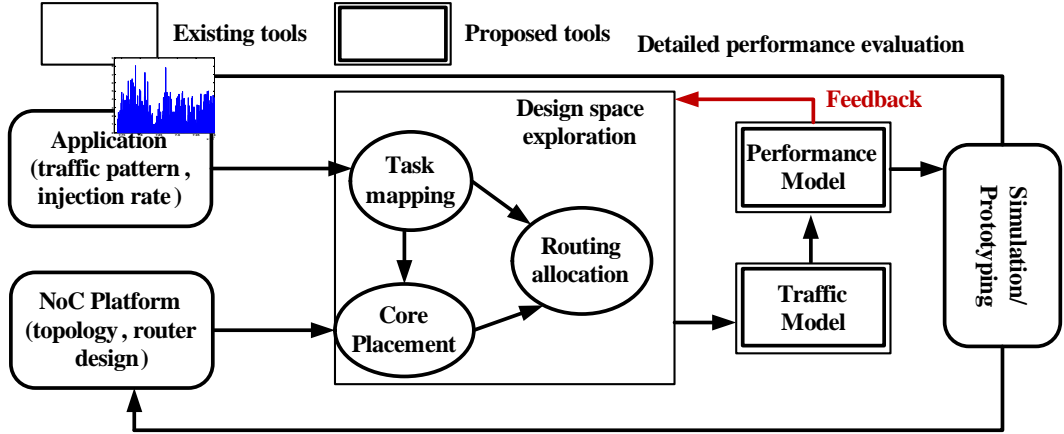}

\caption{\label{fig:The-NoC-performance}The NoC performance model used in
the synthesis inner loop for large design space exploration}
\end{figure}

The NoC complexity as well as its tight requirements on the power, latency,
and throughput have become major challenges in
the design of NoC-based multi-core systems \cite{umit}. As shown in Fig. \ref{fig:The-NoC-performance},
a typical NoC-based system design requires many synthesis steps including
task allocation, mapping, core placement and routing. Specifically,
based on the pre-characterized application traffic, the designers
first need to schedule and map the tasks on the available processor
elements (PEs). After the task scheduling and mapping are done, core
placement and routing path allocation need to be explored. Each of
the above steps can produce numerous design choices. Therefore, a
performance analysis tool is needed to evaluate whether the chosen
NoC configuration for the input application leads a better design over others
while satisfying the design constraints at the same time. Detailed
network simulations can provide performance evaluation results with
high fidelity. However, they suffer from long evaluation times and
so are only suitable for estimating a small subset of alternatives
in the final prototyping stage. Because of this, NoC performance models
are widely adopted to guide the pruning of the design space during
the system synthesis \cite{ICCAD_latency}. 

\begin{table}
\caption{\label{Table-1.-Taxonomy}Summary and comparison of NoC latency
models}
\begin{tabular}{|p{0.12\columnwidth}|p{0.10\columnwidth}|p{0.10\columnwidth}|p{0.10\columnwidth}|p{0.10\columnwidth}|p{0.10\columnwidth}|p{0.12\columnwidth}|p{0.10\columnwidth}|}
\hline 
 & \multicolumn{6}{c|}{\textbf{Queuing theory based analytical models}} & \textbf{Learning based}\tabularnewline
\hline 
{\small Models} & {  \cite{Guz07networkdelays}} & {  \cite{umit,ICCAD_latency}} & {  \cite{TVLSI12}} & {  \cite{Delay_hetergeneous_noc}} & {  \cite{Ge-type}} & {  Proposed } & {  SVR-NoC}\tabularnewline
\hline 
{\small Queue } & { \footnotesize M/M/1} & { \footnotesize M/G/1/K} & { \footnotesize G/G/1/$\infty$} & { \footnotesize M/M/m/K} & { \footnotesize GE/G/1/$\infty$} & { \footnotesize GE/G/1/$K$} & { \footnotesize N/A}\tabularnewline
\hline 
 & \multicolumn{6}{c|}{\textbf{Application related models}} & \tabularnewline
\hline 
{\small Arrival } & { \small Poisson} & { \small Poisson} & { \small General } & { \small Poisson} & { \small GE} & {\small  GE} & { \small General}\tabularnewline
\hline 
{\small Service } & { \footnotesize Memoryless} & { \footnotesize General } & { \footnotesize General } & { \footnotesize Memoryless} & { \footnotesize General } & { \footnotesize General } & { \footnotesize General}\tabularnewline
\hline 
{\small Traffic pattern} & { \small Arbitrary} & { \small Arbitrary} & {\small  Arbitrary} & {\small  Arbitrary} & { \small Uniform } & {\small  Arbitrary} & { \small Arbitrary}\tabularnewline
\hline 
 & \multicolumn{6}{c|}{\textbf{NoC architecture related models}} & \tabularnewline
\hline 
{\small Buffer size} & { \small $1$ flit } & { \small $K$ packets} & { \small $B$ flits } & { \small $1$ flit } & { \small $1$ flit } & { \small Arbitrary} & { \small Arbitrary}\tabularnewline
\hline 
{\small PB ratio}%
\footnote{{  PB ratio is defined as the ratio of average packet
size ($m$ flits) to the buffer depth ($B$ flits)}%
} & {\small  $m$ ($m\gg1)$} & {\small  $<1$} & {\small  Arbitrary} & { \small $m$ ($m\gg1)$} & {\small  $m$ ($m\gg1)$} & { \small Arbitrary} & { \small Arbitrary}\tabularnewline
\hline 
{\small Arbitration } & { \small Round-robin} & { \small Round-robin} & {\small  Round-robin} & { \small Fixed priority} & {\small  Round-robin} & {\small  Round-robin} & {\small  Round-robin}\tabularnewline
\hline 
\end{tabular}
\end{table}

Among all the NoC performance metrics, latency is recognized as one
of the most critical design parameters since it determines the whole
system throughput under specific workloads \cite{TVLSI12}. In this
work, we propose a latency model to predict the average delay of flows
in an NoC-based multi-core system for the design space exploration.
In order to derive a latency model, most previous researches are based
on the queuing-theory formalisms and treat each input channel in the
NoC router as an $M/M/1$ \cite{Guz07networkdelays}, $M/G/1/N$ \cite{ICCAD_latency},
$G/G/1$ \cite{TVLSI12}. Indeed, these models provide accurate performance
estimations when the following assumptions hold: i) The packet length
satisfies an exponential distribution, and therefore the packet service
time in the router is exponentially distributed as well \cite{Guz07networkdelays}.
ii) The traffic inter-arrival time is assumed to follow a Poisson
distribution at all traffic sources \cite{ICCAD_latency,umit}. However,
it has been observed that in many NoC systems, the behavior of the
traffic follows the fractal/long-range-dependent (LRD) pattern \cite{fractal_traffic}
and the distributions of the service time are correlated as well.
Consequently, the accuracy of the queuing theory-based model is compromised
in these cases.

In this chapter, we attempt to develop an NoC latency model which is
suitable for the synthesis inner loop and has higher prediction accuracy
by using a new approach based on the machine learning techniques.
More specifically, we first propose a new queuing-theory-based delay
evaluation methodology which can work for a variety of NoC configurations
and traffic scenarios. The proposed performance queuing (PQ) model
is based on a $G/G/1/K$ queuing formalism and generalizes the previous
NoC PQ models as follows: i) The existing traffic arrival modeling
using Poisson approximations is extended to a generalized exponential
(GE) packet inter-arrival distribution which can account for burst
traffic patterns. ii) The packet service process within
each router is modeled with a general distribution to account for
the service time correlation between routers and traffic flows. iii)
A more general NoC architecture model is used so that routers with
finite buffer depth are accommodated, thus enabling the consideration
of arbitrary buffer depth and packet length combinations. iv) By considering
the link dependencies, the proposed framework is completely generic
and can be applied to any NoC topology with different task mapping
and routing algorithms. Then, to relax some of the assumptions in
the PQ model (such as the GE traffic arrival process) and further
improve the modeling accuracy, we propose SVR-NoC, which is a support
vector regression (SVR) based NoC latency model. In SVR-NoC, the delay
predictions based on enhanced queuing-theory-based PQ model are included
as part of the features in the learning process. In the training stage,
the training data-set is formed by collecting the latency simulation
results of the same NoC platform on various synthetic traffic patterns.
We employ\textit{ Support vector regression} \cite{Vapnik}
techniques to learn the channel queuing and the source queuing models
as functions of their feature sets, respectively. During the learning
process, \textit{cross validation} is used to avoid training data
over-fitting. In the prediction stage, the learned SVR model is used
to estimate the average waiting time at each input buffer channel
and then the overall packet flow latency for the new application patterns.

The rest of the chapter is organized as follows. In Section 2.2, we review
the previous arts and highlight our contributions. In Section
2.3, we present the proposed generic queuing-theory based latency
model. Section 2.4 details the learning-based SVR-NoC latency model.
The experimental results of the proposed NoC latency model on both
synthetic and real applications are shown in section 2.5. Finally, Section
2.6 concludes this chapter.

\section{Background}

The analytical models for evaluating the NoC average latency can be
classified into three groups: probabilistic models \cite{prob_model},
network calculus models \cite{network_calculus_1}, and queuing theory
models \cite{ICCAD_latency,umit,TVLSI12}. In \cite{prob_model},
a probabilistic analysis framework was developed to model a single
wormhole router performance. However, additional effort is needed
to extend to network of routers. In \cite{network_calculus_1}, the
network calculus approach was adopted to characterize the NoC performance.
However, the average delay prediction error is larger than that of
the queuing models \cite{network_calculus_1}. Therefore, most of
the previous efforts are based on queuing models to evaluate the NoC
delay. 
\begin{table}
\caption{\label{tab:parameter} Parameters and notations in NoC latency model}
\begin{tabular}{|p{0.12\columnwidth}|p{0.78\columnwidth}|}
\hline 
Parameters & Description\tabularnewline
\hline 
\hline 
$H_{s}$ & Service time for head flit (including the link transfer)\tabularnewline
\hline 
$m$ & Average packet size (flits)\tabularnewline
\hline 
$B$ & Buffer size in each channel (flits)\tabularnewline
\hline 
$f_{s,d}$ & Communication flow from source $s$ to destination $d$\tabularnewline
\hline 
$d_{f}$ & Length (number of hops) of flow $f$\tabularnewline
\hline 
$l_{ab}$ & Link channel connecting router $a$ and $b$\tabularnewline
\hline 
$P_{f}\,(P_{s,d})$ & Set of links that form the routing path of flow $f_{s,d}$\tabularnewline
\hline 
$F_{l_{ab}}$ & Aggregate set of flows sharing link $l_{ab}$\tabularnewline
\hline 
$l_{i}^{f}$ & Link that resides in the $i^{th}$ hop of flow $f$\tabularnewline
\hline 
$C_{f}^{2}$ & SCV of the packet inter- arrival time of $f$\tabularnewline
\hline 
$\lambda_{l_{ab}}$ & Mean packet arrival rate at link $l_{ab}$\tabularnewline
\hline 
$q_{l_{i}^{f}}$ & Delay for a flit to reach the head of buffer in link $l_{i}^{f}$\tabularnewline
\hline 
$h_{l_{i}^{f}}$ & Delay for a packet head to acquire the link $l_{i}^{f}$\tabularnewline
\hline 
$\eta_{l_{i}^{f}}$ & Header flit transfer time over link $l_{i}^{f}$\tabularnewline
\hline 
$s_{l_{i}^{f}}/s_{l_{ab}}$ & Service time for a packet that travels link $l_{i}^{f}$/ $l_{ab}$\tabularnewline
\hline 
$z_{l_{i}^{f}}$ & Time that a header flit reaches the point where the accumulated buffer
space can hold the whole packet\tabularnewline
\hline 
$v_{s}$ & Source queuing time at time $s$\tabularnewline
\hline 
$L_{f}\,(L_{s,d})$ & Average latency of flow $f_{s,d}$ (cycles)\tabularnewline
\hline 
\end{tabular}
\end{table}

For the general class of queuing-theory-based NoC models, most of
the early works consider the modeling of wormhole (WH) routers under
the assumption of Poisson arrival time distribution and memoryless
packet service time distribution. For example, in \cite{Guz07networkdelays},
an M/M/1 approximation of link delay is used to analyze the capacity
and flow allocation. Although generally tractable, the accuracy of
M/M/1 models can be significantly compromised as the assumption of
exponential arrival and service time distributions may not hold in
many real applications \cite{ICCAD,fractal_traffic}. Several works
have been proposed to improve the estimation accuracy by generalizing
the arrival and service time distributions. In \cite{Hu_ananalytical},
an analytical model based on M/G/1/K queue is proposed to account
for finite size input buffers in local area networks (LANs). However,
this model is based on the Laplace-Stieltjes transform and is too
complicated to be used in the NoC synthesis loop. In \cite{umit},
an M/G/1 based latency model for NoC analysis is proposed. It only
assumes that the arrival rate of the header flits (as opposed to the
entire packet) follows a Poisson distribution. In \cite{TVLSI12},
a fixed-priority G/G/1 based NoC latency model, which attempts to
model the burst arrival times with a 2-state Markov-modulated Poisson
process (MMPP), was proposed. However, this approach targets a specific
priority-based router architecture, while many NoC routers may utilize
a more fair arbitration such as the round robin (RR) scheme. In \cite{Delay_hetergeneous_noc},
an M/M/m/K queue-based analytical model is proposed to analyze the
delay of NoCs with variable virtual channels per link. This approach
assumes negligible flit buffers (\textit{i.e.,} single flit buffer) such that
a packet reaches its destination before its tail leaves the source
host. In \cite{ICCAD_latency}, an $M/G/1/N$ queuing mode is proposed
for both wormhole and virtual channel NoCs. However, this approach
assumes that the granularity of the buffers is in terms of packets
instead of flits and therefore a single channel buffer can hold up
to $N$ packets during the analysis. This may not be the case for
NoCs whose buffers are rather small (only several flits) to save area
and power \cite{TVLSI12}.

Machine learning is a technique that has been extensively used in
the domain of pattern recognition or artificial intelligence where
it is usually difficult to derive the exact mathematical relationship
between the outputs and inputs in these problem formulations \cite{Vapnik}.
In NoC performance modeling and analysis, most of the learning models
focus on using learning-based model to improve the area/power model
accuracy. For example, in \cite{learning-aspdac-area-power}, the
multivariate adaptive regression splines (MARS) technique is used
to develop a non-parametric NoC router power and area regression model.
Compared to the conventional ORION2.0 model \cite{orion2.0}, the
learning-based model improves the prediction accuracy over a variety
of NoC implementations. Later, in \cite{area-power-model-dac}, the
model accuracy is further enhanced by explicitly modeling of control
and data paths in the regression analysis. Besides area/power modeling,
the learning techniques have also been used in optimizing NoC runtime
performance, such as the reinforcement learning based DVFS control
\cite{da-cheng-islped}, the Q-learning based congestion-aware routing
\cite{learning-routing} and the neural network based optimal dynamic
routing \cite{neural-network-routing}. However, for the latency performance
metric analysis, machine learning techniques for improving modeling
accuracy over the conventional queuing model have not been thoroughly
studied yet.

\begin{figure}
\includegraphics[width=1.00\columnwidth]{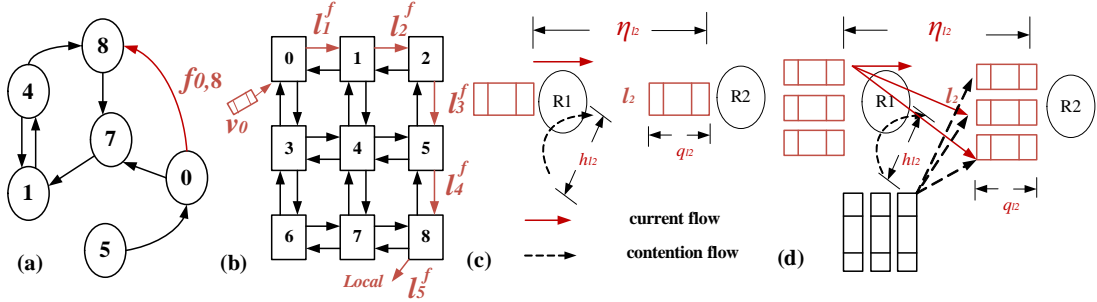}
\caption{\label{fig:An-example-showing}An example showing the flow delay: a) the core communication graph, b) the links involved to calculate the
end-to-end delay of flow $f_{0,8}$ c) the acquisition time $h$ and
transfer time $\eta$ for link $l_{1.2}$ in WH model, d) $h$ and
$\eta$ in VC model}
\end{figure}

In this chapter, we propose a new learning-based NoC latency model which
generalizes the previous work by considering: i) the arrival traffic
burstiness, ii) the general service time distribution, iii) the finite
buffer depth and arbitrary packet length. For clarity purposes, in
Table \ref{Table-1.-Taxonomy}, we summarize and compare our proposed queuing model and the
learning-based SVR-NoC model against other models proposed to date \footnote{{In the table, PB ratio is defined as the ratio of average packet
size ($m$ flits) to the buffer depth ($B$ flits)}%
}. As shown in the table, our proposed model offers a much broader coverage
for various temporal and spatial traffic patterns, as well as NoC
architectures. This provides more flexibility for the designers to
explore the NoC design space.

To the best of our knowledge, this chapter brings the following new
contributions over the previous efforts:
\begin{itemize}
\item We proposed a new queuing-theory based NoC traffic model which is
topology-independent and can be used to analyze a variety of traffic
scenarios as well as arbitrary buffer size and packet length combinations.
\item In addition to the proposed queuing model, we propose and develop
a learning-based framework for NoC latency analysis. The model has
fewer assumptions related to the packet length and traffic distribution
as well as the router architectures. 
\item We show the accuracy and scalability of the proposed SVR model using
both the synthetic traffic and real applications. The speedup of the
learning model over the conventional simulations can significantly
benefit the NoC synthesis and optimization process.
\end{itemize}

\section{Proposed NoC analytical model}

\subsection{Basic assumptions and notations}

We assume that the target applications have been scheduled and mapped
onto the target NoC platform and the source and destination tile addresses
for each specific flow $f$ in the application are known. Also, borrowing
from the idea of modeling bursty traffic in hyper-cube multi-computers
\cite{Ge-model-two}, we assume that the packet inter-arrival
times of flow $f$ have been characterized using a general exponential
(GE) distribution \cite{Ge-type} (discussed later in Section
2.3.4) with mean $\lambda_{f}^{-1}$ and a square coefficient of variation
(SCV) $C_{f}^{2}$. Therefore, the $(\lambda_{f}^{-1},C_{f}^{2})$
characterization of the traffic model is an input to our analytical
framework. Moreover, a deadlock-free and deterministic routing algorithm
is used to guarantee that no cycles are formed by the link dependencies.
Without loss of generality, in this work we use X-Y routing. Other
deadlock-free and deterministic routing schemes can also be used.
We also adopt a wormhole router architecture, where there exists a
single buffer at each input port. For simplicity, we assume that the
packets have a fixed size of $m$ (flits) as in \cite{ICCAD}. However,
this assumption can be relaxed to cover arbitrary packet length distribution.
Other assumptions are that the traffic sources (i.e., the source PEs)
have an infinite queue size and the destinations immediately consume
the arriving flits. To facilitate the discussion, the symbols in Table
\ref{tab:parameter} are used consistently in this chapter, which follows the name conventions in \cite{Delay_hetergeneous_noc,Hu_ananalytical}.

\subsection{End-to-end delay formulation}

In a WH or VC NoC, the end-to-end flow latency $L_{s,d}$
of a specific flow $f_{s,d}$ (shown in Fig. \ref{fig:An-example-showing}-a
and -b) is made up of three parts \cite{Delay_hetergeneous_noc, Hu_ananalytical}:
1) the queuing time at the source $s\:(i.e.,\: v_{s})$, 2) the packet
transfer time $(i.e.,\:\eta_{s,d})$ and 3) the path acquisition time
($i.e.,\: h_{s,d}$). It is expressed as \cite{Delay_hetergeneous_noc}: 

\begin{equation}
L_{s,d}=v_{s}+\eta_{s,d}+h_{s,d}\label{eq:40}
\end{equation}

In order to calculate $h_{s,d}$, we need to consider the path acquisition
time of every link $l_{i}^{f}$ residing in the routing path of $f$ (Fig \ref{fig:An-example-showing}-b
shows the links used for the flow $f_{0,8}$) \cite{Delay_hetergeneous_noc} and therefore: $h_{s,d}=\sum_{i=1}^{d_{f}}h_{l_{i}^{f}}$,
where $h_{l_{i}^{f}}$ is the time for a packet header to contend
a channel (in WH routing) or a VC (in VC routing) with other flows
for link $l_{i}^{f}$ and $d_{f}$ is routing path length (number
of hops) of $f$. 

If we denote the time to transmit the header flit by $\eta_{l_{i}^{f}}$,
then the packet transfer time $\eta_{s,d}$ can be rewritten as \cite{Delay_hetergeneous_noc}:

\begin{equation}
\eta_{s,d}=\sum_{i=1}^{d_{f}}\eta_{l_{i}^{f}}+(m-1)\label{eq:31}
\end{equation}

where the first term denotes the header flit transmission time over
the network, and the second term approximates the packet serialization
time of the body and tail flits. The notations of the queuing delay
$v_{s}$, $h_{l_{i}^{f}}$and $\eta_{l_{i}^{f}}$ are also illustrated
in Fig.\ref{fig:An-example-showing} for both the WH and VC NoCs.

In order to derive $L_{s,d}$, $\eta_{l_{i}^{f}}$ and $h_{l_{i}^{f}}$
for each link, which are the elements of the routing path of $f_{s,d}$,
need to be obtained. Two issues hereby arise. First, it is important
to determine the order of analyzing the links of the path. Second,
the detail model to obtain $\eta$ and $h$ of each link has to be
developed. In particular, we need to differentiate between the VC
routing and the WH routing, which are shown in Fig. \ref{fig:An-example-showing}-c
and -d, respectively. As shown in the figure, the VC router complicates
the $\eta$ and $h$ modeling in the following two aspects: i) For
the link contention time $h$, the packet can randomly choose among
$V$ VCs instead of $1$ and therefore the simple single server queuing
model in WH models can not be used directly. ii) For the flit transfer
time $\eta$, we need to consider the additional time required due
to the flow multiplexing among the VCs over the same physical link as described in \cite{Delay_hetergeneous_noc}. 

In the following subsections, we first present the procedure to determine
the orders of the analysis of the links. After that, we will then
elaborate on the formulation of the WH and VC router models.

\subsection{Link dependency analysis}

To account for the impact of the congestion at the downstream routers
on the blocking of a particular router, it is important to obtain
the dependency among all the links. Fig. \ref{fig:An-example-showing}-b
shows an example. Due to back-pressure, the waiting times $\eta_{2}^{f}$
and $h_{2}^{f}$ of link $l_{12}$ (\textit{i.e.,} $l_{2}^{f}$) will
affect the time to serve a packet in the buffer head of link $l_{01}$
(\textit{i.e.,} $l_{1}^{f}$). Therefore, $l_{01}$ is said to be
dependent on $l_{12}$. Similarly, $\eta_{2}^{f}$ and $h_{2}^{f}$
depend on those downstream links of $l_{12}$ which is determined
by the application mapping and routing. Thus it is required to finish
the analysis of all the downstream links of link $l_{01}$ in the
flow $f$ before we can calculate the queuing service and waiting
time of link $l_{01}$. To obtain the link dependencies, a link dependency
graph (LDG) is built first and the topological sort algorithm as used in \cite{Hu_ananalytical}
is then used to order the links. The detail of link dependency analysis
is presented in Algorithm \ref{alg:Link-dependency-analysis-1}. The
vertices in LDG correspond to the link channels in the NoC while a
directed edge joining two vertices reflects that there is a dependency
between these two links\textit{ (e.g., }$l_{01}$ \textit{and }$l_{12}$\textit{
in }Fig. \ref{fig:An-example-showing}\textit{).} When the routing
or the task mapping solution changes, the LDG needs to be rebuilt.
The LDG is built by checking every flow $f$ in the application communication
set $F$. An edge is added between two vertices of the LDG if there
exists a flow $f_{s,d}\in F$ that the two links corresponding to
the two vertices are two adjacent links on the routing path. The order
of the link analysis are then obtained by applying the topological
sort algorithm on the LDG. 

\begin{algorithm}
\caption{\label{alg:Link-dependency-analysis}Link dependency analysis}
\begin{algorithmic}[1]
\STATE {\textbf{Input:} $F$ the application flow set} 

\STATE {\textbf{Output:} $G$ the ordered list of links for queuing analysis}

\STATE {\textbf{Container:} $LDG=(V,E)$ the link dependency graph}

\FORALL {link $l_{ab}$} 

\STATE {$LDG.addnote(l_{ab})$}

\ENDFOR \COMMENT{initialize the $LDG$}

\FORALL {flow $f \in F$} 

\STATE {$P_{f}=routing\_function(f)$;}

\STATE {$d_{f}=length(P_{f})$;}

\FOR {$i=1:d_{f}-1$} 

\STATE {$l_{up}=P_{f}(i)$;} \COMMENT{the upstream link}

\STATE {$l_{down}=P_{f}(i+1);$} \COMMENT{the downstream link}

\STATE {$LDG.addedge(l_{down},l_{up});$}

\ENDFOR  \COMMENT {considering all the links in the path of $f$}

\ENDFOR \COMMENT {considering all flows in application}

\STATE {return $G=topological\_sort(LDG)$;}
\end{algorithmic}
\end{algorithm}

\subsection{GE-type traffic modeling}

\begin{figure}
\includegraphics[width=0.90\columnwidth]{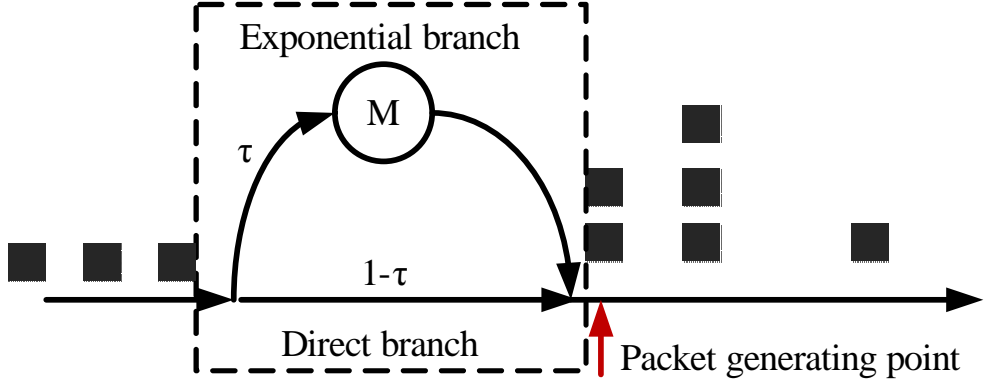}
\caption{\label{fig:Modeling-of-the}Modeling of the GE packet generation process in \cite{Ge-type}}
\end{figure}

Many applications in NoCs show burst patterns of traffic over a wide
range of time scales. Therefore, the Generalized Exponential (GE)
distribution is utilized to model the arrival traffic at the source
PEs and the links \cite{Ge-type}. In the following, we briefly review the GE type distribution proposed in \cite{Ge-type,Ge-model-two}. Under GE distribution, the cumulative distribution
function (cdf) of the inter-arrival time $X$ is given by \cite{Ge-type}: 

\begin{equation}
F(t)=P(X\leq t)=1-\tau e^{-\tau\lambda t},\; t\geq0
\end{equation}

where $\tau=2/(1+C^{2})$ and ($\lambda^{-1},C^{2}$) are the mean
and square coefficient of variation (SCV) of $X$. The GE
packet generation process is shown in Fig. \ref{fig:Modeling-of-the} from \cite{Ge-type,Ge-model-two}. As shown
in the figure, a packet experiences a zero service time to reach the
departure point (\textit{i.e.,} the point to generate the packet) with probability
$1-\tau$. For the rest (\textit{i.e.,} with probability $\tau$), the packet
needs to traverse the system with an exponentially distributed service
time with mean equal to $1/\lambda\tau$. The burst of packets consist
of a packet which comes from the exponential branch (M branch) in addition to
a number of continuous packets arriving through the direct branch \cite{Ge-type}.
The GE distribution is a versatile and simple distribution which helps
to make the queuing formulation analytically tractable \cite{Ge-type}.
Moreover, it has been demonstrated that the GE distribution also provides
efficient approximations for short or long-range dependent traffic
in supercomputers \cite{Ge-type}. 

\subsection{WH router modeling}

In this section, we present the techniques to estimate the three key
components of the WH analytical models, \textit{i.e.,} the flit transfer
time $\eta$, the path acquisition time $h$ and the source queuing
time $v_{s}$. 
\begin{figure}
\includegraphics[width=0.98\columnwidth]{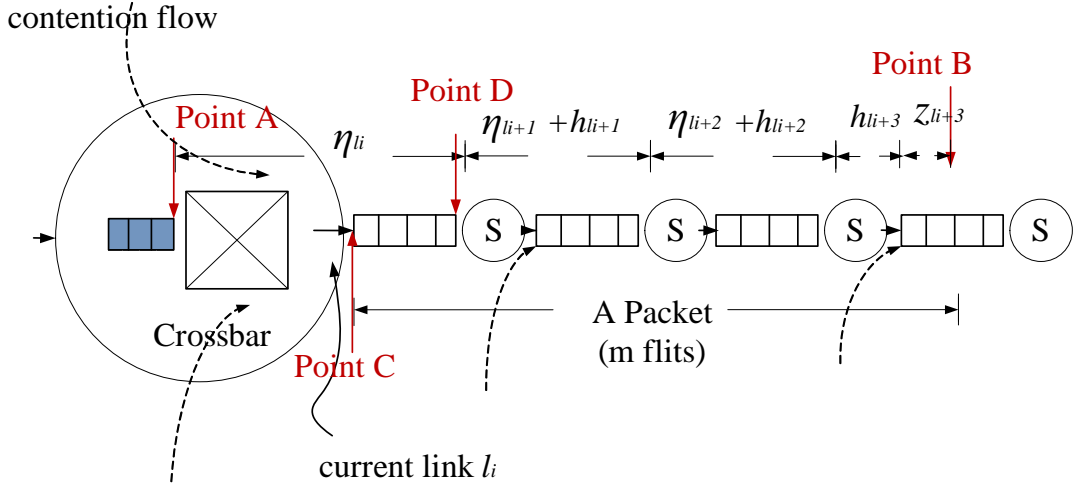}
\caption{\label{fig:An-illustration-of}An illustration of the channel service
time of link $l_{i}$ under WH routing}
\end{figure}

1)\textit{Flit transfer time}: The flit transfer time $\eta$ of
a link $l_{ab}$ is defined as the time taken for the header flit
to leave the buffer head in the upstream node and reach the front
of the buffer in the current link $l_{ab}$ after being granted the
link access. Fig. \ref{fig:An-illustration-of} illustrates this timing
concept. It is equivalent to the time taken from Point A to Point
D. More specifically, it consists of two parts, the first part is
the time to leave the upstream router (\textit{i.e.,} the time from
Point A to Point C in Fig. \ref{fig:An-illustration-of}). For WH
routing, this is a constant value depending on the number of pipeline
stages ($H_{s}$) in the router and the link. The second part accounts
for the time the header flit takes to arrive at the front of the buffer
of link $l_{i}$ ($q_{l_{l}}$). It is illustrated by the time to
travel from Point C to D in Fig. \ref{fig:An-illustration-of}. This
time value can be approximated by the waiting time of a queuing system
(such as the M/M/1/K ) with capacity equals to the buffer size $B$.
The mean flit arrival rate of this queuing system is given by:
\begin{equation}
\lambda_{flit}^{l_{ab}}=m\times\lambda_{packet}^{l_{ab}}
\end{equation}
\begin{equation}
\lambda_{packet}^{l_{ab}}=\sum_{f\in F_{l_{ab}}}\lambda_{f}=\sum_{\forall s}\sum_{\forall d}(\lambda_{sd}\times R(s,d,l_{ab}))
\end{equation}
where $R(s,d,k,l_{ab})$ is a binary value indicating whether the
channel is an element of the routing path set $P_{s,d}$ \cite{umit}:
\begin{equation}
R(s,d,l_{ab})=\begin{cases}
\begin{array}{c}
1\\
0
\end{array} & \begin{array}{c}
if\: l_{ab}\in P_{s,d}\\
otherwise
\end{array}\end{cases}
\end{equation}
The mean time to serve a flit ($s_{flit}^{l_{ab}}$) in this queuing
system is calculated by the weighted average of the service time for
all flows passing through link $l_{ab}$ and is given by: 
\begin{equation}
s_{flit}^{l_{ab}}=\frac{\sum_{\forall f\in F_{l_{ab}}}[\lambda_{f}\times(\frac{h_{l_{i+1}^{f}}}{m}+1)]}{\sum_{\forall f\in F_{l_{ab}}}\lambda_{f}}\label{eq:44}
\end{equation}
Eqn. \ref{eq:44} shows that for a packet with $m$ flits, it takes
$h_{l_{i+1}^{f}}$ cycles for a header in the buffer to win the next
link of flow $f$. Therefore, $h_{l_{i+1}^{f}}+1$ cycles are needed
in total for the service of the header flit. After that, the body
and tail flits are transferred in $1$ cycle without any additional
delay. The mean flit service time over the whole packet is thus computed
as $h_{l_{i+1}^{f}}/m+1$. 

After the mean arrival rate ($\lambda_{flit}^{l_{ab}}$) and the mean
service time ($s_{flit}^{l_{ab}}$) at this queue are obtained, the
M/M/1/K queuing formulation presented in \cite{Guz07networkdelays} can
be applied to approximate $q_{l_{l}}$. $\eta_{l_{ab}}$ is then obtained
as $\eta_{l_{i}^{f}}=q_{l_{i}}+H_{s}$. 

2)\textit{Path acquisition time}: The path acquisition time $h$
of flow $f$ at its $i-th$ hop $l_{ab}$ (\textit{i.e.,} $l_{i}^{f}$)
is defined as the waiting time of the header flit to be granted the access to the buffers
in $l_{ab}$ after contention with other flows routing towards the same output direction \cite{Delay_hetergeneous_noc}. 
It is usually modeled as the waiting time of a queuing system as in \cite{Delay_hetergeneous_noc, Hu_ananalytical}. Examples
of models used are the G/G/1 \cite{TVLSI12}, M/G/1/K \cite{ICCAD_latency,SC_model},
GE/G/1 \cite{Ge-type} and MMPP/G/1 \cite{Geyong-min} queues. For
the fair allocation policies such as the round-robin arbitration \cite{NoC13-latency-model},
each flow has the same priority and takes turns to use the output
link. Therefore, the system capacity $K$ of the queuing system to
derive $h$ is the number of flows that contend for the same link.
In Fig \ref{fig:An-illustration-of}, $k=3$ for $l_{i}^{f}$. The
arrival process of the queuing system is the merging of all flows
that route to $l_{ab}$. More precisely, the mean arrival rate is
$\lambda_{l_{ab}}=\sum_{f\in F_{l_{ab}}}\lambda_{f}$. The higher
moments of the arrival process of this queue are calculated
according to the specific queuing model used. For example, if the
GE/G/1 queuing model is employed to derive $h$, the squared
coefficient of variance (SCV) of the arrival flows are required
in order to apply the queuing formula. The SCV of the traffic to
$l_{ab}$ can be approximated as: 
\begin{equation}
C_{a_{l_{ab}}}^{2}=\sum_{\forall f\in F_{l_{ab}}}(\lambda_{f}\times C_{f}^{2})/\sum_{f\in F_{l_{ab}}}\lambda_{f}\label{eq:45}
\end{equation}
In \cite{Geyong-min,Ge-model-two,KPC-toolbox}, detailed mathematical
manipulations are shown to obtain a more accurate derivation of $C_{a_{l_{ab}}}^{2}$.
Moreover, the method can be used to generate the estimations of other
higher moments that are required by the more generalized queuing models,
such as the MMPP/G/1 or MAP/G/1 queues. However, the computation complexity
of this method is high as more traffic details are considered. Anyway, it provides a way to model more complex traffic phenomena
such as self-similarity and LRD in the queuing models.
The service time of the queue that is used to compute $h$ is
illustrated in Fig.\ref{fig:An-illustration-of} (referenced from \cite{Hu_ananalytical,Ge-model-two}). Without loss of
generality, we assume that a packet has $m$ flits and it spreads
over several adjacent links along the path. Here we assume the input
buffer depth is $B$ flits. The service time accounts for the time
that a packet occupies the link $l_{ab}$ \cite{Hu_ananalytical}. For example, in Fig. \ref{fig:An-illustration-of},
assume a header flit in Point A is granted the current link $l_{i}$
access. Then, the service process of this packet begins when the header
flit leaves A and ends when the tail flit departs A so as to release
$l_{i}$ for other flows. If the downstream links are not congested,
this service time simply equals to the packet length (\textit{i.e.,}
$m$ cycles) because the whole packet can traverse in a continuous
way \cite{Hu_ananalytical}. However, when there is severe blockage along the path, the worst-case
scenario occurs when the packet head reaches Point B (Fig.\ref{fig:An-illustration-of})
and the accumulated buffer spaces from Point C to Point B are just
enough to hold the whole packet \cite{Ge-model-two,Hu_ananalytical}.
The time delay for the packet at point A to reach point B, $x_{l_{i}}^{f}$
can be given by \cite{Hu_ananalytical}:
{ 
\begin{equation}
x_{l_{i}}^{f}=\begin{cases}
\begin{array}{c}
\eta_{l_{i}^{f}}+h_{l_{i+\Lambda_{l_{i}}^{f}}}+\sum_{j=1}^{\Lambda_{l_{i}}^{f}}(\eta_{l_{i+j}^{f}}+h_{l_{i+j}^{f}}+H_{s})+z_{l_{i+\Lambda_{l_{i}^{f}}}}\\
0
\end{array} & \begin{array}{c}
if\;\Lambda_{l_{i}}^{f}\geq1\\
otherwise
\end{array}\end{cases}\label{eq:46}
\end{equation}
}
where $\Lambda_{l_{i}}^{f}$ represents the effective number of hops
that a packet may span \cite{Ge-model-two,Hu_ananalytical}. Specifically, for a flow $f$ with a Manhattan
distance from the source to the destination tile equal to $d_{f}$,
if $m/B$ is smaller than the number of remaining hops $d_{f}-i$
, $\Lambda_{l_{i}}^{f}$ equals to $\left\lfloor m/B\right\rfloor $;
otherwise, it equals to $d_{p}-i$ \cite{Ge-model-two,Hu_ananalytical}. Of note, different from \cite{Hu_ananalytical}, we have included the router pipeline delay $H_{s}$ in Eqn. \ref{eq:46} to reflect a more tight packet transfer time to Point B.

The service time $s_{l_{i}^{f}}$ for the flow $f$ is then bounded by $m$ and $x_{l_{i}}^{f}$ under
different congestion conditions and can be approximated as \cite{Hu_ananalytical,Ge-model-two,SC_model}:
{ 
\begin{equation}
s_{l_{i}^{f}}=\begin{cases}
\begin{array}{c}
[m\times(m+x_{l_{i}}^{f})+2\times x_{l_{i}}^{f}\times m]/(m+2x_{l_{i}}^{f})\\
{}[m\times(m+x_{l_{i}}^{f})+2\times(x_{l_{i}}^{f})^{2}]/(m+2x_{l_{i}}^{f})
\end{array} & \begin{array}{c}
if\: x_{l_{i}}^{f}<m\\
otherwise
\end{array}\end{cases}\label{eq:47}
\end{equation}
}
The rationale of Eqn. \ref{eq:47} is explained as follows \cite{Hu_ananalytical,Ge-model-two,SC_model}: when the
downstream channels along the routing path is not congested\textit{
}(\textit{e.g.,} $x_{l_{i}}^{f}\rightarrow0$), the link service time
$s_{l_{i}^{f}}$ is approximated by the packet length $m$. At the
other extreme, when there exists severe blockage at the subsequent
hops (\textit{i.e.,} $x_{l_{i}}^{f}\rightarrow\infty\gg m$), $s_{l_{i}^{f}}$
is approximated by $x_{l_{i}}^{f}$ as the congestion delay in the
subsequent links dominates the current channel service time. 
\\From Eqn. \ref{eq:47}, once the downstream links transfer time $\eta$
and contention time $h$ are known, the overall link service time
$\bar{s}_{l_{ab}}$ is then the weighted average of the service time
of all flows passing through the link $l_{ab}$, which yields the mean
service time $s_{l_{ab}}$ as: 
\begin{equation}
\bar{s}_{l_{ab}}=\sum_{\forall f\in F_{l_{ab}}}(\lambda_{f}\times s_{l_{i}^{f}})/\sum_{\forall f\in F_{l_{ab}}}\lambda_{f}\label{eq:48}
\end{equation}
For M/G/1/K, G/G/1 and MMPP/G/1 models, the SCV of the service time
for link $l_{ab}$ is also required. It can be approximated in a similar
way as \cite{TVLSI12} and is given by: 
\begin{equation}
C_{s_{l_{ab}}}^{2}=\frac{\overline{s_{l_{ab}}^{2}}}{(\bar{s}_{l_{ab}})^{2}}-1=(\frac{\sum_{\forall f\in F_{l_{ab}}}\lambda_{f}\times s_{l_{i}^{f}}^{2}}{\sum_{\forall f\in F_{l_{ab}}}\lambda_{f}})/(\bar{s}_{l_{ab}})^{2}-1\label{eq:49}
\end{equation}
Based on the above discussions, after obtaining the arrival and the
service process characteristics, in this work, we first use the M/G/1/K queuing
formula to obtain the waiting time $h^*_{l_{ab}}$ and then extend to $h_{l_{ab}}$ by considering the second moment (SCVs) of the GE traffic input of the queuing system:
\begin{equation}
h_{l_{ab}}=\frac{(C_{s}^{2}+C_{a}^{2})}{(1+C_{s}^{2})}\times h_{l_{ab}}^{*}\label{eq:49-1}
\end{equation}
where $h^*_{l_{ab}}$ is the M/G/1/K queue based waiting time, $C_{s}^{2}$ and $C_{a}^{2}$ are the SCVs of the service 
and inter-arrival time, respectively.

3)\textit{ Source }queuing time: The source queue at the network interface
(NI) is modeled as a queuing system with infinite capacity, and hence
an M/M/1/$\infty$ \cite{Delay_hetergeneous_noc} or a GE/G/1/$\infty$
\cite{Ge-type} queuing model can be used. For example, if GE/G/1
model is used, the queuing time is given by \cite{Ge-type}: 

\begin{equation}
v_{s}=\frac{\bar{s_{l_{s}}}}{2}(1+\frac{C_{a}^{2}+\lambda_{a}\times\frac{(\bar{s_{l_{s}}}-m)^{2}}{\bar{s_{l_{s}}}}}{1-\lambda_{a}\times\bar{s_{l_{s}}}})-\bar{s_{l_{s}}}
\end{equation}
where the arrival process ($\lambda_{a}$,$C_{a}^{2}$) and the mean
service time $\bar{s_{l_{s}}}$ at the source node $s$ (link $l_{s}$)
are calculated in similar manners as presented for the router
channels.

\subsection{Discussion on extensions to VC router modeling}

When there are multiple virtual channels sharing a single physical
port, the inputs received from a physical link will be multiplexed among
all the available VCs \cite{VC_multiplexing_dally,Delay_hetergeneous_noc}. In order to model the effect of VC multiplexing,
we need to scale the mean packet waiting time by a factor $\bar{V}$,
which represents the average degree of VC multiplexing that takes
place at a given link channel as in \cite{VC_multiplexing,VC_multiplexing_dally,Europar-vc,Ge-type}.
The mean message delay for the flow $f_{s,d}$ is rewritten as: 
\begin{figure}
\includegraphics[width=1.00\columnwidth]{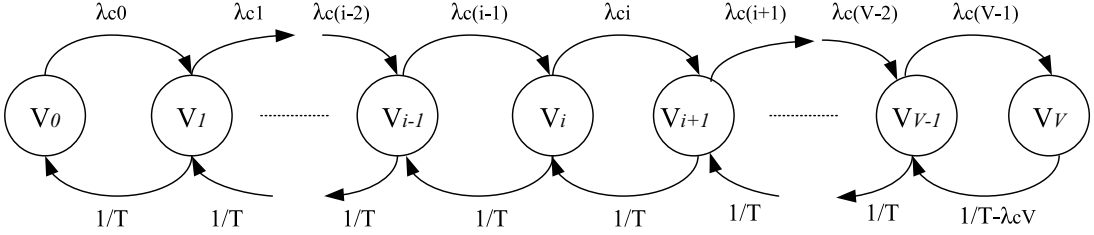}
\caption{\label{fig:The-transition-diagram}The transition diagram to calculate
the average degree of virtual channel multiplexing \cite{Ge-type}}
\end{figure}
\begin{equation}
L_{s,d}=(v_{s}+\eta_{s,d}+h_{s,d})\times\bar{V}\label{eq:51}
\end{equation}
To obtain the value of $\bar{V}$ , the VC state transition diagram
(STD) of a physical channel is used for the analysis \cite{VC_multiplexing}. As shown in
Fig. \ref{fig:The-transition-diagram}, state $V_{i}$ ($0\leq i\leq V$)
in the STD represents that there are $i$ VCs currently being occupied by packets in a physical
channel. The transition rate from state $V_{i}$ to $V_{i+1}$ is
denoted as $\lambda_{ci}$, which reflects the mean packet arrival rate at the given input channel when the STD is at state $V_{i}$ \cite{VC_multiplexing}.
It not only depends on the current state (i.e., the number of VCs being used) in STD, but also
is related to the traffic arrival model. In \cite{Ge-type} and \cite{Geyong-min},
$\lambda_{ci}$ is derived based on the maximum entropy principles
\cite{Maximum_entropy} for the GE and MMPP traffic arrival models.
Similarly, the rate from state $V_{i}$ to $V_{i-1}$ is $1/T$, where
$T$ is the mean service time at each physical channel. Based on the
STD, the state probability, $P_{v}$, ($0\leq v\leq V$), that represents
$v$ VCs are busy at a given channel, can be determined by solving
the steady-state equations of the Markov chain in Fig. \ref{fig:The-transition-diagram}
\cite{Geyong-min,Ge-type}. The average degree of VC multiplexing
that takes place at a given physical channel is then given by \cite{VC_multiplexing_dally}:
$\overline{V}=\frac{\sum_{v=0}^{V}v^{2}P_{v}}{\sum_{v=0}^{V}vP_{v}}$. Consequently, the source queuing time $v_{s}$ and the path acquisition time $\eta_{s,d}$ in the VC channel model will equal to those calculated in the wormhole model multiplying $\overline{V}$ as shown in Eqn. \ref{eq:51}.

\section{SVR-NoC latency model}

To further improve the accuracy of the analytical performance model,
we introduce a learning based performance model \cite{svr-noc}. The main characteristics
of this learning model is that in addition to the using of collected
traffic data statistics as training features, we also include the
prediction results of the queuing model presented in section 2.3 as
part of the features. More precisely, the SVR-NoC model
learns and refines the proposed queuing model based on the typical
training data from the simulation of the target NoC platform. In this
section, we first define two regression functions that need to be
learned and then elaborate on the support vector regression techniques
for obtaining these two models, respectively.

\subsection{Channel and source queuing regression models }

Based on Eqn. \ref{eq:40}, we define two queuing delays that make up of the
traffic flow latency. The first is the source queuing delay $SQ_{s}$
which is the waiting time of the packets at the source queue $s$
before injected into the network (\textit{i.e.,} $SQ_{s}=v_{s}$). The second
is the channel waiting time $CQ_{k,dir}$ which is the total packet
transfer time at the direction $dir$ of router $k$ and equals to
$h_{k,dir}+\eta_{k,dir}$ in Eqn. \ref{eq:40}. In this work, we model these
two components via two regression functions $f_{SQ}$ and $f_{CQ}$.
We denote the feature sets used in learning $f_{SQ}$ and $f_{CQ}$
by two vectors $X_{CQ}=[x_{cq1},x_{cq2},\cdots,x_{cqn}]$ and $X_{SQ}=[x_{sq1},x_{sq2},\cdots,x_{sqn}]$,
respectively. The proposed supervised SVR-NoC learning framework is
applied on the training data set to formulate the following models
in which the estimated queuing delays are specified as functions of
the selected features:

\begin{equation}
\widehat{CQ}_{k,dir}=f_{CQ}(X_{CQ});\;\hspace{4mm}\widehat{SQ_{s}}=f_{SQ}(X_{SQ}) \label{eq:cq_sq}
\end{equation}
Using Eqn. \ref{eq:cq_sq}, given the feature sets, we can estimate $SQ_{s}$ and
$CQ_{k,dir}$ and substitute them in Eqn. \ref{eq:40} to obtain the
latency of each specific flow. Once the $CQ_{k,dir}$ metric is
estimated, the channel buffer utilization and the network throughput
can also be obtained as in \cite{umit}.

\subsection{Overall SVR-NoC methodology }

In SVR-NoC, the step to obtain the estimates of the channel and source
queuing delay using the proposed PQ model is summarized in Algorithm
2. In the training stage, we first apply the link dependency analysis
presented in section 2.3.3 to determine the correct link ordering
for performance analysis. Then, for each link $l_{ab}$ in the ordered
list $G$, we calculate the arrival traffic model ($\lambda_{l_{ab}},C_{al_{ab}}^{2}$)
according to the routing algorithm and applications. As shown
in Eqn. \ref{eq:48}, the link transfer time $\eta_{l_{ab}}$ only depends on
its downstream link contention time $h_{l_{i+1}^{f}}$, which should
have been analyzed in the previous loop. On the other hand, the path
acquisition time $h_{l_{ab}}$depends not only on its downstream link
contention time $h_{l_{i+1}^{f}}$ and transfer time $\eta_{l_{i+1}^{f}}$
but also on the current link's $\eta_{l_{i}^{f}}$ (Eqn. \ref{eq:46} and Eqn. \ref{eq:47}). Therefore,
$\eta_{l_{ab}}$ is calculated first. After $\eta_{l_{ab}}$
is obtained, we then calculate the mean and SCV of the link service
time ($\bar{s_{l_{ab}}},C_{sl_{ab}}^{2}$). If this link
connects two routers, we utilize GE/G/1/K queue to obtain $h_{lab}$. Otherwise, we calculate the source queuing time $v_{a}$. Once all the $(h,v,\eta)$ variables are known, the predicted
channel waiting time $CQ^{*}$ can be obtained as $h+\eta$ while the
source waiting time $SQ^{*}=v$. 

\begin{algorithm}
\caption{\label{alg:Link-dependency-analysis-1}Analytical working flow to
obtain $CQ^{*}$and $SQ^{*}$}
\begin{algorithmic}[1]
\STATE {\textbf{Input:} $F$ the application flow set; $G$ the ordered list
of links for queuing analysis}

\STATE {\textbf{Output:} $CQ_{l_{ab}}^{*}$ the estimated channel waiting time
for link $l_{ab}$; $SQ_{a}^{*}$the estimated source waiting time}

\FORALL {link $l_{ab}\in G$}

\STATE{($\lambda_{l_{ab}},C_{al_{ab}}^{2}$) = traffic\_model($F_{l_{ab}}$)}

\STATE{$\eta_{l_{ab}}=calculate\_transfer\_time(\lambda_{l_{ab}},s_{flit}^{l_{ab}},m,B)$;}

\FORALL {$f\in F_{l_{ab}}$and $l_{i}^{f}=l_{ab}$}

\STATE {$s_{l_{i}^{f}}=calculate\_link\_service\_time()$;}

\ENDFOR

\STATE {($\bar{s_{l_{ab}}},C_{sl_{ab}}^{2}$)=$service\_time()$;}

\IF {$a\neq b$} 

\STATE {$h_{l_{ab}}=GE\_G\_1\_K\_queue(\lambda_{l_{ab}},C_{al_{ab}}^{2},\bar{s_{l_{ab}}},C_{sl_{ab}}^{2},k);$}

\STATE {$CQ_{l_{ab}}^{*}=h_{l_{ab}}+\eta_{l_{ab}}$}

\ELSE 

\STATE {$v_{a}=GE\_G\_1\_queue(\lambda_{l_{ab}},C_{al_{ab}}^{2},\bar{s_{l_{ab}}},C_{sl_{ab}}^{2},);$}

\STATE {$SQ_{a}^{*}=v_{a};$}

\ENDIF

\ENDFOR 
\end{algorithmic}
\end{algorithm}

\begin{figure}[h]
\includegraphics[width=0.95\columnwidth]{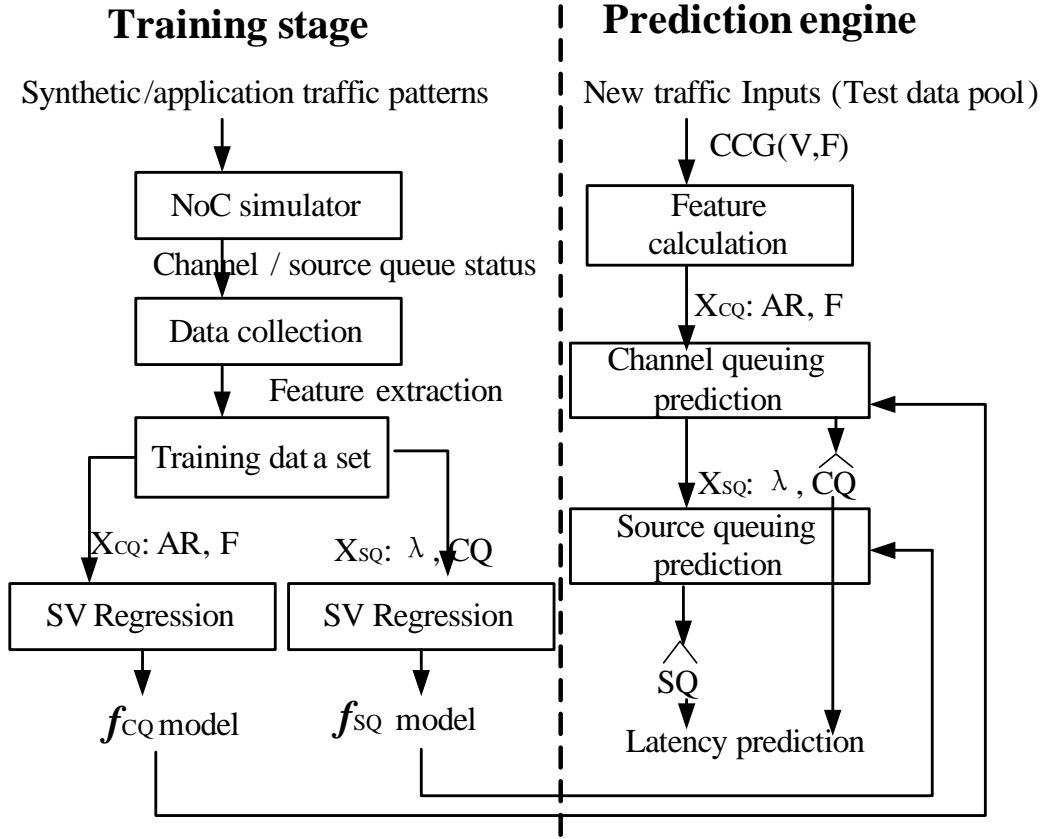}
\caption{\label{fig:SVR-NoC-methodology-overview}SVR-NoC methodology overview}
\end{figure}

Figure \ref{fig:SVR-NoC-methodology-overview} shows the overall SVR-NoC
framework, which consists of two parts, namely the \textit{training
stage} and\textit{ prediction engine}. In the training stage, different
traffic patterns are fed into the system level NoC simulator to collect
the statistics of the training data of the channels in the routers
and the source queues. The channel queue feature set $X_{CQ}$ includes
the packet arrival rates $AR$ and the forwarding probabilities $F$
in the router as well as the calculated channel waiting time $CQ^{*}$
from the proposed analytical model. The source queue feature set $X_{SQ}$
includes the injection rates $\lambda$ at the traffic source, the
neighboring channel waiting times from the analytic model $CQ^{*}$ as well as the calculated
source queuing time $SQ^{*}$. Both $X_{CQ}$ and $X_{SQ}$ are extracted
from the simulation results to form the training data set. Support
vector regression is then carried out to obtain the $f_{CQ}$ and
$f_{SQ}$ models, respectively. 

After the training stage, the obtained SVR models are used in the
prediction engine to estimate the latency performance for different
traffic patterns. The prediction engine can be embedded in the NoC
synthesis framework for the inner-loop design evaluation. Of note,
the features in $X_{CQ}$ and $X_{SQ}$ are computed from the input core
communication graph (CCG) as well as the proposed queuing model. Then,
the $f_{CQ}$ and $f_{SQ}$ functions are used to evaluate
the channel and source queuing delay, respectively. Finally, the traffic flow latency and the
overall latency can be computed according to Eqn. \ref{eq:40}.

\begin{figure}[h]
\includegraphics[width=0.95\columnwidth]{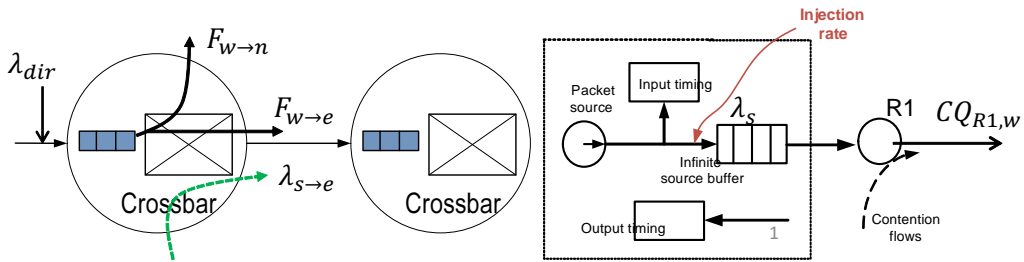}
\caption{\label{fig:Illustraction_of_feature_vector}Illustration of the channel and source queuing feature vectors}
\end{figure}

\subsection{Feature extraction for training data}
\subsubsection{Channel queuing feature vector}

The average waiting time in the channel $dir$ of router $k$ ($CQ_{k,dir}$
) depends not only on its packet arrival rate $AR_{k,dir}$ but also
on the contention among the channels within the same router, as well
as the traffic conditions in the neighboring routers (such as the
congestion in the downstream router). In the analytical models, these
contentions are usually calculated explicitly by computing the contention
matrix using the application traffic arrival rate matrices and the forwarding
probability matrices \cite{umit,ICCAD_latency}. In contrast, we aim
at learning the impact of these contentions implicitly by providing the SVR engine with sufficient samples. Therefore, for the channel queuing feature vector, it is made up of three parts: 1)we first include the arrival rate $\lambda_{dir}$ of the current channel to indicate the traffic workload injected to the channel. 2)The forwarding probability from current channel to different output directions $F_{k}$ and the amount of traffic from other input channels to the same output direction $\lambda_{k}$ ($k\neq dir$) are included to reflect the contentions situations. 3) Finally, the estimated channel service time $S_{dir}$ and the channel waiting time $\eta_{dir}+h_{dir}$ from queuing models are included to provide an estimation for $CQ$. Of note, both the arrival rate matrix $AR_{k}$ and the forwarding probability matrix $F_{k}$ at
node $k$ can be calculated offline based on the application CCG. Specifically, as in \cite{umit},
$AR_{k,dir}$ is given by 

\begin{table}
\caption{\label{tab-training} Training features for regression}
\begin{tabular}{|p{0.12\columnwidth}|p{0.15\columnwidth}|p{0.20\columnwidth}|p{0.35\columnwidth}|}
\hline 
Feature vector & Elements notation & Elements & Description\tabularnewline
\hline 
\hline 
 & $AR_{dir}$ & $X_{CQ}[1]$ & The arrival rate in current channel $dir$ \tabularnewline
\cline{2-4} 
$X_{CQ}$ & $AR_{k}$ & $X_{CQ}[2:5]$ & The aggregated traffic from other input channels in router and contends with $dir$ in the output direction $k$ \tabularnewline
\cline{2-4} 
 & $F_{i,j}$ & $X_{CQ}[6:9]$ & Forwarding probability from current channel $i$ to four output directions\tabularnewline
\cline{2-4} 
 & $CQ^{*}$ & $X_{CQ}[10]$ & The estimated channel queuing time $\eta_{dir}+h_{dir}$\tabularnewline
\cline{2-4} 
 & $S_{dir}$ & $X_{CQ}[11]$ & The estimated channel service time \tabularnewline
\hline 
 & $\Lambda_{s}$ & $X_{SQ}[1]$ & Arrival rate at source $s$\tabularnewline
\cline{2-4} 
$X_{SQ}$ & $F_{s}$ & $X_{SQ}[2:5]$ & Forwarding probability from $s$ to different output directions\tabularnewline
\cline{2-4} 
 & $lamda_{k}$ & $X_{SQ}[6:9]$ & The aggregated traffic that contends with $s$ to the output direction $k$ \tabularnewline
\cline{2-4} 
 & $CQ^{*}_{k}$ & $X_{SQ}[10:13]$ & The predicted waiting time in the downstream channel of output direction $k$\tabularnewline
\cline{2-4} 
 & $s^{*}_{k}$ & $X_{SQ}[14:17]$ & The predicted channel service time in the downstream links of output direction $k$\tabularnewline
\cline{2-4}
 & $SQ^{*}$ & $X_{SQ}[18]$ & The estimated source queuing time from the proposed queuing model\tabularnewline
 \cline{2-4}
 & $S^{*}$ & $X_{SQ}[19]$ & The estimated source service time from the proposed queuing model\tabularnewline
\hline 
\end{tabular}
\end{table}
\begin{equation}
AR_{k,dir}=\sum_{\forall s\in V}\sum_{\forall d\in V}(\lambda_{s,d}\times R(s,d,k,dir))
\end{equation}
where $R(s,d,k,dir)$ indicates whether the channel is part of  the
routing path set $P_{s,d}$ and is equal to (as in \cite{umit}):
\begin{equation}
R(s,d,k,dir)=\begin{cases}
\begin{array}{ccccc}
1 &  & if & (k,dir)\in P_{s,d}\\
0 &  &  & otherwise
\end{array}\end{cases}
\end{equation}
The forwarding probability $F_{k,i,j}$ denotes the portion of traffic
that arrives at channel $i$ of node $k$ and is forwarded to the
output direction $j$. It is given by (as in \cite{umit}):
\begin{equation}
F_{k,i,j}=\frac{\sum_{\forall s\in V}\sum_{\forall d\in V}(\lambda_{s,d}\times g(s,d,k,i,j))}{AR_{k,i}}
\end{equation}
where $g(s,d,k,i,j)$ is a binary indicator that returns 1 if the
routing path $P_{s,d}$ contains the channel $(k,i)$ through the
output direction $j$.  
\\In SVR-NoC, in order to reduce the size of training data that needs to be collected from the simulations, we propose to
include an estimation of the channel queuing time $CQ^{*}$ in the
channel queuing feature set. Of note, $CQ^{*}$ is obtained from the
proposed GE/G/1/K queuing model and is independent of the training
set simulation results. In Table \ref{tab-training}, we list all the elements in the channel queuing feature
vector $X_{CQ}$. In Fig. \ref{fig:Illustraction_of_feature_vector}-a, we show an example of the forwarding probability and aggregated contention traffic. More specifically, as shown in Fig. \ref{fig:Illustraction_of_feature_vector}, for the current west input channel ($dir$), when considering the east output direction, the forward probability $F_{w-e}$ represents the portion of traffic that will be forwarded towards east direction. In addition, $\lambda_{s-e}$ in the figure indicates the traffic that contends with $dir$ to the east output from the south input channel. By including all other input channels, the input feature vector $X_{CQ}$ can be formed as in Table \ref{tab-training}.  

\subsubsection{Source queuing feature vector}

The time that a packet needs to wait in a particular source queue
depends on the traffic generation rate at the processor and the network
congestion status, which are reflected by the arrival rate and waiting time in the attached router, respectively. Therefore, the packet generation rate $\Lambda_{s}$
is first included in $X_{SQ}$ as shown in Table \ref{tab-training}. The source generation rate $\Lambda_{s}$ can be computed by:
\begin{equation}
\Lambda_{s}=\sum_{\forall d\in V}\lambda_{sd}
\end{equation}
As shown in Table \ref{tab-training}, to achieve a better inference on the level
of network congestion and a higher accuracy, we also need to include
the average waiting time $CQ_{k}$ of the channels in the downstream links in $X_{SQ}$, where $k$
is an output direction that the current source queue $s$ has traffic to forward to. Similar to the channel
queuing feature set, we include the calculated source queuing time
$SQ^{*}$ in the feature set (shown in Table \ref{tab-training}) to improve the learning
function accuracy. In Fig. \ref{fig:Illustraction_of_feature_vector}-b, the source traffic rate $\lambda_{s}$ as well as the waiting time $CQ_{R1,W}$ in the west down-streaming link are shown, which corresponds to $X_{SQ}[1]$ and $X_{SQ}[10]$ in Table \ref{tab-training}, respectively. 

\subsection{Support vector regression for $f_{CQ}$ and $f_{SQ}$}

After obtaining the feature vectors $X_{CQ}$ and $X_{SQ}$, we apply
$\epsilon$-SVR \cite{Vapnik} to learn the two nonlinear models $f_{CQ}$
and $f_{SQ}$, respectively. The objective of $\epsilon$-SVR is to
find a function $f(X)$ that deviates from the actual target values
 $y_{i}$ (\textit{i.e.,} $CQ_{k,dir}$ and $SQ_{s}$) in the data-set
by at most $\epsilon$ \cite{Vapnik}. There are three steps in the $\epsilon$-SVR
\cite{ML}: (1) Primal form optimization, (2) Dual problem formulation
and (3) Implicit mapping via kernels. The primal formulation is a
straightforward representation of the regression problem whereas the latter
two steps provide the a practical transformation considering the non-linear extension
via kernel tricks. We present the $\epsilon$-SVR formulation of the
channel average waiting time function $f_{CQ}$  while similar procedures
can be applied for obtaining $f_{SQ}$.

\subsubsection{Primal form formulation}

Without loss of generality, we begin by assuming $f_{CQ}$ is a linear
function. This assumption will be relaxed later to highly non-linear
functions by using the Radial Basis kernels \cite{Vapnik} discussed
in step 3. Given a set of $l$ training data points, \textit{i.e.,}
$\{(X_{CQ1},CQ_{1}),\cdots(X_{CQl},CQ_{l})\}$, where $X_{CQi}\in\mathbb{R}^{d}$
is the $i^{th}$ sample feature vector with a dimensionality of $d$
($d=11$ as shown in Table \ref{tab-training}) . Under the linear model assumption,
$\widehat{CQ}$ can be expressed as \cite{ML}:
\begin{figure}
\includegraphics[width=0.98\columnwidth]{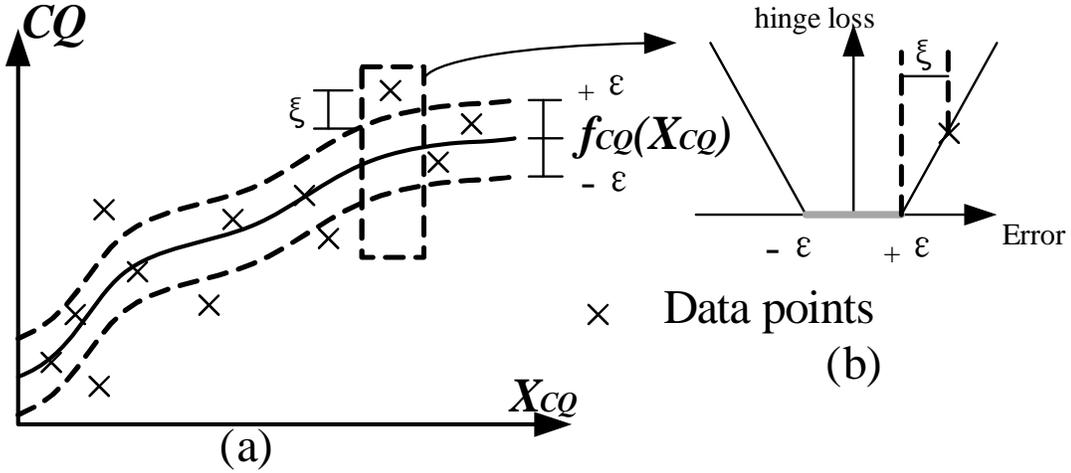}
\caption{\label{soft_margin_loss} The soft margin loss setting in $\epsilon$-SVR \cite{ML}}
\end{figure}

\begin{equation}
\widehat{CQ}=f(X_{CQ})=\sum_{i=1}^{d}w_{i}x_{cqi}+b=\mathbf{w^{T}\mathrm{\mathit{X}_{CQ}}+}b
\end{equation}
where $\mathbf{w}\in\mathbb{R}^{d}$, $b\in\mathbb{R}$ . Let $\epsilon$
be the maximum prediction error (\textit{i.e.}, $||CQ-\widehat{CQ}||$
and $||\bullet||$ represents the L2 norm) that we can tolerate. We
introduce two slack variables, $\xi$ and $\xi^{*}$, to represent
the error larger and smaller than the target value by more than $\epsilon$,
respectively \cite{Vapnik}. They are defined as the \textquotedblleft{}soft margin loss\textquotedblright{}.
Figure \ref{soft_margin_loss}-a illustrates the soft margin loss for $f_{CQ}(X_{CQ})$.
We want to minimize the "soft margin loss" and hence we add
penalty if $\xi$ or $\xi^{*}$ is larger than zero. From Figure \ref{soft_margin_loss}-b, only the
points outside the $\epsilon$region will be penalized in a linear
fashion \cite{Vapnik}. In order to prevent data over-fitting when learning $f(X_{CQ})$
from the training samples, a regularization term proportional to $||\mathbf{w}||^{2}$
is added into the objective function \cite{ML}. To find the optimal
$f_{CQ}$, the problem is formulated as \cite{ML}:
\begin{equation}
minimize\;\hspace{2mm}\frac{1}{2}||\mathbf{w}||^{2}+C\times\sum_{i=1}^{l}(\xi_{i}+\xi_{i}^{*}) \label{eq:obj_eq}
\end{equation}
\begin{equation}
subject\; to\;\begin{cases}
\begin{array}{c}
(\mathbf{w}^{T}X_{CQi}+b)-CQ_{i}\leq\epsilon+\xi_{i}\\
CQ_{i}-(\mathbf{w}^{T}X_{CQi}+b)\leq\epsilon+\xi_{i}^{*}\\
\xi_{i},\xi_{i}^{*}\geq0,\epsilon\geq0
\end{array} & (10)\end{cases} \label{eq:formulation}
\end{equation}

where $C>0$ is the regularization parameter that determines the tradeoff
between the fitting accuracy over the training samples
(the second term in Eqn. \ref{eq:obj_eq}) and the regularization term for preventing
data over-fitting (the first term in Eqn. \ref{eq:obj_eq}) \cite{ML}.

\subsubsection{Dual problem expansion}

The primal form optimization problem in Eqn. \ref{eq:obj_eq} and Eqn. \ref{eq:formulation} is difficult
to solve, therefore we solve its corresponding Lagrangian formulation
$\mathcal{L}$ by introducing a set of Lagrangian multipliers $\alpha_{i}$,
$\alpha_{i}^{*}$,$\eta_{i}$,$\eta_{i}^{*}$ \cite{ML,Vapnik}, where:

\begin{equation}
\mathcal{L}=\begin{array}{c}
\frac{1}{2}||\mathbf{w}||^{2}+C\times\sum_{i=1}^{l}(\xi_{i}+\xi_{i}^{*})-\sum_{i=1}^{l}(\eta_{i}\xi_{i}+\eta_{i}^{*}\xi_{i}^{*})\\
-\sum_{i=1}^{l}\alpha_{i}\times(\epsilon+\xi_{i}-CQ_{i}+\mathbf{w}^{T}X_{CQi}+b)\\
-\sum_{i=1}^{l}\alpha_{i}^{*}\times(\epsilon+\xi_{i}^{*}+CQ_{i}-\mathbf{w}^{T}X_{CQi}-b)\:(11)
\end{array} \label{eq:dual}
\end{equation}

It follows that the partial derivatives of $\mathcal{L}$ with respect
to the primal variables $(\mathbf{w},b,\xi_{i},\xi_{i}^{*})$ are
zero for the optimal point in the primal form \cite{ML}. Hence, the primal variables $(\mathbf{w},b,\xi_{i},\xi_{i}^{*})$
can be represented by $\alpha_{i}$, $\alpha_{i}^{*}$ after setting:
\begin{equation}
\partial_{b}\mathcal{L=}\partial_{\mathbf{w}}\mathcal{L=}\partial_{\xi_{i}}\mathcal{L=\partial}_{\xi_{i}^{*}}\mathcal{L}=0
\end{equation}
By substituting the primal variables in Eqn. \ref{eq:obj_eq} and Eqn. \ref{eq:formulation} with their dual
variable representations, the dual optimization problem can be obtained
\cite{Vapnik} as follows:
\begin{equation}
max\;\hspace{2mm}\begin{cases}
\begin{array}{c}
-\frac{1}{2}\sum_{i,j=1}^{l}(\alpha_{i}-\alpha_{i}^{*})(\alpha_{j}-\alpha_{j}^{*})<X_{CQi},X_{CQj}>\\
-\epsilon\sum_{i=1}^{l}(\alpha_{i}+\alpha_{i}^{*})+\sum_{i=1}^{l}CQ_{i}(\alpha_{i}-\alpha_{i}^{*})\;
\end{array}\end{cases} \label{eq:form1}
\end{equation}
\begin{equation}
subject\; to\;\sum_{i=1}^{l}(\alpha_{i}-\alpha_{i}^{*})=0\; and\;\alpha_{i},\alpha_{i}^{*}\in[0,C] \label{eq:form2}
\end{equation}
where $<\text{\ensuremath{\centerdot}},\text{\ensuremath{\centerdot}}>$
is the dot product of two vectors. This is a quadratic programming
problem and can be solved efficiently in polynomial time. $f_{CQ}(X_{CQ})$
is then obtained as \cite{Vapnik}:
\begin{equation}
f_{CQ}(X_{CQ})=\sum_{i=1}^{m}(\alpha_{i}-\alpha_{i}^{*})<X_{CQi},X_{CQ}>+b \label{eq:solution}
\end{equation}
From Eqn. \ref{eq:solution}, it can been seen that the complexity of the regression function $f_{CQ}$ depends
only on the training samples $X_{CQi}$ that have nonzero $(\alpha_{i}-\alpha_{i}^{*})$
terms (which are defined as the support vectors) \cite{ML}. Hence the SVR model benefits
from keeping only a few training samples for very efficient predictions \cite{ML,Vapnik}.

\subsubsection{Kernel trick for nonlinear extension}

In \cite{ML,Vapnik}, it has been proven that the linear model formulation in Eqn.
\ref{eq:form1} and Eqn. \ref{eq:form2} is still valid if we substitute the dot product operation $<\centerdot,\centerdot>$
with a number of various kernel functions $k$:
\begin{equation}
k(X_{CQ},X_{CQ}^{'})=<\Phi(X_{CQ}),\Phi(X_{CQ}^{'})>
\end{equation}
From the analytical delay model, it is suggested that the router delay
is a non-linear function of the extracted features. Therefore we use
the most common Radial Basis Function (RBF) kernel \cite{ML} ($k(\mathbf{x},\mathbf{x}^{'})=exp(-\gamma\cdot||\mathbf{x}-\mathbf{x}^{'}||^{2})$) in this work,
where $\gamma$ is a tuning parameter to replace the linear dot production
operation in Eqn. \ref{eq:form1}. As the RBF kernel is highly nonlinear \cite{ML},
this kernel trick extends the $f_{CQ}$ formulation from the previous linear assumption into a non-linear
function.

\subsubsection{Cross-validation and grid-search}

\begin{figure}
\includegraphics[width=0.93\columnwidth]{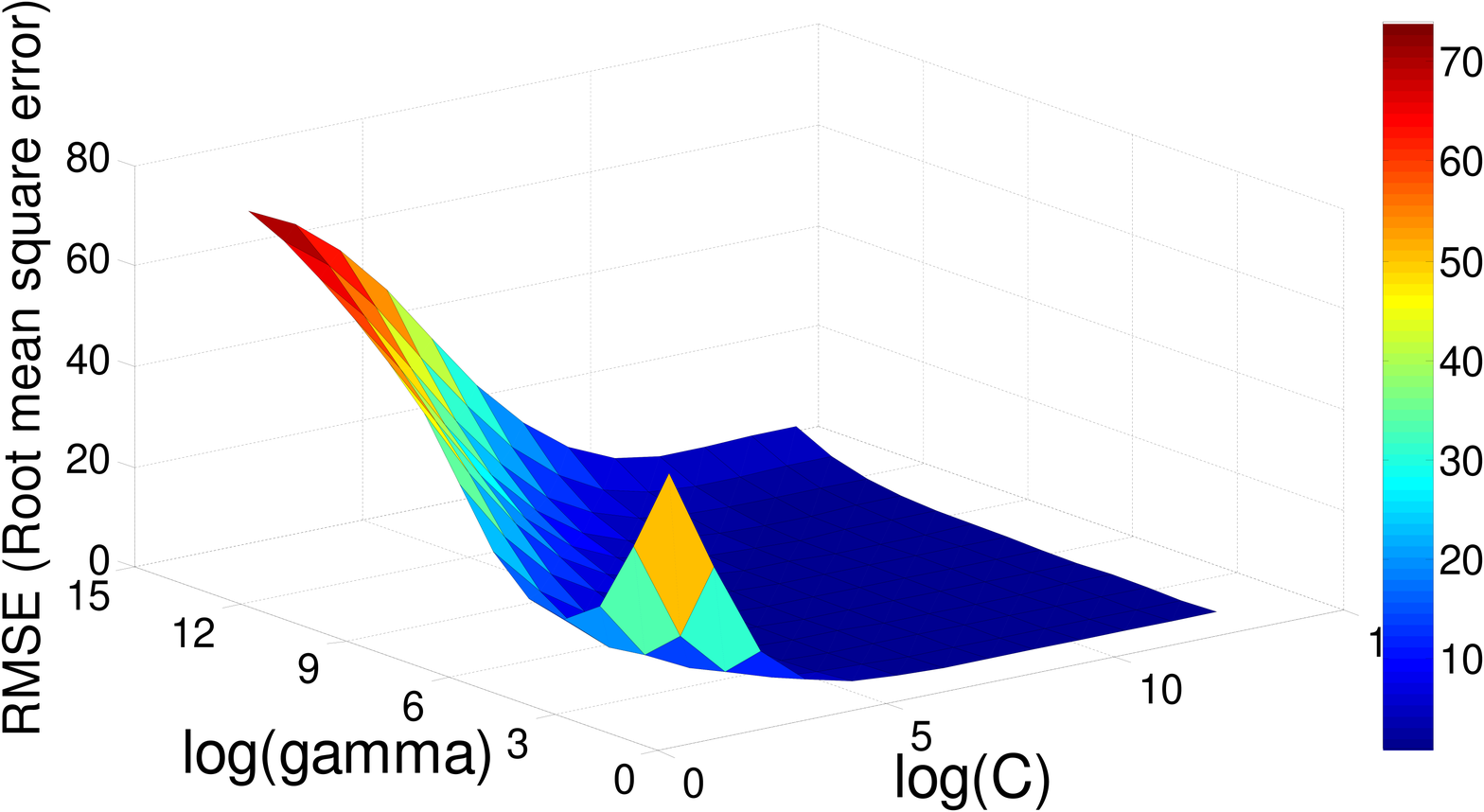}
\caption{\label{surface} An example RMSE surface in the grid search }
\end{figure}

When solving the optimization problem in Eqn. \ref{eq:form1}-\ref{eq:form2}, the tuning parameters
$\epsilon$, $C$, together with the $\gamma$ in the RBF kernel are
assumed to be fixed. They can be set to any arbitrary values. To find out these parameters with highest model accuracy, we
need to search over different combinations of $(\epsilon,C,\gamma)$.
In this work, we adopt a $v$-fold cross-validation approach \cite{ML,Vapnik}.
A $v$-fold cross-validation separates the training data into $v$
subsets. $v-1$ subsets are used in the training regression stage and the remaining
subset is used as the testing set. $v$ runs are carried out with
different subsets as the testing set. The cross-validation accuracy
is then equal to the root mean square error (RMSE) of all the $v$
runs \cite{ML,Vapnik}. In SVR-NoC, we carry out the parameter search on $(\epsilon,C,\gamma)$
using a $5$-fold cross-validation as suggested in \cite{libsvm}.
Fig. \ref{surface} shows an example root mean square error surface for $\epsilon$=0.5
with different $(\gamma,C)$ combinations.
In this work, various combinations of $(\epsilon,C,\gamma)$ values are tried and
the one with the best cross-validation accuracy (\textit{i.e.,} the lowest RMSE)
is selected. The range and resolution of $C$, $\gamma$ and $\epsilon$
are chosen according to \cite{libsvm,lssvm}.

\section{Experimental results}

\subsection{Experimental setup and training data preparation}

We use Booksim2.0 \cite{booksim} to simulate the router and NoC performance
on mesh NoCs. For each mesh size (ranging from $4\times4$ to $12\times12$),
various synthetic traffic patterns including \textit{uniform random},
\textit{transpose} and \textit{shuffle} and \textit{tornado}\cite{Book} are used
as inputs to the target router architecture to obtain the training
data-set. To verify the model, after the training stage, the learned
model is used to predict the delay with different injection rates
for these three patterns. Also, we use this learned model to predict
the delay for other traffic patterns that are different from the training
sets. Here we use \textit{bitreversal} and \textit{bitcomplement} \cite{Book}
traffics. The training data set contains runs with different packet injection rates under each
traffic pattern, which provides sufficient training data for the SVR
learning engine as well as reasonable training time. The SVR-NoC framework
is implemented in MATLAB based on Libsvm and LS-SVM library \cite{libsvm,lssvm}. Both
synthetic and real application traffic patterns are used in the evaluation.
The real application in the experiment includes DVOPD (Dual Object Plane Decoder) \cite{noc_design_65nm} which is a video MPSoC benchmark mapped
onto a $5\times5$ mesh NoC. 

\subsection{Proposed queuing model accuracy }

\begin{figure}
\includegraphics[width=0.99\columnwidth]{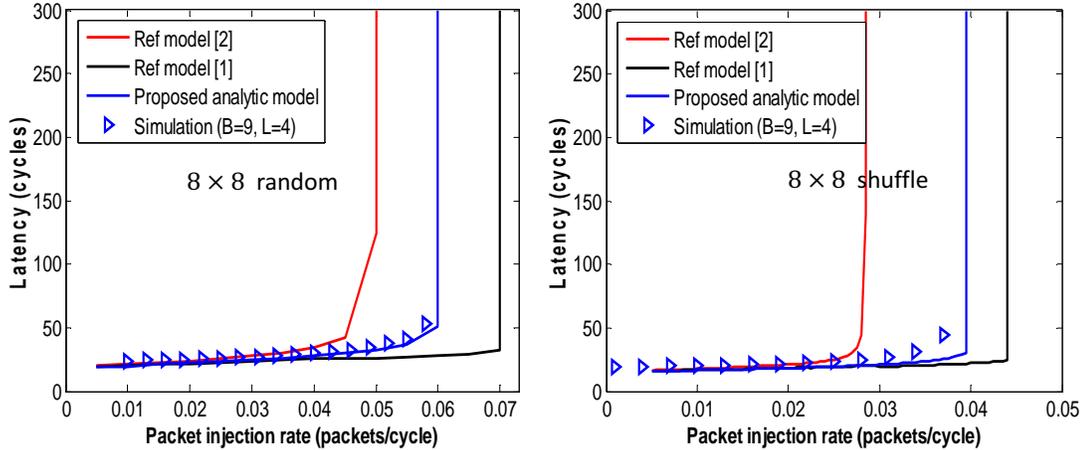}
\caption{\label{fig:Analytical-model-accuracy}Analytical model accuracy comparisons}
\end{figure}

We first compare and evaluate the proposed analytical model using
two synthetic traffic patterns (\textit{i.e.,}, random and shuffle \cite{Book}
traffic) with Poisson packet injection rates at each source node.
The packet length $m$ is assumed to be fixed of $4$ flits. The buffer depth $B$ is adopted to be $9$ flits per input port. We compare the proposed model with two state-of-the-art NoC analtyical models. The reference model $1$ is adopted from \cite{TVLSI12}, where the priority G/G/1 queue has been modified to a generalized G/G/1 queue formula to model the round-robin arbitration in NoC. The reference model $2$ is implemented as proposed in \cite{umit} which is based on a generalized M/G/1/ queue. The comparison results of three analytical models against the simulations under $8\times8$ mesh size are
summarized in Fig. \ref{fig:Analytical-model-accuracy}. 
\\As shown in Fig. \ref{fig:Analytical-model-accuracy}, compared to the two reference models, which over- and under-estimate the network saturation injection points by $10\%-15\%$, the proposed model achieves less than $10\%$ error in predicting the network saturation
point (\textit{i.e.,} the injection rate where the overall latency starts increasing
dramatically) for both the random and shuffle traffic patterns. Consequently, the proposed analytic model provides a better estimation of the channel and source queuing time and helps to relieve the burden of the learning process.
\subsection{SVR prediction accuracy}

We show the $f_{CQ}$ regression accuracy for the training
data-set in Figure \ref{fig:Regression-accuracy-for}-a. As shown in
the figure, our SVR predictor is very accurate for fitting the channel regression curve.
The root mean square error (RMSE) between the predicted and the actual
values is less than $0.5$ and the squared correlation coefficient
is higher than $0.99$.
\subsubsection{Average latency prediction}
\begin{figure}
\includegraphics[width=0.98\columnwidth]{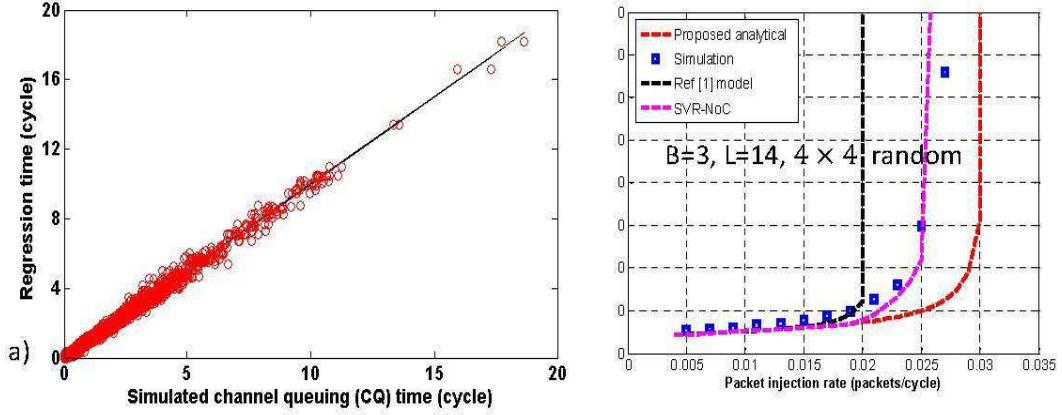}
\caption{\label{fig:Regression-accuracy-for} a) Regression curve of the channel queuing time (CQ) b)Comparisons of the learning and queuing model}
\end{figure}

\begin{figure}
\includegraphics[width=0.99\columnwidth]{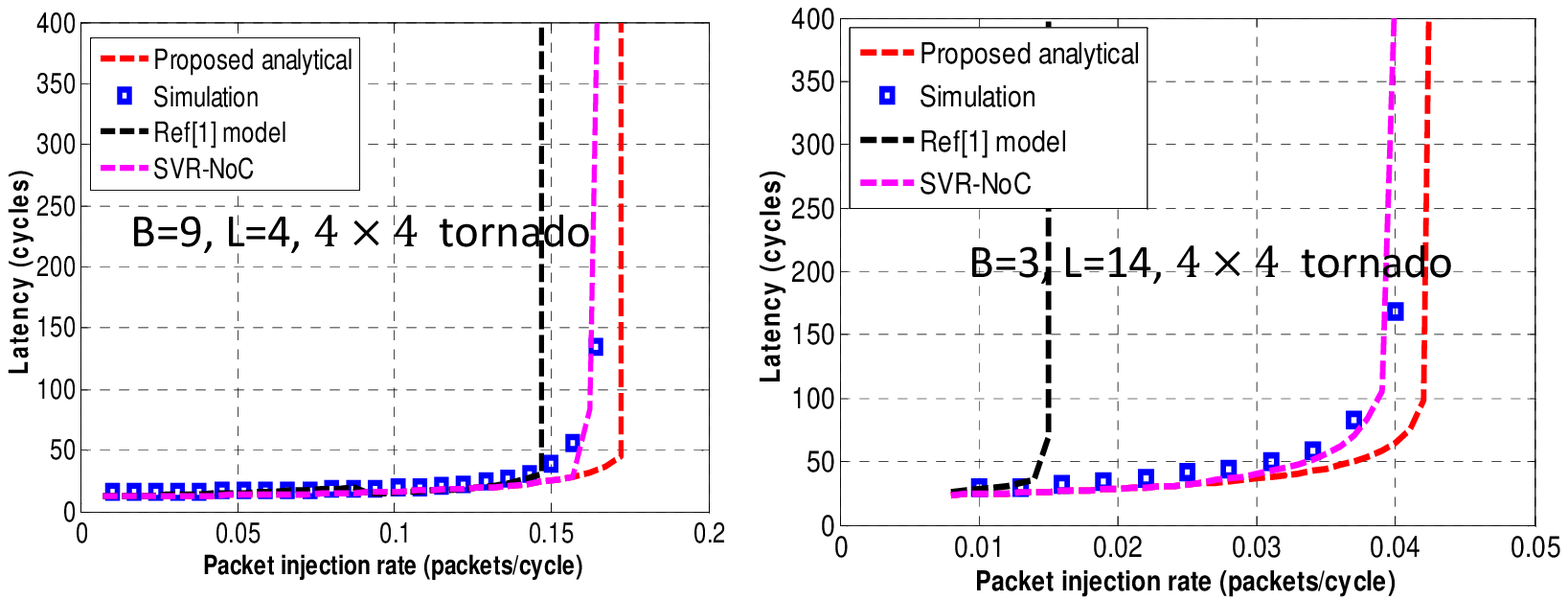}
\caption{\label{fig:Comparison-with-analytical}Comparisons with the queuing
model under different traffic pattern and NoC configurations}
\end{figure}

\begin{figure}
\includegraphics[width=0.99\columnwidth]{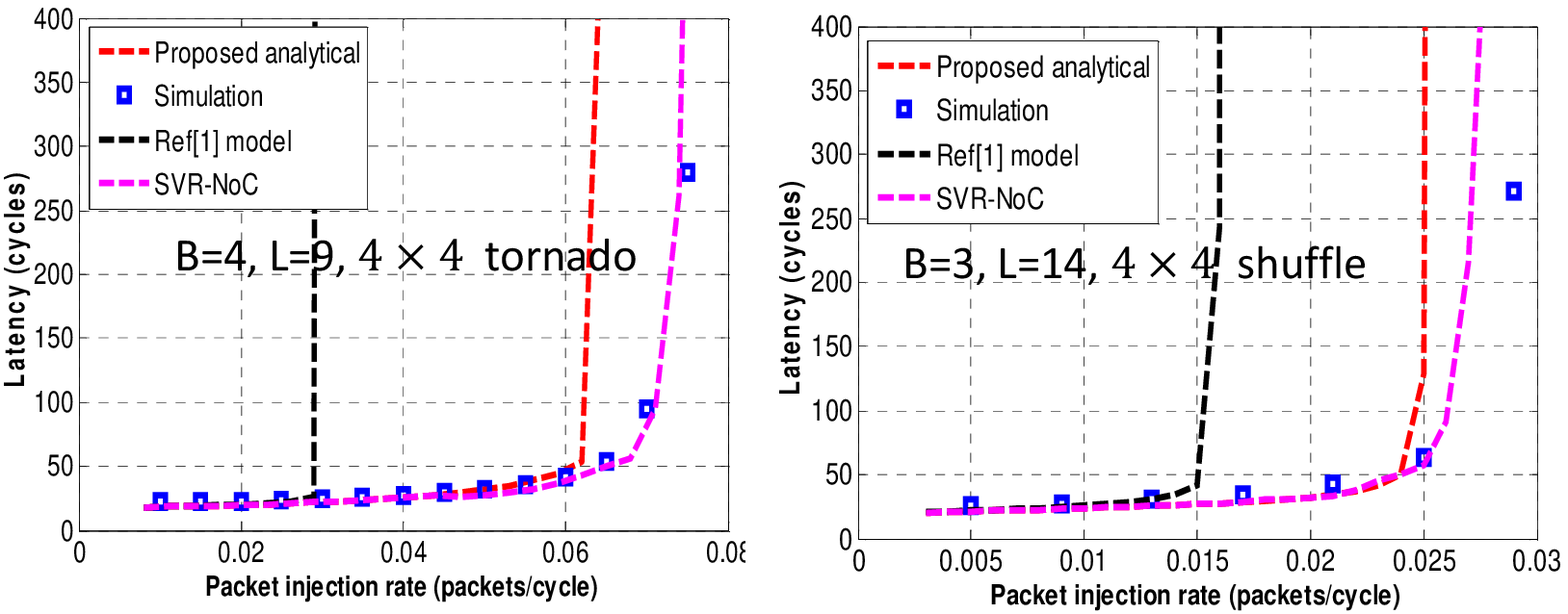}
\caption{\label{fig:Comparison-with-analytical-2}Comparisons with the queuing
model under different traffic pattern and NoC configurations -2 }
\end{figure}

Although the proposed queueing model provides a comprehensive estimation of different NoC configurations and improves the accuracy under a variety of traffic patterns. From our experimental results, it is still observed the accuracy of the queueing model decreases under some traffic patterns due to the approximations and assumptions made in the derivation. To illustrate this, we first compare the SVR-NoC with the proposed analytical model. In
this comparison, $4\times4$ mesh NoC with random, tornado and shuffle traffic are considered. We consider different buffer depth $B$ and packet length $L$ combinations in the evaluation. The average latencies obtained by
different methods are summarized in Fig. \ref{fig:Regression-accuracy-for}-b, Fig. \ref{fig:Comparison-with-analytical} and Fig .\ref{fig:Comparison-with-analytical-2}.
As shown in Figure \ref{fig:Comparison-with-analytical},  the analytical
model incurs  more than $10\%$ error in predicting the network criticality.
Especially for the uneven traffic such as tornado, the error is extremely large
for all the packet length and buffer depth combinations. The inaccuracy
is due to the assumption and simplification made in the derivation
of the PQ model. On the other hand, the SVR-NoC can predict the network
saturation point very accurately, with less than $3\%$ error for
both the random and transpose traffic patterns.

\begin{figure}
\centering
\subfigure[Synthetic traffic used in the training-1]{
                \includegraphics[width=0.95\columnwidth]{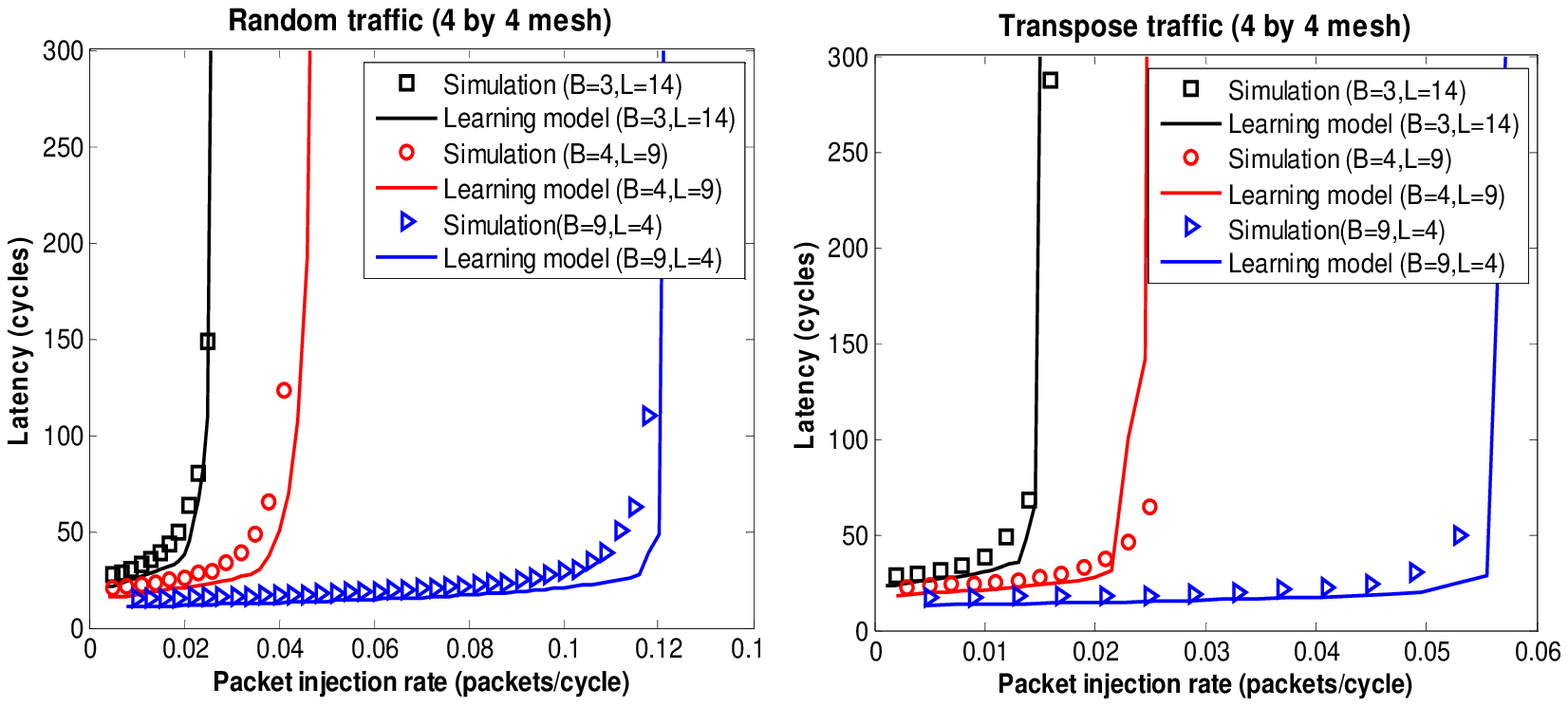}}
\subfigure[Synthetic traffic used in the training-2]{
                \includegraphics[width=0.95\columnwidth]{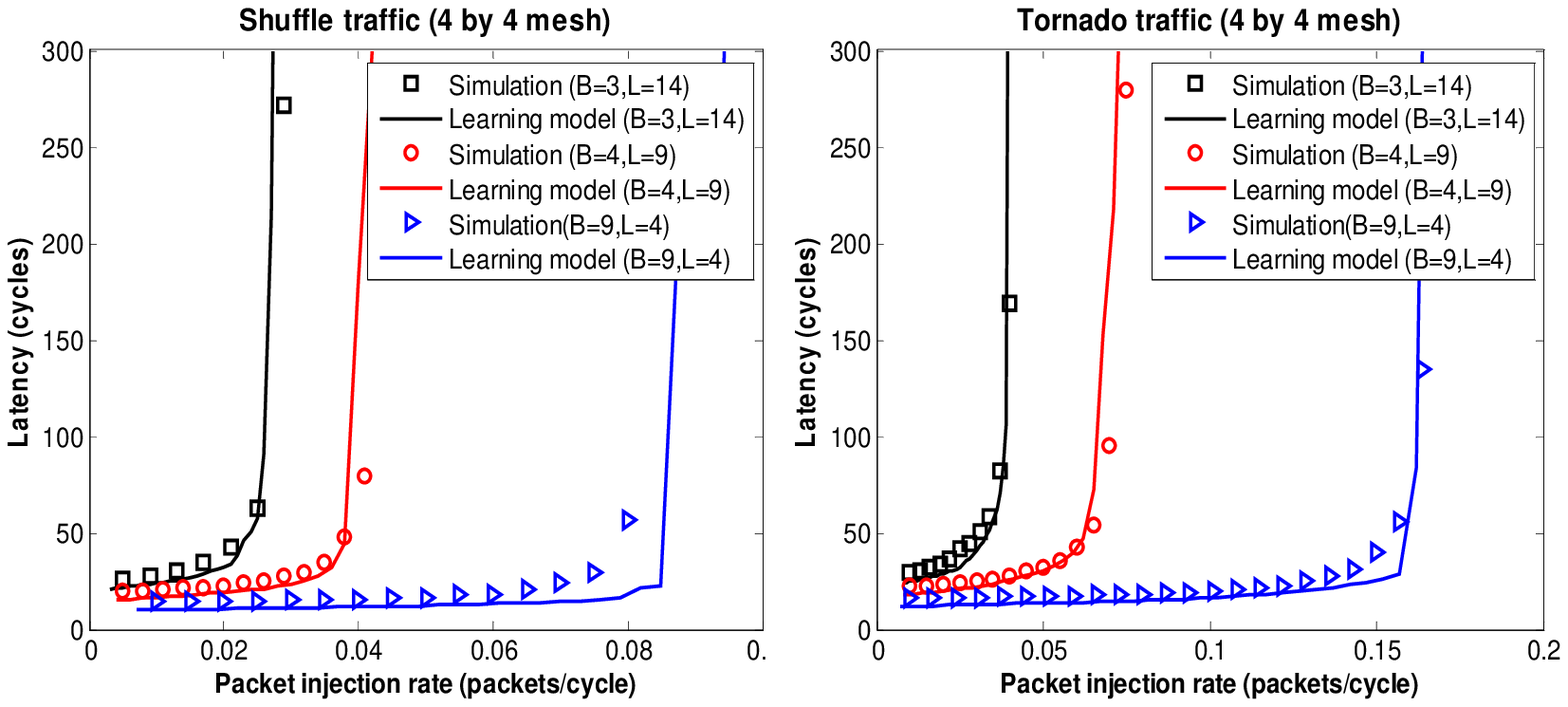}}
\subfigure[Synthetic traffic used in the training-3]{
                \includegraphics[width=0.95\columnwidth]{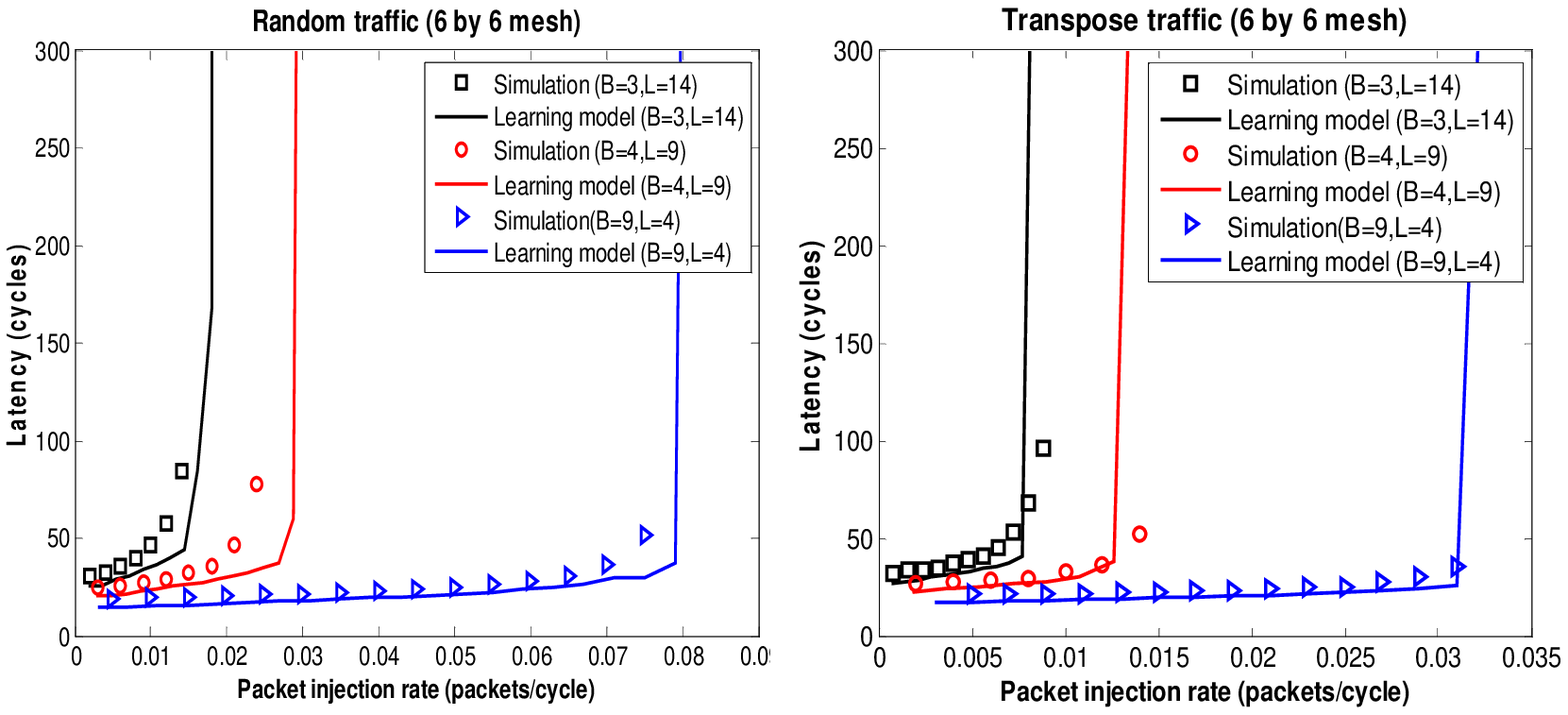}}
\caption{\label{fig:Latency-prediction-performance}Latency prediction performance
under trained traffic patterns}
\end{figure}

\begin{figure}
\subfigure[Synthetic traffic used in the training-4]{
                \includegraphics[width=0.95\columnwidth]{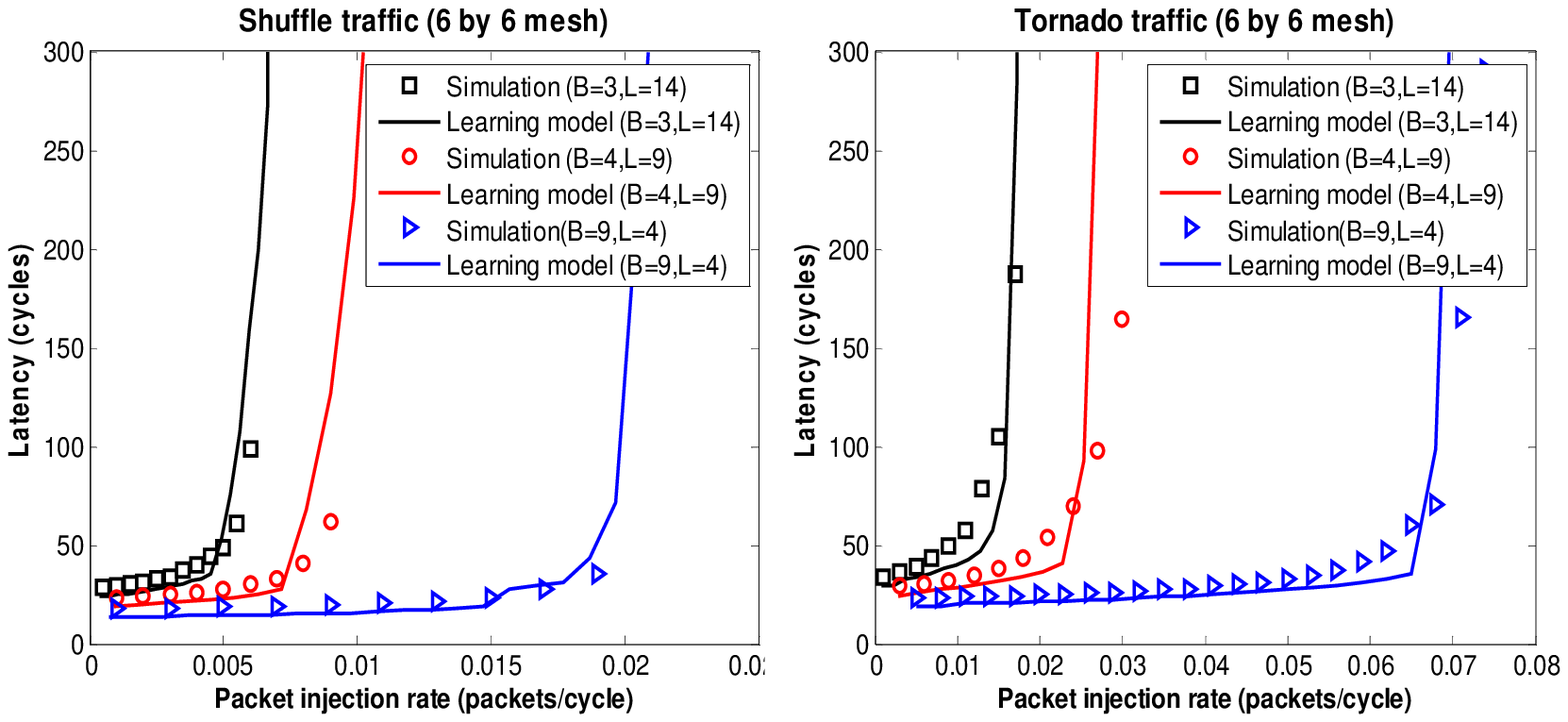}}
                \subfigure[Synthetic traffic used in the training-5]{
                \includegraphics[width=0.95\columnwidth]{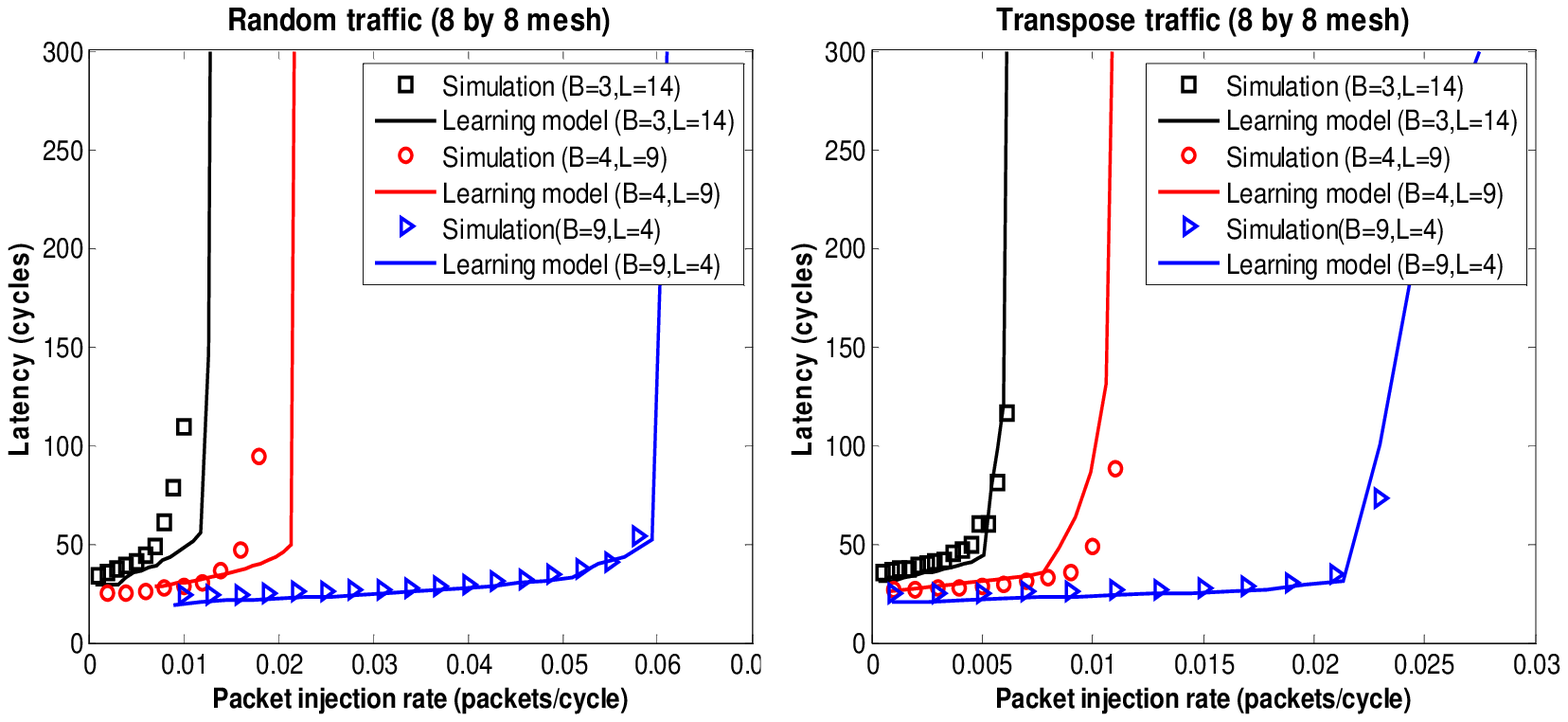}}
                \subfigure[Synthetic traffic used in the training-6]{
                \includegraphics[width=0.95\columnwidth]{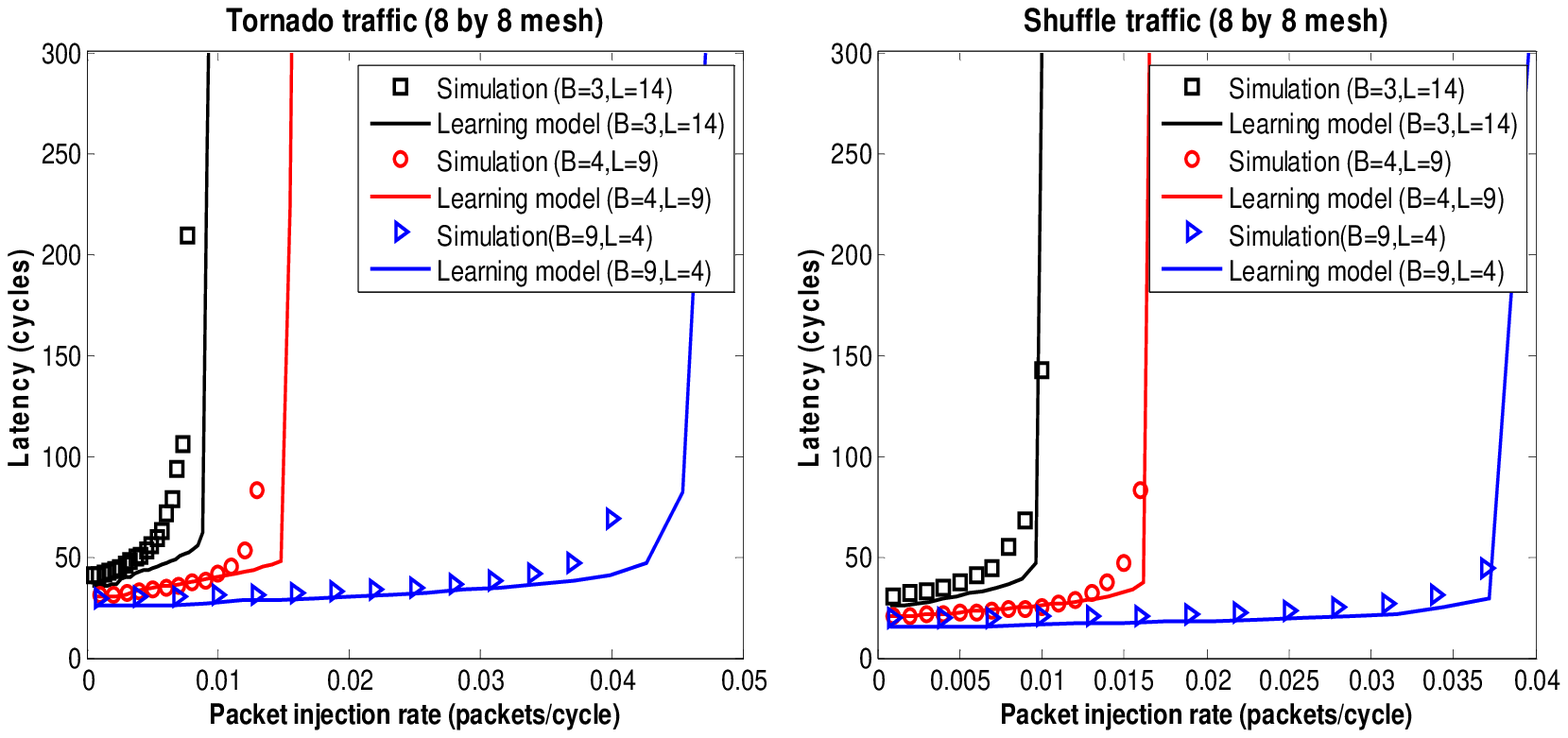}}
\caption{\label{fig:Latency-prediction-for}Latency prediction for trained
synthetic traffics -2 }
\end{figure}

\begin{figure}
\subfigure[Synthetic traffic used in the training-7]{
                \includegraphics[width=0.95\columnwidth]{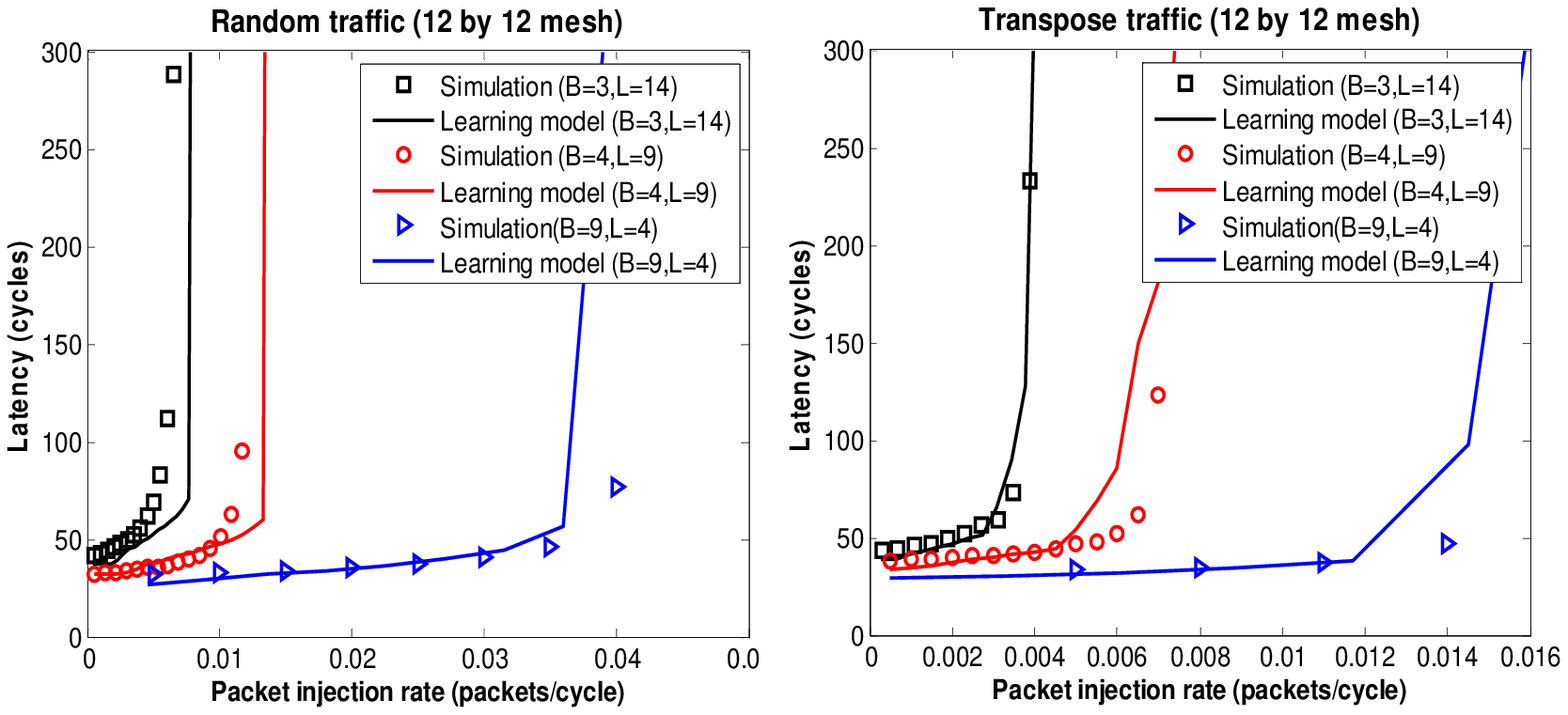}}
                \subfigure[Synthetic traffic used in the training-8]{
                \includegraphics[width=0.95\columnwidth]{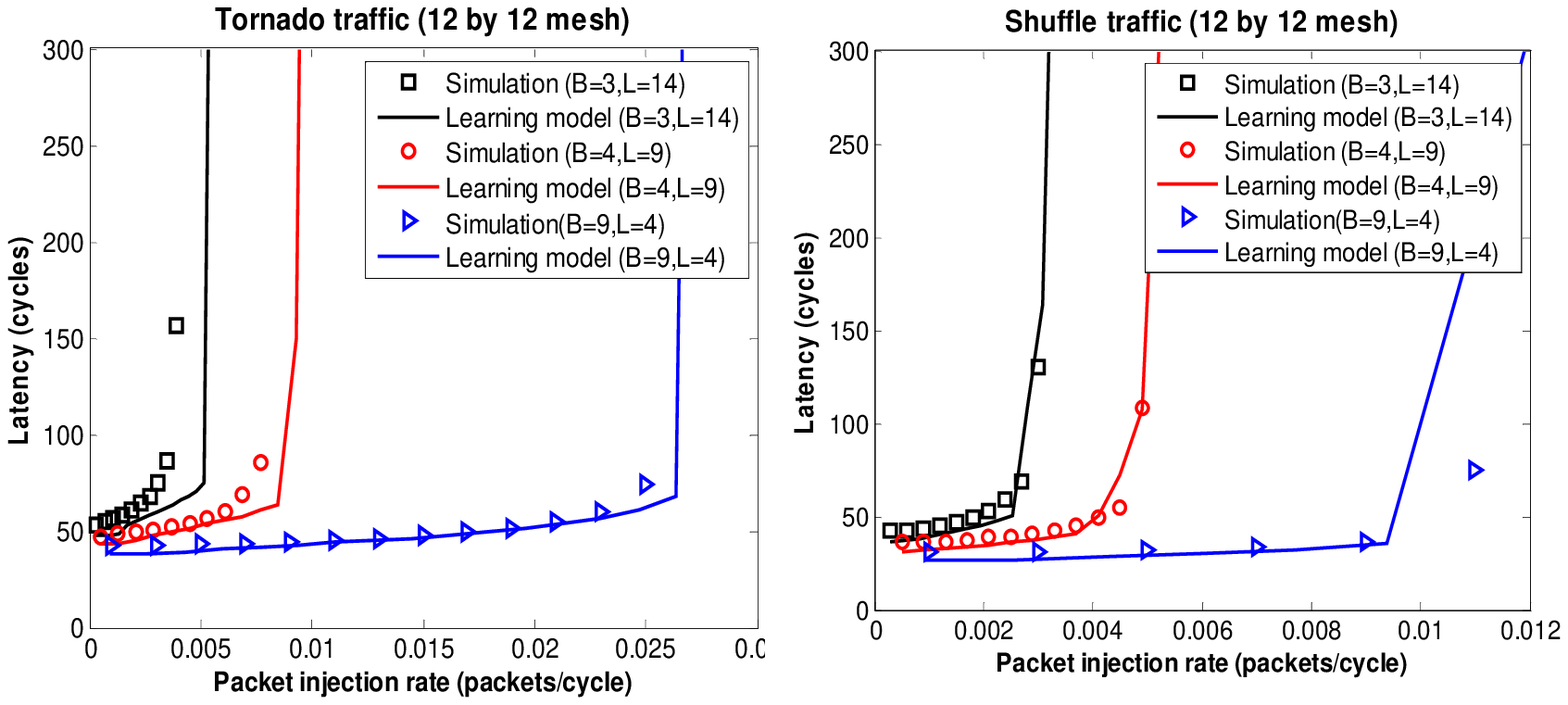}}
\caption{\label{fig:Latency-prediction-for-2}Latency prediction for trained
synthetic traffics -3 }
\end{figure}

Figure \ref{fig:Latency-prediction-performance}-\ref{fig:Latency-prediction-for-2} show the prediction
results for various traffic patterns and router architectures. The traffic patterns in these figures are used in the training stage. There
are three router architectures considered. They are wormhole routers
with: 1) Packet length $L=4$(\textit{flits}) and buffer depth $B=9$ \textit{flits}; 2) $L=9$ and
$B=4$ and 3)$L=14$ and $B=3$. As can be seen in Figure \ref{fig:Latency-prediction-performance}-\ref{fig:Latency-prediction-for-2},
the errors of SVR-NoC in predicting the network saturation point are
within $10\%$ for all the traffic patterns and router architectures.

\begin{figure}
\subfigure[Test the untrained synthetic traffic patterns-1]{
                \includegraphics[width=0.95\columnwidth]{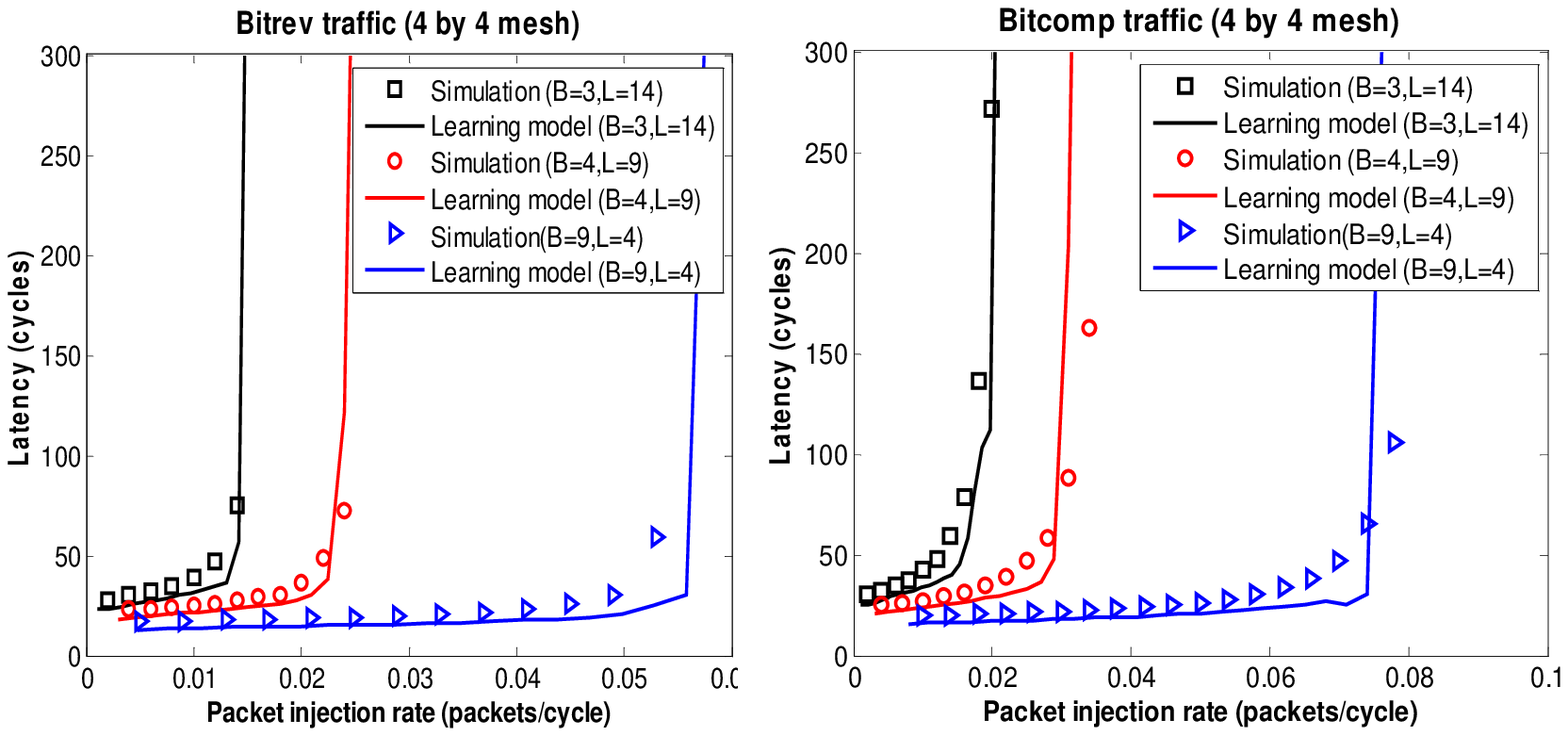}}
                \subfigure[Test the untrained synthetic traffic patterns-2]{
                \includegraphics[width=0.95\columnwidth]{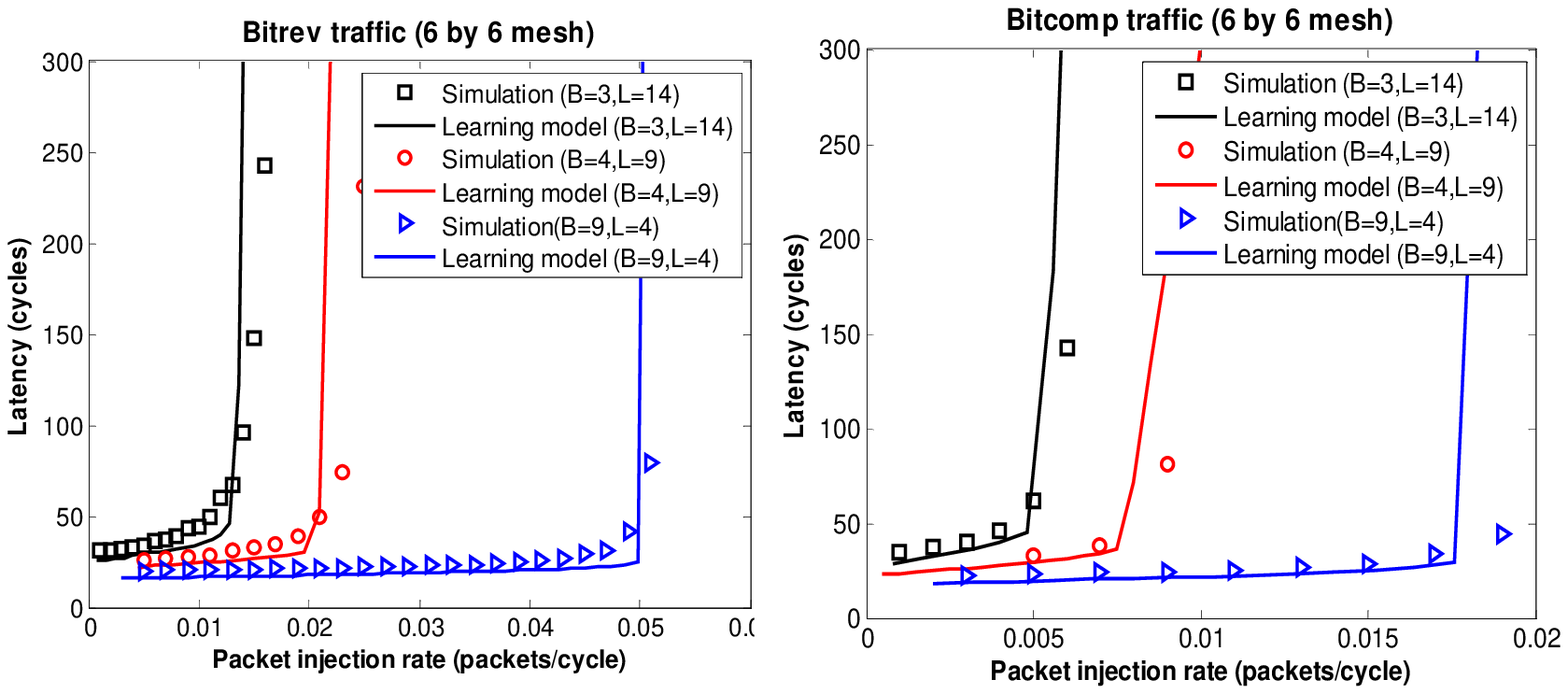}}
                \subfigure[Test the untrained synthetic traffic patterns-3]{
                \includegraphics[width=0.95\columnwidth]{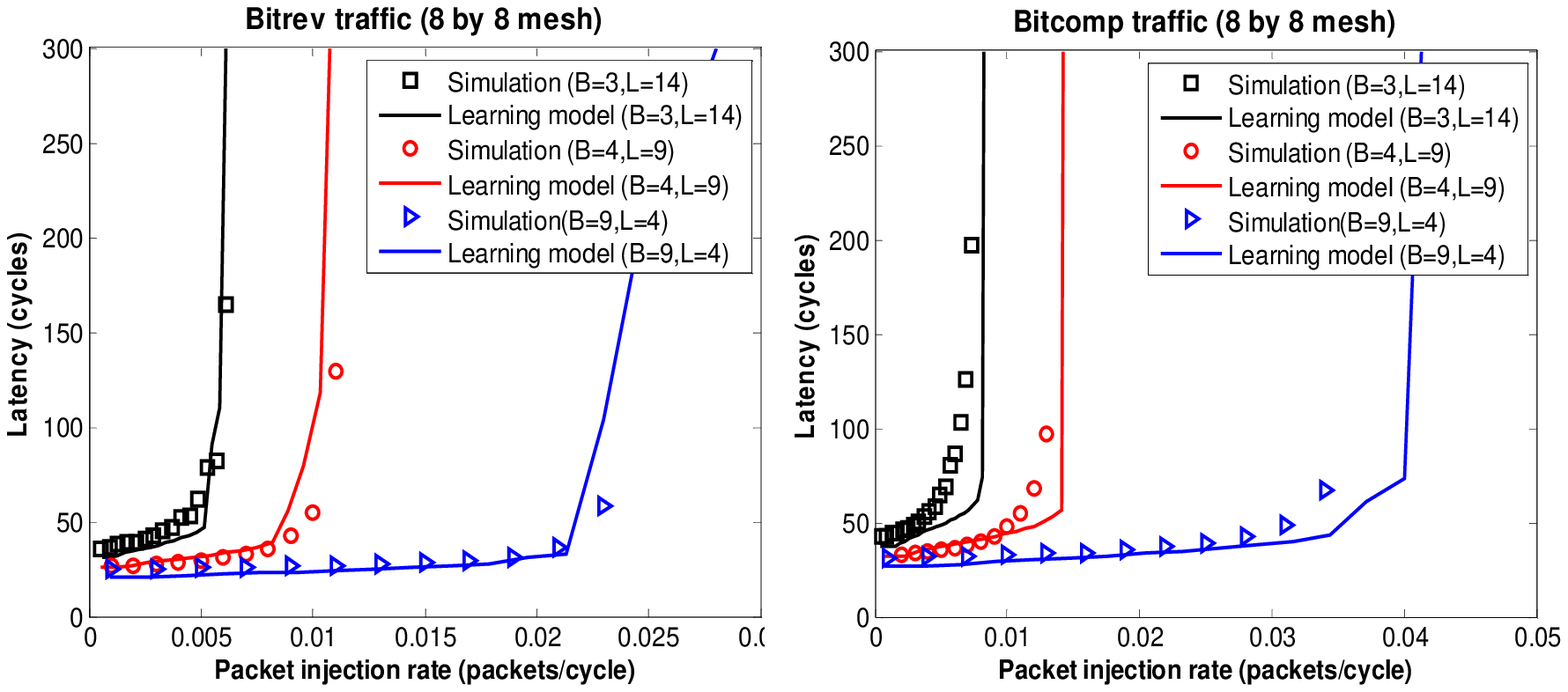}}
\caption{\label{fig:Latency-prediction-for-1}Latency prediction for untrained
synthetic traffics}
\end{figure}

\subsubsection{SVR-NoC for untrained traffic and real applications }

\begin{figure}
                \includegraphics[width=0.95\columnwidth]{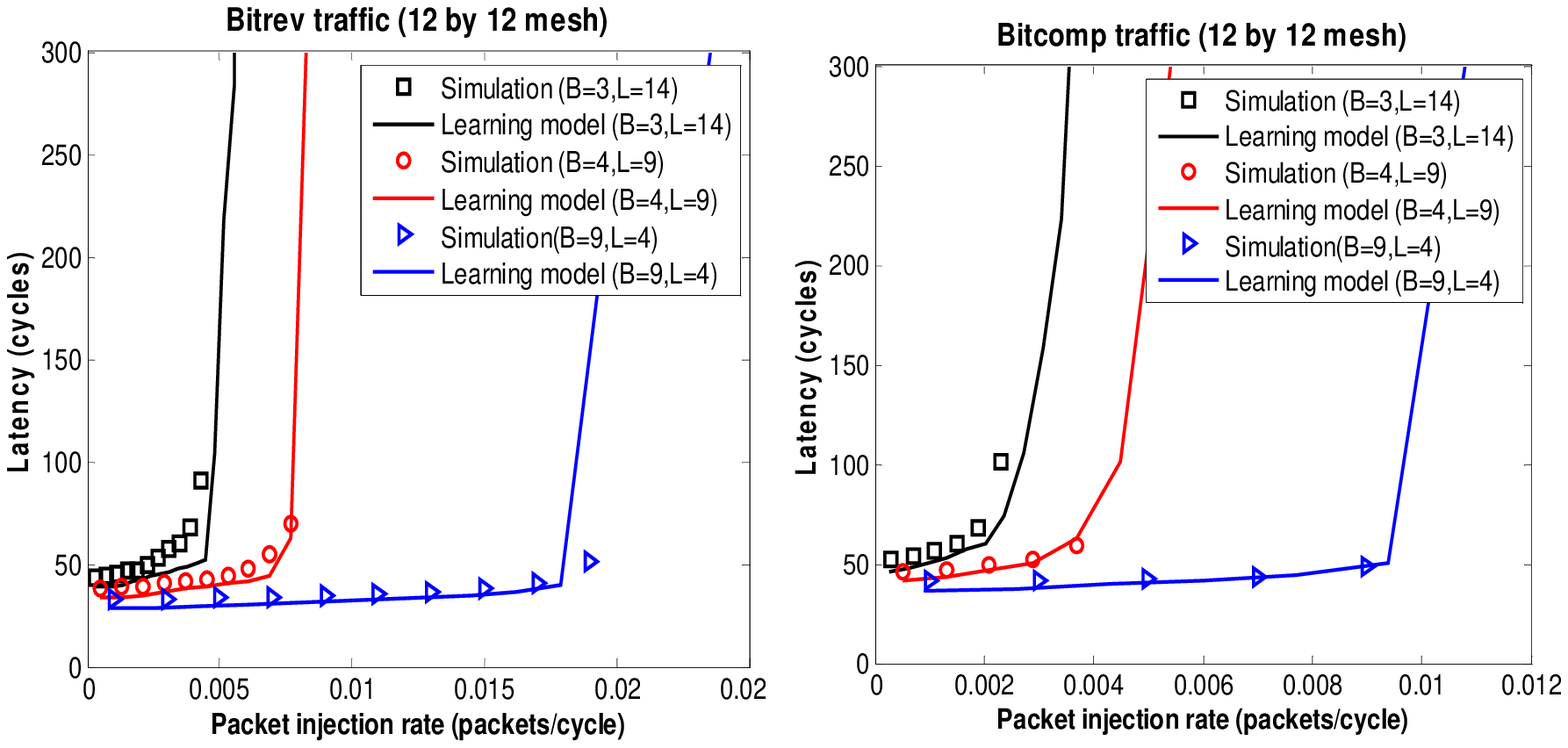}
\caption{\label{fig:Latency-prediction-for-4}Latency prediction for untrained
synthetic traffics -2}
\end{figure}

\begin{figure}
\includegraphics[width=0.9\columnwidth]{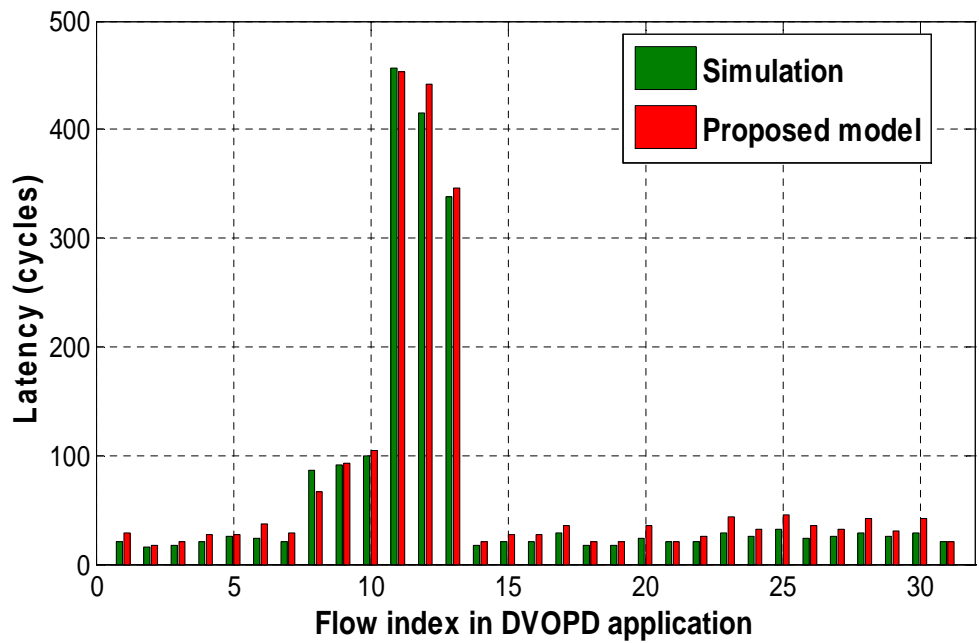}
\caption{\label{Flow_latency_comparison} Flow latency comparison for DVOPD benchmark}
\end{figure}

Next, we show the latency prediction performance of the SVR-NoC for
the other traffic patterns that are not included in the training data
set. Figure \ref{fig:Latency-prediction-for-1}-\ref{fig:Latency-prediction-for-4} show the comparison
results for $4\times4$, $8\times8$ and $12\times12$ meshes. As shown
in the figures, the SVR-NoC still demonstrates
high accuracy in the bit-reversal and bit-complement unseen patterns with less than $5\%$ error
in predicting the network criticality status. We also use a real benchmark,
Dual Object Plane Decoder (DVOPD) \cite{noc_design_65nm} for the comparison and
demonstrate the accuracy of SVR-NoC. Figure \ref{Flow_latency_comparison} shows the comparison
of the latency obtained by the simulation and that predicted by SVR-NoC
for each traffic flow. As shown in Figure \ref{Flow_latency_comparison}, the maximum difference between the simulation results
and the SVR-NoC regression results is less than $15\%$. Similar observations can be obtained for
several different mappings as well as other application benchmarks. 

\subsection{Runtime comparison}

\begin{figure}
\includegraphics[width=0.9\columnwidth]{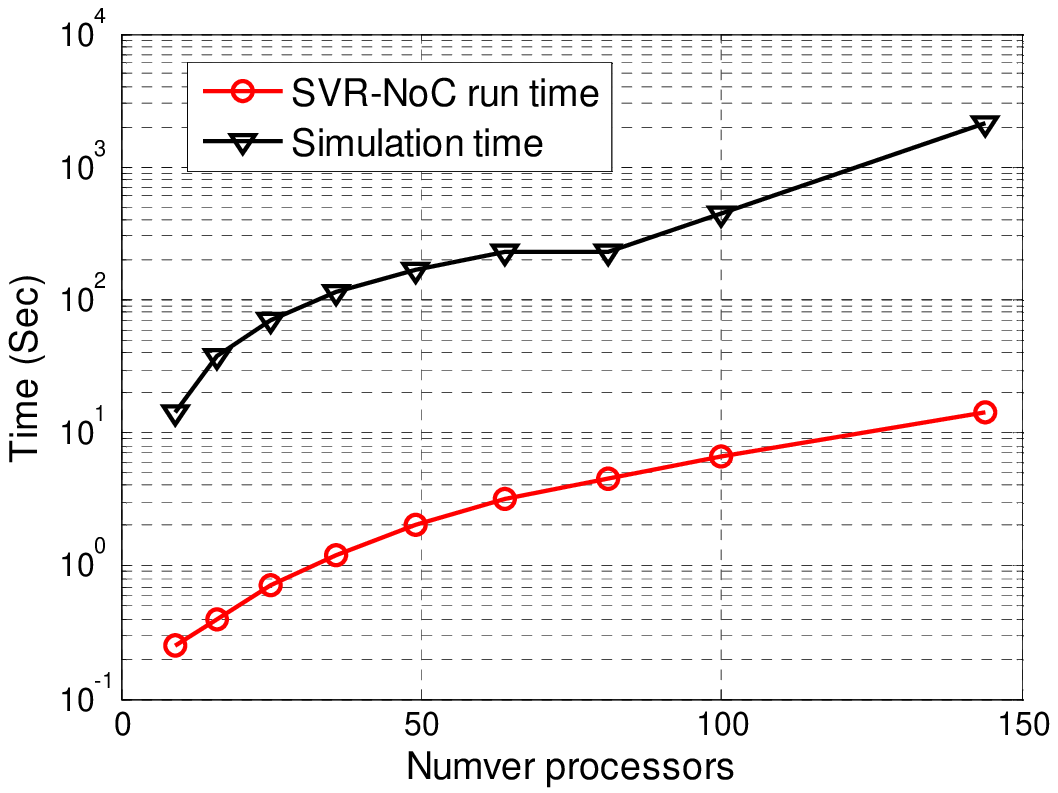}
\caption{\label{run_time_comparison} Run time comparisons of the learning model and simulation}
\end{figure}

The training time of SVR-NoC for an $8\times8$ mesh with $5\times10^{4}$
samples takes about $6.8$ hours. However, the training process is
a one-time effort and it does not incur additional overhead during
the inner-loop performance prediction.  The average simulation time
for an $8\times8$ mesh for $1\times10^{6}$ cycles is about fifteen
minutes while the SVR-NoC predictive framework takes about eight seconds.
The latency estimation method using an analytical model \cite{ICCAD_latency}
takes about six seconds. We can see that the SVR-NoC has a similar
run-time with the analytical model and has a $120X$ speedup over
the simulation-based prediction. In Fig. \ref{run_time_comparison}, we show the run time comparisons of the simulations and learning models under different mesh sizes. As can be seen from the figure, the learning model achieves orders of magnitude improvement in the running time, which can benefit the exploration of a large design space in the design phase.

\section{Conclusion}

In this chapter, we propose a machine-learning based approach for NoC
performance prediction. We demonstrate the modeling accuracy by using
independent test data sets. The SVR-NoC model achieves less than $10\%$
error for various traffic patterns with about $120X$ speedup compared
to the simulation-based estimation.

\chapter{Thermal-aware and Application-specific Routing Algorithm Design}

\textit{In this chapter, we propose a routing algorithm to reduce
the hotspot temperature for application-specific Network-on-chip (NoC).
Using the traffic information of the applications as well as the mapping
of the tasks onto the NoC, we develop a routing scheme which can achieve
a higher adaptivity than the generic ones and at the same time distribute
the traffic more uniformly. A set of admissible paths which is deadlock
free for all the communications is first obtained. To reduce the hotspot
temperature, we find the optimal distribution ratio of the communication
traffic among the set of candidate paths. The problem of finding this
optimal distribution ratio is formulated as a linear programming (LP)
problem and is solved offline. A router microarchitecture which supports
our ratio-based selection policy is also proposed. From the simulation
results, the peak energy reduction considering the energy consumption
of both the processors and routers can be as high as 20\% for synthetic
traffic and real benchmarks.}

\section{Introduction}

For the design of sophisticated MPSoC on high performance NoC, power
and temperature have become the dominant constraints \cite{thermal_noc}.
Higher power consumption leads to higher temperature and at the same
time, the uneven power consumption distribution across the chip will
create thermal hotspots. These thermal hotspots have adverse effect
on the carrier mobility, the meantime between failure (MTBF), and
also the leakage current of the chip. As a result, they will degrade
the performance and reliability dramatically. Consequently, it is
highly desirable to have an even power and thermal profile across
the chip \cite{hotspot_elimination}. This imposes
as a design constraint for NoC to avoid uneven power consumption profile
so as to reduce the hotspot temperature.

In a typical NoC-based MPSoC design, we need to allocate and schedule
the tasks on the available processors and map these processors onto
the NoC architecture first. After the task and processor mapping,
routing algorithm is developed to decide the physical paths for sending
the packets from the sources to the destinations. Each phase of the
NoC design affects the total power consumption and also the power
profile across the chip. Previously, task mapping and processor core
floorplan algorithms \cite{thermal_aware_IP} have been proposed
to achieve a thermal balanced NoC design. However, the routing algorithm
is rarely exploited for this purpose. Since communication network
(including the routers and the physical links) consumes a significant
part of the chip\textquoteright{}s total power budget ($39\%$ of total
tile power \cite{router_example}), the decision on
the routing path of the packets will greatly affect the power consumption
distribution and hence the overall chip hotspot temperature \cite{thermal_noc}.
Therefore it is important to consider the thermal constraint in the
routing phase of the NoC design.


In this chapter, we tackle the routing problem to achieve an even
temperature distribution for an application-specific MPSoC. Given
an application described by the task flow graph and a target NoC topology,
we assume that the tasks are already scheduled and allocated to the
processors and the processor mapping is also done. We then utilize
the traffic information of the application specified in the task flow
graph, which can be obtained through profiling \cite{1411933,Noc_synthesis},
to decide how to split the traffic among different physical routes
for an even power profile. 


Similar to \cite{antnet_routing}, we use the peak
power (or the peak energy under a given time window) metrics to evaluate
the effectiveness in reducing the hotspot temperature. Through simulation-based
evaluation, we demonstrate that the proposed algorithm can reduce
the peak energy of the tiles by $10\%-20\%$ while improving or maintaining
the throughput and latency performance.
\section{Related Work}
\label{sec:2} 

In the area of temperature-aware NoC design, many previous works focus
on the power consumption distribution of the processor cores. In \cite{hotspot_prevention},
a dynamic task migration algorithm was proposed to reduce the hotspot
temperature due to the processor core (\textit{i.e.,} PE). In \cite{inducing_thermal_wareness},
a thermal management hardware infrastructure was implemented to adjust
the frequency and voltage of the processing elements according to
the temperature requirements at run time.

Since the power consumption of the routers is as significant as the
processor core, the thermal constraint should also be addressed in
the routing algorithm design. There have been a lot of works on NoC
routing algorithms for various purposes including low power routing
\cite{inducing_thermal_wareness}, fault-tolerant routing \cite{vicis}
and congestion avoidance routing to improve latency \cite{fluidity_concept_noc}.
However, there are only a few works \cite{antnet_routing,thermal_noc,routing_traffic_migration}
taking temperature issue into account. 

In \cite{antnet_routing}, an ant-colony-based
dynamic routing algorithm was proposed to reduce the peak power. Heavy
packet traffics are distributed on the chip based on this dynamic
routing algorithm to minimize the occurrence of hot spots. However
this dynamic routing algorithm is generic in nature and does not take
into account the specific traffic information of the applications.
Therefore it may not be able to achieve an optimal path distribution.
Special control packets are sent among the routers to implement the
algorithm which increases the power overhead. Also, two additional
forward and backward ant units are needed in the router which results
in a large area overhead. Moreover, this work only minimizes the peak
power of the routers but does not consider the effect of the processor
core power on the temperature. In \cite{thermal_noc},
a run-time thermal-correlation-based routing algorithm is proposed.
When the peak temperature of the chip exceeds a threshold, the NoC
is under thermal emergency and the dynamic algorithm will throttle
the load or re-route the packets using the paths that have the least
thermal correlation with the run time hottest regions. However, the
algorithm also does not consider the specific traffic information
of the applications. It may be inefficient if multiple hotspots occur
at the same time. Also it does not clearly describe how to do the
re-routing while still guaranteeing the deadlock free property. In
\cite{routing_traffic_migration}, a new routing-based traffic migration
algorithm VDLAPR and the buffer allocation scheme are proposed to
trade-off between the load balanced and the temperature balanced routing
for 3D NoCs. In particular, the VDLAPR algorithm is designed for 3D
NoCs by distributing the traffic among various layers. For routing
within each layer, a thermal-aware routing algorithm such as the one
introduced in this chapter is still needed.

In this chapter, we focus on the thermal-aware routing for application
specific NoC \cite{my_asp_dac_2011}. To guarantee deadlock free property, generic routing
schemes use algorithms such as X-Y routing, odd-even routing or forbidden
turn routing \cite{NoCbook_Peh}. However, this
will limit the flexibility of re-distributing the traffic to achieve
an even power consumption profile. Here, we utilize the characterized
application traffic information to achieve a larger path set for routing
and at the same time provide deadlock avoidance. Higher adaptivity
and hence better performance can be achieved since more paths can
be used for the re-distribution of the traffic. Given the set of possible
routing paths, we formulate the problem of allocating the optimal
traffic among all paths as a mathematical programming problem. At
run time, the routing decisions will be made distributively according
to the calculated traffic splitting ratios. We demonstrate the effectiveness
of the proposed routing strategy on peak energy reduction using both
synthetic and real application traffics. 
\section{Methodology overview}
\subsection{Assumptions and preliminaries}

\label{sec:3} 
We aim to achieve an even power consumption distribution and reduce
the hotspot temperature for application specific MPSoC using NoC connection.
We assume that the given application is specified by a task flow graph
which characterizes the communication dependencies and the bandwidth
requirements among the tasks of the application. In this chapter,
we use a tile-based mesh topology for the NoC due to its popularity
and simplicity. However, our methodology can be easily extended to
other irregular topologies. Based on the task flow graph and the target
topology, task allocation and processor mapping are determined before
the routing phase. The energy consumption of each processor core is
then estimated according to the tasks mapped to it. Taking all the
above information as input, we address the issue of reducing the peak
temperature via routing algorithm design. We adopt an adaptive routing
strategy instead of static routing scheme because it provides path
diversity to distribute traffic among the sources and the destinations
\cite{NoCbook_Peh} and achieves better performance
in terms of latency, throughput and congestion avoidance. We also
assume minimum path routing is used.

To facilitate the discussion of the proposed methodology and algorithms,
we borrow some of the definitions presented and discussed  in \cite{application_specific_plaesi}.
\begin{definition}
A Task graph $TG=T(T,C)$ is a directed graph where the vertex $t_{i}\in T$
represents a task in the application and the edge $c_{ij}=(t_{i},t_{j})\in C$
represents the communication from $t_{i}$ to $t_{j}$.
\end{definition}
\begin{definition}
A core communication graph $CCG=(P,E)$ is a directed graph where
the vertex $p_{i}\in P$ represents the $i^{th}$ processor in the
mesh and the edge $e_{ij}=(p_{i},p_{j})\in E$ represents the communication
from $p_{i}$ and $p_{j}$. Given an application-specific task graph
and its corresponding mapping function, $CCG$ can be generated accordingly.

\end{definition}
\begin{definition}
A channel dependency graph $CDG=(L,R)$ is a directed graph where
the vertex $L_{mn}\in L$ represents a physical link $(m,n)$ in the
mesh topology pointing from tile $m$ to tile $n$. An edge $R_{\mu\nu}=(L_{\mu},L_{\nu})\:(\mu=mn,\,\nu=np)\in R$
exists if there is a dependency from $L_{\mu}$ to $L_{\nu}$ which
indicates a path passing through the two links $(m,n)$ and $(n,p)$consecutively.

\end{definition}
From Duato\textquoteright{}s theorem \cite{deadlock_duato}, a routing
algorithm is deadlock free for an application if there are no cycles
in its channel dependency graph (CDG).\vspace{-4mm}
\begin{definition}
Strongly connected components (SCC): A directed graph is strongly
connected if there is a path from each vertex to every other. The
strongly connected components of a graph are its maximally connected
sub-graphs.\vspace{-6mm}

\end{definition}
\begin{figure}
\includegraphics[width=1\linewidth]{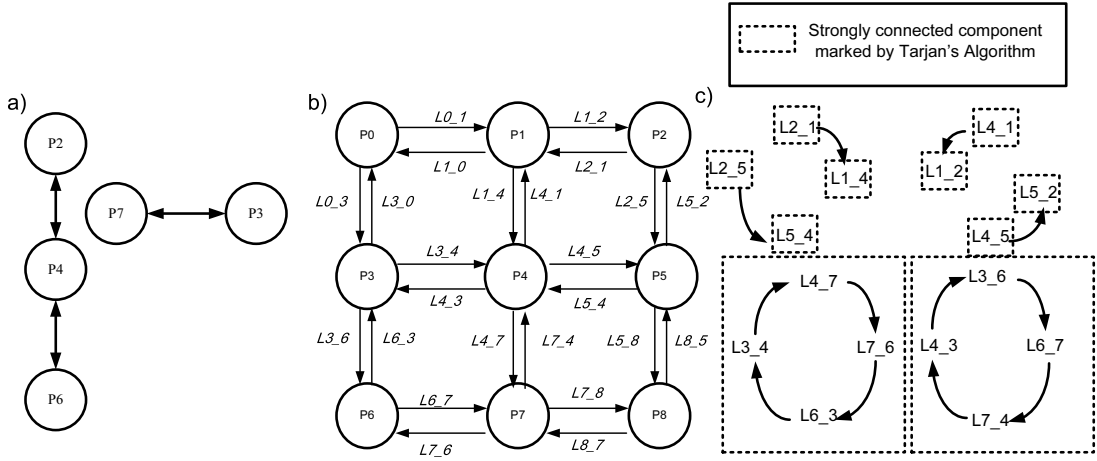}

\caption{\label{fig:an-example-of}An example of: a) Core communication graph
(CCG), b) NoC architecture, c) Channel dependency graph (CDG) and its
strongly connected components(SCC) after mapping}
\end{figure}

We illustrate the definitions of CCG, CDG and SCC by an example shown
in Figure \ref{fig:an-example-of}. Here we use a $3\times3$ mesh
NoC (Figure \ref{fig:an-example-of}-b as the communication backbone
for an application characterized by the core communication graph shown
in Figure \ref{fig:an-example-of}-a. For each communication pair
$e_{ij}=(p_{i},p_{j})$ in the CCG, assume the Manhattan distance
(i.e. the minimum number of hops) between the source and destination
tile pair $(i,j)$ is $d(i,j)=|i_{x}-j_{x}|+|i_{y}-j_{y}|$ and the
distance between the two tiles in the $x$-direction is $d_{x}(i,j)=|i_{x}-j_{x}|$,
the total number of minimum length paths connecting the two tiles
is given by $N_{ij}=C_{d(i,j)}^{d_{x}(i,j)}$ %
\footnote{$C_{n}^{r}$ is the combination notation, where $C_{n}^{r}=\frac{n!}{r!(n-r)!}$ %
} since we need to traverse along the $x$ direction $d_{x}(i,j)$
times among total $d(i,j)$ number of hops. Let $l(i,j,k)\,(k\leq N_{ij})$
denote the $k^{th}$ path connecting the two tiles. If $l(i,j,k)$
traverses two network links $(m,n)$ and $(n,p)$ consecutively, then
an edge is added to connect the two nodes  $L_{mn}$ and $L_{np}$
in the CDG. By inspecting all the feasible paths, we can construct
the whole channel dependency graph. In Figure \ref{fig:an-example-of},
taking the communication pair $(2,4)$ as an example, two minimum
length paths exist: $(2\rightarrow5\rightarrow4)$ and $(2\rightarrow1\rightarrow4)$.In
the CDG, we add two edges connecting $(L_{2-5,}L_{5-4})$ and $(L_{2-1},L_{1-4})$
, respectively. As shown in Figure\ref{fig:an-example-of}-c, there
are in total 10 strongly connected components in the resulting CDG
which can be found by the Tarjan's algorithm \cite{tarjan_algorithm}.
We can find two circles in Figure \ref{fig:an-example-of}-c, \textit{i.e.,
$L_{3-4}\rightarrow L_{4-7}\rightarrow L_{7-6}\rightarrow L_{6-3}\rightarrow L_{3-4}$
}and\textit{ $L_{4-3}\rightarrow L_{3-6}\rightarrow L_{6-7}\rightarrow L_{7-4}\rightarrow L_{4-3}$
, }where some edges are required to be removed by the path set finding
algorithm in Section 4.1 to avoid deadlocks.
\begin{definition}
Routing adaptivity: Routing adaptivity is defined as the ratio of
the total number of available paths provided by a routing algorithm
to the number of all possible minimum length paths  between the source
and destination pairs. Higher adaptivity will improve the capability
of avoiding congestion and redistributing the traffic to reduce thermal
hotspot. However, we need to meet the deadlock-free constraints while
improving the routing adaptivity.\vspace{-2mm}

\end{definition}
For a given application, we first use the task graph $TG$ (Definition
1) to characterize its traffic patterns as well as the communication
volume between the tasks. Then based on the mapping algorithm on the
target NoC platform, the task graph $TG$ is transformed to the core
communication graph $CCG$ (Definition 2). The channel dependency
graph $CDG$ (Definition 3) is built based on the routing paths assigned
to each edge in $CCG$. In this chapter, we propose to find the strongly
connected components $SCC$ (Definition 4) of the underlying $CDG$
to remove the circular dependency among the channel resources to avoid
potential deadlock. Finally, routing adaptivity (Definition 5) is
the metric that reflects the capability of redistributing traffic
for different cycle breaking algorithms in $SCC$.

\subsection{Motivation for thermal-aware routing}

\begin{figure}
\centering{}\includegraphics[width=1\textwidth]{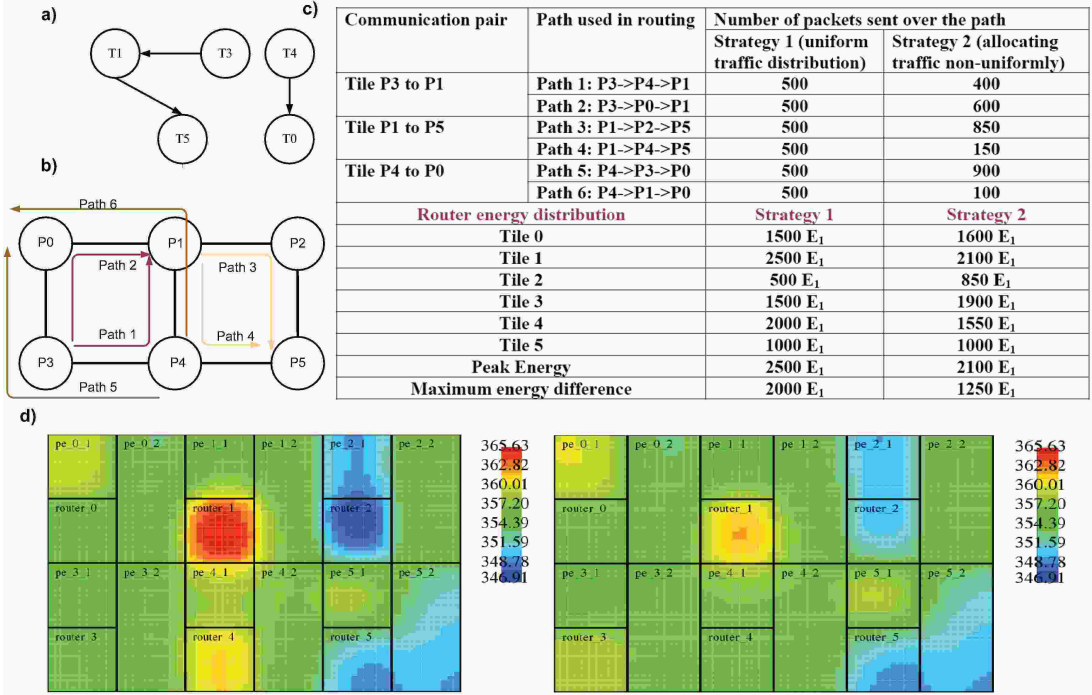}\caption{\label{fig:A-motivation-example-1}A motivation example of allocating
traffic among paths: a) the core communication graph(CCG), b) routing
paths allocation on $2\times3$ NoC, c) two strategies of using the
routing paths, d) thermal profile comparison (left: strategy one, right:
strategy two) }
\end{figure}

For adaptive routing, normally there will be more than one path available
for every communication pair. If traffic is distributed equally on
all the paths, some of the routers may have more paths passing through
them and hence more packets to receive and send.  We show the need
to allocate traffic properly among the paths by an example illustrated
in Figure \ref{fig:A-motivation-example-1}. 
 In Figure \ref{fig:A-motivation-example-1}, three communication
pairs occur concurrently: from tile P3 to P1, tile P1 to P5 and tile
P4 to P0. We assume 1000 packets are generated for each pair within
a time window and sent over the network in this example. The energy
consumption of processing a single packet in a router is denoted as
$E_{1}$. Two routing strategies are compared. Strategy 1 uses uniform
traffic distribution among all paths between the source and the destination
nodes for routing. Strategy 2 allocates traffic non-uniformly among
the candidate paths. The total number of packets handled by each router
is different in these two strategies. The router energy distribution
is shown in Figure \ref{fig:A-motivation-example-1}-c. By properly
allocating the traffic, we can reduce the peak energy of the tile
by 16\% and the energy difference among the tiles by 37.5\%. In this
simple example, we assume the average energy consumption of the processor
core and the router are about the same. We use Hotspot \cite{hotspot}
to simulate and evaluate the thermal profile. As shown in Figure \ref{fig:A-motivation-example-1}-d,
strategy 2 indeed makes the thermal profile more uniform and reduces
the hotspot temperature.
\subsection{Thermal-aware routing design flow}

From the above example, we can see that we need to find a set of paths
for routing the packets for every communication pair and allocate
the traffic properly among these paths so as to achieve a uniform
power consumption profile. One critical issue of determining the path
set is to provide deadlock avoidance. In generic routing algorithms,
deadlock is prevented by disallowing various turns \cite{NoCbook_Peh}.
For application specific NoCs, it will be too conservative and unnecessarily
prohibit some legitimate paths to be used \cite{application_specific_plaesi}
as some disallowed turns can actually be used because the application
does not have traffics interacting with these turns to form circular
dependencies in the CDG. By using the information specified by the
task flow graph and the derived communication graph, we can find more
paths available for routing and increase the flexibility of distributing
traffic among the sources and the destinations. Here we use a similar
approach as \cite{application_specific_plaesi}
to find the set of admissible paths for each communication pair while
still satisfying the deadlock free requirement.

After obtaining the admissible path set for routing, we find the optimal
traffic allocation to each path based on the bandwidth requirement
to achieve an even power distribution profile. The problem is formulated
as a mathematical programming problem and solved by a LP solver. These
phases of design are done offline at the design time. After that,
the optimal distribution ratio of each path is obtained. For each
router and a particular source-destination communication pair, the
ratios of the paths passing through it are combined into the probabilities
of using each port to route to the destination. At run time, these
probability values are stored into the routing table in each router.
For each incoming packet, the router will inquiry its routing table
and return the candidate output ports for this packet according to
the input direction and the destination. The final output port will
be chosen according to the probability value of each candidate. 
\begin{figure}
\includegraphics[width=1.02\linewidth]{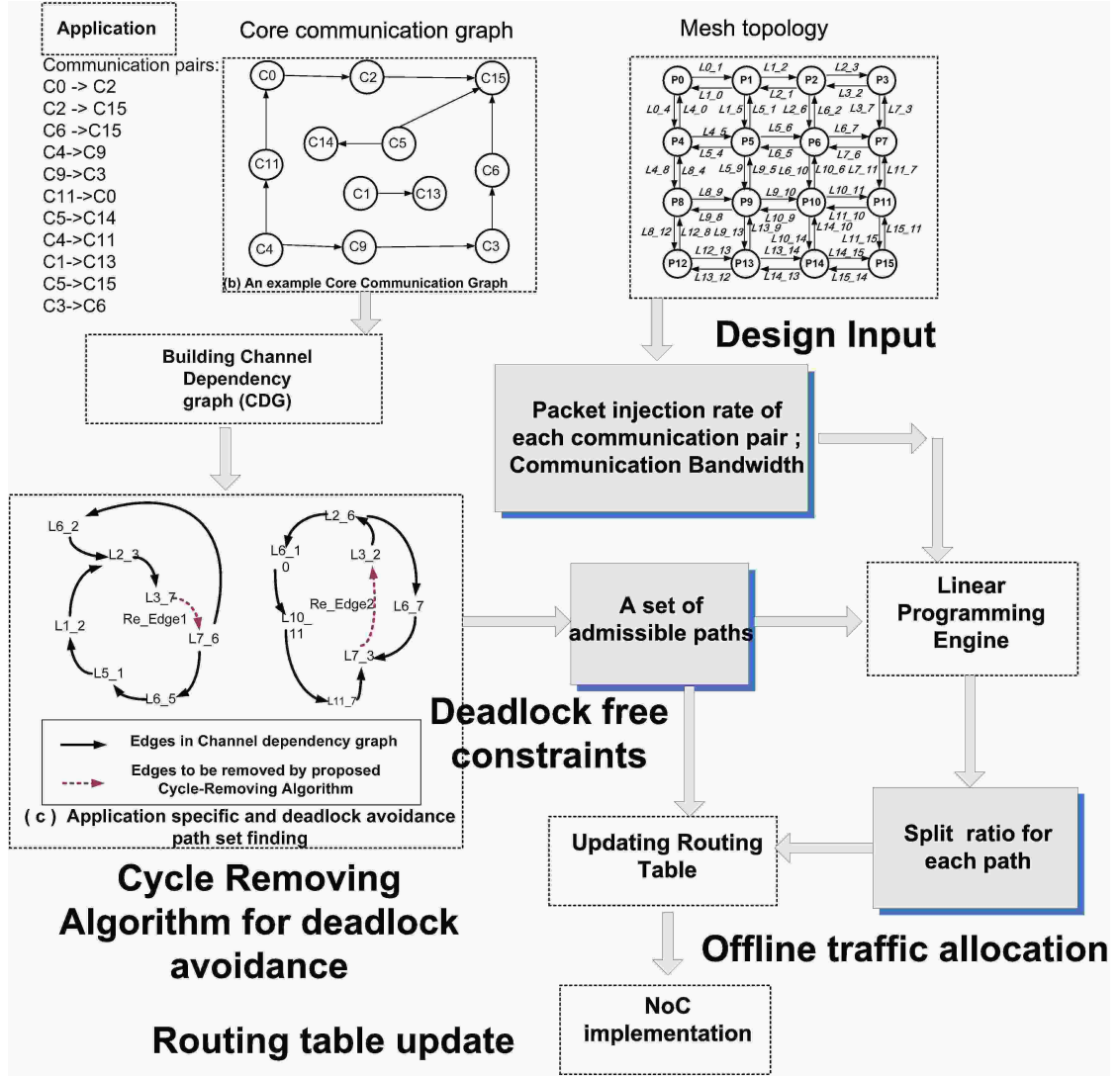}

\caption{\label{fig:Proposed-design-methodology}Proposed design flow for the
thermal-aware routing}
\end{figure}

Figure \ref{fig:Proposed-design-methodology} summarizes the whole
design flow of our proposed methodology. In the next section, the
details of the algorithm will be presented.

\section{Main algorithm}

In this section, we present the details of the offline routing algorithms.
We first discuss the algorithm of finding the set of admissible paths
for each source-destination communication pair. The admissible paths
avoid the circular dependency among any paths and hence provide the
deadlock free property. Then we present the optimal traffic allocation
problem formulation.

\subsection{Application-specific path set finding algorithm}

In \cite{application_specific_plaesi},
a dynamic routing algorithm that increases the average routing adaptivity
while maintaining deadlock free is proposed. The average routing adaptivity
is often used to represent the degree of adaptiveness and flexibility
of a routing algorithm. Here we use a similar approach. Instead of
maximizing the average routing adaptivity, we aim to maximize the
flexibility to re-divert the traffic to even out the power consumption
distribution. Therefore we have to consider the bandwidth requirement
of each communication also since the amount of packet processed will
directly affect the energy consumption.
\begin{figure}
\includegraphics[width=1.01\linewidth]{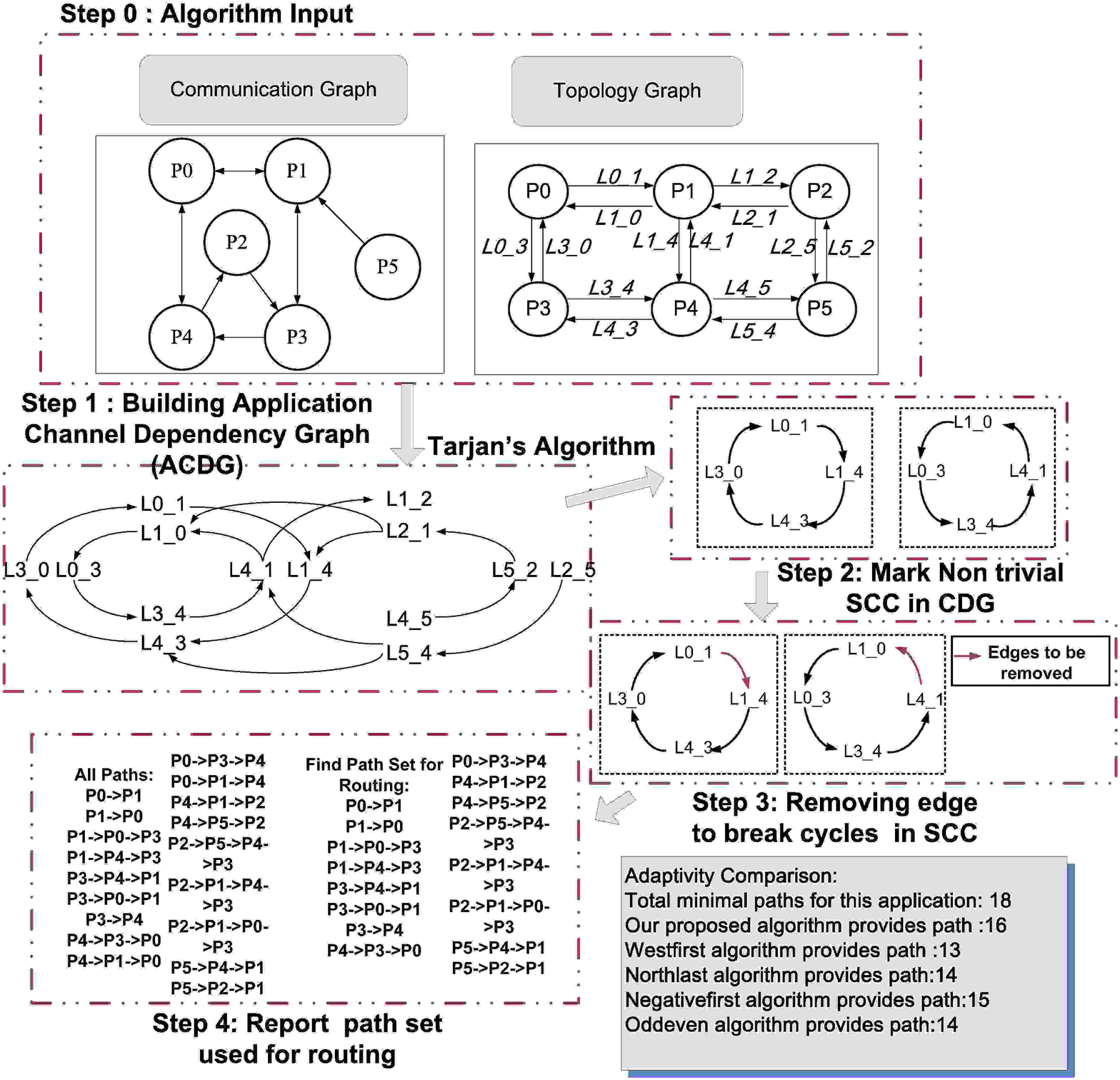}

\caption{\label{fig:Application-specific-and}Application specific and deadlock
free path finding algorithm}
\end{figure}

Figure \ref{fig:Application-specific-and} shows the main flow of
our path finding algorithm. Similar to \cite{application_specific_plaesi},
based on the application's task flow graph, we examine all the paths
between the source and destination pairs to build the application
channel dependency graph (ACDG). Most likely, there will be cycles
in the ACDG so that some edges have to be removed to break these cycles
to guarantee deadlock free. In \cite{application_specific_plaesi}, a branch and bound algorithm is used to select the set of edges
to remove all the cycles while maintaining all the connectivity and
maximizing the average adaptivity. The average adaptivity $\alpha$
is defined as:
\begin{equation}
max\,\alpha=max\,\frac{1}{|C|}\sum_{c\in C}\alpha_{c}\,;\;\alpha_{c}=\frac{|\Phi_{S_{edge}}(c)|}{|\Phi(c)|}
\end{equation}
where
\begin{equation}
\Phi_{S_{edge}}(c)=\Phi(c)\setminus\{p|p\in\Phi(c)\bigwedge p\in P_{\mu\nu}\:\forall R_{\mu\nu}\in S_{edge}\}
\end{equation}

\begin{table}
\caption{\label{tab:Notations-of-cycle}Notations of application specific path
set finding algorithm }
\begin{minipage}[c]{8cm}%
\global\long\def\arraystretch{1.5}
\tabcolsep 2pt \global\long\def\thefootnote{a}
\begin{tabular}{l@{~~~}l}
\hline 
\parbox[c]{11mm}{%
Notations%
}  & Description\tabularnewline
\hline 
$p$ & a path from tile $src(p)$ to tile $dst(p)$\tabularnewline
$c$ & a communication edge in core communication graph\tabularnewline
$W(c)$ & bandwidth of communication $c$\tabularnewline
$C$ & whole set of communication $c$ in application \tabularnewline
$\Phi(c)$ &  set of minimum paths for communication $c$\tabularnewline
$S_{edge}$ & set of edges to be removed for breaking cycles\tabularnewline
$R_{\mu\nu}$ & an edge joining two link vertex $\mu,\,\nu$ in CDG\tabularnewline
$P_{\mu\nu}$ & set of paths which introduce $R_{\mu\nu}$in CDG\tabularnewline
$\Phi_{S_{edge}}(c)$ & set of paths for $c$ after edges in $S_{edge}$ removed\tabularnewline
\hline 
\end{tabular}%
\end{minipage}
\end{table}
The notations used in the above equations are summarized in Table
\ref{tab:Notations-of-cycle}. Connectivity is guaranteed for every
communication $c$ by making sure that at least one path exists between
the source and destination nodes. So we have $|\Phi_{S_{edge}}(c)|\geqslant1;\:\forall c\in C$.
In this chapter, instead of finding all cycles in the ACDG and breaking
each cycle respectively, as done in \cite{application_specific_plaesi},
we apply Tarjan\textquoteright{}s algorithm \cite{tarjan_algorithm}
to find all the strongly connected components (SCC) and try to eliminate
cycles within each nontrivial components (components containing more
than one vertex). One important feature of SCC is that cycles of a
directed graph are contained in the same components. We then eliminate
cycles within each nontrivial components to achieve deadlock free.
Tarjan\textquoteright{}s algorithm is used because the complexity
is lower ($O(|V|+|E|)$). In many cases, several edges are shared
among different cycles (as illustrated in Figure \ref{fig:Proposed-design-methodology},the
two edges $(L_{3-7},L_{7-6})$ and $(L_{7-3},L_{3-2})$ are shared
among several cycles). If we inspect each cycle separately, we may
consider these edges more than once. On the other hand, when we use
Tarjan's algorithm, cycles with common edges are in the same component
and hence decision can be made more efficiently if we remove some
shared edges to break these cycles simultaneously. When we select
edges to break the cycle, instead of optimizing the average routing
adaptivity, we maximize the following objective function:
\begin{equation}
max\:\frac{1}{|C|}\sum_{c\in C}\alpha_{c}W_{c}=max\:\frac{1}{|C|}\sum_{c\in C}W(c)\times\frac{|\Phi_{S_{edge}}(c)|}{|\Phi(c)|}
\end{equation}

Here we weight the routing adaptivity of each communication with its
corresponding bandwidth requirement ($W(c)$ is the bandwidth of communication
$c$). The rationale is that for the communications that have large
bandwidth requirement, more packets need to be processed and routed,
the impact on the power consumption distribution is higher. Therefore
we should have higher flexibility for these communications to re-divert
the traffic and hence higher adaptivity.

\subsection{Optimal traffic allocation }

In the following, we use the notations summarized in Table \ref{tab:Notations-of-cycle-1}
and the energy model described in section 3.4.2 to obtain the linear
programming (LP) formulation of the optimal traffic allocation problem.

\subsubsection{Energy consumption model}

We assume the energy consumption of each processor $i$ ($E_{p-i}$)
is available after task mapping. Wormhole routing is used in our routing
scheme. In wormhole routing, each packet is divided into several flits
which are the minimum units for data transmission and flow control.
For every data packet, the head flit sets up the path directions for
the body and the tail flits. Thus, $E_{rc}$, $E_{sel}$ and $E_{vc}$
only incur when the head flit is processed by the router. Total energy
consumption for processing a single packet in router $i$ is given
by: 
\begin{equation}
\triangle E_{r-i}=(E_{buffer-rw}+E_{forward}+E_{sw})\times S_{packet}+E_{rc}+E_{sel}+E_{vc}
\end{equation}
Let $n_{i}$ denote the number of packets that received by router
$i$, then the total router energy consumption is equal to $E_{r-i}=\triangle E_{r-i}\times n_{i}$.
The total energy of each tile $i$ ($E_{i}$) is equal to $E_{p-i}+E_{r-i}$.

\begin{table}
\caption{\label{tab:Notations-of-cycle-1} Parameters and notations in LP formulation }

\begin{minipage}[c]{8cm}%
\global\long\def\arraystretch{1.5}
\tabcolsep 2pt \global\long\def\thefootnote{a}
\begin{tabular}{l@{~~~}l}
\hline 
\parbox[c]{11mm}{%
Notations%
}  & \textbf{Network on chip parameters}\tabularnewline
\hline 
$N$ & total number of tiles in NoC\tabularnewline
$C_{ij}$ & link capacity between tiles $i$ and $j$\tabularnewline
$L_{ij}$ & number of minimum paths between tiles $i$ and $j$ \tabularnewline
$l(i,j,k)$ &  the $k^{th}$ path joining tiles $i$ and $j$ ; $1\leqslant k\leqslant L_{ij}$\tabularnewline
$T_{i}$ & set of paths pass tiles $i$ $T_{i}=\{l(m,n,t)|i\in l(m,n,t)\}$\tabularnewline
$p(i,j)$ & packet injection rate between tiles $i$ and $j$\tabularnewline
$S_{packet}$ & number of flits in a packet\tabularnewline
$S_{bit}$ & number of bits in a packet\tabularnewline
\hline 
\hline 
\parbox[c]{11mm}{%
Notations%
}  & \textbf{Tile energy parameters} \tabularnewline
\hline 
$E_{p-i}$ & average processor energy of the $i^{th}$ tile\tabularnewline
$E_{forward}$ & energy of forwarding a flit in the router \tabularnewline
$E_{buffer-rw}$ &  buffer read and write energy for a flit in the router\tabularnewline
$E_{sw}$ &  switch allocation energy for a flit in the router\tabularnewline
$E_{rc}/E_{sel}$ &  routing computation/output port selection energy\tabularnewline
$E_{vc}$ &  virtual channel allocation energy\tabularnewline
\hline 
\end{tabular}%
\end{minipage}
\end{table}

\subsubsection{LP problem formulation}

Given the set of admission paths for every communication pair which
is deadlock free, we want to obtain the ratio of traffic allocated
to each path so as to minimize the maximum energy consumption among
all the tiles. We formulate the following linear programming problem
to solve the optimal path ratios:

\textbf{1) Variables $r(i,j,k)$:} ratio of traffic allocated to the
$k^{th}$ path $l(i,j,k)$ between tiles $i$ and $j$ among all the
$L_{ij}$ paths, where $1\leq i\leq N_{tile,\,}1\leq j\leq N_{tile},\,1\leq k\leq L_{ij}$.

\textbf{2) Objective functions :} The energy consumption of the $i^{th}$
tile, $E_{i}$, is given by 
\begin{equation}
E_{i}=E_{p-i}+\triangle E_{r-i}\times n_{i}
\end{equation}
where $n_{i}$ is the total number of packets received by router $i$
per unit time and is equal to the summation of the number of the packets
from all paths that pass through tile $i$ , \textit{i.e.,} $T_{i}$
. $n_{i}$ is given by:

{\small 
\begin{equation}
n_{i}=\sum_{\forall l(a,b,k)\in T_{i}}r(a,b,k)\times p(a,b)
\end{equation}
}In order to balance the tile energy $E_{i}$, our objective function
is written in a min-max form as follows:

\begin{equation}
obj:\; min(max(E_{i}))\; for\,1\leq i\leq N_{tile}
\end{equation}
It is equivalent to:

\begin{equation}
min(E)\; s.t.\; E\geq E_{i}\; for\,1\leq i\leq N_{tile}
\end{equation}

\textbf{3) Problem constraints : }The objective function is optimized
subject to the following constraints:

3-1) Traffic split constraints : summation of all the traffic allocation
ratios between a given pair $(i,j)$ should equal to 1.
\begin{equation}
\sum_{k=1}^{L_{i,j}}r(i,j,k)=1\;\forall(i,j)\in C
\end{equation}
\begin{equation}
r(i,j,k)\geq0\:\forall i,j\in[1,N_{tile}],\: k\leq L_{ij}
\end{equation}

3-2) Bandwidth constraints : the aggregate bandwidth used for a specific
link should not exceed the link capacity.

The communication bandwidth = packet injection rate $\times$packet
size $\times$clock frequency. Assume $T$ is the cycle time and $(i,j)$
is a physical link in the mesh NoC, a path $l(a,b,k)$ will traverse
this link if $l(a,b,k)\in T_{i}\cap T_{j}$. So we have{\small 
\begin{equation}
\sum_{\forall l(a,b,k)\in T_{i}\cap T_{j}}\frac{r(a,b,k)\times p(a,b)\times S_{bit}}{T}\leq C_{ij}
\end{equation}
}{\small \par}

Given the application's task flow graph and the task mapping, the
packet injection rate $p(a,b)$ can be calculated by summing the bandwidth
requirement from all the communication pairs where the source tasks
are mapped onto tile $a$ and destination tasks are mapped onto tile
$b$. 

The above formulation is a typical linear programming problem and
can be solved efficiently using MATLAB CVX optimization toolbox \cite{matlab_cvx}. 

\subsubsection{Using the path ratios in the routers}

After solving the LP problem, we obtain a set of admissible paths
and their corresponding traffic allocation ratios. To use these information
in the implementation of the NoC routing, we can use two schemes.
The simplest way is source routing. In the source processor, the header
flit of each packet contains the entire path information. The source
processor $i$ decides which path to use to send the packet to destination
$j$ according to the traffic allocation ratios $r(i,j,k)$ of the
set of admissible paths. Intel Teraflops chip \cite{NoCbook_Peh}
uses this source-based routing scheme as the router will be simpler.
However, this will create a large overhead on the effective bandwidth
as the packet needs to contain additional payload to record the entire
routing path.

A more efficient way of implementing the thermal aware routing is
to use routing tables stored in each router. One of the major advantages
of table-based routing is that it can be dynamically reconfigured
or reloaded \cite{application_specific_plaesi}
to allow modifications in the communication requirements. However
since the routing decision is made within each router without knowing
the entire path information, the path traffic-allocation ratios can
not be directly used. In the following, we will show how to convert
these ratios into local probability values for the router to select
which output port to send out the packet. 

To support the thermal-aware routing, the routing table is organized
as follows: for each router $t$ in the mesh topology and each of
its input direction $d\in D_{in}(t)$, there is a routing table $RT(t,d)$
\cite{application_specific_plaesi}. For each
output port $o$, there is a set of corresponding entries in the routing
table. Each entry consist of the tile id numbers of the source ($s$),
and destination ($b$) pair, and the probability values $p(o)$ of
using $o$ to route to $b$. Formally, $RT(t,d)=\{(s,b,o,p(o)|o\in D_{out}(t),0\leq p(o)\leq1\}$. 

When using routing tables within the routers, the final routing path
for a packet of a specific traffic flow is composed by the output
port selected in each router along the path distributively. One problem
may hereby arise: will the selected output ports in the routers form
a new path that is not included in the admissible path set and hence
introduce the possibility of deadlock? In the following, we prove
that path generated at run-time using the table-based routing is indeed
deadlock free provided the path set used for generating the routing
table $RT(t,d)$ contains no circles.
\begin{lemma}
\textbf{(deadlock-free property in distributive routing):} For the
table-based routing, after generating the distributed routing table
for each router according to the deadlock free path set $P$, the
actual route generated at run time will not contain a disallowed path
not included in $P$. Deadlock-free property is hence inherited from
the path set $P$.
\begin{figure}
\includegraphics[width=0.92\columnwidth]{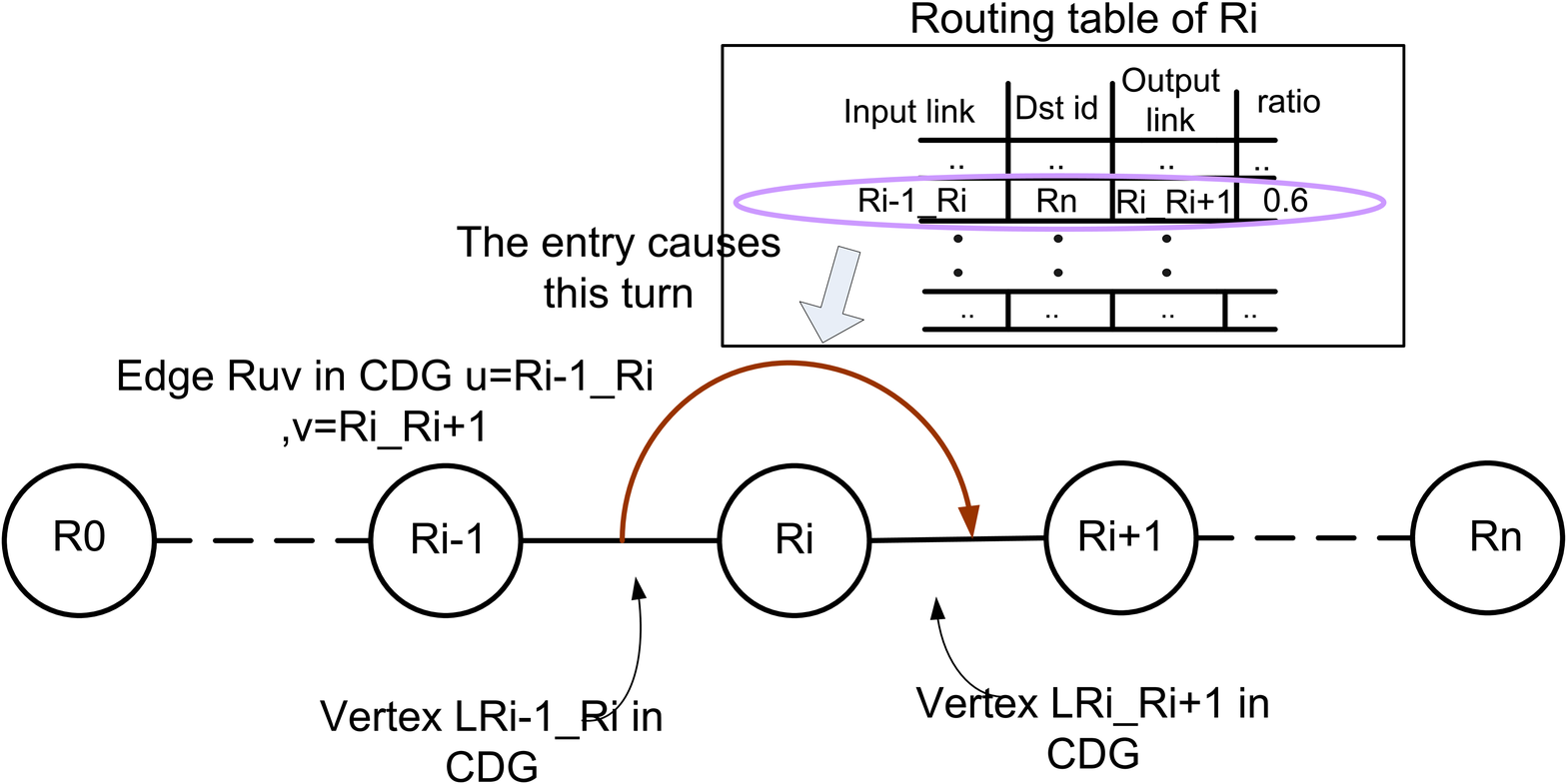}

\caption{\label{fig:proof-of-deadlock}Proof of deadlock free inheritance}
\end{figure}
\end{lemma}
\textbf{(by contradiction): }We prove that a disallowed path $R$
can not exist if all the entries of the routing table are created
according to $P$. Assume $R=<r_{0},\cdots,r_{i-1},r_{i},r_{i+1}\cdots r_{n}>$
is an actual path used to route a packet from node $r_{0}$ to $r_{n}$
at run time (Figure \ref{fig:proof-of-deadlock}). $R$ is obtained
by sequentially looking up the routing tables from routers $r_{0}$
to $r_{n-1}$. Assume this path $R\notin P$ is a disallowed path
which may cause circular dependencies with the other routing paths
in $P$, then at least one edge $R_{\mu\nu}(\mu=r_{i-1}r_{i},\;\nu=r_{i}r_{i+1})$
is an element of $S_{edge}$ in Table-\ref{tab:Notations-of-cycle}
in order to generate the circular dependencies. However, since $R$
is built from the routing tables in router $r_{i}$, which means there
is an entry in the routing table of $r_{i}$ using link $r_{i-1}r_{i}$
as input and link $r_{i}r_{i+1}$as output. It is already known the
entries of the routing tables are created according to the paths in
the admissible path set $P$. Therefore, there is another path $p\in P$
while $p\neq R$ traversing through $r_{i}$ from the input link $r_{i-1}r_{i}$
to the output link $r_{i}r_{i+1}$, hence $p$ also contains the edge
$R_{\mu\nu}(\mu=r_{i-1}r_{i},\;\nu=r_{i}r_{i+1})$ in its CDG. On
the other hand, since $R_{\mu\nu}\in S_{edge}$, it will create cycles
in the CDG and $P$ is actually not a deadlock-free path set. This
contradicts with the assumption. 

Now the issue is how to obtain the probability values $p(o)$ for
each output $o$ from the path traffic allocation ratios. For each
router $t$ and each of its input port $d$ , we obtain the subset
of admissible paths ( \textit{i.e.,}$T_{t,d}$) that pass through
$t$ using input port $d$. For a given destination tile $b$, using
minimum path routing, only two candidate output ports ($o_{1}$and
$o_{2}$) are feasible. The probabilities of selecting ports $o_{1}$
and $o_{2}$ are calculated by comparing the aggregate traffic of
the paths inside $T_{t,d}$ that use ports $o_{1}$ and $o_{2}$,
respectively, to route to the tile $b$. Using the notations in Table
\ref{tab:Notations-of-cycle}, let paths $l(s,b,k)\in T_{t,d}(o_{1},s,b)$
and $l(s,b,l)\in T_{t,d}(o_{2},s,b)$, we have: {\small 
\begin{equation}
p_{t,d}(o_{1},s,b)+p_{t,d}(o_{2},s,b)=1
\end{equation}
\begin{equation}
\frac{p_{t,d}(o_{1},s,b)}{p_{t,d}(o_{2},s,b)}=\frac{\sum_{\forall l(s,b,k)\in T_{t,d}(o_{1},s,b)}p(s,b)\times r(s,b,k)}{\sum_{\forall l(s,b,l)\in T_{t,d}(o_{2},s,b)}p(s,b)\times r(s,b,l)}\:
\end{equation}
}{\small \par}
After considering all the routers and the communication pairs, the
port probability values are obtained and stored into the routing tables
offline.

In order to maintain the global path traffic allocation ratios, each
entry in the routing table $RT(t,d)$ needs to distinguish the source
tile location $s$ of the packet. Thus the number of entries in the
routing table is increased. We can reduce the table size by grouping
the entries of different sources but the same destinations together.
The new format of the routing table becomes $RT(t,d)=\{(b,o,p(o)|o\in D_{out}(t),0\leq p(o)\leq1\}$.
In this case, the port probability value $p(o,b)$ in $RT(t,d)$)
is calculated as: {\small 
\begin{equation}
p_{t,d}(o_{1},b)+p_{t,d}(o_{2},b)=1
\end{equation}
\begin{equation}
\frac{p_{t,d}(o_{1},b)}{p_{t,d}(o_{2},b)}=\frac{\sum_{s}\sum_{\forall l(s,b,k)\in T_{t,d}(o_{1},s,b)}p(s,b)\times r(s,b,k)}{\sum_{s}\sum_{\forall l(s,b,l)\in T_{t,d}(o_{2},s,b)}p(s,b)\times r(s,b,l)}\:
\end{equation}
}{\small \par}

Although the exact path traffic allocation ratios cannot be maintained,
the traffic distribution on each router and each link can still be
maintained which is more important for uniform energy distribution
purpose. From the simulation results which will be presented in Section
6, this routing table implementation achieves similar improvement
in peak energy reduction and latency performance compared to the routing
table using the source-destination pair. 
\begin{figure}
\includegraphics[width=1\linewidth]{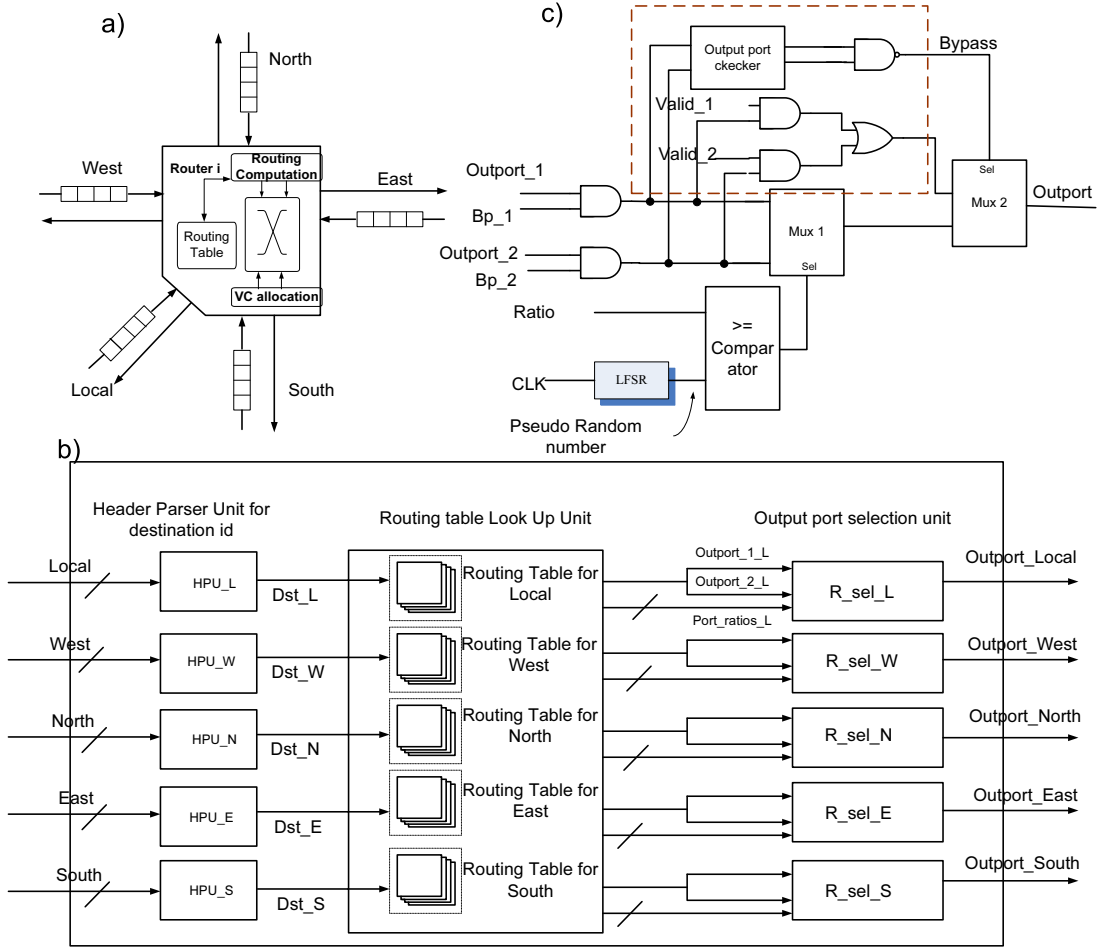}

\caption{\label{fig:block-diagram-of}Block diagram of the router supporting
ratio-based routing: a)Router microarchitecture, b)Routing computation
unit, c)Ratio-based output port selection }
\end{figure}

\section{Router microarchitecture}

The block diagram of our proposed router to support thermal aware
routing is illustrated in Figure \ref{fig:block-diagram-of}-a. For
minimum path routing, if the input direction and the destination are
fixed, there are at most two candidate output ports, $o_{1}$ and
$o_{2}$ , and $p(o_{1})+p(o_{2})=1$. The routing selection unit
in the router selects the output port $O_{dir}$ by comparing the
probability value $p(o_{1})$ with a random number $\tau$ in $[0,1]$.
It will select port $O_{1}$ $if\:\tau\leq p(o_{1})$, otherwise port
$O_{2}$ will be chosen.

A pseudo random number generator using linear feedback shift register
(LFSR) is employed to generate the random number. 
 The header flits from each input port are first decoded by a parser
(the HPU module shown in Figure \ref{fig:block-diagram-of}-b) to
extract the destinations. Then by checking $RT(t,d)$, two candidate
output ports are returned with the corresponding probability values.
The output port selection unit then make a decision on the output
port. In addition to the probability selection, backpressure information
( Bp\_1 and Bp\_2 in Figure \ref{fig:block-diagram-of}-c) from downstream
routers are also used. If one candidate output port is not available
due to limited buffer space, the backpressure signal will disable
this output port from the selection. 

\section{Experimental results}

\subsection{Simulation environment setup}

We implemented the proposed thermal-aware routing strategy for mesh-based
NoC architecture. A C++ program is developed to analyze the traffic
parameters automatically and generate the corresponding LP formulation
for a given application. The LP problem is then solved by CVX \cite{matlab_cvx}
which is a toolbox embedded in MATLAB for convex optimization. In
order to evaluate the NoC performance, a systemC based cycle-accurate
simulator was developed combining the features of several widely adopted
academic tools (including Noxim \cite{the:Noxim-simulator-User},
Nirgam \cite{nigram}
and Booksim \cite{booksim}).
We used both synthetic traffic and real benchmarks to evaluate the
performance and compare with other deadlock free routing algorithms
(westfirst, northlast, negativefirst, oddeven and X-Y routing \cite{NoCbook_Peh}).
The real benchmarks include PIP (Picture-In-picture)\cite{Noc_synthesis},
MWD (Multi-Window Display) \cite{Noc_synthesis},
MPEG4 \cite{xpipe_luca_benini}, VOPD (Video
Object Plane Decoder) \cite{Noc_synthesis},
MMS\_1(Multimedia system mapping to 16 cores) \cite{1411933},
MMS\_2(Multimedia system mapping to 25 cores) \cite{bandwidth_aware_routing}
and DVOPD (Dual Video Object Plane Decoder) \cite{noc_design_65nm}.
For synthetic traffic, we use 8 different traffic scenarios, namely
Uniform random, Transpose-1, Transpose-2 \cite{turn_model}, Hotspot in center
(Hotspot-C) \cite{isca_routing}, Hotspot
in top-right corner (hotspot-Tr) \cite{application_specific_plaesi},
Hotspot in right-side (Hotspot-Rs) \cite{application_specific_plaesi},
Butterfly \cite{the:Noxim-simulator-User} and Bursty traffic
\cite{the:Noxim-simulator-User}. $3\times3$, $4\times4$ and
$5\times5$ NoC meshes are used for different benchmarks and $4\times4$
meshes are used for all synthetic traffics.

\subsection{Adaptivity improvement}

\begin{figure}
\includegraphics[width=1\linewidth]{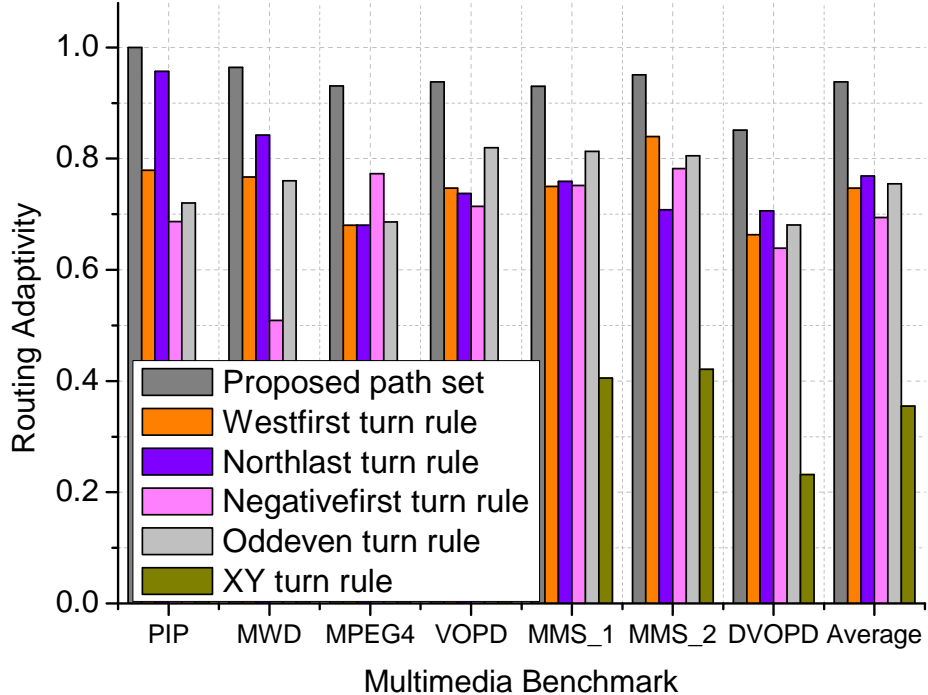}

\caption{\label{fig:adaptivity-comparisons-of}Adaptivity comparisons for different
benchmarks}
\end{figure}

First we compare the routing adaptivity with other routing algorithm
under several real benchmarks. The results are shown in Figure \ref{fig:adaptivity-comparisons-of}.
It can be seen that 20\%-30\% more paths are available if we take
the application traffic into consideration.

\subsection{Peak energy simulation results}

\begin{figure}
\includegraphics[width=1\linewidth]{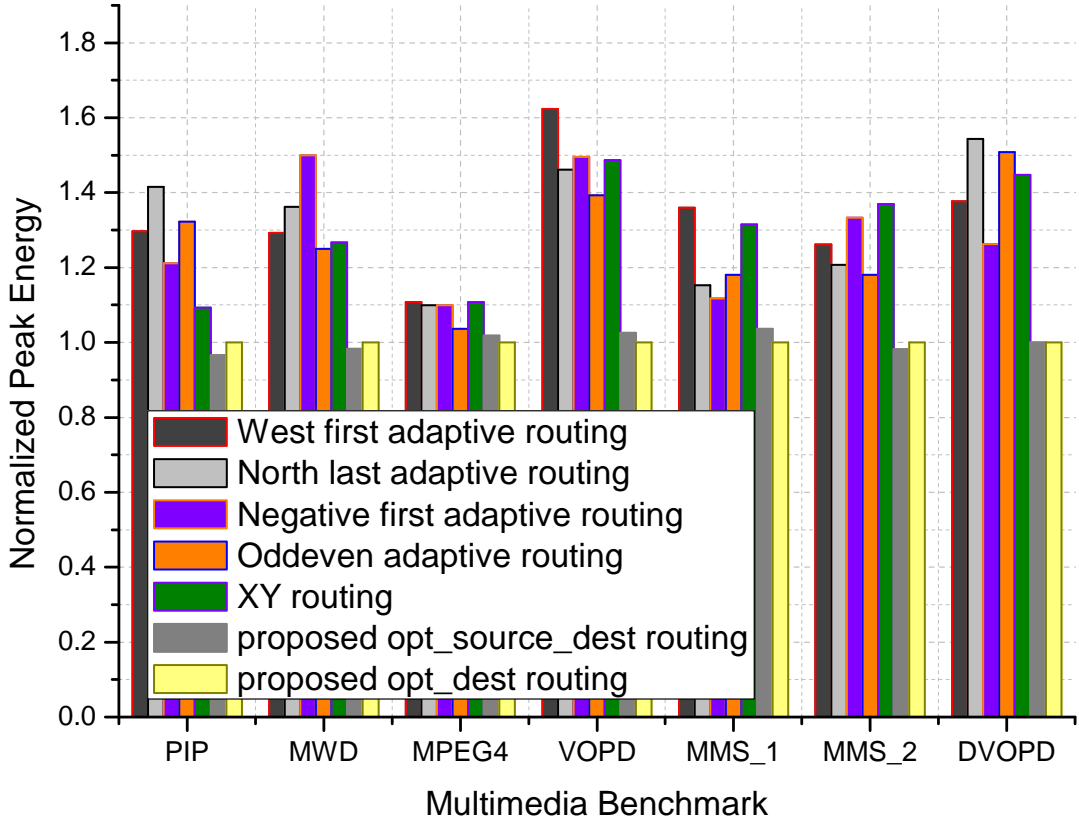}

\caption{\label{fig:Peak-energy-simulation}Peak energy simulation for real
benchmarks}
\end{figure}

\begin{figure}
\subfigure[Random traffic]{\includegraphics[width=1\linewidth]{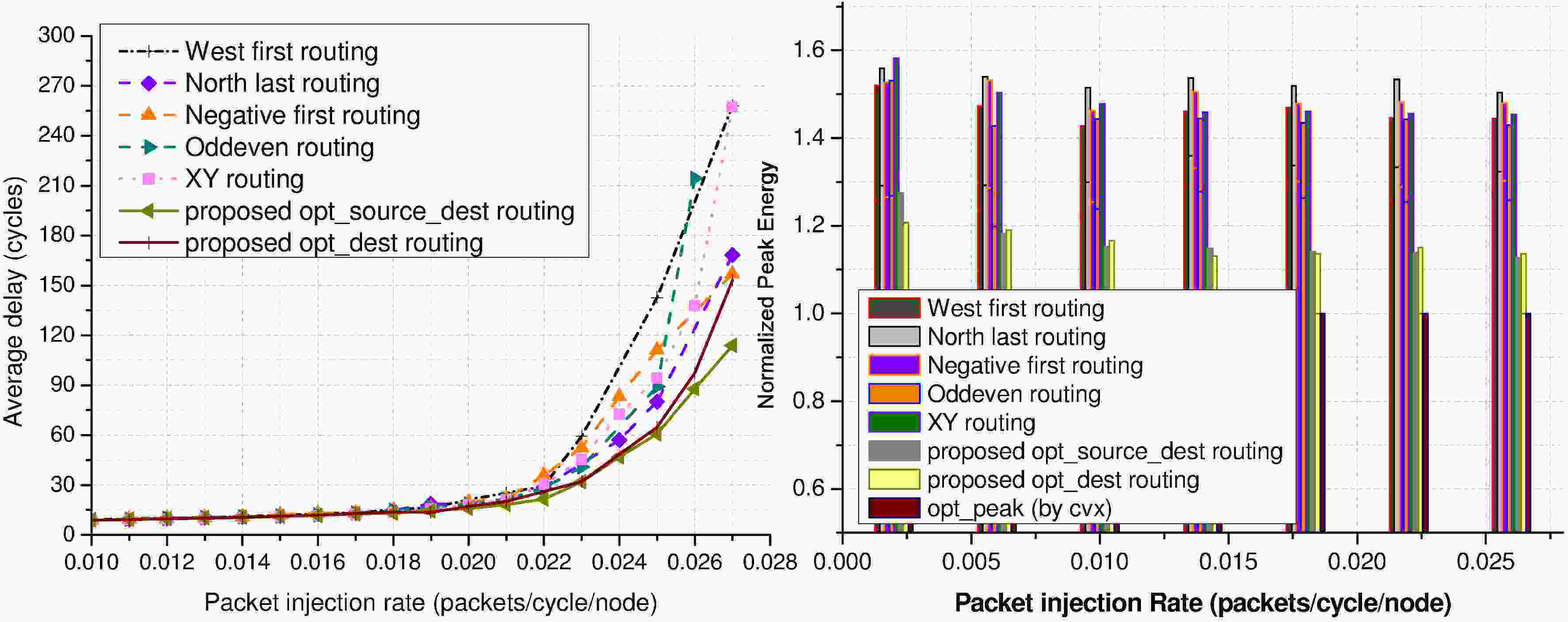}

}

\subfigure[Transpose-1 traffic]{\includegraphics[width=1\linewidth]{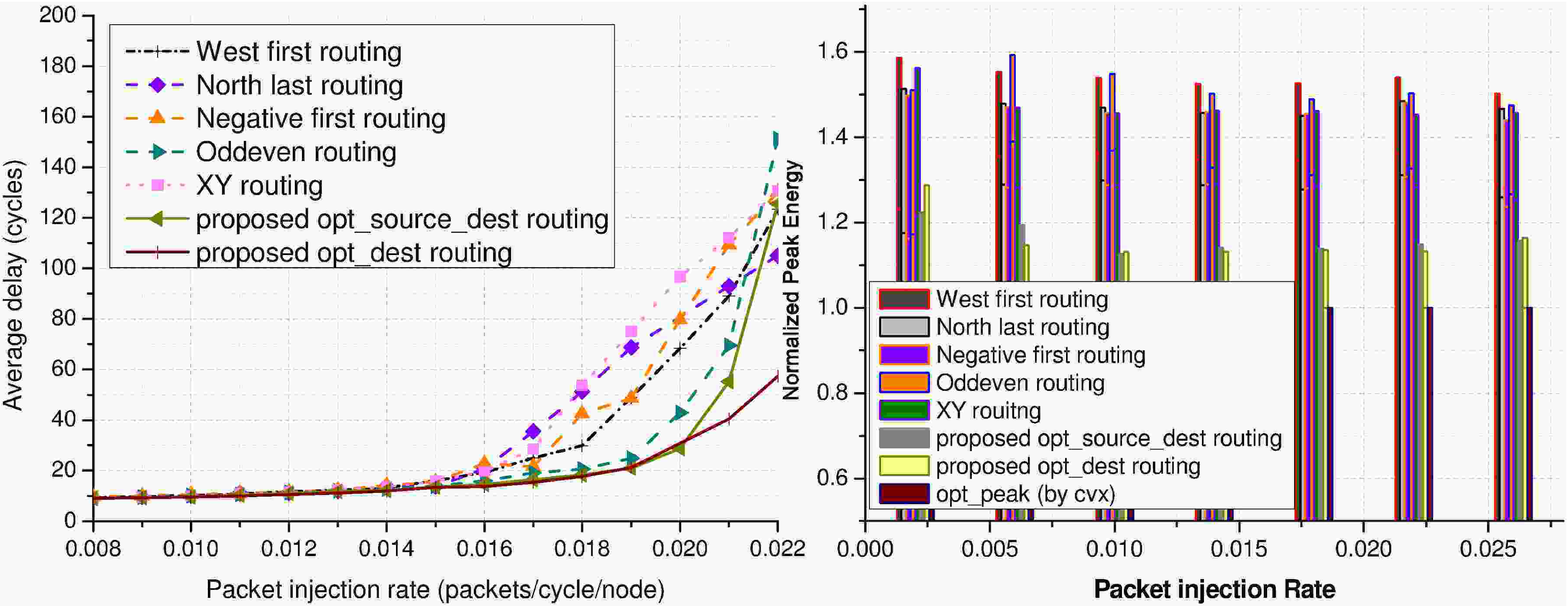}

}

\subfigure[Transpose-2 traffic]{\includegraphics[width=1\linewidth]{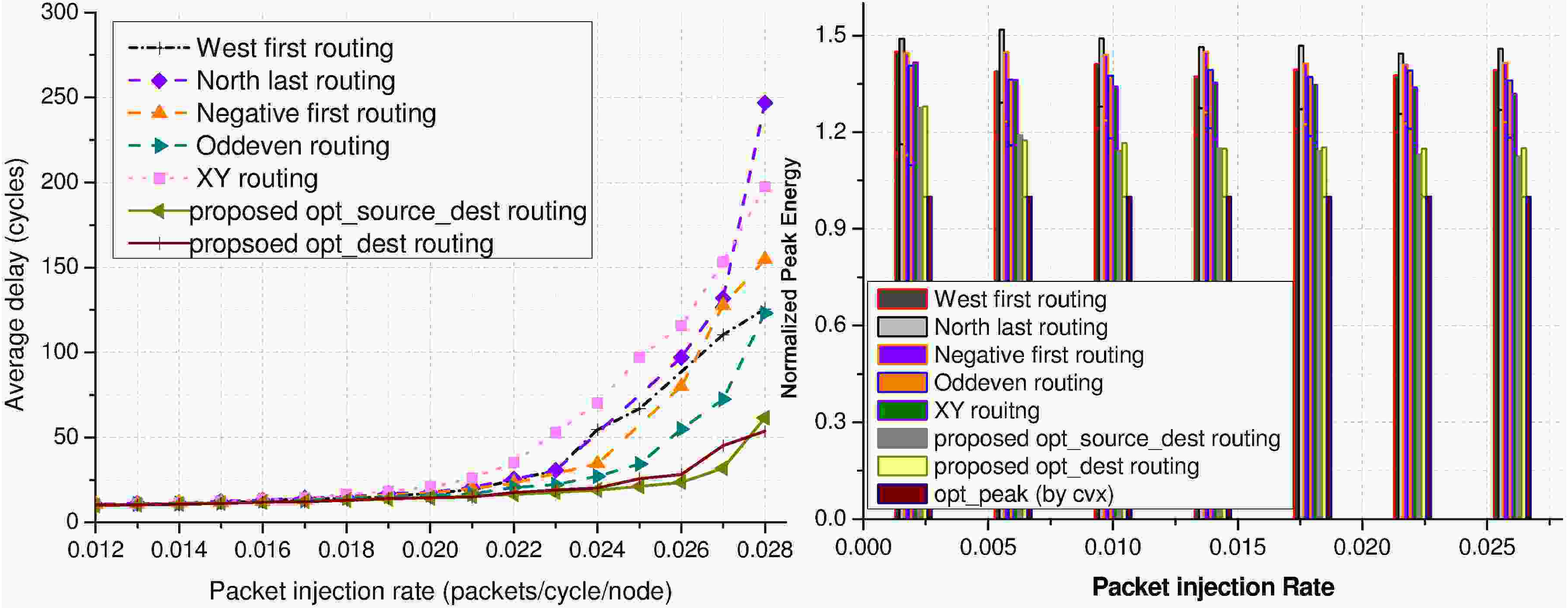}

}

\raggedleft{}\caption{\label{fig:Latency-and-peak}Latency and peak energy simulation for
a) Random, b) Transpose-1 and c) Transpose-2 traffic}
\end{figure}

\begin{figure}
\subfigure[Hotspot in right-side (Hotspot-Rs) traffic]{\includegraphics[width=1\linewidth]{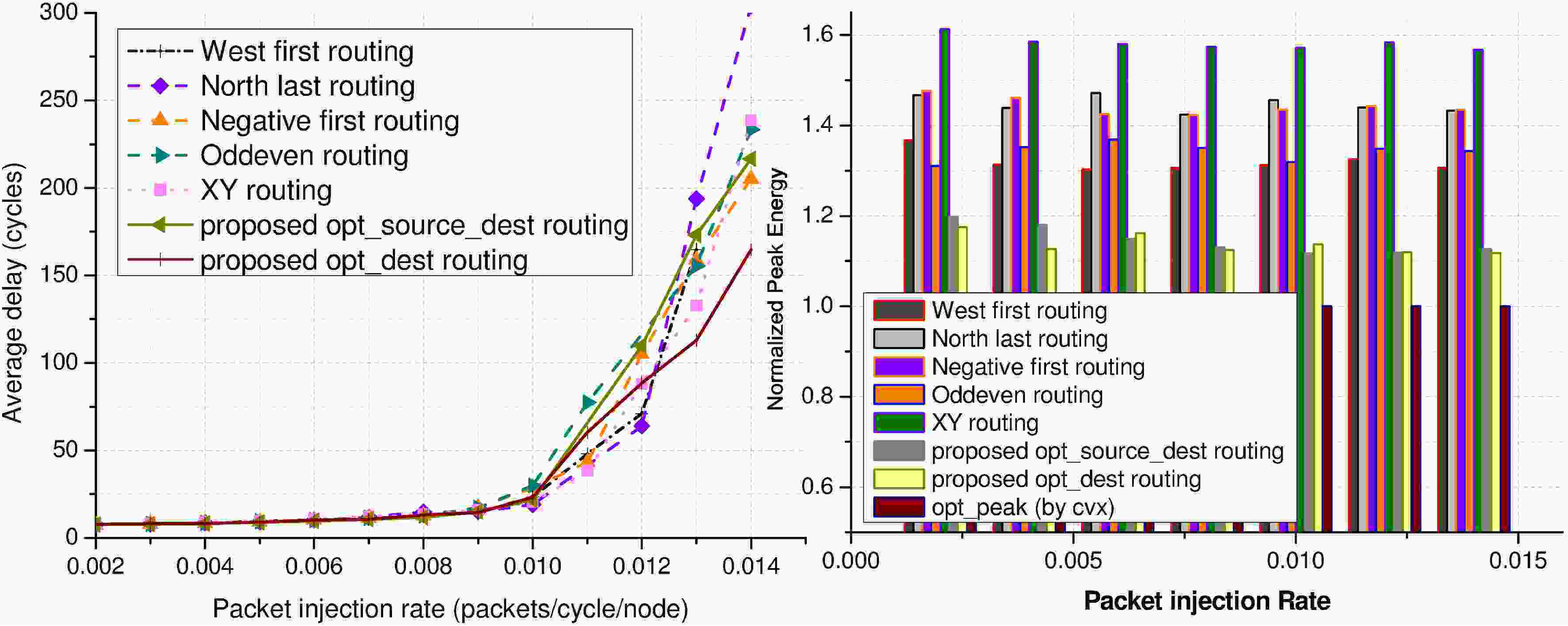}

}

\subfigure[Hotspot in top-right corner (Hotspot-Tr) traffic]{\includegraphics[width=1\linewidth]{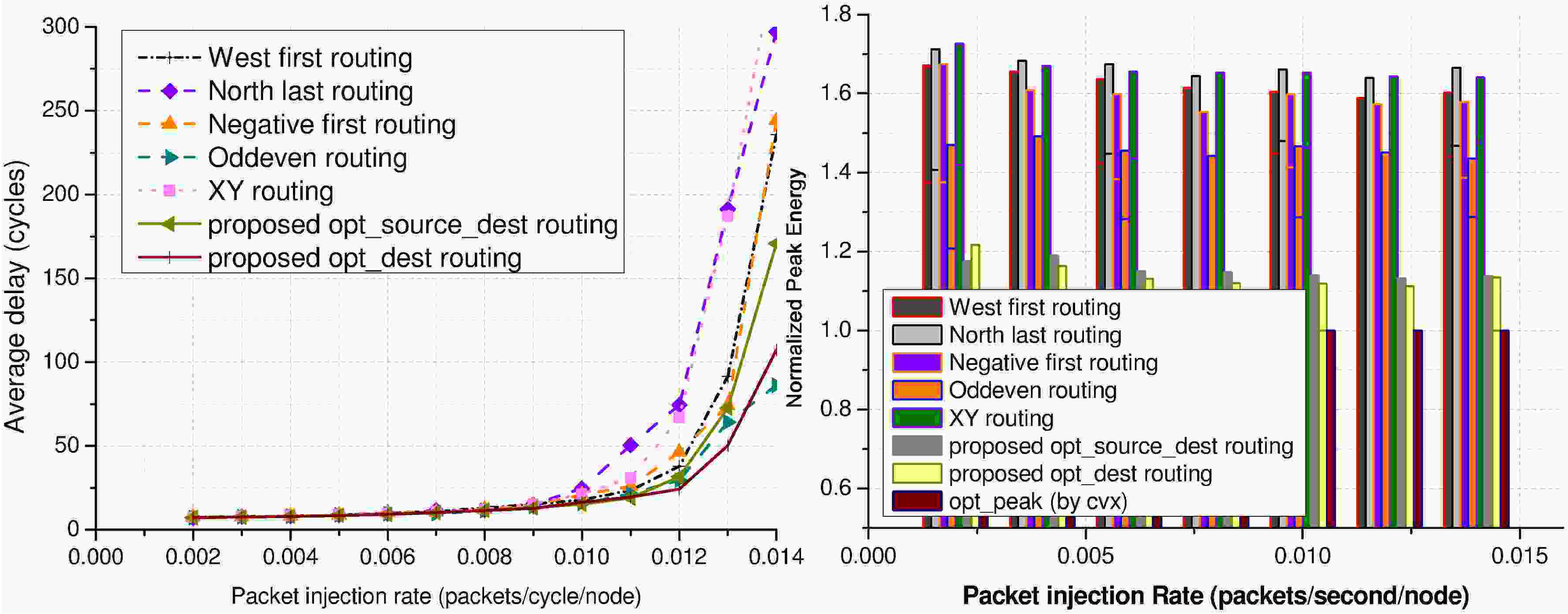}

}

\subfigure[Bursty traffic]{\includegraphics[width=1\linewidth]{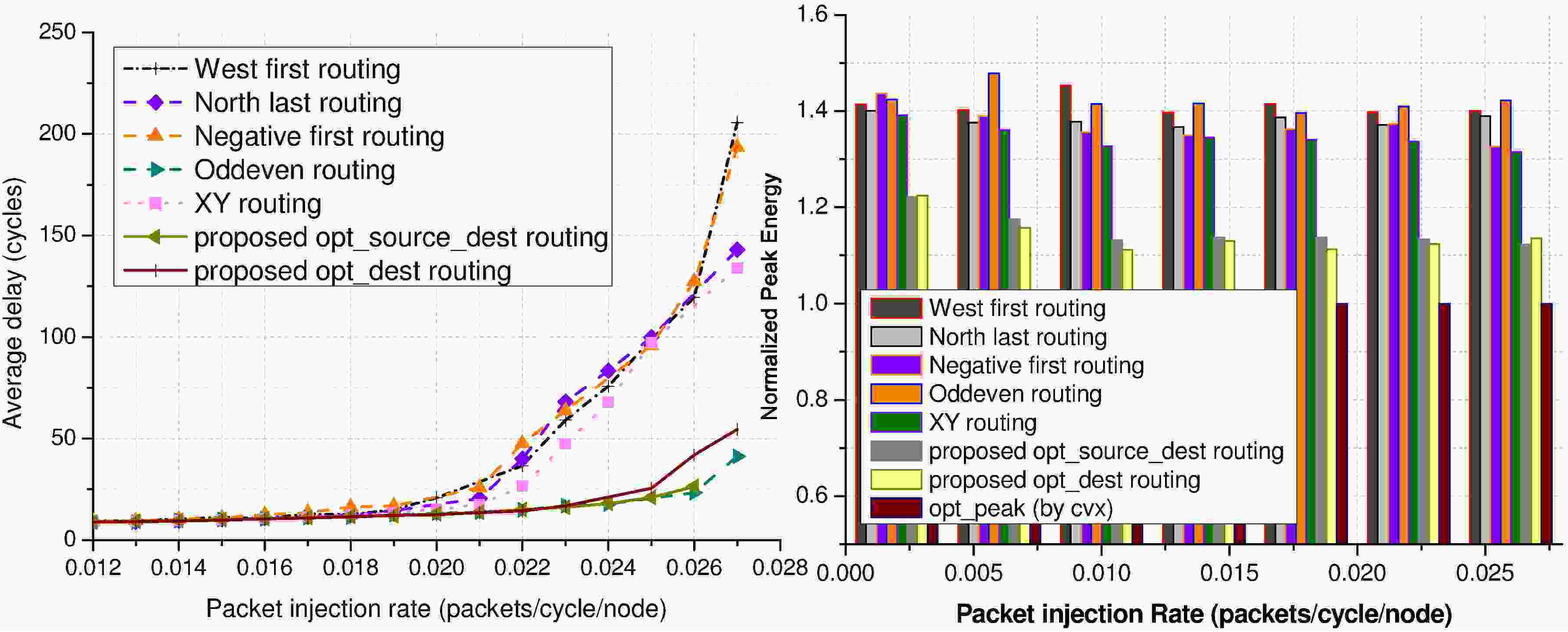}

}

\raggedleft{}\caption{\label{fig:Latency-and-peak-1}Latency and peak energy simulation
for a) Hotspot-Rs, b) Hotspot-Tr and c) Bursty traffic}
\end{figure}

\begin{figure}[ht]
\subfigure[Hotspot in center (Hotspot-C) traffic]{\includegraphics[width=1\linewidth]{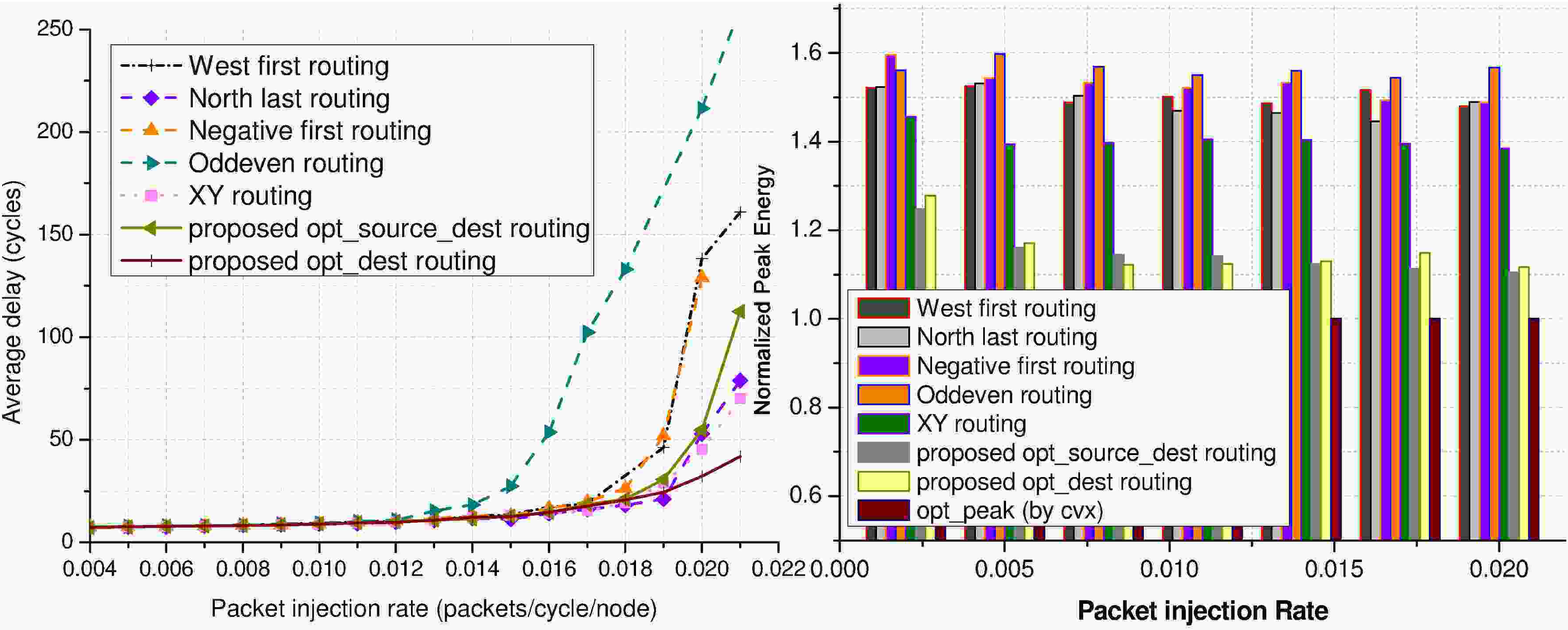}

}

\subfigure[Butterfly traffic]{\includegraphics[width=1\linewidth]{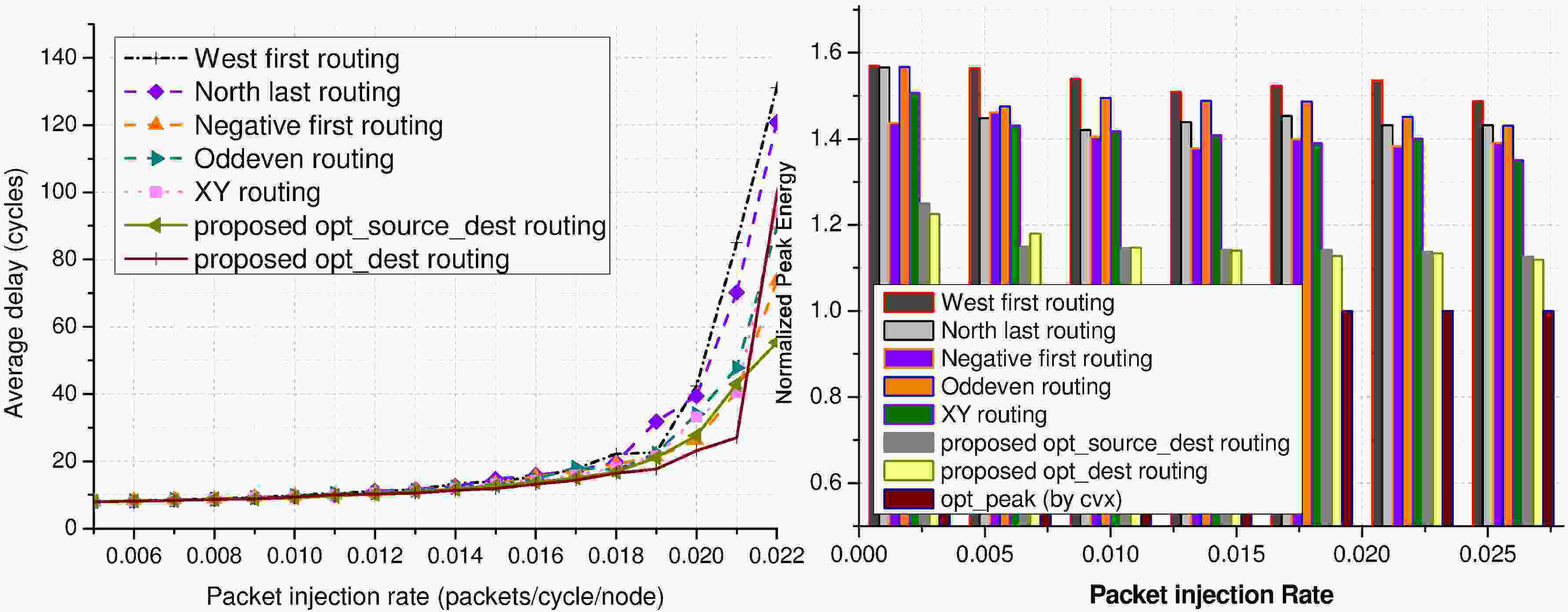}

}

\raggedleft{}\caption{\label{fig:Latency-and-peak-1-1}Latency and peak energy simulation
for a) Hotspot-C and b) Butterfly traffic}
\end{figure}

Next we evaluate the performance of the proposed algorithm in peak
energy reduction. In the experiment, energy parameters ($E_{buffer-rw},\: E_{rc}$
etc.) are adopted from Noxim and Booksim simulator. After $10,000$
warm-up cycles, we carry out energy simulation using a fixed time
window for all cases. Different simulations were done for each
packet injection rate to obtain the average peak energy consumption.
The simulation results for the real benchmarks and the synthetic benchmarks
are shown in Figure \ref{fig:Peak-energy-simulation} and Figure \ref{fig:Latency-and-peak}-\ref{fig:Latency-and-peak-1-1},
respectively.

In the figures, the proposed source-destination routing table and
destination-only routing table schemes are denoted as opt\_source\_dest
and opt\_dest, respectively. The theoretical optimal energy consumption
which is obtained by solving the LP formulation using CVX is also
included as a reference and is denoted as opt\_peak.

From Figures \ref{fig:Peak-energy-simulation} and \ref{fig:Latency-and-peak},
we can see that comparing with other adaptive routing algorithms,
the proposed routing algorithm can achieve more than 20\% peak energy
reduction in most traffics. At the same time, the latency performance
can also be improved or maintained for all the cases. 
\begin{figure}[h]
\includegraphics[width=1\linewidth]{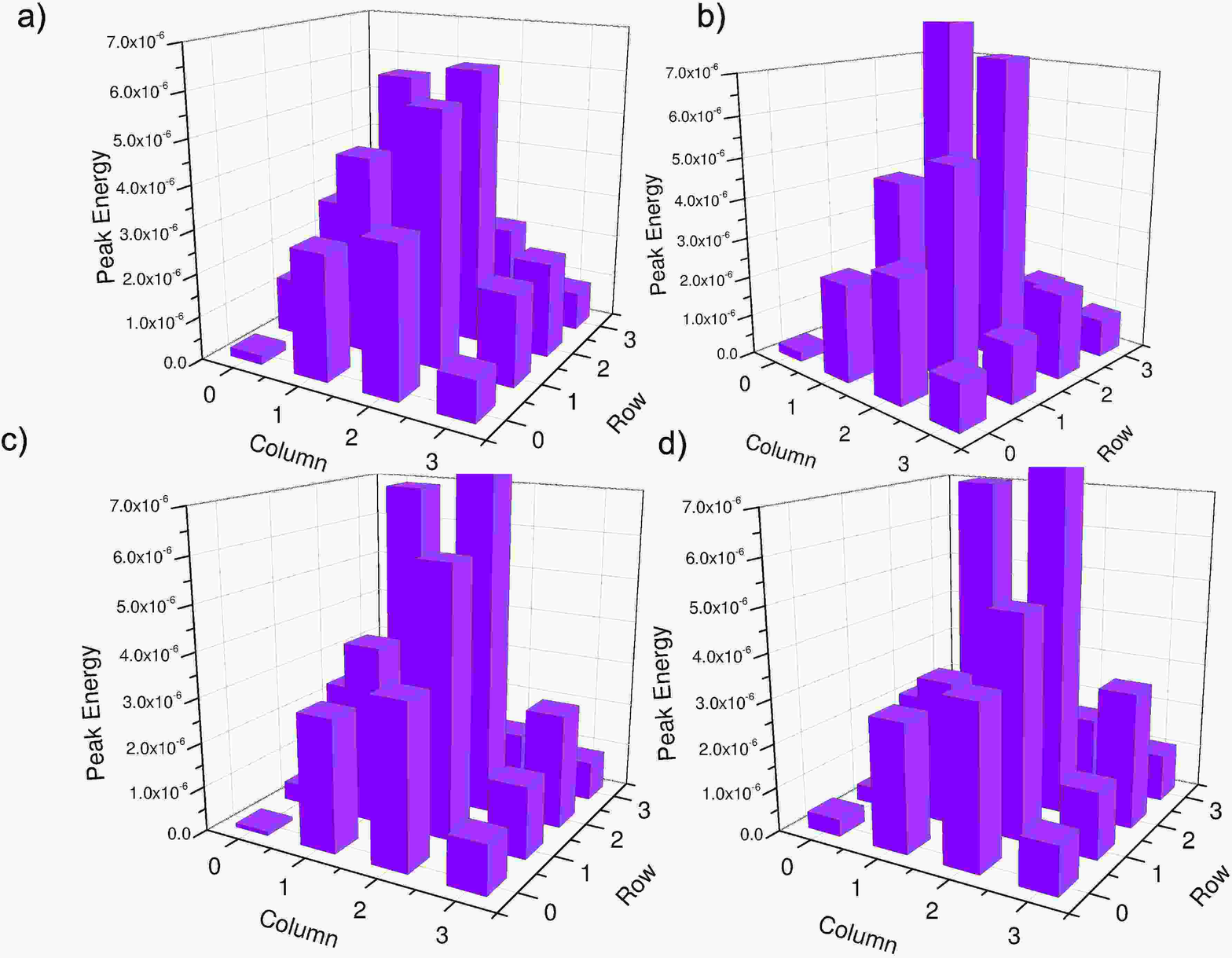}

\caption{\label{fig:An-example-of}An example of the NoC energy profile under
hotspot-4c traffic: a) proposed routing b) oddeven routing c) negativefirst
routing d) XY routing}
\end{figure}

In Figure \ref{fig:An-example-of}, we illustrate a scenario of the
tile energy distribution under Hotspot-C traffic (\textit{i.e.,} four center
cores are traffic hotspots). Figure \ref{fig:An-example-of}-a is
the energy distribution using the proposed thermal-aware routing algorithm,
and Figures \ref{fig:An-example-of}-b, -c and -d show the energy
distribution using odd-even, negative-first and XY routing, respectively.
For all the four cases, the total energy consumption of the NoC is
same, \textit{i.e.,} $4.55\times10^{-5}J$. However, the peak tile
energy of the proposed, odd-even, negative-first and XY routing are
$5.95\times10^{-6}J$, $8.09\times10^{-6}J$, $7.53\times10^{-6}J$
and $7.72\times10^{-6}J$ respectively. From the figures, we can see
that our proposed scheme indeed leads to a more uniform energy profile
across the NoC chip.

In Table \ref{tab:Improvement-in-peak-1-1}, we summarize the execution
time of our off-line routing algorithm (including path generation
and LP solving) under various mesh size and communication density
$\rho$ ($\rho$ is defined as the ratio of the total number of communications
pairs to the number of processors in mesh). It can be seen that the
execution time is reasonable. For larger mesh size ($7\times7$ or
more) and more communication pairs (100 or more), the number of minimum
paths increase dramatically. It takes longer time (2 -3 hours in average)
to obtain the traffic allocation ratios. Since the traffic allocation
is determined offline, in most cases, the time cost is still affordable.
In case we want to reduce the execution time, we can restrict the
number of minimum-length paths to a smaller subset.

\begin{table}
\caption{\label{tab:Improvement-in-peak-1-1}Execution time for different mesh
sizes}
\begin{tabular}{|p{0.20\linewidth}|p{0.08\columnwidth}|p{0.05\columnwidth}|p{0.08\columnwidth}|p{0.05\columnwidth}|p{0.08\columnwidth}|p{0.05\columnwidth}|}
\hline 
 & \multicolumn{6}{p{0.65\linewidth}|}{Running time (s)}\tabularnewline
\cline{2-7} 
Mesh Size & \multicolumn{2}{p{0.25\linewidth}|}{$3\times3$} & \multicolumn{2}{p{0.25\linewidth}|}{$4\times4$} & \multicolumn{2}{p{0.13\columnwidth}|}{$5\times5$}\tabularnewline
\hline 
Com density ($\rho$) & 2 & 4 & 2 & 4 & 2 & 4\tabularnewline
\hline 
\hline 
Time (s) & 5 & 6 & 18 & 20 & 227 & 243\tabularnewline
\hline 
\end{tabular}
\end{table}

\subsection{Simulation results with different processor/router energy ratio}
\begin{table} [t]
{\caption{\label{tab:Improvement-in-peak}Peak energy reduction under various
$r_{e}$(for Random, Transpose-1, Transpose-2 and Hotspot-C traffic)}
}{\par}

\begin{tabular}{|p{0.15\columnwidth}|p{0.07\columnwidth}|p{0.07\columnwidth}|p{0.07\columnwidth}|p{0.07\columnwidth}|p{0.07\columnwidth}|p{0.07\columnwidth}|p{0.07\columnwidth}|p{0.07\columnwidth}|}
\hline 
 & \multicolumn{8}{p{0.55\columnwidth}|}{{ Peak energy reduction}}\tabularnewline
\cline{2-9} 
{ Benchmark} & \multicolumn{2}{p{0.2\linewidth}|}{{ Random}} & \multicolumn{2}{p{0.2\columnwidth}|}{{ Transpose-1}} & \multicolumn{2}{c|}{{Transpose-2}} & \multicolumn{2}{c|}{{Hotspot-C}}\tabularnewline
\hline 
{ Average Energy Ratio ($r_{e}$)} & { vs. XY} & { vs. OE} & { vs. XY} & { vs. OE} & { vs. XY} & { vs. OE} & { vs. XY} & { vs. OE}\tabularnewline
\hline 
\hline 
{ 0.67} & { 17.4\%} & { 12.8\%} & { 15.6\%} & { 17.9\%} & { 9.5\%} & { 12.1\%} & { 15.7\%} & { 17.6\%}\tabularnewline
\hline 
{  1.00} & {  15.7\%} & {  15.3\%} & {  13.3\%} & {  14.2\%} & {  10.3\%} & {  10.3\%} & {  14.4\%} & {  16.2\%}\tabularnewline
\hline 
{  1.33} & {  13.3\%} & {  9.5\%} & {  12.5\%} & {  14.5\%} & {  7.2\%} & {  9.4\%} & {  13.2\%} & {  14.9\%}\tabularnewline
\hline 
{  1.67} & {  10.6\%} & {  8.6\%} & {  11.2\%} & {  13.8\%} & {  6.4\%} & {  9.4\%} & {  12.3\%} & {  14.0\%}\tabularnewline
\hline 
{  2.00} & {  11.9\%} & {  9.3\%} & {  11.1\%} & {  13.8\%} & {  6.5\%} & {  9.1\%} & {  11.6\%} & {  15.0\%}\tabularnewline
\hline 
{  2.33} & {  10.5\%} & {  8.2\%} & {  8.5\%} & {  10.7\%} & {  5.5\%} & {  7.0\%} & {  10.8\%} & {  13.9\%}\tabularnewline
\hline 
{  2.67} & {  9.4\%} & {  7.7\%} & {  8.9\%} & {  11.1\%} & {  4.3\%} & {  6.9\%} & {  10.2\%} & {  13.0\%}\tabularnewline
\hline 
{  3.00} & {  8.8\%} & {  7.0\%} & {  8.8\%} & {  10.6\%} & {  3.7\%} & {  5.6\%} & {  9.6\%} & {  10.9\%}\tabularnewline
\hline 
{  3.33} & {  7.9\%} & {  5.9\%} & {  7.1\%} & {  9.9\%} & {  4.2\%} & {  6.6\%} & {  9.1\%} & {  10.5\%}\tabularnewline
\hline 
{  3.67} & {  8.1\%} & {  6.5\%} & {  7.8\%} & {  9.2\%} & {  4.1\%} & {  5.8\%} & {  8.7\%} & {  9.9\%}\tabularnewline
\hline 
{  4.00} & {  7.9\%} & {  6.1\%} & {  7.0\%} & {  9.0\%} & {  4.6\%} & {  6.4\%} & {  8.3\%} & {  9.4\%}\tabularnewline
\hline 
{  Average} & {  11.04\%} & {  8.81\%} & {  10.16\%} & {  12.25\%} & {  6.03\%} & {  8.05\%} & {  11.26\%} & {  13.21\%}\tabularnewline
\hline 
\end{tabular}
\end{table}

\begin{table}
{\caption{\label{tab:Improvement-in-peak-2}Peak energy reduction under various
$r_{e}$(for Butterfly, Hotspot-Tr,Hotspot-Rs and Bursty traffic)}
}{\par}

\begin{tabular}{|p{0.15\columnwidth}|p{0.07\columnwidth}|p{0.07\columnwidth}|p{0.07\columnwidth}|p{0.07\columnwidth}|p{0.07\columnwidth}|p{0.07\columnwidth}|p{0.07\columnwidth}|p{0.07\columnwidth}|}
\hline 
 & \multicolumn{8}{p{0.55\columnwidth}|}{{ Peak energy reduction}}\tabularnewline
\cline{2-9} 
{ Benchmark} & \multicolumn{2}{p{0.2\linewidth}|}{{ Butterfly}} & \multicolumn{2}{p{0.2\columnwidth}|}{{ Hotspot-Tr}} & \multicolumn{2}{c|}{{ Hotspot-Rs}} & \multicolumn{2}{c|}{{ Bursty}}\tabularnewline
\hline 
{ Average Energy Ratio ($r_{e}$)} & { vs. XY} & { vs. OE} & { vs. XY} & { vs. OE} & { vs. XY} & { vs. OE} & { vs. XY} & { vs. OE}\tabularnewline
\hline 
\hline 
{ 0.67} & {  14.7\%} & {  20.2\%} & {  16.6\%} & {  9.9\%} & {  21.4\%} & {  11.2\%} & {  12.6\%} & {  16.1\%}\tabularnewline
\hline 
{  1.00} & {  13.1\%} & {  19.1\%} & {  15.4\%} & {  9.2\%} & {  19.9\%} & {  10.5\%} & {  10.4\%} & {  13.0\%}\tabularnewline
\hline 
{  1.33} & {  10.8\%} & {  15.2\%} & {  13.6\%} & {  7.9\%} & {  18.5\%} & {  9.7\%} & {  8.1\%} & {  11.3\%}\tabularnewline
\hline 
{  1.67} & {  9.8\%} & {  14.4\%} & {  12.5\%} & {  7.3\%} & {  17.2\%} & {  8.9\%} & {  7.5\%} & {  9.9\%}\tabularnewline
\hline 
{  2.00} & {  9.9\%} & {  13.7\%} & {  11.6\%} & {  6.7\%} & {  16.3\%} & {  9.4\%} & {  5.6\%} & {  9.1\%}\tabularnewline
\hline 
{  2.33} & {  8.9\%} & {  13.0\%} & {  11.0\%} & {  6.5\%} & {  15.5\%} & {  8.2\%} & {  7.1\%} & {  8.7\%}\tabularnewline
\hline 
{  2.67} & {  8.8\%} & {  12.8\%} & {  10.1\%} & {  5.8\%} & {  14.6\%} & {  7.8\%} & {  5.3\%} & {  7.9\%}\tabularnewline
\hline 
{  3.00} & {  8.7\%} & {  11.2\%} & {  9.5\%} & {  5.4\%} & {  13.8\%} & {  7.2\%} & {  5.5\%} & {  7.6\%}\tabularnewline
\hline 
{  3.33} & {  7.4\%} & {  9.8\%} & {  8.9\%} & {  5.1\%} & {  13.1\%} & {  6.9\%} & {  4.9\%} & {  7.2\%}\tabularnewline
\hline 
{  3.67} & {  5.8\%} & {  9.4\%} & {  8.4\%} & {  4.8\%} & {  12.5\%} & {  6.6\%} & {  4.3\%} & {  6.0\%}\tabularnewline
\hline 
{  4.00} & {  6.0\%} & {  8.8\%} & {  8.0\%} & {  4.6\%} & {  12.0\%} & {  6.3\%} & {  4.5\%} & {  6.4\%}\tabularnewline
\hline 
{  Average} & {  9.45\%} & {  13.42\%} & {  11.42\%} & {  6.65\%} & {  15.89\%} & {  8.43\%} & {  6.89\%} & {  9.38\%}\tabularnewline
\hline 
\end{tabular}
\end{table}

\begin{table} [t]
\caption{\label{tab:Improvement-in-peak-1}Peak energy reduction under various
$r_{e}$ (for real benchmark traffic)}
{\small
}{\small \par}
\begin{tabular}{|p{0.25\columnwidth}|p{0.12\columnwidth}|p{0.08\columnwidth}|p{0.12\columnwidth}|p{0.08\columnwidth}|p{0.12\columnwidth}|p{0.08\columnwidth}|}
\hline 
 & \multicolumn{6}{p{0.55\columnwidth}|}{{ Peak energy reduction}}\tabularnewline
\cline{2-7} 
{Benchmark} & \multicolumn{2}{p{0.25\linewidth}|}{{ MMS-1}} & \multicolumn{2}{p{0.16\columnwidth}|}{{MPEG4}} & \multicolumn{2}{p{0.25\linewidth}|}{{VOPD}}\tabularnewline
\hline 
{Average Energy Ratio ($r_{e}$)} & {vs. XY} & {vs. OE} & {vs. XY} & {vs. OE} & {vs. XY} & {vs. OE}\tabularnewline
\hline 
\hline 
{  0.67} & {  17.7\%} & {  11.8\%} & {  7.8\%} & {  10.2\%} & {  16.5\%} & {  28.9\%}\tabularnewline
\hline 
{  1.00} & {  15.7\%} & {  10.4\%} & {  6.9\%} & {  9.0\%} & {  12.3\%} & {  23.6\%}\tabularnewline
\hline 
{  1.33} & {  14.5\%} & {  9.8\%} & {  6.1\%} & {  8.0\%} & {  10.3\%} & {  20.5\%}\tabularnewline
\hline 
{  1.67} & {  12.5\%} & {  8.1\%} & {  5.5\%} & {  7.3\%} & {  9.9\%} & {  19.0\%}\tabularnewline
\hline 
{  2.00} & {  10.6\%} & {  6.5\%} & {  5.4\%} & {  7.1\%} & {  9.3\%} & {  17.7\%}\tabularnewline
\hline 
{  2.33} & {  9.3\%} & {  5.6\%} & {  5.0\%} & {  6.5\%} & {  8.3\%} & {  16.1\%}\tabularnewline
\hline 
{  2.67} & {  10.4\%} & {  7.0\%} & {  4.6\%} & {  6.0\%} & {  7.6\%} & {  14.8\%}\tabularnewline
\hline 
{  3.00} & {  9.7\%} & {  6.5\%} & {  4.3\%} & {  5.6\%} & {  7.6\%} & {  14.2\%}\tabularnewline
\hline 
{  3.33} & {  8.5\%} & {  5.6\%} & {  4.0\%} & {  5.5\%} & {  7.7\%} & {  13.8\%}\tabularnewline
\hline 
{  3.67} & {  7.9\%} & {  5.2\%} & {  4.2\%} & {  5.4\%} & {  7.1\%} & {  13.0\%}\tabularnewline
\hline 
{  4.00} & {  7.3\%} & {  4.7\%} & {  3.9\%} & {  5.0\%} & {  6.5\%} & {  12.0\%}\tabularnewline
\hline 
{  Average} & {  11.28\%} & {  7.38\%} & {  5.25\%} & {  6.87\%} & {  9.37\%} & {  17.6\%}\tabularnewline
\hline 
\end{tabular}
\end{table}
The energy consumption of a tile consists of that of the processor
and the router. For different applications, the power contribution
of the router and the processor varies greatly. Our thermal-aware
routing algorithm can only re-distribute the router power. We want
to evaluate the effectiveness of our algorithm on the peak energy
reduction when the ratio of the energy contribution from the router
and the processor varies. Let $r_{e}=\frac{Average\, processor\, energy}{Average\, router\, energy}$
. The experimental results shown in the previous sections assume $r_{e}=1$.
In this sub-section, we simulate the peak energy reduction using our
routing scheme for different $r_{e}$ values. Tables \ref{tab:Improvement-in-peak},
\ref{tab:Improvement-in-peak-2} and \ref{tab:Improvement-in-peak-1}
summarize the results for synthetic traffic and real benchmarks, respectively.
From the results we can see that when $r_{e}$ increases, the peak
energy reduction is smaller as the relative contribution of the router
energy reduces. Overall, we can achieve an average 6\%-17\% peak energy
reduction over other two existing routing schemes when $r_{e}$ ranges
from $0.67$ to $4$ across all the benchmarks.\vspace{5mm}
\section{Conclusion}

NoC has been widely adopted to handle the complicate communications
for future MPSoCs. As temperature becomes one key constraint in NoC,
in this chapter, we propose an application-specific and thermal-aware
routing algorithm to distribute the traffic more uniformly across
the chip. A deadlock free path set finding algorithm is first utilized
to maximize the routing adaptivity. A linear programming (LP) problem
is formulated to allocate traffic properly among the paths. A table-based
router is also designed to select output ports according to the traffic
allocation ratios. From the simulation results, the peak energy reduction
can be as high as $20\%$ for both synthetic traffic and real benchmarks.\\


\chapter{Fault-tolerant Routing Scheme for NoCs
using Dynamic Reconfiguration of Partial-Faulty Routing Resources}

\textit{In this chapter, we propose a fault-tolerant framework for Network-on-Chips
(NoC) to achieve maximum performance under fault. A fine-grained fault
model is first introduced. Different from the traditional link or
node NoC fault models which assume the faulty resource is totally
unfunctional, we distinguish the faulty components and handle them
according to their fault classes. By doing so, we can avoid unnecessary
partitioning of the network and hence achieve a higher connectivity
under high fault rate. In particular, two new dynamic re-configuration
schemes at the router level, namely Dynamic Buffer Swapping (DBS)
and Dynamic MUX Swapping (DMS), are proposed to deal with the buffer
and crossbar faults, which are the main sources of failure in the
router. In these schemes, the healthy resources in the router are
maximally utilized to mitigate the faults. A deadlock recovery scheme
is designed to handle the potential deadlock hazard due to the proposed
DBS operation. Experimental results show that we can achieve higher
packet acceptance rate as well as lower latency compared
with state-of-the-art fault-tolerant routing schemes}

\section{Introduction}

For NoC-based Multiprocessor-System-on-Chip (MPSOC), the complex system
is highly susceptible to the prominence of faults \cite{resilient_routing}. A
significant error tolerance to both permanent and temporary faults
is thus strongly desired. Temporary errors are mainly due to the on-chip
crosstalks and coupling noise while permanent faults are mostly caused
by the worn-out devices, manufacturing defects and accelerated aging
effects \cite{5467330}. The NoC architecture offers a good opportunity of achieving
high reliability since it provides multiple paths between the source-destination
communication pairs and allows reconfiguration under faulty conditions.%

\begin{figure}
\includegraphics[width=0.95\columnwidth]{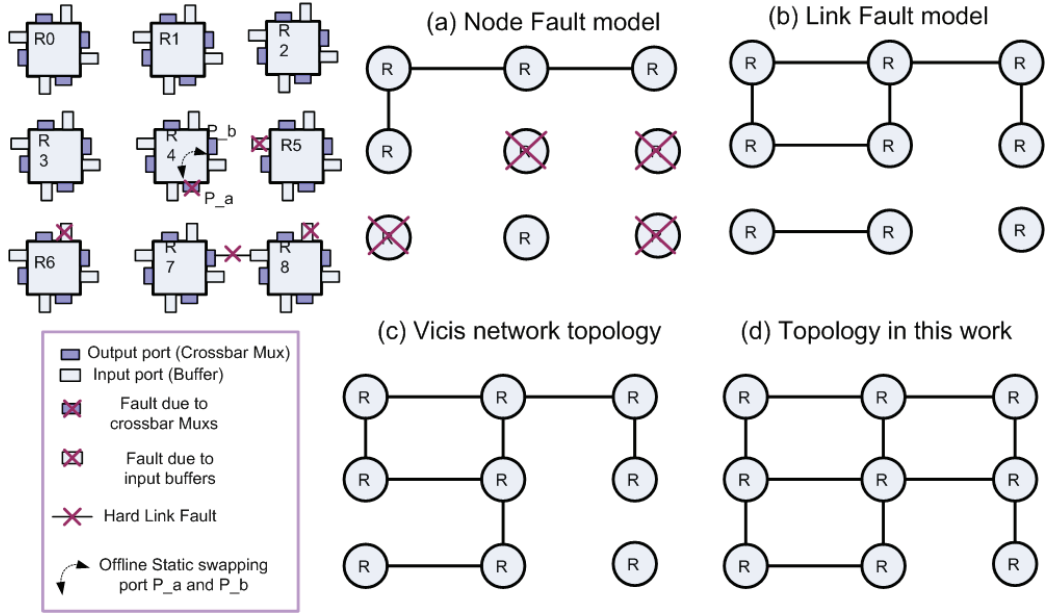}
\caption{\label{fig:A-Comparison-of}A Comparison of various fault model }
\end{figure}

To achieve fault tolerance in NoC, a proper fault model is required
for the clear identification and isolation of the defective components
so that the routing algorithm can maximally utilize the remaining
functionality to achieve the optimal performance. Traditional NoC
fault model treats the router fault as a node fault \cite{4555858} or
a link fault \cite{resilient_routing,5158098,vicis} and assumes total outage of
the faulty resource. However, a fine-grained functional fault model
will be more beneficial because graceful degradation of network operation
by exploiting the partially functional resources can be achieved \cite{5467330}.
In this chapter, we identify the major fault components in the NoC and
propose a fine-grained permanent fault model, which distinguishes
the partial router faults (buffer or crossbar faults) from the hard
link faults. Hard link faults may lead to a partitioned network. On
the other hand, a faulty router with an appropriate reconfiguration
strategy, may be able to tolerate the buffer or crossbar errors and
still supports the full connectivity in the NoC with reduced capacity.
This motivates a need to develop an integrated fault handling methodology
for different fault classes. An example comparing different fault
models given the same fault pattern is shown in Fig. \ref{fig:A-Comparison-of}.
As shown in the figure, if we adopt the generic node or link fault
model (Fig. \ref{fig:A-Comparison-of}-a and -b) which do not consider
whether the fault is caused by the buffer or the crossbar, we may
isolate many nodes from the network. To increase the connectivity
under faulty situation, \cite{vicis} adopts a static offline port
swap algorithm to reduce the number of faulty links (replacing port
P\_a with port P\_b in this example). However, the resultant network
is still partitioned, because the port fault is treated as a hard
link fault under this algorithm. 

In this chapter, we propose a dynamic swapping scheme which shares the
healthy resource in the router for different ports at run-time. As
shown in Fig. \ref{fig:A-Comparison-of}-d, by leveraging the other
healthy buffers or multiplexers (Muxes) and dynamically re-allocating the resources,
the network connectivity can remain intact which result in better performance in terms of packet acceptance rate. Two new techniques (DBS and DMS) are proposed to
handle the buffer and crossbar faults, respectively. An approach that
combines the turn model based routing and an efficient deadlock recovery
scheme is also proposed to resolve the potential deadlock issue when
the routers are working in the DBS mode.

\section{Background}

The growing concern on reliable NoC has prompted extensive researches
in this area, from self-detect and self-diagnosis routers \cite{5158098} to
fault-tolerant routing algorithm design\cite{resilient_routing}. In \cite{5158098},
a router with built-in-self-test (BIST) and self-diagnose circuits
is proposed to detect and locate the faults in the FIFOs and the crossbar
MUXes. After identifying the faults, the FIFO faults and the MUX faults
are treated as input and output link faults, respectively. This fault
categorization method reduces the connectivity of the network and
hence degrades the performance if there are a large number of faults.
In \cite{5467330}, a detailed online fault diagnosis framework is proposed
to diagnose the link, crossbar and arbitration/allocation logic faults.
The remaining healthy paths are identified from the faulty paths and
the routing algorithm utilizes these healthy paths to achieve graceful
degradation. However, the faulty paths can still lead to network partitioning
as they are determined statically. Moreover, this work is designed
for buffer-less router and may not be suitable for buffered router
architecture, which is required for high throughput and low latency
design.

There have been many fault-tolerant routing algorithms proposed to
deal with faulty components in NoC. In \cite{4555858}, the faulty router
is treated as a node fault and the routing algorithm can be dynamically
reconfigured to re-route the packets following the contour around
the faulty nodes. This region-based routing algorithm is only suitable
for one-faulty-router topology. Also, the node fault model may cause
the faulty router totally isolated from the network, while in real
situation, the faults may only affect a certain part of the router
and the rest of it can still function correctly. In \cite{resilient_routing},
the NoC faults are modeled as link level hard failures and a highly
resilient routing algorithm is designed to reconfigure the routing
table in an offline process. The algorithm is static in nature and
there exists a trade-off between the network connectivity and the
deadlock avoidance by selectively removing the turn rules. In \cite{4542002},
a deflection routing algorithm is proposed to dynamically re-route
the packets towards the destinations based on the stress factors which
reflect the traffic condition. The NoC faults are treated as the link
faults and it does not clearly describe how to avoid deadlock and
livelock based on the stress factors.

\section{Fault-tolerant NoC design }

In the context of NoC, the most common source of failures includes
link errors and single-event upsets within the router\cite{1633499}.
Since the input buffers and the crossbar occupy the majority of the
router area ($67.68\%$ for the FIFO buffer and $11.14\%$ for the crossbar
MUX in \cite{5158098}) , most of the faults in the router may occur
there\cite{vicis}. In this chapter, first we propose a hybrid fine-grained
fault model to distinguish the nature and consequence of different
faults and briefly describe the fault-diagnosis technique to distinguish
the types of fault \cite{my_vlsi_soc}. Then, we introduce the DBS and DMS techniques
which dynamically reconfigure the routers to handle buffer faults
and crossbar faults while still maintaining the maximum service of
the router. After that, the scheme to deal with link faults is discussed.
Finally, we introduce the deadlock recovery mechanism used in our
framework.%
\begin{figure}
\includegraphics[width=0.95\columnwidth]{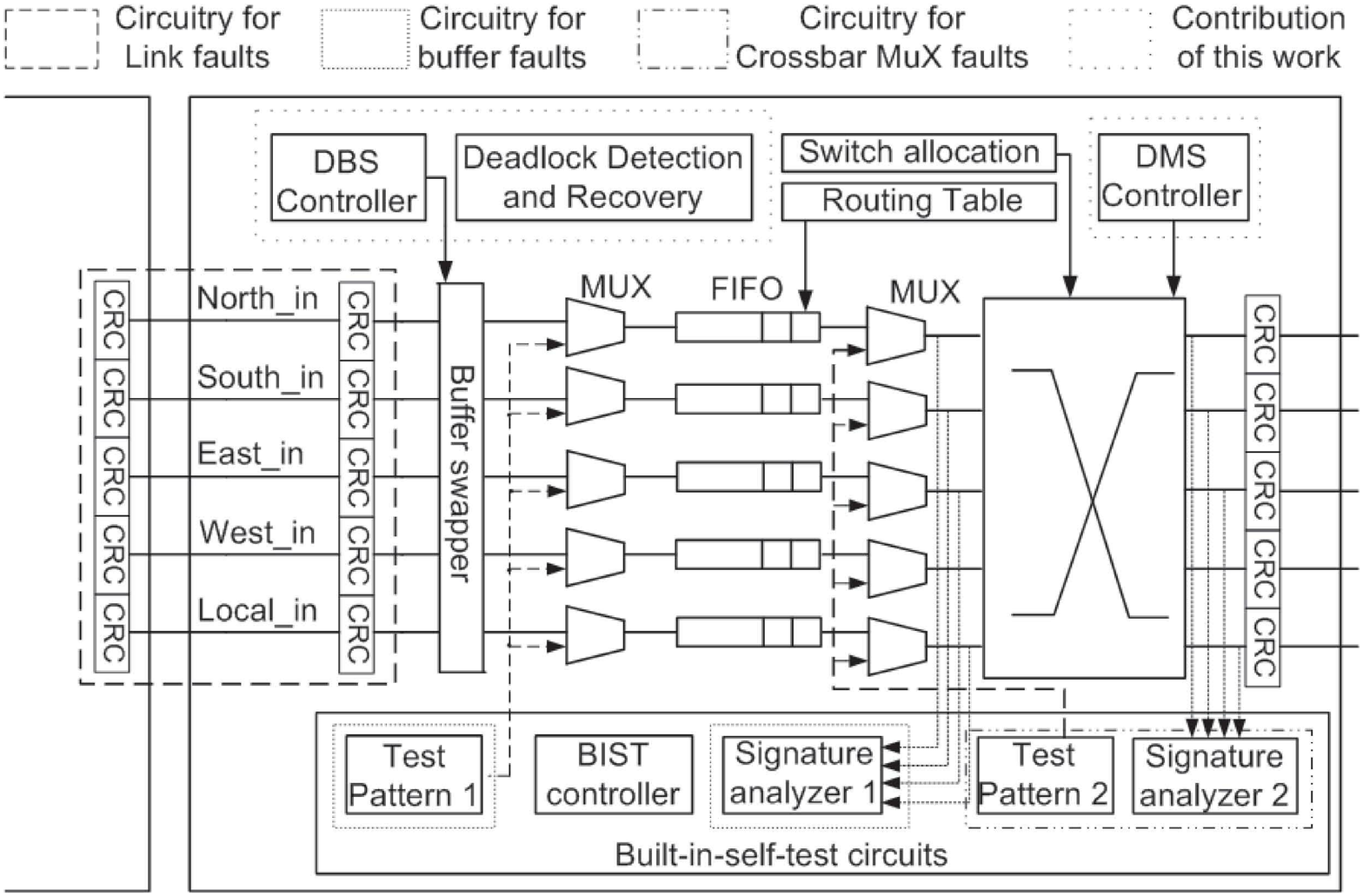}

\caption{\label{fig:Architecture-of-fault-tolerant}Architecture of fault-tolerant
NoC \cite{5158098,5467330}}
\end{figure}

\subsection{Fault-diagnosis and hybrid fault model}

\subsubsection{Fault-diagnosis infrastructure}

The fault diagnosis techniques not only should detect the occurrence of
faults but also should indicate precisely the type as well as the position of
the faults. In this work, we assume the NoC fault diagnostic technique proposed in \cite{5158098,5467330} 
is used (shown in Fig. \ref{fig:Architecture-of-fault-tolerant}), which provides the detail information of the faults in router. In the following, we briefly review the state-of-the-art fault-diagnosis structures in NoC \cite{5158098,5467330, vicis}.

As shown in Fig. \ref{fig:Architecture-of-fault-tolerant}, the
cyclic redundancy check (CRC) units are used for the link fault diagnosis.
The packet correctness is verified by the CRC modules associated with
each input port. The link transmission errors is characterized by
a corresponding checksum mismatch \cite{5467330}. Once a switch-to-switch transmission
error is detected, the BIST controller in the upstream router will
start the link diagnosis by sending the test patterns to the downstream
neighbor and find out the specific link wires that have faults. Similarly, the intra-router error will be discovered if
the packets pass the CRC check at the input of the router but fail at the output
side \cite{5467330}. On the detection of the internal fault, the router will switch
to the test mode to diagnose the specific faulty components such as the input buffers or switch crossbars. As illustrated
in Fig \ref{fig:Architecture-of-fault-tolerant}, the Test Pattern
Generator (TPG) will generate the test vectors for the buffers (TPG1)
and crossbar MUX (TPG2) respectively, while the Signature Analyzer
will compare the results and identify the specific faulty components in either FIFO buffers
or the crossbar. The diagnosis results are finally forwarded to the
DBS and DMS units for router reconfiguration based on the proposed
hybrid fault model.

\subsubsection{A novel hybrid fault model}

\begin{table}
\caption{\label{The fault model} The proposed hybrid model for permanent faults}
\begin{tabular}{|p{0.15\columnwidth}|p{0.17\columnwidth}|p{0.28\columnwidth}|p{0.25\columnwidth}|}
\hline 
{ Category} & { Component} & { Handling scheme} & { Impacts}\tabularnewline
\hline
\hline 
{ Static } & { Link} & { fault-tolerant routing to avoid faulty links} & { Partitioned network}\tabularnewline
\hline 
{ Dynamic} & { Buffer} & { time-multiplexed reuse other buffers } & { Reduced bandwidth / deadlock}\tabularnewline
\cline{2-4} 
 & { Crossbar} & { time-multiplexed reuse other MUXs } & { Reduced bandwidth}\tabularnewline
\hline
\end{tabular}
\end{table}

Table \ref{The fault model} summarizes our proposed hybrid fault model. The permanent
faults are divided into two classes: the static and the dynamic fault.
Faults occurred at the interconnect wires is denoted as link fault
and is classified as a static fault, which physically disconnects
two nodes. Of course in real situation, not all the wires are faulty
and we can still exploit the healthy wires to maintain the connectivity
similar to the approaches proposed in \cite{4669214}. In this chapter,
we mainly focus on the recovery of the router faults and hence we
assume a totally-failure link fault. To handle the static fault, a
fault-aware routing algorithm is exploited to bypass the error links. On the other hand, if the
fault occurs within the buffer or the crossbar, we can mitigate them
by reusing the other healthy resources in the router. All the datapath
will remain intact with a small penalty of reduced bandwidth. This
kind of fault is categorized as dynamic faults in our model.

\subsection{Dynamic buffer swapping (DBS) }

Dynamic buffer swapping (DBS) algorithm is proposed to handle the
input buffer faults. For traditional fault model, the buffer fault
is treated as a hard link fault and data cannot be transmitted even
the link is healthy. In our approach, through proper swapping of another
substitute buffer to hold the packets destined for the faulty buffer,
a graceful performance degradation can be achieved by fully utilizing
the link. By adopting DBS algorithm, no extra buffers or virtual
channels are needed in the router, which satisfies the stringent area
constraint of NoC \cite{4555858}.

\subsubsection{DBS operation}

\begin{figure}[h]
\includegraphics[width=0.95\columnwidth]{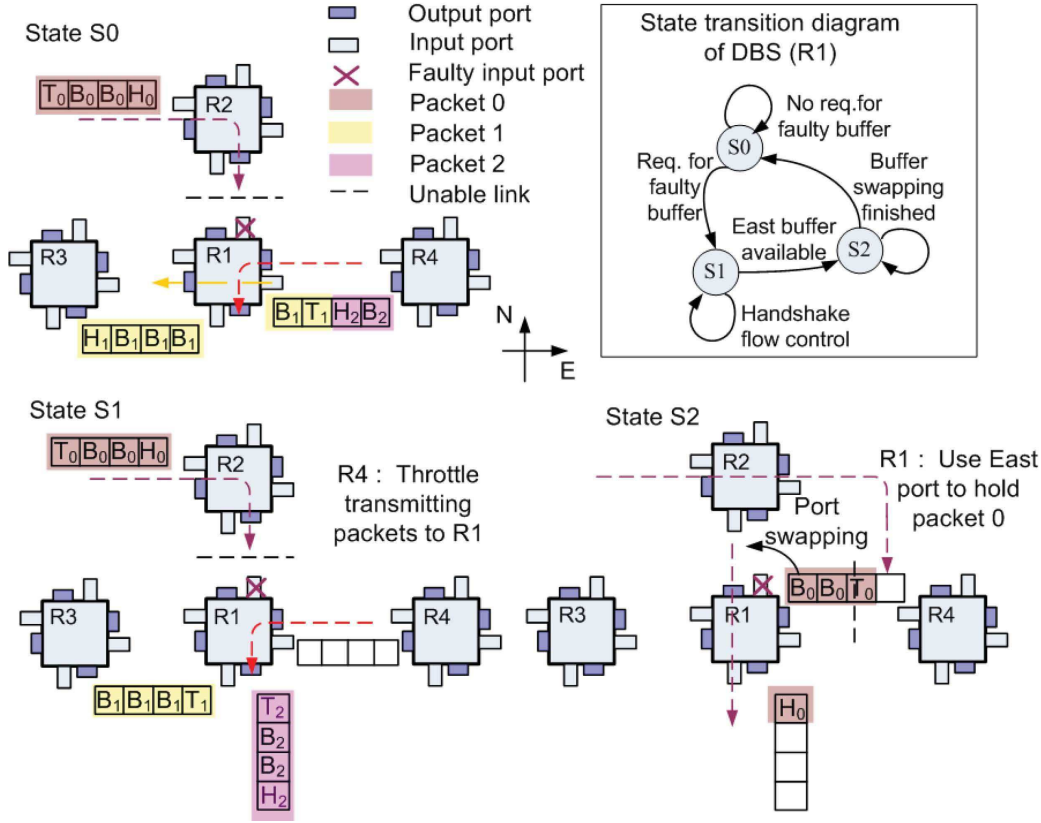}

\caption{\label{fig:Dynamic-Buffer-swapping}Dynamic Buffer Swapping (DBS)
technique}
\end{figure}

DBS technique operates with three states. As shown in Figure \ref{fig:Dynamic-Buffer-swapping},
assume fault occurs at the north input buffer of router R1, packet
0 is blocked in router R2. We denote this situation as state S0. After
receiving the transmission request token, R1 enters the port-swap
scheduling state S1. In S1, R1 will arrange another input port buffer
to substitute the north buffer based on a predefined swapping strategy. Assume the east input buffer
is chosen in this example. Then a traffic control token is propagated
to router R4 indicating that the transmission link (R1-R4) will be
shut down in the next state for DBS operation. Upon the reception
of this flow control token, R4 will stop allocating and granting new
requests to traverse through its west output port after the tail flit
of packet 2 finishes routing. R1 will also continuously monitor its
east input buffer. Once the buffer serves the tail flit of packet
2, it will enter the buffer swapping state S2. In this state, the
communication between R2-R1 link is enabled while link R4-R1 is currently
disabled. The packet 0 can thus be transmitted to R1's east input
buffer and further be forwarded towards its destinations. After packet
0 departs R1, R1 returns to normal state S0 and link R4-R1 is re-enabled
again. The router will keep running at state S0 until another transmission
request is asserted to use R1's north input buffer.%

\subsubsection{Avoid packet interleaving in DBS}

\begin{figure}
\includegraphics[width=0.95\columnwidth]{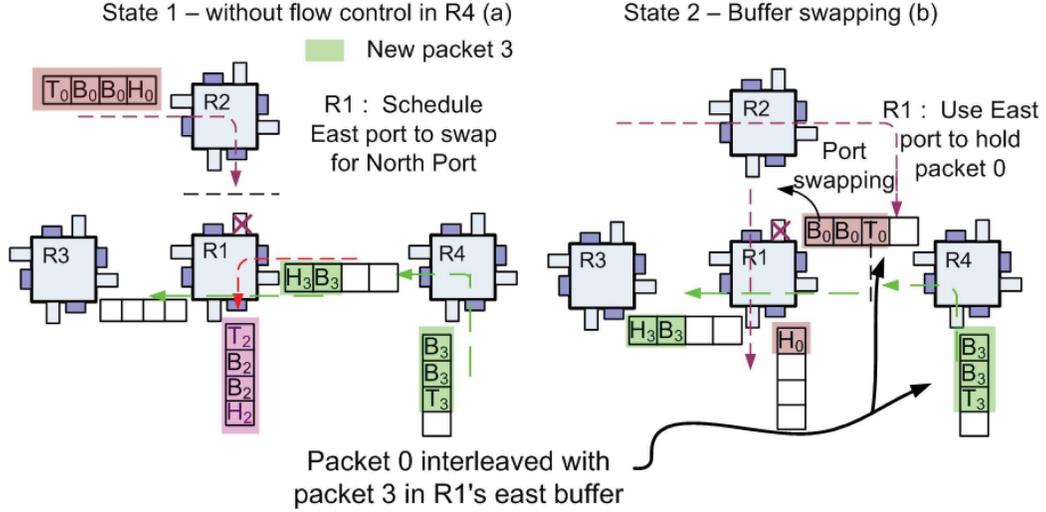}

\caption{\label{fig:A-pathological-case}A pathological case for packets interleaving}
\end{figure}

One situation needs to be addressed and avoided is the packet-level
interleaving when DBS operation is performed. In the previous example,
in S1, R4 will control its traffic towards its west output port temporarily.
Without this control, if R4 continues its transmission for the following
packets (packet 3 in Fig. \ref{fig:A-pathological-case} for instance)
and R1 jumps to state S2 for buffer swapping operation, the east buffer
of R1 will store part of the flits from packet 3 and the whole packet
0 from R2. The remaining flits of packet 3 are still in R4 waiting
to be routed. Thus packet 0 will interleave with packet 3 in the east
input buffer of R1. The header of packet 0 will erase the output reservation
information created for packet 3. After packet 0 is sent and R1 returns
to the initial state S0, the residual flits of packet 3 will enter
R1 from R4. These flits lose the routing information in R1 which is
previously reserved by their header and have to be discarded.%

Figure \ref{fig:A-pathological-case} illustrates the above situation. In order to avoid interleaved packets due to the DBS operation, a
simple but effective flow control scheme is developed. Two handshake
signals, current transmit status (CTS) and current port status (CPS),
are associated with each port. Both CTS and CPS signals are sent by
the faulty routers and read by their neighbors. The CTS (1bit) token
manages the link transmission. The router can send the packets across
the link only when it receives a '1' in the CTS. The CPS (2 bits)
token is responsible for intra-router transmission. If {}``00''
is read from one port, the port can be used as an normal output port
for all the packets. When {}``01'' is read, the router ( R4 in Fig.
\ref{fig:Dynamic-Buffer-swapping}) will begin monitor this port and
grant it only to the packets which have been partially transmitted
across the link (packet 1 and 2 in Fig. \ref{fig:Dynamic-Buffer-swapping})
to avoid the packet-level interleaving. Once the residual tail flit
($T_{2}$ in Fig. \ref{fig:Dynamic-Buffer-swapping}) is received,
the faulty router enters buffer swapping state, and sends CPS {}``10''
to its swapped port (east output port of R1 in Fig. \ref{fig:Dynamic-Buffer-swapping})
and the corresponding neighbor (R4 in Fig. \ref{fig:Dynamic-Buffer-swapping})
will not transmit packets to this link during the whole state. Within
each router, three registers are used to record the faulty port (FP\_R),
the scheduled swapping port (SP\_R) and the disabled port (DP\_R),
respectively. The status of these tokens for the example shown in
Fig. \ref{fig:Dynamic-Buffer-swapping} are summarized in Table
\ref{handshake}. 

{\small }%
\begin{table}
\caption{\label{handshake} Handshake flow control for DBS}
{\small }\begin{tabular}{|p{0.1\columnwidth}|p{0.4\columnwidth}|p{0.4\columnwidth}|}
\hline 
{\small States} & {\small Signals status (North port of R1 is faulty)} & {\small Operations / representation on tokens}\tabularnewline
\hline
{\small S0} & {\small R1 North writes: CPS {}``10''; CTS {}``0'' } & {\small R2 blocks packet 0 }\tabularnewline
 & {\small R1 Reg: FP\_R: {}``N''; SP\_R: {}``$\phi$'';DP\_R: {}``N''} & {\small R1 north port is faulty}\tabularnewline
\hline 
{\small S1} & {\small R1 North writes: CPS {}``10''; CTS {}``0''} & {\small R2 blocks packet 0}\tabularnewline
 & {\small R1 East writes: CPS {}``01'';CTS {}``1''} & {\small R4 monitors request to the west }\tabularnewline
 & {\small R1 Reg: FP\_R {}``N''; SP\_R: {}``E''; DP\_R: {}``N''} & {\small R1 prepares east to substitute north }\tabularnewline
\hline 
{\small S2} & {\small R1 North writes: CPS {}``00''; CTS {}``1''} & {\small R2 transmits packet 0 to east }\tabularnewline
 & {\small R1 East writes: CPS {}``10'';CTS {}``0''} & {\small R4 blocks transmission to R1}\tabularnewline
 & {\small R1 Reg: FP\_R {}``N''; SP\_R: {}``E''; DP\_R: {}``E''} & {\small R1 east port buffer is used for north link}\tabularnewline
\hline
\end{tabular}{\small \par}
\end{table}
{\small \par}

\subsubsection{buffer swapping strategy design}

In DBS we need to schedule a healthy port to swap with the faulty
port. As shown in Fig. \ref{fig:Dynamic-Buffer-swapping},
a buffer swapper is needed to match the physical link connections
with the swapped healthy FIFO input. We can use a five-by-five port
swapper to connect each physical link with every input buffer, similar
to the swapper used in \cite{vicis}. This architecture provides the
highest flexibility for buffer swap scheduling. A round-robin DBS
scheduling can be adopted by dynamically using different buffers to
swap with the faulty port so to even out the bandwidth penalty for
each input. For example, if the north port is faulty and in the previous
swapping state, we use the east port to hold the packets from the
north, then we will choose to use the south port to handle the next
request for the north port.

A more area-efficient implementation for buffer swapping is to use
a dedicated swapping strategy. This will reduce the area overhead
of the port swapper as we do not need to have a crossbar-like connection
architecture. The faulty port can only choose its predefined neighbor
for swapping operation. From the simulation results, which will be
presented in the next section, it is shown that this swapping strategy
achieves only a little performance penalty compared with the round-robin
DBS.

\subsection{Dynamic MUX swapping (DMS)}

\begin{figure}[h]
\includegraphics[width=0.95\columnwidth]{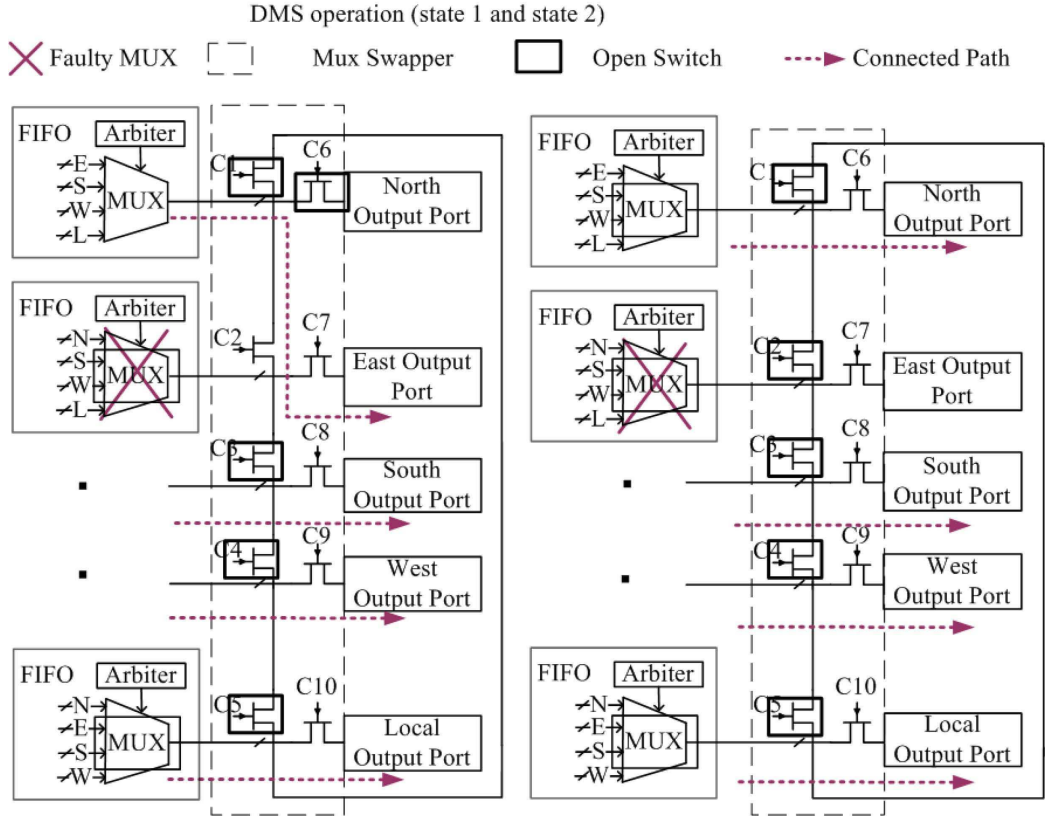}

\caption{\label{fig:DMS-hardware-implementation}DMS hardware implementation}
{\small
}
\end{figure}

The crossbar in the router is usually implemented in the form of five
$4\times1$ MUXs \cite{5158098}. As shown in Fig. \ref{fig:DMS-hardware-implementation},
if one of the MUX malfunctions, the data will not be presented to
the output link correctly. Here we present a dynamic MUX swapping
(DMS) technique to mitigate the MUX faults and make all the output
links available for routing. The proposed ring-based crossbar swapper
is shown in Fig. \ref{fig:DMS-hardware-implementation}. The ring
topology allows each MUX to be shared among its neighboring output
ports. If the MUX corresponding to the east output port is faulty,
the MUX for the north port can be time shared with the north and east
ports by turning on-off the switches in MUX swapper accordingly. By
doing so, all the output links can be maintained.

\subsection{Handling hard link faults}

For the link wire faults, since the physical connectivity between
the routers has been damaged, we need to use a robust routing algorithm
to bypass the faulty links. In this work, we use a similar approach
as proposed in \cite{resilient_routing,vicis} to route the packets under the
existence of hard link faults.

\subsection{Deadlock issue and handling strategy}

\subsubsection{Deadlock in proposed framework}

To ensure deadlock free routing, in normal operation, Oddeven (OE)
routing algorithm \cite{OE} is used in our framework. However
when faults occur, due to the DBS or the re-routing, a deadlock may
be introduced.%
\begin{figure}
\includegraphics[width=0.95\columnwidth]{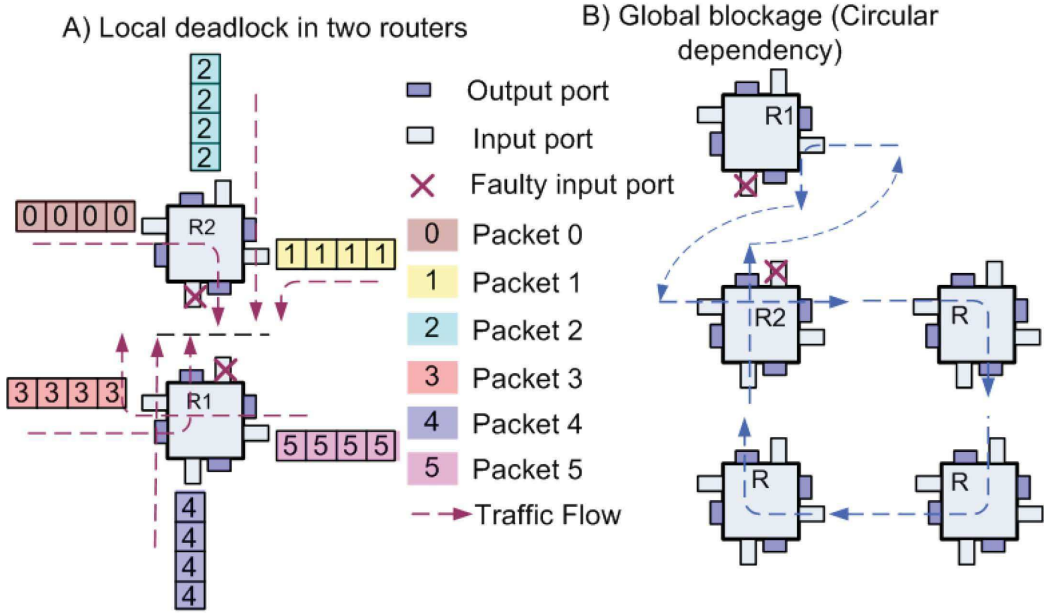}

\caption{\label{fig:Deadlock-situation-for}Deadlock situation for DBS operation}
{\small \vspace{-4mm}
}
\end{figure}

Figure \ref{fig:Deadlock-situation-for} gives an example of two
deadlock situations. In Fig. \ref{fig:Deadlock-situation-for}-a,
a local deadlock will occur between the two nodes R1 and R2 if the
south port of R2 and north port of R1 are faulty and all the packets
in other ports choose these two ports as outputs. In this case, R1
(R2) will be stuck at state S1 waiting for the original packets in
the substitute port of R2 (R1) being completely transmitted. In Fig.
\ref{fig:Deadlock-situation-for}-b, a global blockage situation is
illustrated assuming the south to east turn is forbidden. If the scheduler
in R2 arranges its west input buffer to hold packets from north and
the scheduler in R1 arranges its east input buffer to substitute the
faulty south buffer, a circular dependency on buffer resources can
thus be generated as in the figure. According to \cite{250114}, this
kind of circular dependency on the NoC resources may bring deadlocks
and stall the whole system. 

For DMS operation, we only time-share the MUX between two output ports
and do not change the dependency of the buffers in the routing algorithm.
Hence it will not create deadlock if the original routing algorithm
is deadlock-free.

\subsubsection{Strategies to relieve deadlock}

We employ a detection scheme, similar to the DISHA approach in computer
network \cite{524561} to detect the blockage in the network. When
the router receives the header flit of a packet, an embedded counter
is triggered to count the number of cycles it stays in the buffer.
The counter continues to increment until the tail flit has been successfully
transmitted or its value exceeds a predefined threshold $T_{sh}$.
In the latter case, the router will enter the deadlock handling mode.
After that the counter would be reset and waits for next packet.
\\In this chapter, we propose a set of strategies to relieve the deadlock
situation. We mainly deal with the packets in the faulty routers in
DBS mode because DMS and the routing algorithm will not generate circular
dependency. First, we set the SP\_R register to the next port so as
to initialize another swapping operation using a different substitute
port. In Fig. \ref{fig:Deadlock-situation-for}-b, by using the
east input buffer of R2 to swap with the north faulty buffer, we can
break the original dependency and avoid the blockage. If all the ports
in the router are not available for swapping as the case in Fig.
\ref{fig:Deadlock-situation-for}-a, we will reset the routing computation
(RC), switch allocation (SA) information of these packets to request
a different output port. In figure \ref{fig:Deadlock-situation-for}-a,
if packet 1 chooses the west instead of south as output direction
and leaves R2 successfully, then the deadlock problem can be solved
accordingly.

If the re-computation step fails to relieve the blockage, and the
local processing element (PE) has available memory spaces in network
interface (NI), next, we will send these blocked packets to the local
PE, and make the PE re-send these packets to the destination. In order
to avoid creating a new circular dependency between the PE and other
routers, this operation is carried out only when NI has free memory
slots. If the memory in NI is full, we need to drop the current packet
and require a higher level end-to-end re-transmission protocol to re-send
the packets\cite{5467330}. For hard link faults, as mentioned in\cite{resilient_routing},
in order to improve network connectivity, we need to selectively remove
the turn rules of some routers at the cost of increasing the hazard
of deadlock. The scheme discussed above can also be applied to the
packets that violate these turn rules so as to return to normal status.

\section{Simulation Results}

\subsection{Simulation environment setup}

In order to evaluate the NoC performance, a systemC based NoC simulator
was developed. The energy parameters are adopted from Noxim\cite{the:Noxim-simulator-User}.
The mesh topology is used for the simulation due to its popularity
and simplicity. $8\times8$ meshes are used in the simulation for
all the synthetic traffics. Several real benchmarks are also used for evaluating
the latency effects. They are MMS (Multimedia system mapped on $5\times5$
mesh) in \cite{4492729}, DVOPD (Dual Video Object Plane Decoder mapped
on $3\times4$ mesh) and MPEG4 (MPEG4 codec) in \cite{Noc_synthesis}. The simulation is carried
out for $150000$ cycles after $10000$ warm up cycles.

\subsection{Comparison of connectivity of the network}

\begin{figure}
\includegraphics[width=0.95\columnwidth]{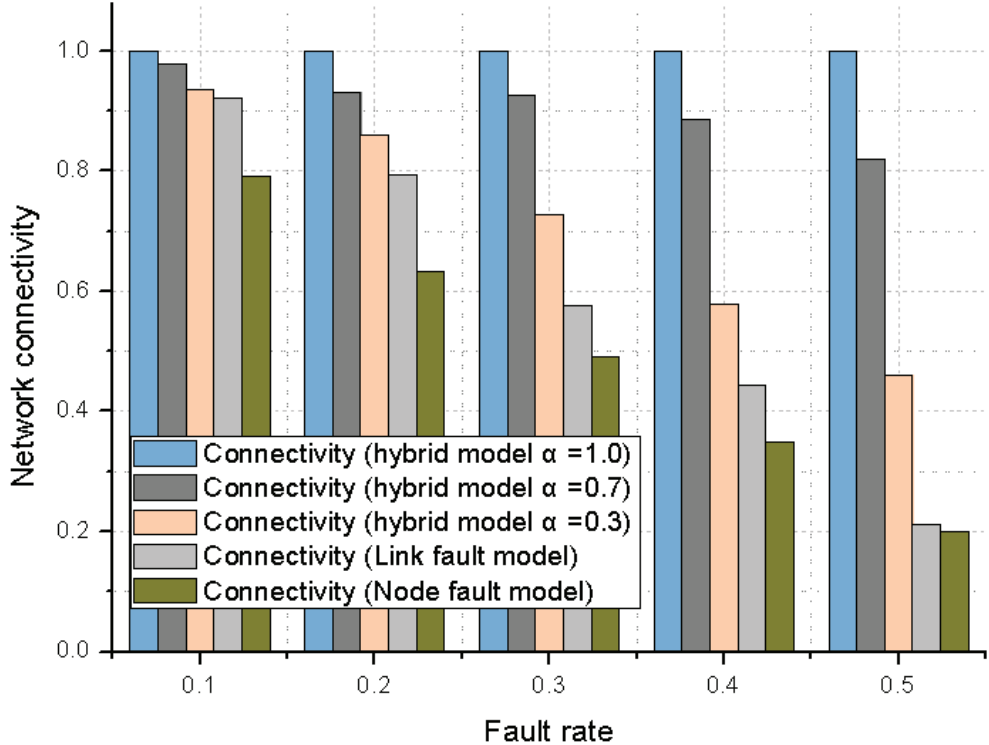}
\caption{\label{fig:Connectivity-comparison-of}Connectivity comparison of
the hybrid fault model}
{\small
}
\end{figure}

We define fault rate $f$ as follows: $f=\ulcorner\frac{Number\: of\: injected\: faults}{Mesh\: size}\urcorner$;
and the parameter $\alpha$ indicates the portion of the partial router
faults in the injected fault pattern: $\alpha=\frac{Number\: dynamic\: faults}{Number\: injected\: faults}$;
the connectivity of a network is defined as the ratio of number of
node pairs that can communicate to total number of node pairs in the
network. Figure \ref{fig:Connectivity-comparison-of} is the analytical
results for the connectivity for $8\times8$ mesh. %

For each fault rate, we randomly generate 30 fault patterns. Then
the connectivity is analyzed by checking whether each node pair can
communicate under the given pattern. From the simulation, we demonstrate
by distinguishing the dynamic faults with static faults, network connectivity
can be improved more than 20\% compared to the link fault model.

\subsection{Improvement in packet acceptance rate}

In table \ref{packet_accpetance_rate}, we show the simulation results for the packet acceptance
rate compared with two state-of-art works in \cite{resilient_routing,4542002}.
We randomly generate 20 fault patterns for each fault rate and use
random traffic to evaluate this merit. The packets between two nodes
which can not communicate anymore will be counted as dropped packets
as well. Under high fault rate, the connectivity drops dramatically
in \cite{resilient_routing,4542002} and hence reduces the packet acceptance rate.
On the contrary, by performing the DBS and DMS operations, the probability
of a successful transmission between any two nodes is greatly improved.
Hence, $16.6\%$-$21.9\%$ higher acceptance rate can be achieved.

\begin{table}
\caption{\label{packet_accpetance_rate} Packet acceptance rate comparison}
\begin{tabular}{|c|p{0.05\textwidth}|p{0.06\textwidth}|p{0.06\textwidth}|p{0.06\textwidth}|p{0.06\textwidth}|p{0.06\textwidth}|p{0.06\textwidth}|p{0.06\textwidth}|p{0.06\textwidth}|p{0.06\textwidth}|}
\hline 
{\footnotesize Mesh} & \multicolumn{1}{p{0.06\textwidth}}{} & \multicolumn{1}{p{0.06\textwidth}}{} & \multicolumn{1}{p{0.06\textwidth}}{{\footnotesize }%
\begin{minipage}[t]{1\columnwidth}%
{\footnotesize $4\times4$ mesh}%
\end{minipage}} & \multicolumn{1}{p{0.06\textwidth}}{} &  & \multicolumn{1}{p{0.06\textwidth}}{} & \multicolumn{1}{p{0.06\textwidth}}{} & \multicolumn{1}{p{0.07\textwidth}}{{\footnotesize }%
\begin{minipage}[t]{0.95\columnwidth}%
{\footnotesize $8\times8$ mesh}%
\end{minipage}} & \multicolumn{1}{p{0.06\textwidth}}{} & \tabularnewline
\hline 
 & \multicolumn{1}{p{0.06\textwidth}}{} & \multicolumn{1}{p{0.06\textwidth}}{{\footnotesize }%
\begin{minipage}[t]{0.95\columnwidth}%
{\footnotesize Previous work}%
\end{minipage}} &  & \multicolumn{1}{p{0.06\textwidth}}{{\footnotesize }%
\begin{minipage}[t]{0.95\columnwidth}%
{\footnotesize This work}%
\end{minipage}} &  & \multicolumn{1}{p{0.06\textwidth}}{} & \multicolumn{1}{p{0.06\textwidth}}{{\footnotesize }%
\begin{minipage}[t]{0.95\columnwidth}%
{\footnotesize Previous work}%
\end{minipage}} &  & \multicolumn{1}{p{0.06\textwidth}}{{\footnotesize }%
\begin{minipage}[t]{0.95\columnwidth}%
{\footnotesize This work}%
\end{minipage}} & \tabularnewline
\hline 
{\footnotesize Fault Rate} & {\footnotesize OE} & {\footnotesize \cite{4542002}} & {\footnotesize \cite{resilient_routing}} & {\footnotesize }%
\begin{minipage}[t]{0.6\columnwidth}%
{\footnotesize $\alpha=0.5$}%
\end{minipage} & {\footnotesize }%
\begin{minipage}[t]{0.95\columnwidth}%
{\footnotesize $\alpha=1.0$}%
\end{minipage} & {\footnotesize OE} & {\footnotesize \cite{4542002}} & {\footnotesize \cite{resilient_routing}} & {\footnotesize }%
\begin{minipage}[t]{0.95\columnwidth}%
{\footnotesize $\alpha=0.5$}%
\end{minipage} & {\footnotesize }%
\begin{minipage}[t]{0.95\columnwidth}%
{\footnotesize $\alpha=1.0$}%
\end{minipage}\tabularnewline
\hline 
{\footnotesize 0\%} & {\footnotesize 100\%} & {\footnotesize 100\%} & {\footnotesize 100\%} & {\footnotesize 100\%} & {\footnotesize 100\%} & {\footnotesize 100\%} & {\footnotesize 100\%} & {\footnotesize 100\%} & {\footnotesize 100\%} & {\footnotesize 100\%}\tabularnewline
\hline 
{\footnotesize 4\%} & {\footnotesize 100\%} & {\footnotesize 100\%} & {\footnotesize 100\%} & {\footnotesize 100\%} & {\footnotesize 100\%} & {\footnotesize 62.3\%} & {\footnotesize 100\%} & {\footnotesize 100\%} & {\footnotesize 100\%} & {\footnotesize 100\%}\tabularnewline
\hline 
{\footnotesize 12\%} & {\footnotesize 55.4\%} & {\footnotesize 95.8\%} & {\footnotesize 100\%} & {\footnotesize 100\%} & {\footnotesize 100\%} & {\footnotesize 98.2\%} & {\footnotesize 100\%} & {\footnotesize 100\%} & {\footnotesize 100\%} & {\footnotesize 100\%}\tabularnewline
\hline 
{\footnotesize 20\%} & {\footnotesize 33.9\%} & {\footnotesize 83.1\%} & {\footnotesize 98.3\%} & {\footnotesize 100\%} & {\footnotesize 100\%} & {\footnotesize 25.3\%} & {\footnotesize 80.3\%} & {\footnotesize 95.9\%} & {\footnotesize 100\%} & {\footnotesize 100\%}\tabularnewline
\hline 
{\footnotesize 28\%} & {\footnotesize 29.8\%} & {\footnotesize 64.3\%} & {\footnotesize 91.0\%} & {\footnotesize 100\%} & {\footnotesize 100\%} & {\footnotesize 13.2\%} & {\footnotesize 56.3\%} & {\footnotesize 89.6\%} & {\footnotesize 93.9\%} & {\footnotesize 100\%}\tabularnewline
\hline 
{\footnotesize 36\%} & {\footnotesize 29.9\%} & {\footnotesize 48.9\%} & {\footnotesize 74.0\%} & {\footnotesize 99.8\%} & {\footnotesize 100\%} & {\footnotesize 14.2\%} & {\footnotesize 32.1\%} & {\footnotesize 71.8\%} & {\footnotesize 90.4\%} & {\footnotesize 96.0\%}\tabularnewline
\hline 
{\footnotesize 40\%} & {\footnotesize 16.4\%} & {\footnotesize 26.0\%} & {\footnotesize 69.7\%} & {\footnotesize 98.9\%} & {\footnotesize 100\%} & {\footnotesize 9.7\%} & {\footnotesize 23.7\%} & {\footnotesize 71.0\%} & {\footnotesize 87.7\%} & {\footnotesize 92.9\%}\tabularnewline
\hline
\end{tabular}
\end{table}

\begin{figure}
\includegraphics[width=0.95\columnwidth]{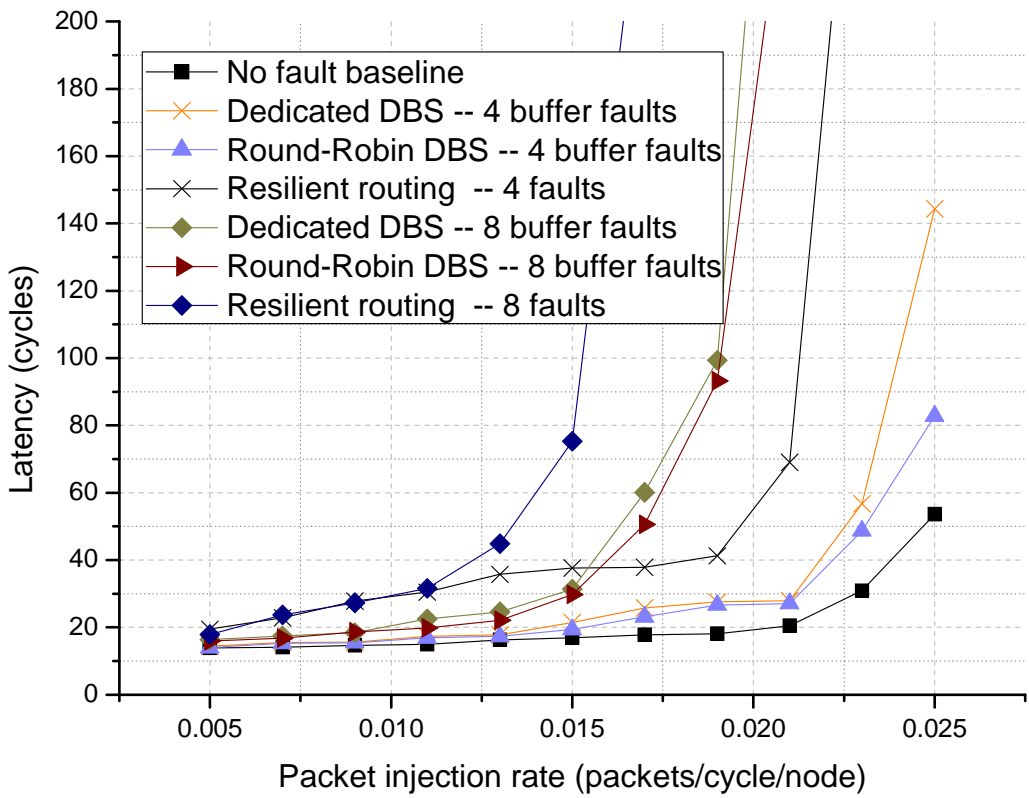}
\caption{\label{fig:Comparison-between-buffer}Comparison between buffer swapping
strategy}
{\small
}
\end{figure}

\subsection{Effects on latency performance}

In Fig. \ref{fig:Comparison-between-buffer}, we make a comparison
of the proposed round-robin and dedicated DBS operation for $8\times8$
mesh. Both two strategies outperform the resilient routing in \cite{resilient_routing}
and a 1.15$\times$ improvement in the saturation point can be observed.%
\begin{figure}
\includegraphics[width=0.95\columnwidth]{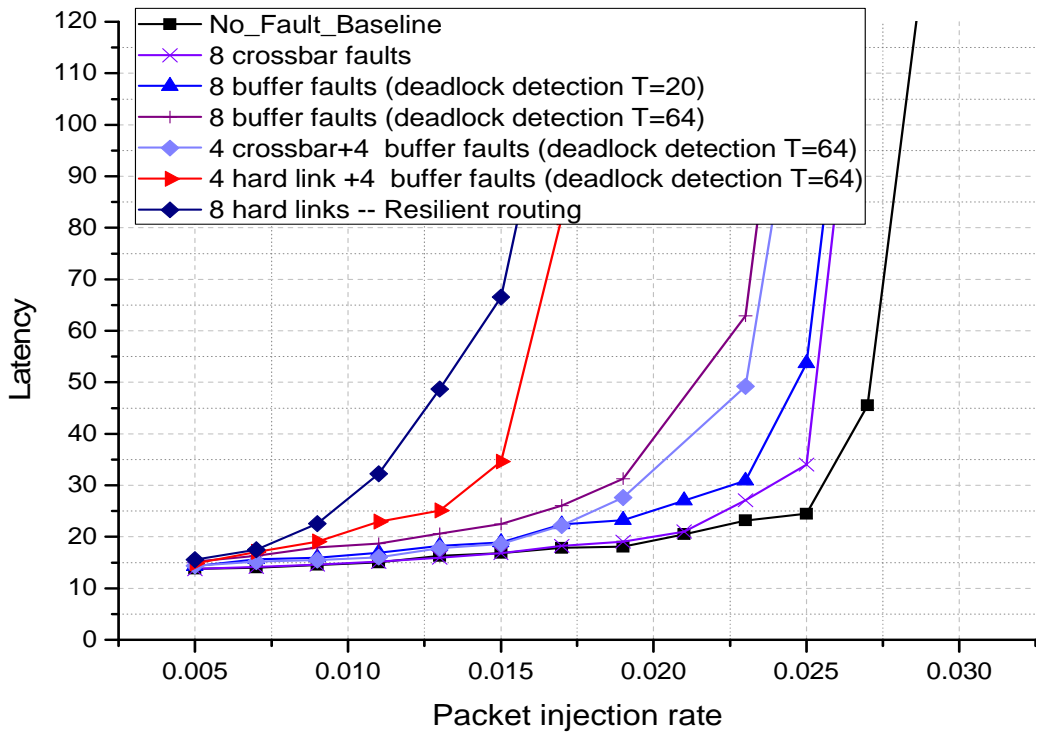}
\caption{\label{fig:Latency-comparison-of}Latency comparison of various fault
model and schemes}
{\small 
}
\end{figure}

Fig. \ref{fig:Latency-comparison-of} is the simulation results
for $8\times8$ mesh under the hybrid fault situation. As discussed
earlier, DMS will not generate deadlocks so the performance of DMS
operation is the closest to the baseline routers. While the DBS operation
may bring deadlocks and the latency depends on the predefined threshold
of the deadlock detection scheme ($T=20$ and $64$) . It is noticeable
that the performance in our work is consistently better than \cite{resilient_routing}
when faults can be categorized as dynamic faults.%

\begin{table}
\begin{tabular}{|p{0.10\linewidth}|p{0.09\linewidth}|p{0.09\linewidth}|p{0.09\linewidth}|p{0.09\linewidth}|p{0.09\linewidth}|p{0.09\linewidth}|p{0.09\linewidth}|p{0.09\linewidth}|}
\hline 
 & \multicolumn{2}{p{0.07\linewidth}|}{{\footnotesize }%
\begin{minipage}[t]{0.2\columnwidth}%
{\footnotesize No fault }%
\end{minipage}} & \multicolumn{2}{c|}{{\footnotesize }%
\begin{minipage}[t]{0.2\columnwidth}%
{\footnotesize Work in \cite{resilient_routing}}%
\end{minipage}} & \multicolumn{2}{c|}{{\footnotesize }%
\begin{minipage}[t]{0.2\columnwidth}%
{\footnotesize This work}%
\end{minipage}} & \multicolumn{2}{c|}{{\footnotesize }%
\begin{minipage}[t]{0.16\columnwidth}%
{\footnotesize Improvement}%
\end{minipage}}\tabularnewline
\hline 
{\footnotesize Traffic} & {\footnotesize Latency}{\footnotesize \par}

{\footnotesize (cycle)} & {\footnotesize Accept Rate} & {\footnotesize Latency}{\footnotesize \par}

{\footnotesize (cycle)} & {\footnotesize Accept Rate} & {\footnotesize Latency}{\footnotesize \par}

{\footnotesize (cycle)} & {\footnotesize Accept Rate} & {\footnotesize Latency}{\footnotesize \par}

{\footnotesize (cycle)} & {\footnotesize Accept Rate}\tabularnewline
\hline 
\hline 
{\scriptsize DVOPD-4 } & {\footnotesize 18.70} & {\footnotesize 100\%} & {\footnotesize 34.24} & {\footnotesize 100\%} & {\footnotesize 22.11} & {\footnotesize 100\%} & {\footnotesize 35.4\%} & {\footnotesize --}\tabularnewline
\hline 
{\scriptsize DVOPD-8 } & {\footnotesize 18.70} & {\footnotesize 100\%} & {\footnotesize 56.22} & {\footnotesize 92.9\%} & {\footnotesize 30.91} & {\footnotesize 100\%} & {\footnotesize 45.0\%} & {\footnotesize 7.1\%}\tabularnewline
\hline 
{\scriptsize MPEG4-4 } & {\footnotesize 24.33} & {\footnotesize 100\%} & {\footnotesize 70.12} & {\footnotesize 100\%} & {\footnotesize 45.51} & {\footnotesize 100\%} & {\footnotesize 35.1\%} & {\footnotesize --}\tabularnewline
\hline 
{\scriptsize MPEG4-8} & {\footnotesize 24.33} & {\footnotesize 100\%} & {\footnotesize 82.77} & {\footnotesize 88.6\%} & {\footnotesize 47.35} & {\footnotesize 100\%} & {\footnotesize 42.8\%} & {\footnotesize 11.4\%}\tabularnewline
\hline 
{\scriptsize E3S-4} & {\footnotesize 23.54} & {\footnotesize 100\%} & {\footnotesize 27.17} & {\footnotesize 100\%} & {\footnotesize 24.87} & {\footnotesize 100\%} & {\footnotesize 8.5\%} & {\footnotesize --}\tabularnewline
\hline 
{\scriptsize E3S-8} & {\footnotesize 23.54} & {\footnotesize 100\%} & {\footnotesize 45.33} & {\footnotesize 93.2\%} & {\footnotesize 26.20} & {\footnotesize 100\%} & {\footnotesize 42.2\%} & {\footnotesize 6.8\%}\tabularnewline
\hline 
{\scriptsize MMS-4} & {\footnotesize 39.74} & {\footnotesize 100\%} & {\footnotesize 58.18} & {\footnotesize 100\%} & {\footnotesize 51.30} & {\footnotesize 100\%} & {\footnotesize 11.8\%} & {\footnotesize --}\tabularnewline
\hline 
{\scriptsize MMS-8 } & {\footnotesize 39.74} & {\footnotesize 100\%} & {\footnotesize 87.81} & {\footnotesize 94.7\%} & {\footnotesize 64.21} & {\footnotesize 100\%} & {\footnotesize 26.9\%} & {\footnotesize 5.3\%}\tabularnewline
\hline 
{\scriptsize TGFF-4 } & {\footnotesize 47.06} & {\footnotesize 100\%} & {\footnotesize 66.48} & {\footnotesize 100\%} & {\footnotesize 52.15} & {\footnotesize 100\%} & {\footnotesize 21.6\%} & {\footnotesize --}\tabularnewline
\hline 
{\scriptsize TGFF-8 } & {\footnotesize 47.06} & {\footnotesize 100\%} & {\footnotesize 285.48} & \centering{}{\footnotesize 100\%} & {\footnotesize 64.17} & {\footnotesize 100\%} & {\footnotesize 77.5\%} & {\footnotesize --}\tabularnewline
\hline 
\multicolumn{7}{|c|}{%
\begin{minipage}[t]{0.8\columnwidth}%
{Average improvement}%
\end{minipage}} & {\footnotesize 34.7\%} & {\footnotesize 7.7\%}\tabularnewline
\hline 
\end{tabular}

\caption{\label{Effects_on_latency}Latency and packet acceptance rate comparison for various traffics}
{\small \vspace{-4mm}
}
\end{table}

In Table \ref{Effects_on_latency}, we summarize the latency and acceptance rate performance
for the traffic scenarios other than the uniform random case. Two
different fault numbers, $4$ and $8$ are used for each traffic. We can
see that for different traffic scenarios, an average $34.7\%$ reduction
in latency and $7.7\%$ improvement in reliability can be achieved.



\section{Conclusion}

In this chapter, we propose a fine-grained fault model for network on
chips to maximally maintain the network connectivity. Based on the
fault model, we distinguish with different fault components and use
specific strategies to handle these faults. DBS and DMS techniques
are proposed to make the partial router still functional and hence
improve the system robustness. Simulation results demonstrate higher packet acceptance rate and better latency performance can be achieved.

\chapter{FSNoC: A Flit-level Speedup Scheme For NoCs Using Self-Reconfigurable Bi-directional Channels}

\textit{In this chapter, we tackle the problem of optimizing the bandwidth utilization
of the Network-on-Chips (NoC). More specifically, we propose a flit-level
speedup scheme to enhance the NoC performance utilizing self-reconfigurable
bidirectional channels. For the NoC intra-router bandwidth, in addition
to the traditional efforts on allowing flits from different packets
using the idling internal bandwidth of the crossbar, our proposed
flit-level speedup scheme also allows  flits within the same packet
to be transmitted simultaneously. For inter-router transmission, a
novel distributed channel configuration protocol is developed to dynamically
control the link directions. In this way, the bisection bandwidth
between the routers can adapt to the changing network status. We describe
the implementation of the proposed flit-level speedup NoC on a two-dimensional
mesh. An input buffer architecture which supports reading and writing
two flits from the same virtual channel at one time is proposed. The
switch allocator is also designed to support flit-level parallel arbitration.
Extensive simulations on both the synthetic traffic and real application
benchmarks, show performance improvement in throughput and latency
over the existing architectures using bi-directional channels.}

\section{Introduction}

\begin{figure}
\includegraphics[width=0.98\columnwidth]{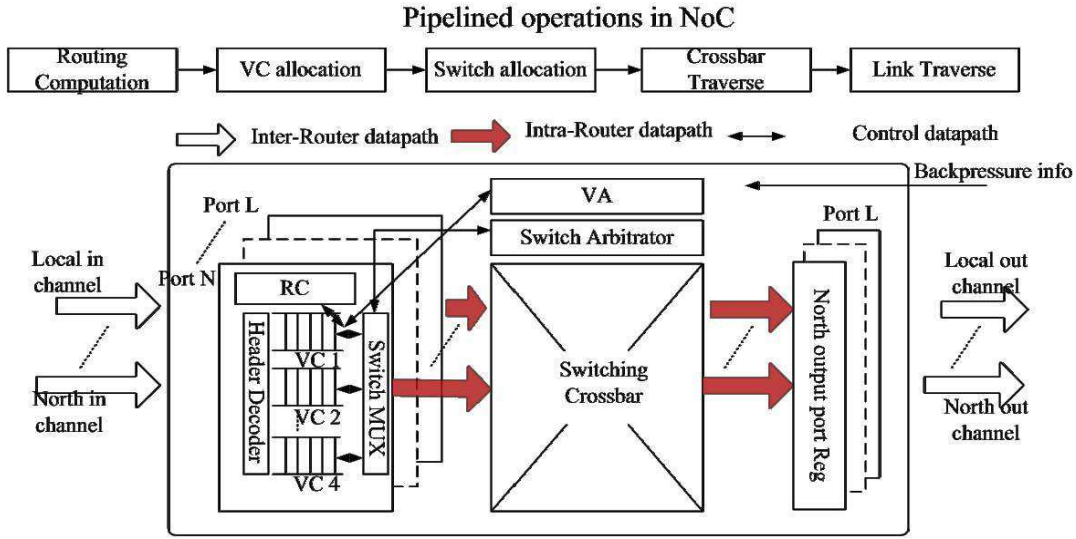}
\caption{\label{fig:Typical-four-stage-pipelined}Typical four-stage pipelined
router micro-architecture using VC-based flow-control }
\end{figure}

NoCs utilize the interconnected routers instead of buses or point-to-point
wires to send and receive packets between processor elements (PE).
Due to the high communication bandwidth provided, it overcomes the
scalability limitations of the buses and bring significant performance
improvement in terms of communication latency, power consumption and
reliability \textit{etc} \cite{976921}. A typical NoC consists of
the processing elements (PEs), network interfaces (NIs) and the routers.
The latter two make up of the communication infrastructures in order
to support high-bandwidth communication. The NIs are used to packetize
the message and transform each packet from the clients (PE cores)
into fixed-length flow-control units (\textit{i.e.,} flits). The flits
are then sent to the attached router, whose function is to route packets
from the input ports to an appropriate output port. In a typical four-stage
pipelined, virtual-channel (VC) based router \cite{NoCbook_Peh} architecture
(shown in Fig. \ref{fig:Typical-four-stage-pipelined}), the router
needs to compute the next hop direction (RC stage) and allocates a
virtual channel (VA stage) for every incoming header flit. Then the
body and tail flits follow the path settled and only need to contend
for the switch entries (SA stage). The winning flits will traverse
the router to the output port (ST stage) and move to the next hop
routers (LT stage). In general, the bandwidth provided by NoC and
its utilization have a significant impact on the overall performance
\cite{Book}. As shown in Fig. \ref{fig:Typical-four-stage-pipelined},
the NoC bandwidth can be categorized into two types: the inter-router
bandwidth (channel bandwidth) and the intra-router bandwidth (switch
bandwidth). The inter-router bandwidth is a set of physical interconnection
wires between the routers (or router and NI) to handle either the
incoming or outgoing traffic from the neighboring routers (or PEs).
The intra-router bandwidth is dictated by the datapath within the
router (\textit{e.g.,} the input buffers, the crossbar fabric and
the output registers).

In order to improve the usage of internal bandwidth, in \cite{Book},
NoC with input speedup is proposed by providing some excess bandwidth
in the cross-point to relieve the contentions in the routers. In \cite{Vichar},
a fine-grained buffer utilization scheme ViChaR is proposed which
target dynamic adaption of the router buffers by providing unified
buffer resources across all input ports. Due to the efficient usage
of the buffers, a 25\% performance increase can be achieved over generic
routers. 

In addition to the efforts on optimizing the internal bandwidth utilization,
several approaches have been proposed recently to dynamically adjust
the inter-router channel resources to match the run time workload
requirements. In \cite{reconfigurable-links-GLVLSI}, based on the
observation that NoC link resources are generally over-provisioned,
a dynamic channel width configuration algorithm is proposed to adjust
the link bandwidth for the current application to optimize the energy-performance
trade-offs. In \cite{PACT}, an adaptive physical channel regulator
(APCR) is proposed to provide finer granularity of using link channels.
The flit size in APCR router is less than the physical channel width
(phit). Therefore, the APCR scheme allows flits from different packets
to share the same physical bandwidth at run time. However, this concurrent
transmission of multiple flits from different packets are implemented
at the cost of increased physical link bandwidth. In \cite{ISCA},
a heterogeneous NoC architecture is proposed where wider links are
allocated to routers that require more resources. At run time, two
flits are combined when possible and then sent through the wider link
in order to reduce the delay. 

Besides the above approaches which are based on the uni-directional
links, new progresses have been made by utilizing the self-reconfigurable
bi-directional channels. This is motivated by the observation that,
between two routers, quite often one uni-directional link may be overflowed
with heavy traffic while the link for the opposite direction remains
idle\cite{5090667,5715603,BiNoC1}. The inefficient link bandwidth
usage in uni-directional NoC may limit the system latency and throughput.
In \cite{5090667}, the link directions are configured in the architecture
level with route re-allocation  to meet the bandwidth and QoS requirement.
In \cite{5715603,BiNoC1},  NoC architectures with run time reconfigurable
bi-directional channels were proposed. The direction of each link
is decided at run time depending on the traffics of the two routers
using either an external bandwidth arbiter \cite{BiNoC1} or a channel
direction control (CDC) protocol \cite{5715603} . For both approaches,
the link direction is reversed if there are more than two packets
requesting to traverse along one direction and no packets requesting
in the opposite direction. Besides the performance improvement in
latency and throughput, NoCs with bi-directional channels have also
demonstrated the advantages in mitigating the static and dynamic channel
failures \cite{5982012}, providing QoS guarantees \cite{5470359}
as well as better energy-performance trade-offs \cite{phit-noc}.

In this chapter, based on the self-reconfigurable bi-directional channels,
we propose a flit-level speedup scheme to further enhance the NoC
performance. More specifically, by allowing flits within the same
packet to be transmitted simultaneously using the inter- and intra-
router bandwidth, the channel utilization is improved and therefore
better latency/throughput performance can be achieved. Moreover, a
new channel direction control method is proposed which masks inter-router
signaling delays within the router pipeline depth and is efficient
for configuring the directions of long wire links using repeaters
or pipelined registers. The input buffers which support reading and
writing two flits from the same virtual channel is proposed. The switch
allocator is also designed to support flit-level parallel arbitration.
Both the synthetic and real-world traffic patterns have been used
to evaluate the performance of proposed flit-level speedup scheme.
The hardware overhead is also compared to the conventional NoC and
BiNoC designs. The results demonstrate better latency and throughput
performance can be achieved with moderate power and area overhead
than the traditional NoC and BiNoC schemes.

Compared to the previous works, this chapter makes the following key
contributions:
\begin{enumerate}
\item A fine-grained, flit-level transmission scheme is proposed which
provides more flexibility in using NoC bi-directional channels. We
demonstrate the performance enhancement using both synthetic traffic
and application benchmarks.
\item A new channel direction control protocol for bi-directional channels
is proposed. It is shown this protocol can be used for long wire links
with repeaters or pipeline registers.
\item The FSNoC router architecture that supports flit-level parallel transmission
is proposed. The performance overhead in terms of area and power are
analyzed and compared with state-of-the-art BiNoC implementations. 
\end{enumerate}
The remainder of this chapter is organized as follows. Section 5.2 gives
the motivation of adding more flexibility in BiNoC and compares the
single router throughput under different transmission schemes. Then,
in Section 5.3, we present the proposed flit-level speedup scheme.
We elaborate on the channel directional control protocols to support
parallel flit transmission in FSNoC, followed by a in-depth discussion
of the hardware implementations. Section 5.4 extends the FSNoC design
for the existence of long wire links made with repeaters or pipeline
registers. Our simulation results are presented in Section 5.5. Finally,
section 5.6 concludes the whole work. 

\section{Motivations of FSNoC}

\subsection{\label{sub:Motivations-of-flit-level}A motivational example }

\begin{figure}
\includegraphics[width=1.04\columnwidth]{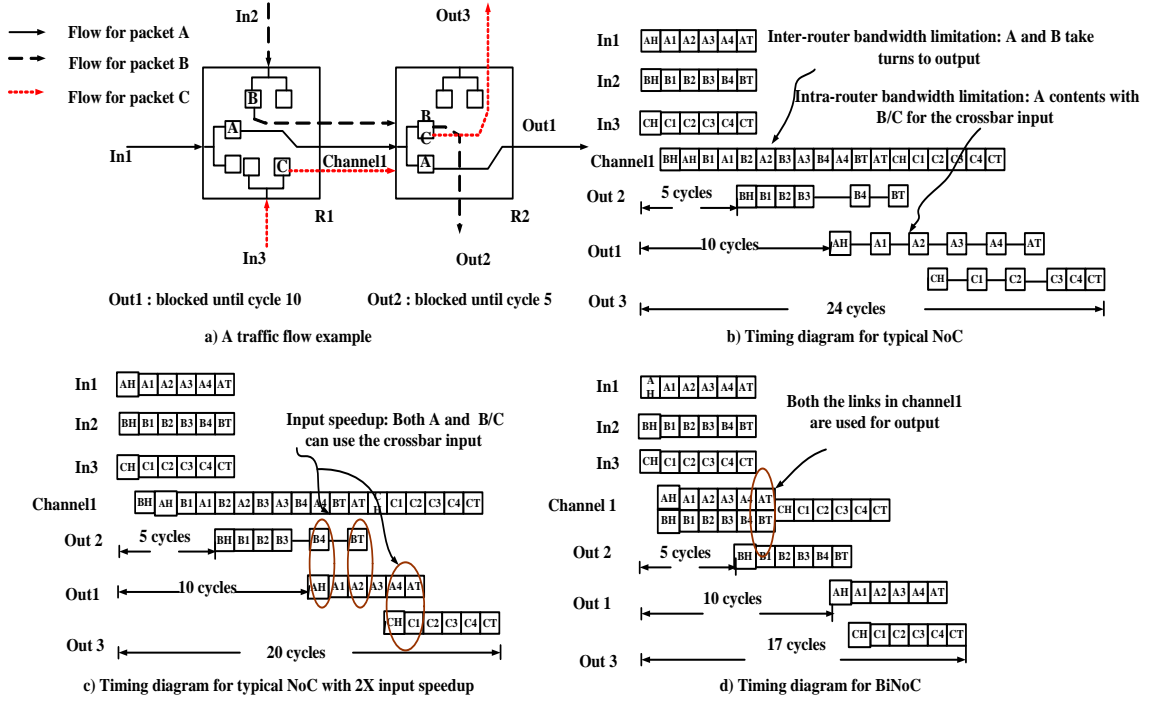}
\caption{\label{fig:A-motivational-example}The comparisons of different architectures
a) example traffic flows b) timing diagram under typical uni-directional
NoC c) timing diagram for typical NoC with 2X internal bandwidth d)
timing diagram for BiNoCs}
\end{figure}

To illustrate the difference in bandwidth utilization under different
NoC architectures, we consider a case of three packets A, B, C traversing
from a router R1 to a router R2 as shown in Fig. \ref{fig:A-motivational-example}-a.
We denote the physical bandwidth of \textit{Channel1} pointing from
router the R1 to R2 as $B$ (flits) and the number of input ports
in the routers as $N$. The packets in flow A, B and C have fixed
length (in this example it is equal to $6$ flits) including the header
and the tail flit. A fair round-robin policy is adopted in the virtual
channel and switch arbitration. At the beginning, the packets A, B
and C need to contend for the VCs in R2. As there are only two VCs
available at each port, we assume packets A and B successfully acquire
the virtual channels while packet C has to wait until A or B finishes
transmission and releases the occupied VC. We assume the packet stalling
\cite{Book} occur in R2 during the transmission of packet A and B.
The blockage times for packets A and B in ports Out1 and Out2 are
$10$ and $5$ cycles, respectively. Moreover, for the sake of simplicity,
in this example, we assume the processing time of a packet transfer
is only one cycle in the routers R1 and R2. For the state-of-the-art
NoC architectures, there are three ways to utilize the NoC inter-
and intra-router bandwidth:

1) \textit{Typical uni-directional NoC switching}: In a typical uni-directional
NoC router \cite{NoCbook_Peh}, the internal crossbar size is $N\times N$
while the inter-router bandwidth $B=1$. Therefore, two kinds of bandwidth
limitations exist in the transmission as shown in Fig. \ref{fig:A-motivational-example}-b.
The first limitation is the inter-router bandwidth limitation, where
as there is only one unidirectional link between R1 and R2 ($B=1$),
packets A and B have to alternately use this single capacity link
(\textit{i.e., Channel1}). The second is the intra-router bandwidth
limitation where in R2, the two virtual channels in the west input
need to compete one of the $N$ entries in the crossbar during the
SA phase. Therefore packet A and B/C can not be transmitted at the
same time.

2) \textit{Uni-directional NoC with input speedup switching}: The
input speedup scheme provides $2X$ internal bandwidth in the input
crossbar side and increases the crossbar size to $2N\times N$. The
intra-router bandwidth limitation can be overcome as shown in Fig.
\ref{fig:A-motivational-example}-c, where both the VCs in R2 can
use the crossbar entries simultaneously. Thus the latency of packet
C is reduced to 20 cycles. However, the inter-router bandwidth limitation
still exists in the transmission.

3)\textit{Bi-directional NoC switching}: In BiNoCs \cite{5715603},
since there are no traffic flows from R2 to R1, the link originally
pointing from R2 to R1 can be reverted which doubles the inter-router
bandwidth from R1 to R2 (\textit{i.e.,} $B=2$). Consequently, as
shown in Fig. \ref{fig:A-motivational-example}-d, both packets A
and B can be transmitted at the same time and the delay of packet
C can be further reduced to 17 cycles.

\begin{figure}[h]
\includegraphics[width=0.90\columnwidth]{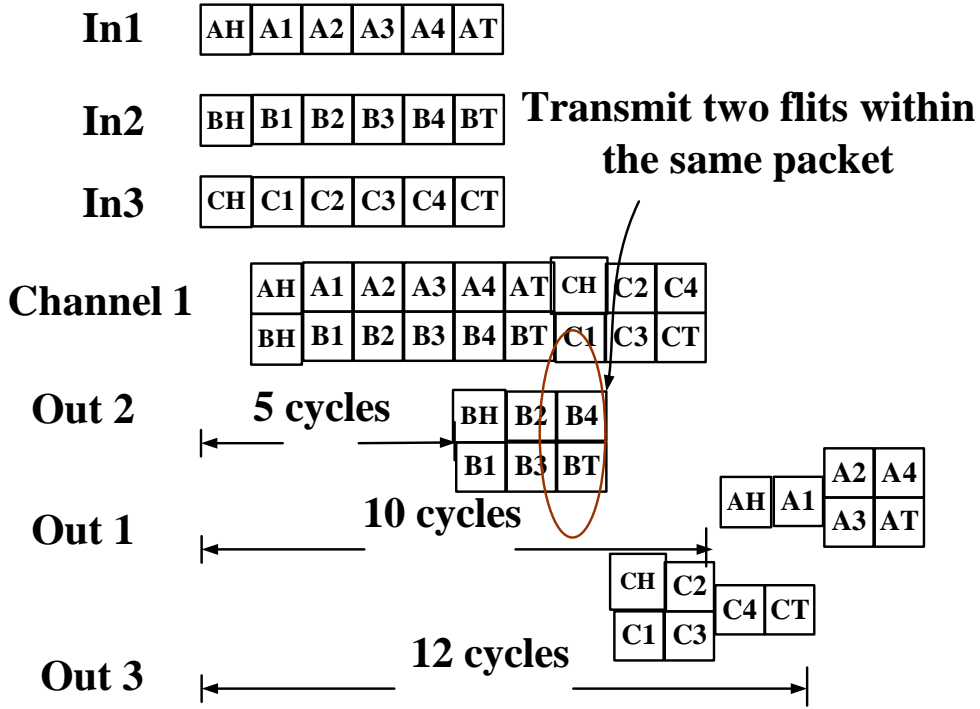}
\caption{\label{fig:The-proposed-flit-level}The proposed flit-level speedup
switching timing diagram for the case in Fig. \ref{fig:A-motivational-example}-a }
\end{figure}

In BiNoC switching, after packets A and B using \textit{Channel1 }simultaneously,
the transmission of flits in packet C only needs one flit-width channel
(shown in Fig. \ref{fig:A-motivational-example}-d). In other words,
half of the effective bandwidth between R1 and R2 remains unused.
Further performance gain can be achieved if we increase the inter-router
bandwidth utilization under this case. Towards this end, we propose
a flit-level speedup scheme which allows flits within the same packet
to use the doubled bandwidth in addition to those from different packets.
Fig. \ref{fig:The-proposed-flit-level} depicts the timing diagram
that allows flits within the same packet (A,B,C) to be transmitted
simultaneously. As can be seen from the figure, the latency of packet
A/B is reduced due to two flits are sent together towards output port
\textit{Out2/Out3}. On the other hand, the latency of packet C is
improved because its waiting time is greatly reduced as the packet
B in front of it leaves and releases the VC in R2 earlier than in
the BiNoC case. In this example, these two effects  further improve
the latency of the three packets by 12.5\%, 27.2\% and 29.4\%, respectively.
\begin{figure}
\includegraphics[width=0.95\columnwidth]{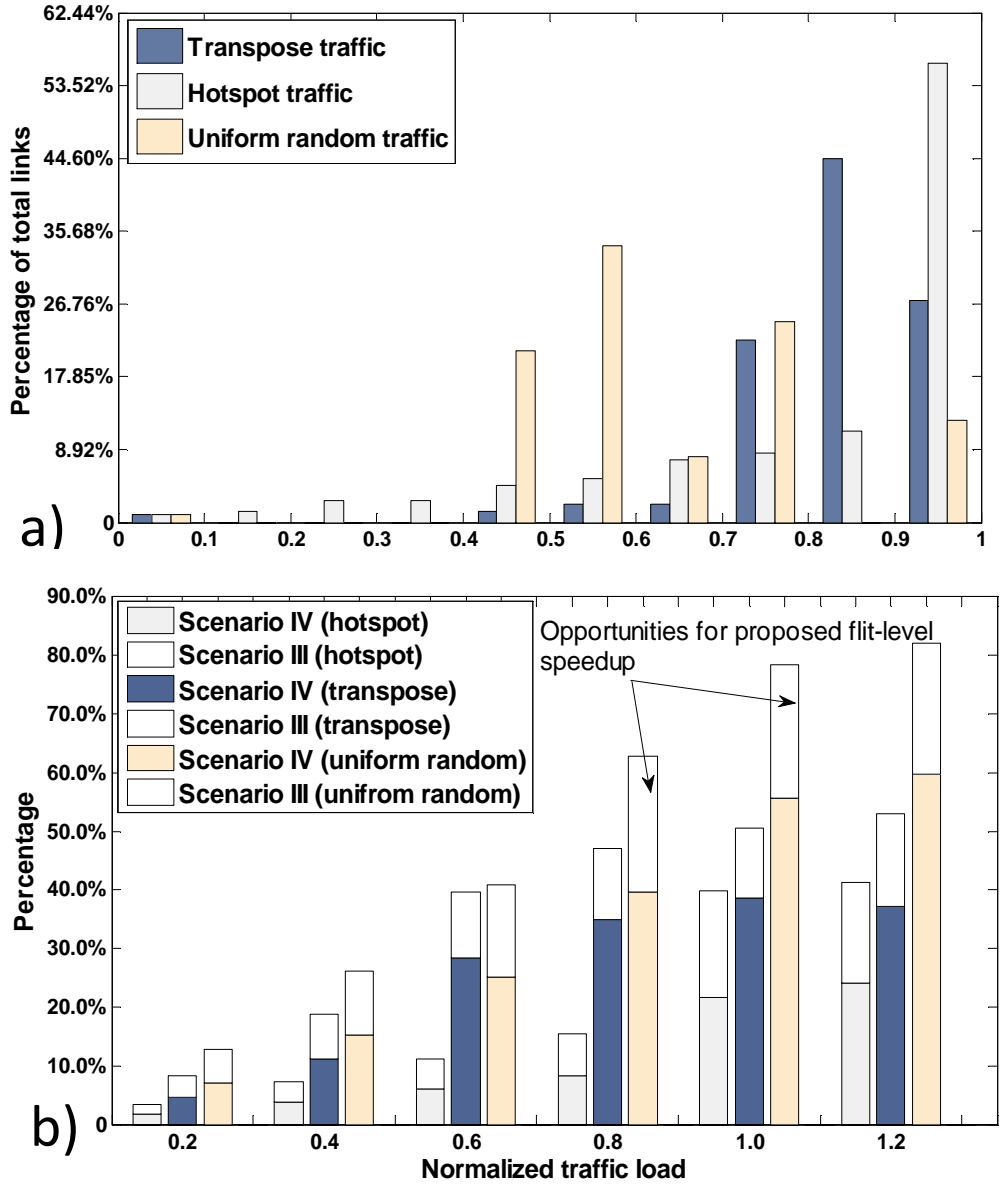}
\caption{\label{fig:Proportion-of-two-1} Profiling results on an $8\times8$
mesh under different traffic patterns: a) Histogram of percentage
of times that no flows in the opposite link direction at run time
b) Breakdown of scenarios III and IV under the condition that there
is no contention in the opposite link }
\end{figure}

The profiling results in Fig. \ref{fig:Proportion-of-two-1} confirms
the high opportunity of using the proposed flit-level speedup scheme.
More specifically, this is based on the following observations:

1) \textit{The uneven traffic between neighboring routers}: For a
channel between two routers, while there is packet sending from one
end to the other, there is a significant portion of time that no packet
is coming from the opposite direction. Fig. \ref{fig:Proportion-of-two-1}-a
shows the profiling results on an $8\times8$ mesh NoC architecture
under $80\%$ of the saturation injection rate of three synthetic
traffic patterns (\textit{i.e.,} uniform random, hotspot and transpose).
As can be seen from the figure, in uneven traffic patterns (\textit{e.g.}
transpose), for more than 80\% of the links, the probability that
there is no packets in the opposite direction is larger than $0.7$
at run time. Therefore, for these links, more than $70\%$ of time,
the effective bandwidth during transmission can be doubled via alternating
the opposite link directions. Even for uniform random traffic, most
of the links (67.50\% in Fig. \ref{fig:Proportion-of-two-1}-a) have
more than $0.5$ possibility to double the bandwidth.

2) \textit{The breakdown of the scenarios when there are no flows
in the opposite direction:} Under this condition, we can further distinguish
into four scenarios for the router under consideration.  They are:
I) the router has no packet to send through the link, II) the router
has only one packet with a single flit to be sent through the link,
III) the router has one packet with multiple flits to be sent through
the link, IV) the router has multiple packets requesting the output
link. For scenarios I and II, we do not need to alternate the opposite
link as the original link bandwidth is sufficient for the transmission.
However, for the last two scenarios, where scenario III and IV corresponds
to flit-level speedup and bi-directional switching, respectively,
we can increase bandwidth and hence the performance if we can reverse
the direction of the opposite link and send two flits through the
two links at the same time. Figure \ref{fig:Proportion-of-two-1}-b
shows the portion of time that scenario III and IV occur among all
of the above four scenarios under various injection rates and traffic
patterns for an $8\times8$ mesh NoC architecture.  Under low injection
rate, since there is little congestion in the network, scenarios I
and II dominate the occurrence. On the other hand, when the injection
rate increases, the probability that we can use the double capacity
of the bi-directional link increases. As shown in the figure, scenario
III occurs frequently in run time. However the state-of-art bidirectional
switching schemes will only work for scenario IV \cite{5715603}\cite{BiNoC1}.

\begin{figure}
\includegraphics[width=0.95\columnwidth]{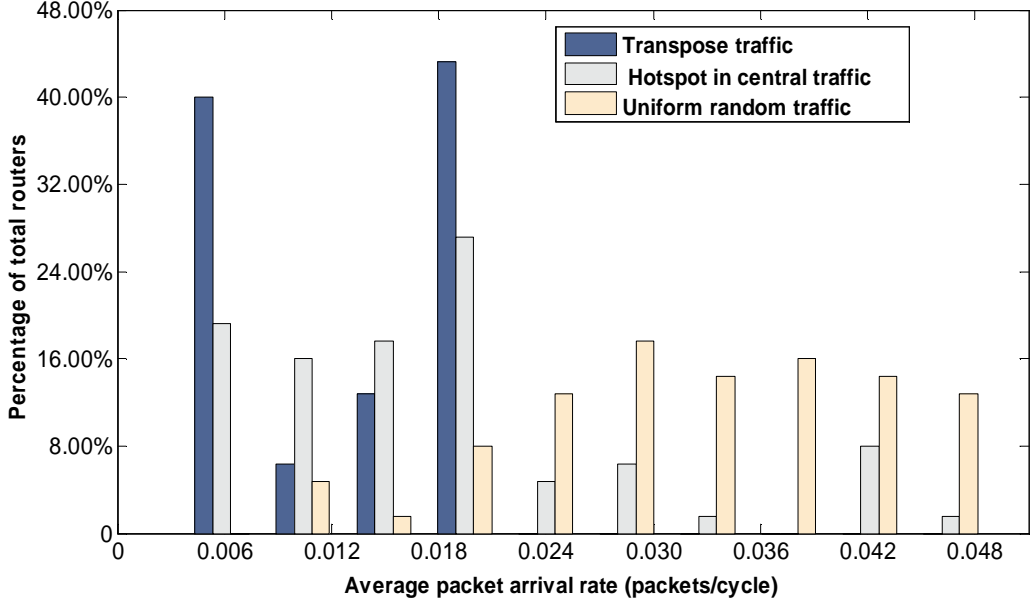}
\caption{\label{fig:The-histogram-of}The histogram of $\beta$ on an $8\times8$
mesh under $80\%$ saturation injection rate}
\end{figure}

\subsection{Single router throughput comparisons }

FSNoC is further motivated by the throughput comparisons of a single
router. We design an event-driven simulator for a single router to
model four different switching schemes: 1) Typical uni-directional
NoC, 2) Typical NoC with 2X input speedup, 3) Bi-directional NoC and
4) Proposed FSNoC. The simulator is based on two parameters $\alpha$
and $\beta$. The\textit{ $\alpha$} parameter represents the probability
that there is no flows in the opposite link (as shown in Fig. \ref{fig:Proportion-of-two-1}-a)
and therefore both two link channels of that direction are enabled
for sending data. The $\beta$ parameter describes the average packet
arrival rate at each crossbar input%
\footnote{In a VC-based router architecture, the internal physical bandwidth
of a router is mainly dictated by its crossbar input bandwidth. Therefore
we abstract the packet arrival process $\beta$ with respect to each
crossbar input entry instead of each VC in the input port.%
}. Both $\alpha$ and $\beta$ can be obtained by profiling the target
applications on specific NoC architectures and are treated as input
to the simulator. For example, let $\beta_{r}$ denote the mean packet
arrival rate of router $r$ over all crossbar inputs in $r$. For
three traffic patterns (random, transpose and hotspot) and under $80\%$
saturation injection rate, the histogram of $\beta_{r}$ is shown
in Fig. \ref{fig:The-histogram-of}. As can be seen from the figure,
the overall average packet rate $\beta$ for random, hotspot and transpose
traffic patterns are $0.033$, $0.013$ and $0.018$ (packets/cycle),
respectively. Similarly, the $\alpha$ values under the same configurations
for these three traffics are $0.67$, $0.84$ and $0.85$. These $\alpha$
and $\beta$ values can be passed to the simulator to evaluate the
single router throughput under a specific traffic pattern. Although
being simplified, the single router throughput analysis can provide
some useful insights on the performance impacts of FSNoC and other
architectures. 

\begin{figure}
\includegraphics[width=0.97\linewidth]{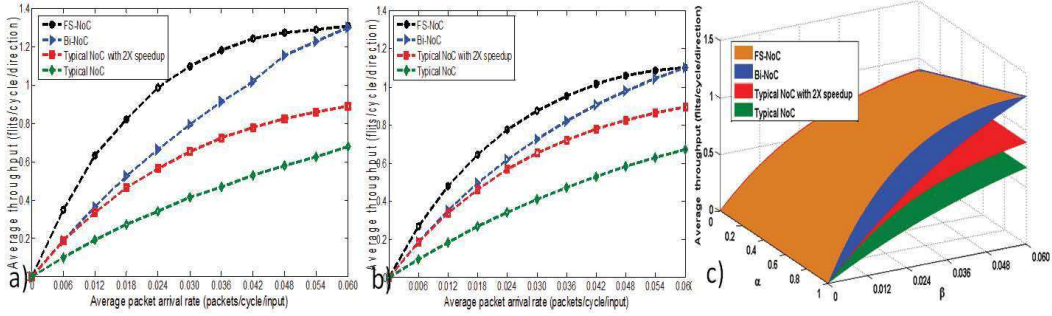}
\caption{\label{fig:Analytical-model-validation}Single router throughput comparisons
of different router architectures: a) Throughput versus packet arrival
rate $\bar{\beta}$ under $\bar{\alpha}=1.0$. b) Throughput versus
packet arrival rate $\bar{\beta}$ given $\bar{\alpha}=0.5$ c) The
throughput comparisons of four NoC architectures under a variety
of $(\bar{\alpha},\bar{\beta})$ combinations.}
\end{figure}

In Fig. \ref{fig:Analytical-model-validation}-a and -b, The simulated
throughputs versus different packet arrival rates $\bar{\beta}$ are
shown with $\bar{\alpha}=1.0$ and $0.5$, respectively. It is observed
that when the packet arrival rate approximates a very large value
(\textit{i.e.,} $\bar{\beta}\rightarrow0.06$), the BiNoC and FSNoC
have the same saturation throughput. This is because under this situation,
all the crossbar entries tend to be occupied by different packets
and scenario III discussed in sub-section \ref{sub:Motivations-of-flit-level}
dominates the time that can use $2X$ bandwidth. However, as profiled
in Fig. \ref{fig:The-histogram-of}, many routers tend to have a smaller
$\bar{\beta}$ (\textit{e.g.,} $0.018-0.033$) at run time, which
provides the opportunities for FSNoC to enhance the system performance.
The throughput under a variety of $\bar{\alpha}$
and $\bar{\beta}$ combinations can be obtained by fitting the simulation curves in Fig. \ref{fig:Analytical-model-validation}-a and -b as shown in Fig. \ref{fig:Analytical-model-validation}-c.
As can be observed from the figure, the FSNoC achieves about $10\%-25\%$
improvement over the BiNoCs for a wide range of $\alpha$ and $\beta$
values(\textit{e.g.,}$\alpha=[0.6,1.0]$ and $\beta=[0.012,0.045]$). 

Based on the above observations, we propose FSNoC which improves the
NoC performance by adding more flexibility in using the bandwidth.
For the inter-router transmission, bidirectional links are employed
to provide a double bandwidth at run time. For the intra-router transmission,
a new input buffer organization and a switch allocator are designed
to allow flits within the same packet to participate in the routing
pipelines. \vspace{-1mm}

\section{Implementation of FSNoC\label{sec:Implementation-of-FSNoC}}

In this section, we present the details of our proposed flit-level
speedup scheme. The channel direction control protocol which adapts
the link directions based on the run time traffic conditions will
be described first. Then we present the router datapath design to
support flit-level parallel transmission. \vspace{-2mm}

\subsection{\label{sub:Inter-router-channel-direction}Inter-router channel direction
control }

\begin{figure}
\includegraphics[width=0.98\columnwidth]{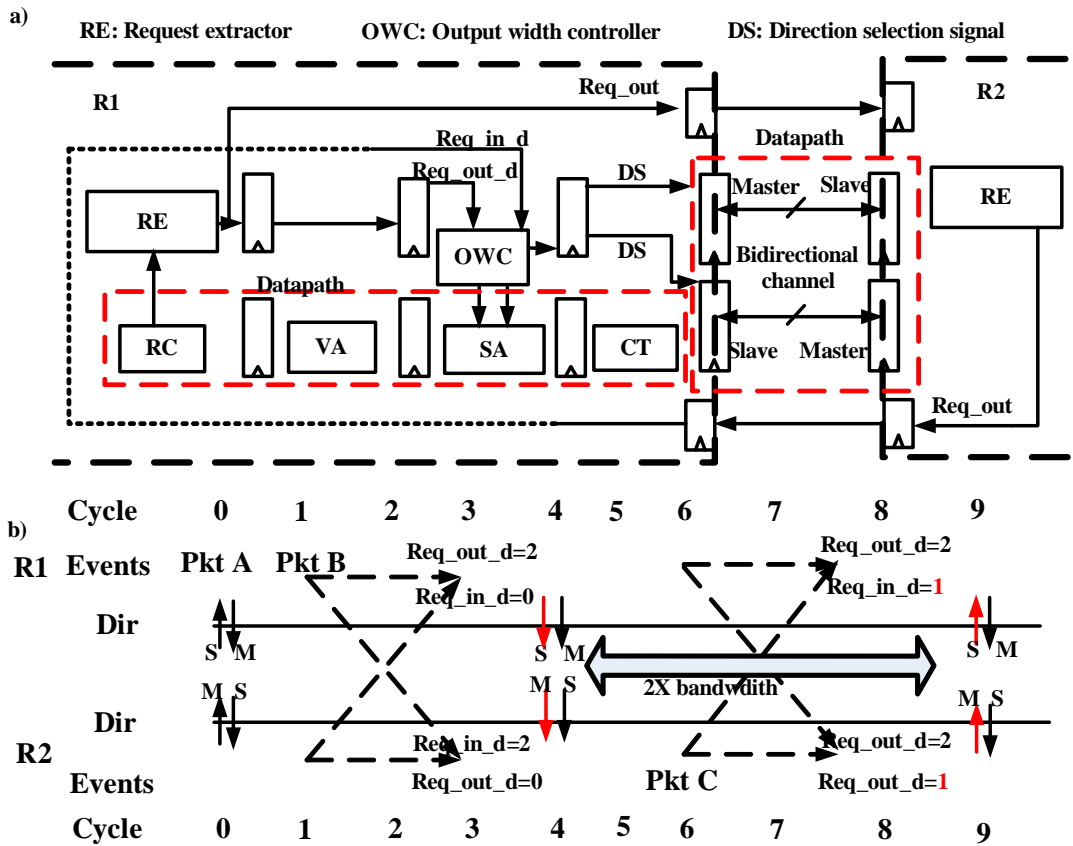}
\caption{\label{fig:The-channel-direction}a) The channel direction control
(CDC) module b) The timing diagram of link direction switching under
the arrival of three packets (A,B,C) at router R1 and R2}
\end{figure}

In FSNoC with bi-directional channels, both links connecting to one
direction of the router can be reconfigured as sending or receiving.
However, in order to keep the original order of the flits within the
same packet, we define one link as the master and the other as the
slave. As shown in Fig. \ref{fig:The-channel-direction}-a, the master
link of a router is also the slave link of its neighbor and vice versa.
If there is traffic in both directions between two neighboring routers,
the master and slave links acts as the sending and receiving links,
respectively. In order to keep the order of flits within the same
packet during the flit-level parallel transmission, when writing two
flits from the same VC into the output channel, the output controller
will always put the first flit on the master link and the second flit
on its slave link. On the receiving side, we always assemble the flit
appeared on the slave link ahead of the flit on the master link into
the VC buffer.

Based on the master and slave link definition, we develop a new channel
direction control (CDC) protocol to configure the links at run time
so as to adapt for various traffic conditions. Previously, there are
two ways to do the direction control, \textit{i.e.,} the centralized
and distributed control. For example, \cite{BiNoC1} employs a separate
bandwidth allocator between two adjacent routers so as to determine
the link directions simultaneously for both routers, while \cite{5715603}
uses a distributed control mechanism without the need of an extra controller.
In this work, the proposed CDC protocol is also based on a distributed
scheme since it is more modular and easier to implement. Therefore,
we mainly compare the proposed CDC protocol with that used in \cite{5715603}. 

\begin{algorithm}
\caption{\label{alg:Operation-of-request}Operation of the RE module}
\begin{algorithmic}[1]
\STATE {\textbf{Input:} Virtual channel status $VS[N][V]$}

\STATE {\textbf{Output:} Pressure signal $Req\_out[N]$ and local pressure register}

\STATE {bank $Req[N]$}

\IF {$clock.reset(\,)$}

\FORALL {each input direction $i\in[1,N]$}

\STATE {$Req[i]=0$}

\FOR {each VC channel $v\in[1,V]$}

\STATE {$VS[i][v].status=idle$;}

\ENDFOR

\ENDFOR

\ENDIF \COMMENT {initialize the pressure counter}

\STATE {always $clock.posedge(\:)$}

\FORALL {input direction $i\in[1,N]$}

\FORALL {VC channel $v\in[1,V]$}

\IF {$VS[i][v].status$==$assigned$}

\STATE {$k=VS[i][v].Get\_output(\:);$}

\STATE {$Req[k].add\_request(\:);$}

\STATE {$VS[i][v].status$==$active$;}

\ELSIF {$VS[i][v].status$==$released$}

\STATE {$k=VS[i][v].Get\_output(\:);$}

\STATE {$Req[k].decrease\_request(\:);$}

\STATE {$VS[i][v].status$==$idle$;}

\ENDIF \COMMENT {add or decrease the request}

\ENDFOR

\ENDFOR

\FOR {each output direction $o\in[1,N]$} 

\IF {$Req[o].size(\:)\geq2$}

\STATE  {$Req\_out[o]=2$;}

\ELSE 

\STATE {$Req\_out[o]=Req[o]$;}

\ENDIF

\STATE {$Write\_out(Req\_out[o]);$}

\ENDFOR
\end{algorithmic}
\end{algorithm}

For the CDC proposed in \cite{5715603}, there are two Finite State
Machines (FSMs) in each router, one for the master link with high
priority (\textit{i.e.,} HP-FSM) and the other for the slave link
with low priority (\textit{i.e.,} LP-FSM). The direction of each link
depends on the state of the corresponding FSM. Since the incoming
request signal from the neighboring router takes two cycles to arrive
at the current router, a "wait" state is required when the FSM
transits from a "receiving" state to a "sending" state
which will introduce $1\sim4$ dead cycles in the channel direction
reversal process \cite{5715603}.

In this work, the proposed channel control scheme is shown in Fig.
\ref{fig:The-channel-direction}-a, which embeds the inter-rouer request
signal delays within the four-stages pipeline depth and uses a single
module to decide the output width ($1$ or $2$ flits) for each direction
instead of two separate FSMs. Two modules, namely Request extractor
(RE) and Output width controller (OWC), are introduced to work together
with the conventional four pipeline stages in the datapath. The RE
module monitors the input channel status and generate the pressure
signals to the corresponding OWC modules in the current and the neighboring
router. The pressure signal of an output direction, which is named
as Req\_out in Fig. \ref{fig:The-channel-direction}-a, represents
the number of virtual channels (VCs) that requests to send out towards
this direction after being granted a VC in the downstream router.
Similar to \cite{5715603}, in order to provide clean signals from
router to router in the transmission, we assume both the data and
control signals are doubly registered (shown in Fig. \ref{fig:The-channel-direction}-a).
Therefore, it takes two cycles for Req\_out to arrive at the neighboring
router and become the input request signal (\textit{e.g.,} Req\_in\_d
shown in Fig. \ref{fig:The-channel-direction}-a) of the OWC module
in the neighboring router. The OWC module within each router then
works with the switch allocator to process the local pressure signal
Req\_out\_d as well as the received neighbor pressure signal Req\_in\_d
and determines the width of each output direction ($2$ flits, $1$
flit or $0$ flit) accordingly. During the crossbar traversal (CT)
and link traversal (LT) stage, the direction control signals (DS)
of the master and slave link are generated based on the output width
calculated in the SA stage and are used to configure the write and
read operations. 

In the proposed CDC protocol, the pressure signal needs two cycles
to arrive at the neighboring router and is required to be used in
the SA/OWC stage. In order to guarantee both the OWC modules in router
R1 and R2 receive the Req\_in\_d and Req\_out\_d signals generated
at the same time, the RE module extracts the pressure signals (\textit{i.e.,}
Req\_out) at the RC stage instead of the VA stage. The effect of generating
the signal in the RC stage is that the VC allocation result is updated
one clock cycle later when a packet first enters the virtual channel
buffer. However, it only affects the performance when the neighboring
router requests to send two flits, while the newly arrived packet
in the current router wins the VC allocation after the RC stage. In
the next cycle, during the switch allocation (SA) stage, the Req\_out\_d
signal received by the OWC module may still be $0$ as it was generated
two cycles ago and the latest VA results has not yet been updated.
In this case, the current router may treat this as there is no request
for sending and the link will then be configured as receiving for
both channels while it should be configured as one sending and one
receiving. The correctness of the router operation is not compromised,
only some of the priority of sending out a flit in current router
is lost in this special case. From our simulation results, it is shown
that there is no significant difference in the latency and throughput
when the pressure signal is generated at the routing computation stage
compared with that generated at the VC allocation stage.

Fig \ref{fig:The-channel-direction}-b shows a typical timing diagram
of the link direction switching under the proposed CDC protocol. Assume
two packets $A$ and $B$ arrived at $R1$ are granted downstream VCs in cycle $0$ and $1$,
respectively. They are forwarded towards router $R2$. Also, we assume
a packet $C$ arrives at router $R2$ and is routed to $R1$ after being allocated a VC in cycle
$6$. In cycle $1$, the RE module in $R1$ extracts a pressure signal
$Req\_out=2$ and sends it to the OWC modules in $R1$ and $R2$.
At the same time, since there are no requests from $R2$ to $R1$,
the pressure signal generated at $R2$ is still $0$. In cycle $3$,
the OWC module in $R1$ receives the local pressure signal $Req\_out\_d=2$
from the RE module and the neighbor pressure signal $Req\_in\_d=0$
from $R2$. The OWC in $R1$ then enables both channels corresponding
to the master and slave link during the switch arbitration (SA) for
packet $A$ and $B$. On the other hand, the OWC module in $R2$ disables
its master and slave channel during the SA stage to prepare for receiving
two flits at one time. After cycle $5$, each flit from $A$ and $B$
can traverse the crossbar and the link using the $2X$ bandwidth together.
In cycle $8$ when the neighboring pressure signal $Req\_in\_out$
received by $R1$ turns to be $1$ due to packet $C$ being granted VC
in cycle $6$, the slave link channel in $R1$ is disabled in the
$SA$ stage and re-configured to the direction of read from cycle
$9$. In contrast, the master link channel in $R2$ is enabled again
during the SA stage for packet $C$. Therefore, flits from packet
$C$ can begin to traverse the crossbar from cycle $9$ smoothly.

The operation of the request extractor is described in Algorithm \ref{alg:Operation-of-request}.
For each VC, its status consists of four states: 1) $idle$: the channel
is waiting for the VC allocation in the corresponding output direction.
2) $assigned$: the packet has just been granted the VC in the downstream
router and has not been included in the $Req[N]$ yet. 3) $active$:
the VC occupies the downstream VC for transmission. 4) $released$:
The tail flit has left the buffer in the last cycle and released the
downstream VC. In the beginning, all the channel status are reset
to $idle$ and the request registers $Req[N]$ are cleared to $0$
(line 3-10). At run time, for a specific channel $v$ of direction
$i$, if its status is $assigned$, then we need to get the output
direction $k$ of this VC and increased the pressure value $Req[k]$
by one (line 14-17). On the other hand, if the VC status is $released$,
we need to decrease the pressure value of direction $k$ by one (line
18-22). Finally, the pressure values in $Req[N]$ are passed to the
OWC modules in both current and neighbor router (line 31). 

Algorithm \ref{alg:Operation-of-the} describes the operation of the
output width controller. It dynamically configures the data width
of each output direction (0, 1 or 2 flits) by enabling/disabling the
master link ($MLA[N]$) and slave link ($SLA[N]$), respectively.
By default, the master link is enabled for write operations while
the slave link is disabled for read. At run time, if the local pressure
signal ($Req\_out\_d[o]$ ) indicates that there are $2$ requests
from different input VCs towards the same direction $o$ and there
are no flows in the opposite direction (dictated by $Req\_in\_d[o]$),
the controller will enable both the master and the slave link channels
corresponding to output direction $o$ to participate in the switch
arbitration (line 18-19). Similarly, if the local pressure equals
to $1$ while the neighbor pressure signal equals to $0$, the slave
link will be enabled to support flit-level parallel transmission (line
14-16). On the other hand, if $Req\_out\_d[o]$ equals to $0$ and
the $Req\_in\_d[o]$ is larger than $0$, both the master and slave
links will be disabled to let the neighboring router output using
the $2X$ bandwidth (line 12-13). In addition to generate $MLA$ and
$SLA$ signals, the OWC module also determines the input switch arbitration
mode $SA\_in[N][V]$. By default, $SA\_in[N][V]$ for each VC is set
to be $normal$ (line 4-8), which means only one flit from this VC
can participate in the SA stage every cycle. However, under the case
of flit-level speedup, the VC can send two flits at the same time.
Therefore, under this condition, the switch arbitration mode of the
requesting VC is set to be $flit\_speedup$ (line 17), which allows
two flits within the same packet to participate in the arbitration
for both the master and slave links of the corresponding output direction.

\begin{algorithm}
\caption{\label{alg:Operation-of-the}Operation of the OWC module}
\begin{algorithmic}[1]
\STATE {\textbf{Input:} Pressure signal $Req\_out\_d[N]$ and $Req\_in\_d[N]$}

\STATE {\textbf{Output:} Master/slave link enable $MLA[N]$ and$SLA[N]$, Switch allocation
mode $SA\_in[N][V]$}

\STATE {always $clock.posedge(\:)$}

\FORALL {input direction $i\in[1,N]$}

\FORALL {VC channel $v\in[1,V]$} 

\STATE {$SA\_in[i][v]=normal;$}

\ENDFOR 

\ENDFOR

\FORALL {output direction $o\in[1,N]$} 

\STATE {$MLA[o]=enable;$}

\STATE {$SLA[o]=disable;$}

\IF {$Req\_out\_d[o]$ == 0\,\&\&$Req\_in\_d[o]>0$}

\STATE {$MLA[o]=disable;$}

\ELSIF {$Req\_out\_d[o]$== 1\,\&\&\, $Req\_in\_d[o]$ == $0$}

\STATE {$SLA[o]=enable;$}

\STATE {$(i,v)=get\_input\_vc(o);$}

\STATE {$SA\_in[i][v]=flit\_speedup;$}

\ELSIF {$Req\_out\_d[o]$ == 2 \,\&\&\, $Req\_in\_d[o]$ == $0$}

\STATE {$SLA[o]=enable;$}

\ENDIF

\ENDFOR
\end{algorithmic}
\end{algorithm}

\subsection{Router datapath design of FSNoC}

Besides the RE and OWC modules added in the control path, the router
datapath also needs to be modified. In FSNoC, the routing computation
(RC) module and virtual channel allocator are identical to those in
the typical NoC \cite{1310774} and BiNoC architecture \cite{5715603}
because the output directions/VCs are computed/reserved in a same
way in all architectures. However, the FSNoC has a number of additional requirements in the input buffer and switch allocator design to support
efficient flit-level parallel transmission.

\subsubsection{Input buffer organization}

\begin{figure}[h]
\includegraphics[width=0.99\columnwidth]{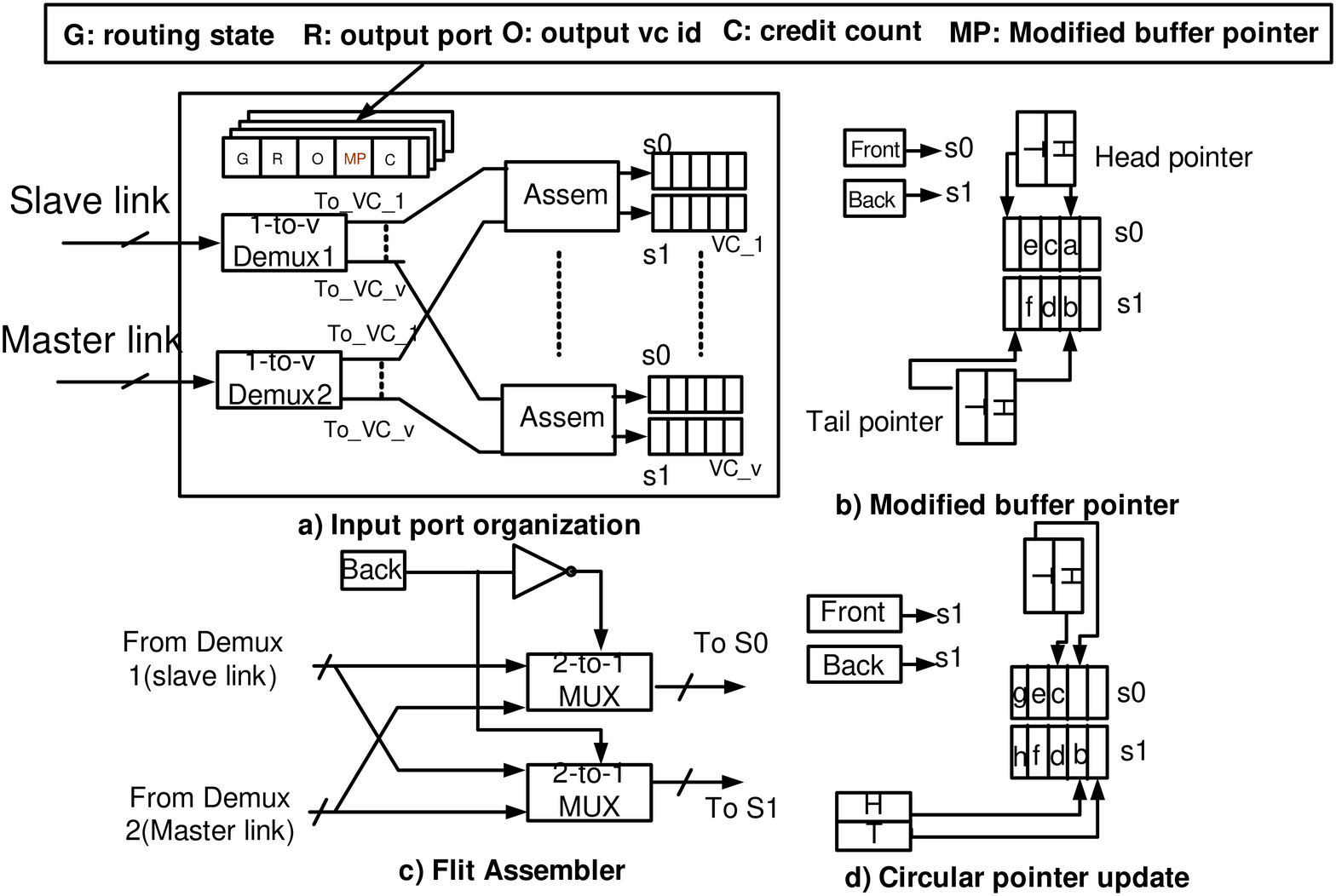}
\caption{\label{fig:Proposed-buffer-organization}Proposed buffer organization }
\end{figure}

In the proposed flit-level speedup scheme, it is possible that two
incoming flits are for the same VC buffer. Therefore, it is required
to support reading/writing two flits from/to the same VC buffer at
the same time. One way is to adopt multiple ports memory. However,
this causes significant overhead in terms of area and memory access
time \cite{Book}. Here we propose a novel input buffer organization,
which is shown in Figure \ref{fig:Proposed-buffer-organization},
to satisfy this requirement.

Figure \ref{fig:Proposed-buffer-organization}-a shows the input port
configuration with $V$ virtual channels. Two $1-to-V$ demux and
$V$ flit assemblers are needed for each port. The original virtual
channel buffer is spitted into two sub-buffers, s0 and s1, respectively.
Both sub buffers share the same buffer state information (the GROPC
vector shown in Figure \ref{fig:Proposed-buffer-organization}-a are defined as in \cite{Book}).
The incoming flits are written into s0 and s1 alternatively so that
two consecutive flits are stored in different sub buffers. Figure
\ref{fig:Proposed-buffer-organization}-b shows the modified buffer
pointer in each virtual channel. For s0 and s1, each  has a head-tail
pointer pair pointing to the start and end addresses of the flits
of the packets.  In addition, two 1-bit registers, denoted as the
Front and Back registers, are added to indicate which sub-buffer holds
the first and the last flit (\textit{e.g.,} flits "a" and "f"
in Figure \ref{fig:Proposed-buffer-organization}-b) of the  VC channel.
\begin{figure}[h]
\includegraphics[width=0.98\columnwidth]{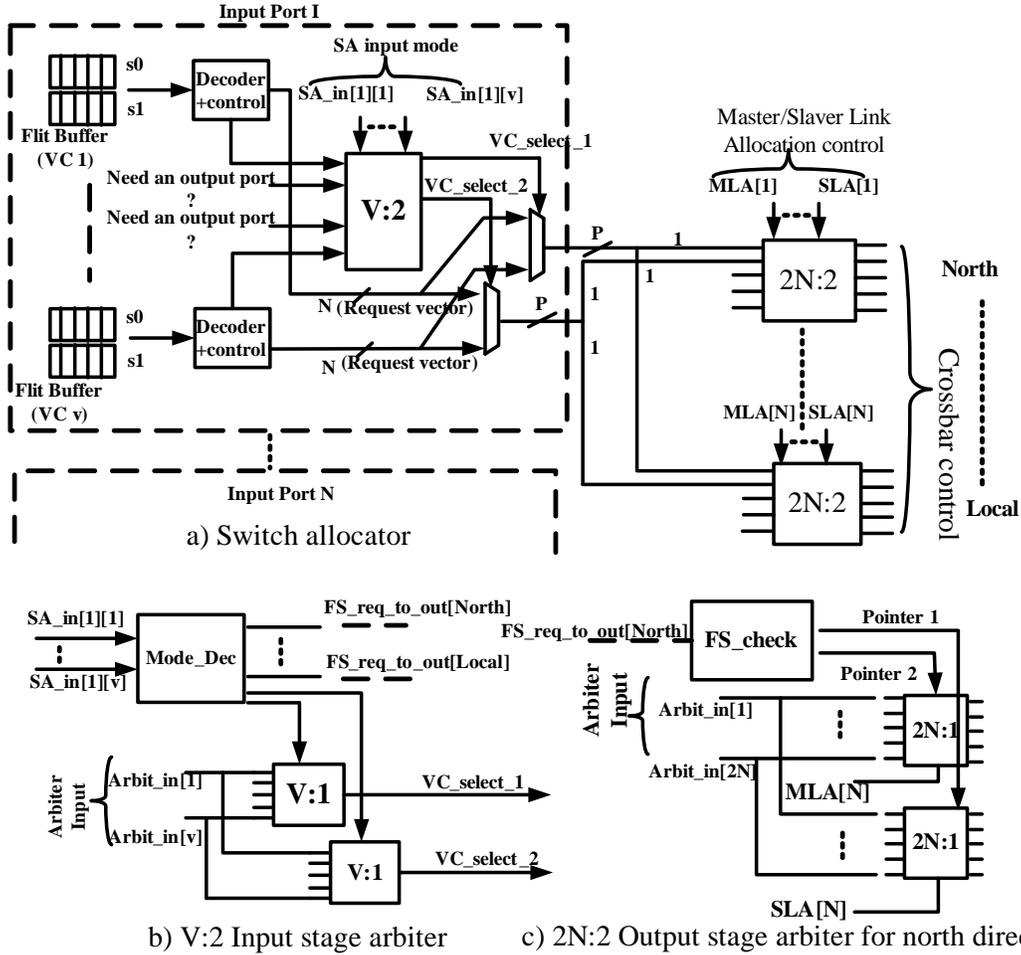}
\caption{\label{fig:Switch-allocator-logic}Switch allocator logic (modified from \cite{1310774})}
\end{figure}

The details of the flit assembler is described in Figure \ref{fig:Proposed-buffer-organization}-c.
It is responsible for assembling the flits into the sub-buffer of
the VC in the right order. When  two flits are sent in the reconfigurable
links, the first flit is always connected to the master link of the
sender (\textit{i.e.,} the slave link of the receiver).  To ensure
the correct assembling of the flits, the slave link is always connected
to the  sub-buffer that is not pointed by the Back register.  Figure
\ref{fig:Proposed-buffer-organization}-b shows an example. The Back
register points to s1 since the last flit "f" is stored in  s1.
The assembler will connect the slave link to s0 and the master link
to s1 in the next operation. Similar operation is executed when only
1 flit is written into the buffer from the input link.  The read operation
is similar to the write operation based on the value of the Front
register. Figure \ref{fig:Proposed-buffer-organization}-d illustrates
the updated pointers after reading a flit "a" from the buffer
and writing flits "g" and "h" into it. As shown in the
figure, the head and tail pointers are updated in a circular manner
while the Front and Back registers are updated based on the number
of flits reading from or writing into the virtual channels.

\subsubsection{Switch allocator design }

The switch allocator assigns the master and slave link channel of
each output direction to the input VCs such that the VCs with granted
switch access can move flits to the corresponding links in the crossbar
and link traversal stage. Different from BiNoC, to support the flit-level
speedup in the NoC, the switch allocator should not only support granting
two requests from different VCs, but also allow a single VC to arbitrate
for both the master and slave links. Figure \ref{fig:Switch-allocator-logic}-a
shows the building block of our switch allocator. This allocator is
accomplished with a two-stage arbitration architecture \cite{Book}, which is
shown in Fig. \ref{fig:Switch-allocator-logic}-a. At the input
stage, a $V:2$ arbiter in each input direction is used to select
$2$ requests out of many requests from the $V$ virtual channels.
Of note, the $2$ winners can come from different VCs or from the
same VC. This is determined by the SA arbitration mode vector ($SA\_in[N][V]$)
discussed in section \ref{sub:Inter-router-channel-direction}. If
all the VCs in the current input direction $i$ have the same mode
$normal$ (\textit{i.e.,} $SA\_in[i][v]=normal$, $\forall v\in[1,V]$),
then the two winning requests come from different VCs. On the other
hand, if some of the VCs have the SA mode $flit\_speedup$, the two
winning requests are combined together and returned to one of these
flit-level speedup VCs. At the output side of the switch allocator,
$N$ $2N:2$ arbiters are utilized to decide the actual number of
the requests that is finally granted.

Figure \ref{fig:Switch-allocator-logic}-b shows the detail design
of the $V:2$ arbiter for the input stage allocation. It consists
of two $V:1$ round-robin arbiters and a mode decision module named
as $Mode\_Dec$. The $Mode\_Dec$ module of direction $i$ reads in
the SA mode vector (\textit{ i.e.,} $SA\_in[i][v],\, v=[1,V]$) and
determines the selection mode for the current input port. If all the
entries of SA vector equal to $normal$, it means that there is no
request that requires two flits to be transmitted from the same VC
and hence the two $v:1$ arbiters will work independently to select
two VCs using different priority pointers . Otherwise, the $Mode\_Dec$
will chooses one VC with the $flit\_speedup$ mode and generate the
priority pointer exactly pointing to that VC for both $V:1$ arbiters. 

Figure \ref{fig:Switch-allocator-logic}-c shows the $2N:2$ arbiter
used in the output side. It consists of two $2N:1$ arbiters. The
link allocation control signals ($MLA[N]$ and $SLA[N]$ discussed
in Algorithm \ref{alg:Operation-of-the}) of each output direction
enable/disable the arbiters accordingly. Moreover, a module named
as $FS\_check$ (shown in Figure \ref{fig:Switch-allocator-logic}-c)
is used to grant two requests to the same input port so  to allow
parallel transmission of two flits from the same VC.

\section{FSNoC for long-wire links}

For many NoC designs, there exist long wire links due to the irregular
topology \cite{long-wire-link} or the short-cut paths created by
long-range links \cite{long-range-link}. In order to address the
issue of high latency and power consumption, repeaters \cite{repeaters-noc}
or pipeline registers \cite{long-range-link,long-wire-link} are usually
inserted between the routers to break the long wires. In this section,
we extend the FSNoC design to support flit-level parallel transmission
over the long wires.
\begin{figure}
\includegraphics[width=0.98\linewidth]{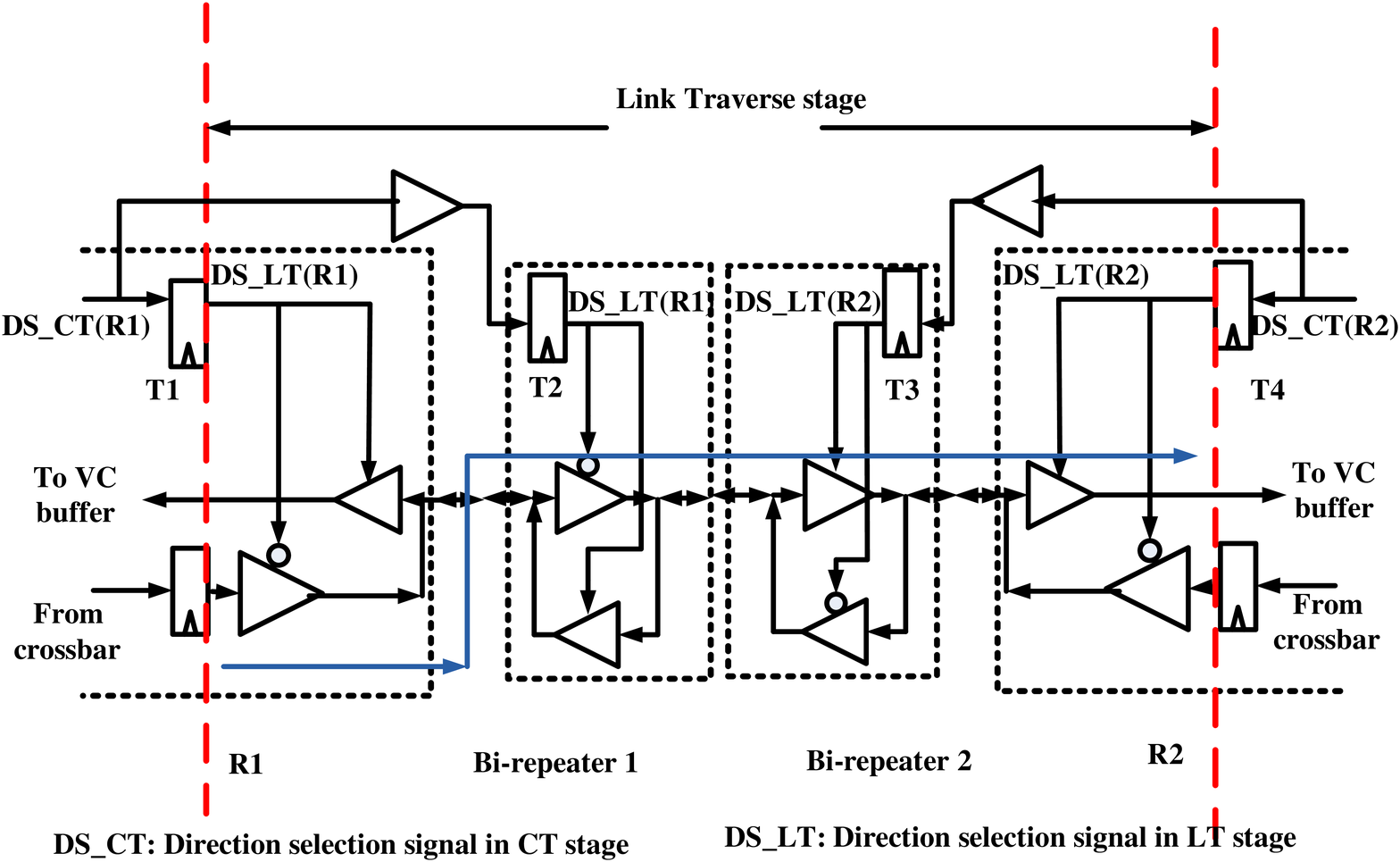}
\caption{\label{fig:FSNoC-design-for}FSNoC design for long wire link with
repeaters}
\end{figure}

\subsection{Long-wire links with repeaters}

For long-wire link with repeaters, it still takes one cycle to move
a flit from one router to the neighbor. In FSNoC, the control channels
(\textit{e.g., $Req\_out$} in Fig. \ref{fig:The-channel-direction}-a)
between two routers are uni-directional. Therefore, the repeaters
can be inserted directly between the sender and receiver as in the
case of typical NoCs. On the other hand, for the data channels, bi-direction
repeaters (Bi-repeater) are needed to allow the data transmission
over both directions. Moreover, the direction selection signal (\textit{i.e.},
$DS$ in Fig. \ref{fig:The-channel-direction}-a) should be propagated
to configure the intermediate Bi-repeaters properly. Fig. \ref{fig:FSNoC-design-for}
shows the detail design of a data channel between two routers $R1$
and $R2$. Within each router, the OWC module determines the output
channel width as well as the $DS$ signal in the SA stage based on
the CDC protocol. Then, the generated $DS$ signal is pipelined together
with the flits in the crossbar traverse (\textit{CT}) and link traverse
(\textit{LT}) stages before reaching the output port to configure
the inout directions. As shown in Fig. \ref{fig:FSNoC-design-for},
the $DS$ signal during the CT and LT stages are represented as $DS\_CT$
and $DS\_LT$, respectively. If there are $n$ Bi-repeaters need to
be inserted between router $R1$ and $R2$ to address the long wire
delays (Seen Fig. \ref{fig:FSNoC-design-for} for a case of two Bi-repeaters),
$\left\lfloor \frac{n+1}{2}\right\rfloor $of them close to $R1$
will be driven by the $DS$ signal from $R1$ while the remaining
are driven by the $DS$ from $R2$. For example, in Fig. \ref{fig:FSNoC-design-for},
Bi-repeater 1 and 2 are driven by $R1$ and $R2$, respectively. The
one-cycle ahead signal $DS\_CT$ from $R1$ and $R2$ are connected
to the registers in the Bi-repeater $1$ and $2$. Every cycle, at
the clock rising edge, the $DS\_CT$ signals will be read into the
registers $T1-T4$ simultaneously. They become the $DS\_LT$ signals
for the current cycle. Then, the $DS\_LT$ signals configure the Bi-repeater
to ensure the write/read directions are the same as the driving router.
As the CDC control protocol discussed in Section \ref{sec:Implementation-of-FSNoC}
guarantees $DS\_LT(R1)$ and $DS\_LT(R2)$ are conflict-free, the
correct data transfer direction can be maintained in the Bi-repeaters.
Also, as the number of pipeline stages do not change, the FSNoC with
repeaters has the same latency as the original design for data transmission.
\begin{figure}[h]
\includegraphics[width=0.98\linewidth]{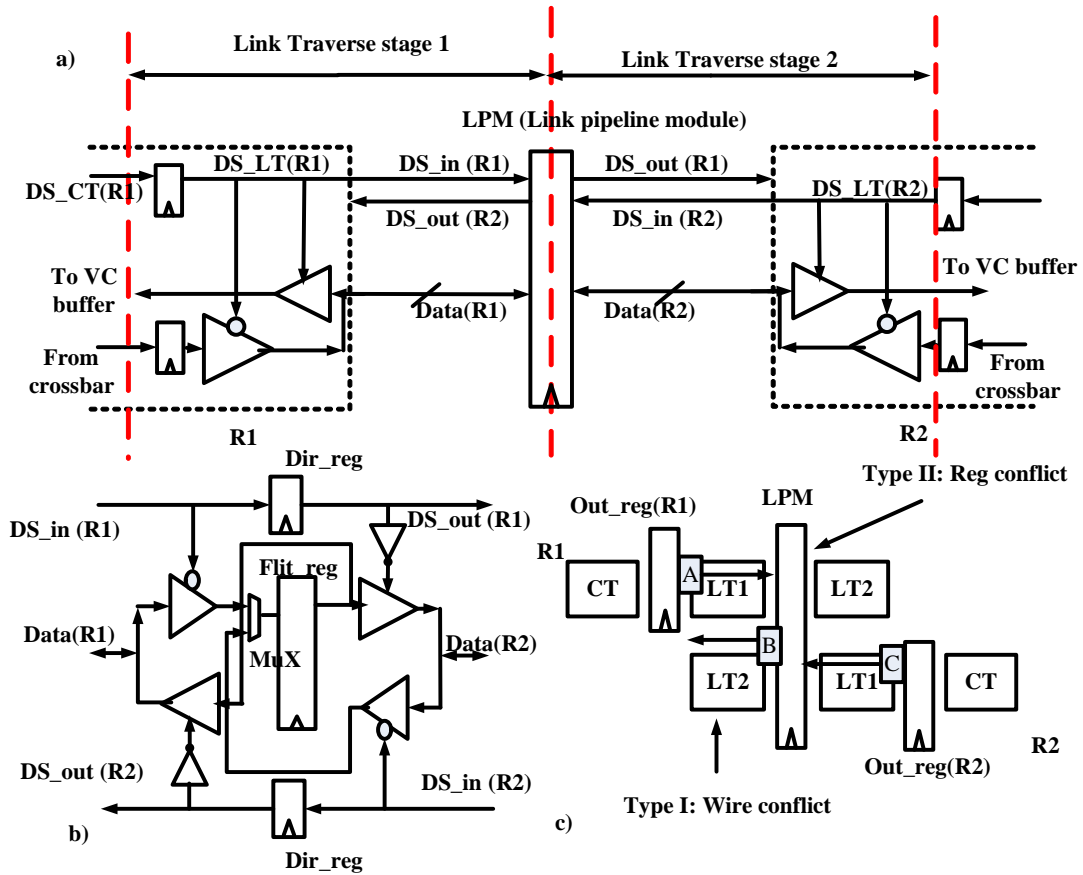}
\caption{\label{fig:FSNoC-design-for-1}a) FSNoC design for long wire links
with pipeline registers, b) the detail schematics for data link pipeline
registers, c) two types of conflicts during the the link traversal
(LT) stage}
\end{figure}

\subsection{Long-wire links with pipeline registers\label{sub:Long-wire-links-with}}

Besides repeaters, pipeline registers can also be used to break long
wire links which achieves better timing properties at the cost of
longer link traversal latency. 

For the control channels, conventional registers can be added directly
as the signals are uni-directional. On the other hand, for the data
channels, the direction selection signals ($DS$ signals) are pipelined
together with the data flits to ensure the proper read and write direction.
This is shown in Fig. \ref{fig:FSNoC-design-for-1}-a, where a link
pipeline module (LPM) is inserted between routers $R1$ and $R2$
for a data link. Two direction selection signals from $R1$ and $R2$,\textit{
i.e.,} $DS\_in(R1)$ and $DS\_in(R2)$, are also connected to the
LPM. The detail schematic of the pipeline module is shown in Fig.
\ref{fig:FSNoC-design-for-1}-b. In LPM, a flit register ($Flit\_reg$)
and two one-bit registers ($Dir\_reg$) are used to buffer the data
flits and direction signals, respectively. Every cycle, based on $DS\_in(R2)$
and $DS\_in(R1)$, a $MuX$ selects whether the flit on $Data(R1)$
or $Data(R2)$ should be written into the internal $Flit\_reg.$ Similarly,
at the output side of $Flit\_reg$, the previous one-cycle pipelined
$DS$ signals, \textit{i.e.,} $DS\_out(R1)$ and \textit{$DS\_out(R2)$,
}determine the link segment ($Data(R1)$ or $Data(R2)$) that current
flit in $Flit\_reg$ needs to be forwarded to. 

Compared to FSNoC in section \ref{sec:Implementation-of-FSNoC},
the additional link traversal cycles due to link pipelining complicate
the channel direction control in two aspects. First, it brings the
\textit{synchronization} issue between the neighboring and local pressure
signals. As shown in Fig. \ref{fig:The-channel-direction}-a, originally,
it takes two cycles for both the neighbor pressure ($Req\_in$) and
local pressure signal ($Req\_out$) to arrive at the $OWC$ module.
As the $OWC$ modules in two adjacent routers make the decision based
on same $(Req\_out\_d,\, Req\_in\_d$) pair every cycle, it is ensured
that no write conflicts (\textit{i.e.,} write from both directions)
occur by the proposed distributed CDC protocol in Algorithm \ref{alg:Operation-of-the}.
On the other hand, here, for the case of $K$-stage pipelined link,
the pressure signal $Req\_out$ takes additional $K$ cycles (\textit{i.e.,}
$K+2$ cycles) to arrive at the $OWC$ in $R1$ while it still needs
two cycles to reach $OWC$ in $R2$. The asynchronization of $OWC$
modules in $R1$ and $R2$ may create incorrect output width combinations
(\textit{e.g.,} an output width of $2$ flits for $R1$ and $1$ flit
for $R2$) and overwrite useful flits on the link. Therefore, in order
to provide correct direction decisions, $K$ additional pipeline registers
should be added between the $RE$ and $OWC$ modules inside each router
to keep the local pressure signals synchronized with the neighbor
pressures. 

The second hurdle due to link pipeline is that \textit{more types
of conflicts} need to be resolved in the CDC protocol. More precisely,
two types of conflicts on a pipelined link are identified, namely
\textit{wire conflict} and \textit{register conflict}. Taking a one-stage
pipelined link in Fig. \ref{fig:FSNoC-design-for-1}-a and -b as
an example, the \textit{wire conflict} happens if both ends try to
write data on the same wire segment. As shown in Fig. \ref{fig:FSNoC-design-for-1}
-a and -c, the output register of router $R1$ will write a flit $A$
on the link segment $Data(R1)$ if the direction selection signal
$DS\_in(R1)$ equals to $OUT$ (\textit{i.e.,} 0). On the other hand,
the $LPM$ will write a flit $B$ on the same link segment if the
direction selection signal $DS\_out(R2)$ equals to $OUT$. Therefore,
a wire conflict happens if both $DS\_out(R2)$ and $DS\_in(R1)$ equal
to $OUT$ at the same time. The\textit{ register conflict} refers
to the operation that try to write flits into the same $LPM$ in one
cycle. For example, if $DS\_in(R1)$ equals to $OUT$, flit $A$ on
$Data(R1)$ will be written into $LPM$ at the next cycle. Similarly,
if $DS\_in(R2)$ equals to $OUT$, flit $C$ on $Data(R2)$ will be
written into $LPM$. Thus, the register conflict occurs if $DS\_in(R1)$
and $DS\_in(R2)$ equal to $OUT$ at the same time. As discussed in
section \ref{sec:Implementation-of-FSNoC}, in FSNoC, the direction
selection signal $DS\_in(R1)$ and $DS\_in(R2)$ are generated in
the same cycle by the $OWC$ module of $R1$ and $R2$, respectively.
Therefore, they cannot be $OUT$ simultaneously under the CDC protocol
in Algorithm \ref{alg:Operation-of-the}, which means the \textit{register
conflict }will not occur. On the other hand, as $DS\_out(R2)$ is
generated and pipelined one cycle before $DS\_in(R2)$, they may both
be equal to $OUT$ at run time and produce the \textit{wire conflicts}
as a result. In order to resolve the wire conflicts in one-stage pipelined
link, the $OWC$ module needs to store the direction selection results
in the previous cycle and ensure the $DS$ signal generated in current
cycle not conflict with that of the neighbor router in both current
and previous cycles. More generally, for a $K$-stage pipelined link
with $K\geq2$, more than one $LPM$s are inserted between two routers.
In order to read and write correctly, we need to guarantee there are
no wire conflicts on the $K+1$ link segments as well as no register
conflicts in the $K$ $LPM$ modules. Therefore, the $OWC$ module
in each router needs to store the $DS$ signals of the previous $K$
cycles. For the master link, its direction will be reversed for the
other router if the current router have not written flits into this
link during the past $K$ cycles. Similarly, for the slave link,
its direction can be reversed for write operations if the current
router receives the neighboring pressure signal $Req\_in=0$ continuously
during the past $K$ cycles.

\section{\label{sec:Simulation-results}{Experimental results}}

\subsection{Simulation setup}

We evaluate and compare the proposed FSNoC architecture with other
designs using a cycle-accurate, SystemC-based simulator extended from
Noxim \cite{the:Noxim-simulator-User}. We assume the mesh-based topology for all the
comparisons. Table \ref{tab:NoC-architectures-used} shows four router architectures used for comparison.
Specifically, we compared the performance of the proposed scheme with
three different architectures, namely the Typical NoC, Typical NoC
with 2X input speedup, and BiNoC. For all architectures, each input
direction of the router has $4$ VCs and the buffer depth of each
VC is $16$ flits as in \cite{5715603}. We assume the packet has
a constant length of $16$ flits. As shown in Table \ref{tab:NoC-architectures-used},
for conventional unidirectional NoC, a $5\times5$ crossbar is used
to support the $5$-in and $5$-out data transmission. The input speedup
scheme provides $2X$ excess bandwidth in the input side and therefore,
a $10\times5$ crossbar is needed. For the BiNoC and FSNoC architectures,
as we need to provide flexible data transmission over the $10$ inout
ports, a larger $10\times10$ crossbar is required. In order to make
a fair comparison over the four architectures, each router architecture
is assumed to operate under the maxium frequency, which is obtained
from the synthesis results by Synopsys Design Compiler under TSMC
$65$nm technology. The simulated latency is then calculated in the
unit normalized to the cycle $T$ of Typical NoC. 
\begin{table}
\caption{\label{tab:NoC-architectures-used}NoC architectures used in experiments}
\begin{tabular}{|p{0.17\columnwidth}|p{0.10\columnwidth}|p{0.10\columnwidth}|p{0.14\columnwidth}|p{0.10\columnwidth}|p{0.10\columnwidth}|p{0.10\columnwidth}|}
\hline 
{ Architecture} & { Buf. /Dir. (flits)} & { Buf. /VC (flits)} & { Ch./Dir.} & { Crossbar size} & { Freq. (MHz)} & { Norm. $T$}\tabularnewline
\hline 
\hline 
\textit{ Typical NoC} & \textit{ $64$} & \textit{ $16$} & \textit{ $1$-in $1$-out} & \textit{ $5\times5$} & $1265$  & $1.0$ \tabularnewline
\hline 
\textit{ 2X speedup} & \textit{ $64$} & \textit{ $16$} & \textit{ $1$-in $1$-out} & \textit{ $10\times5$} & $1219$ & $1.04$ \tabularnewline
\hline 
\textit{ BiNoC} & \textit{ $64$} & \textit{ $16$} & \textit{ $2$-inout} & \textit{ $10\times10$} & $1020$ & $1.24$ \tabularnewline
\hline 
\textit{ FSNoC} & \textit{ $64$} & \textit{ $16$} & \textit{ $2$-inout} & \textit{ $10\times10$} & $1020$  & $1.24$ \tabularnewline
\hline 
\end{tabular}
\end{table}

For four architectures, the Dimension ordered XY routing is adopted
in order to avoid deadlocks (except Fig. \ref{fig:Histogram-of-the-4}-a,
which evaluates the effects of oddeven routing algorithm). Various
traffic patterns were used in the evaluation, including synthetic
traffic and real benchmarks. More specifically, two types of real
benchmarks are used in the evaluation. The first type is task-graph-based
MPSoC applications, including MWD (Multi-Window Display) \cite{Noc_synthesis},
MMS (Multimedia system) \cite{1411933}, MPEG4 (MPEG4 codec)\cite{Noc_synthesis}
and DVOPD (Dual Video Object Plane Decoder) \cite{noc_design_65nm} as well
as three E3S \cite{E3S} applications named auto\_indust, telecom
and consumer. For these application-specific benchmarks, we need to map the task
graphs onto the mesh architecture first. The second type of applications
are extracted from SPEC-web \cite{Spec} benchmarks which model the memory access patterns for CMP architectures.
For SPEC-web applications, five 16-node multithreaded workloads for
IBM and Oracle database server (\textit{i.e., DB2V} and\textit{ Oracle}),
the Apache HTTP server (\textit{i.e., Apache}), scientific workloads
of matrix factorization (\textit{i.e.,Sparse}) and the ocean dynamic
simulation (\textit{i.e., Ocean}) are extracted on a $4\times4$ mesh.
Moreover, application consolidations \cite{DBAR} are evaluated by
choosing four applications and randomly mapping them onto an $8\times8$
mesh (\textit{i.e.,} \textit{Mixed}). 
In the following, we first evaluate the performance of the NoC architectures without link pipeline stages in sub-section
\ref{sub:Simulation-results-on}-\ref{sub:Simulation-results-on-1}. Then, we discuss the simulation results for FSNoC designs link pipelines.
\begin{figure}
\includegraphics[width=0.99\columnwidth]{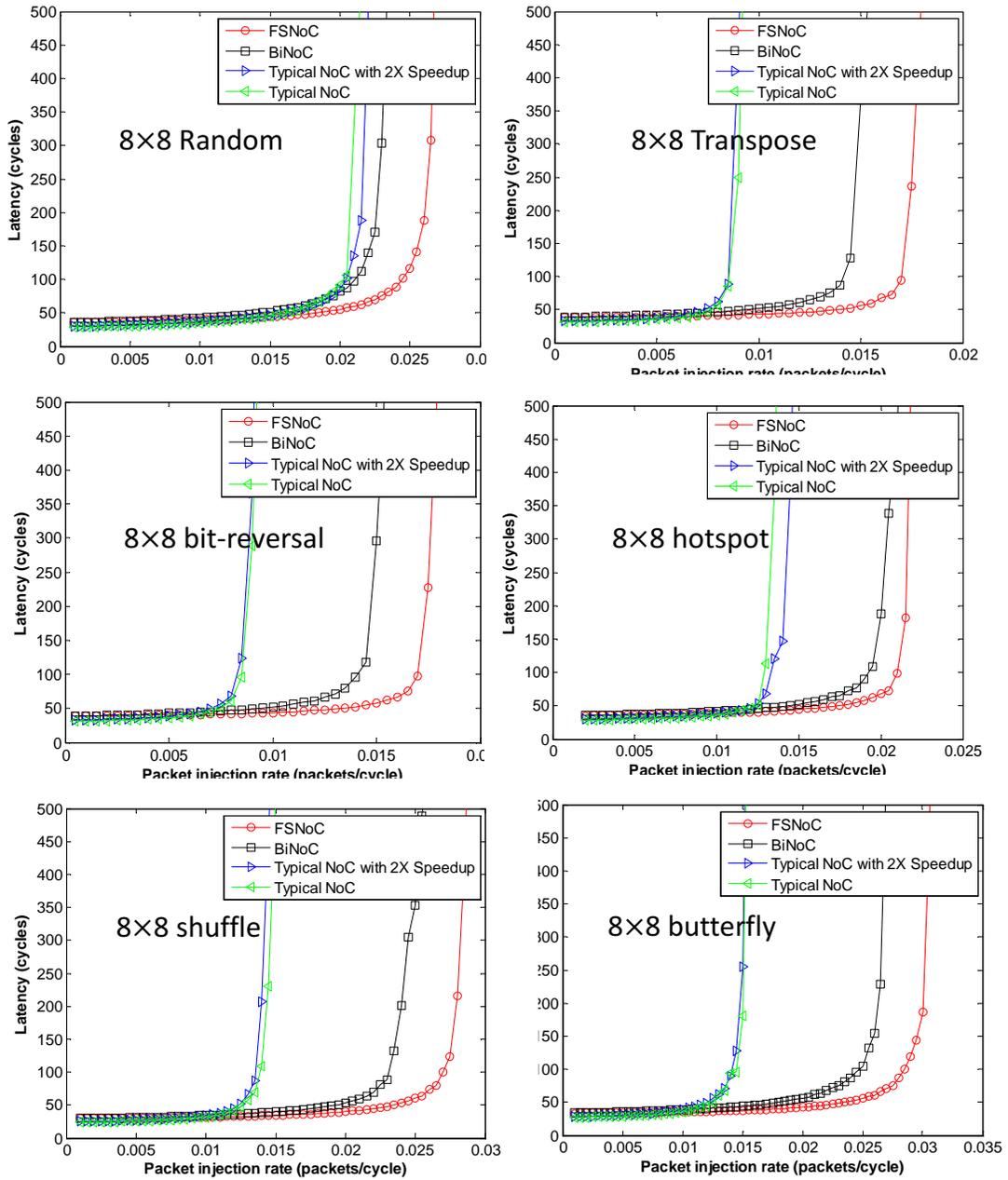}
\caption{\label{fig:Latency-simulation-results}Simulation results for six
synthetic traffics on an $8\times8$ mesh under XY routing}
\end{figure}

\subsection{\label{sub:Simulation-results-on}Simulation results on synthetic
traffics\vspace{-0.9mm}
}

\begin{figure}
\includegraphics[width=0.85\linewidth]{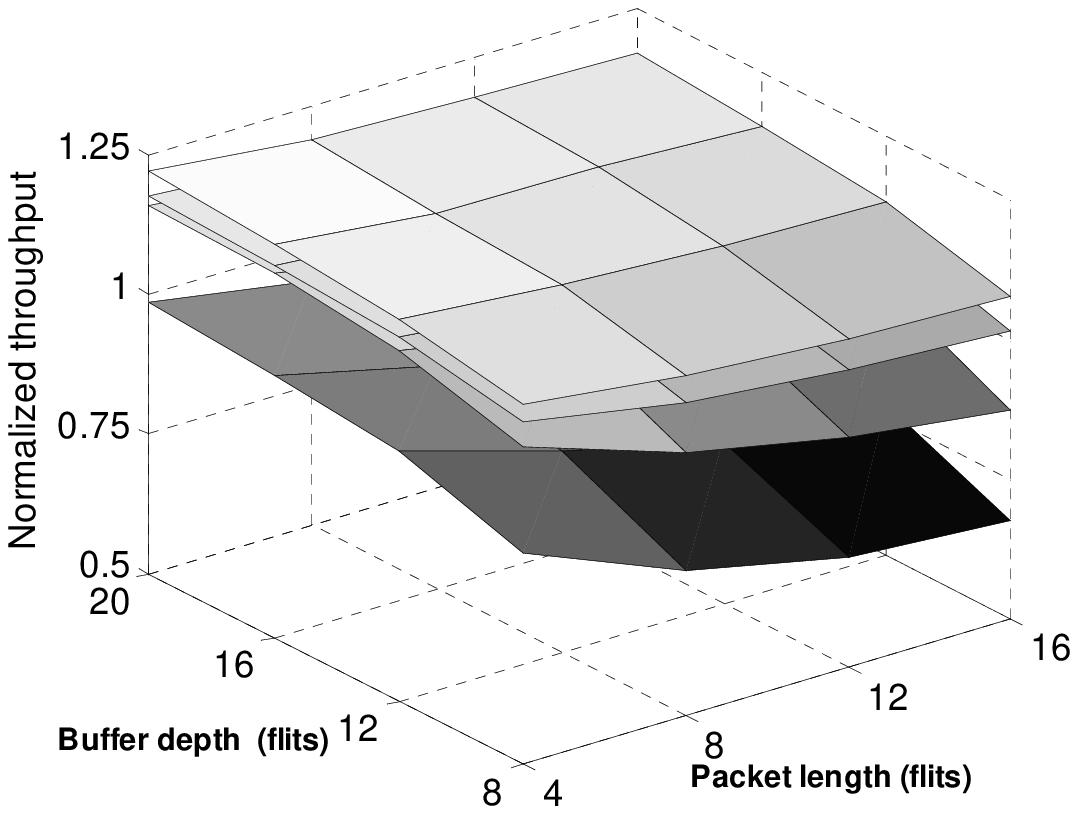}
\caption{\label{fig:Latency-simulation-results-1}The impacts of buffer depths
and packet size on overall throughput }
\end{figure}

\begin{figure}
\includegraphics[width=1\columnwidth]{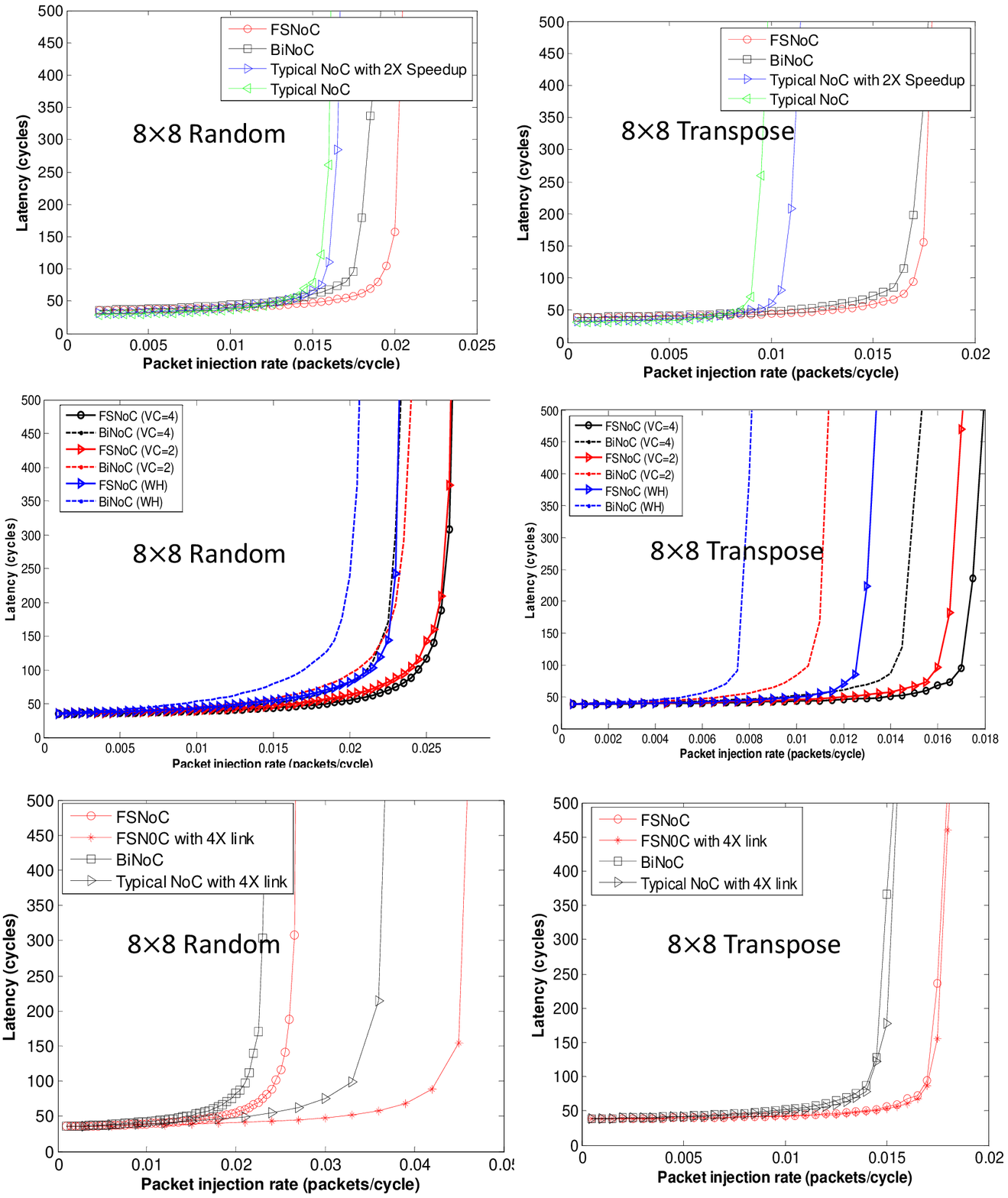}
\caption{\label{fig:Histogram-of-the-4}Latency comparisons for a) four architectures
under oddeven (OE) routing algorithm b) FSNoC and BiNoC with WH router
architecture and VC router architecture with different VC numbers
c) NoC architectures with 4L links }
\end{figure}

For the synthetic traffic comparison, six traffic patterns were used,
\textit{i.e.,} random, transpose, hotspot, shuffle, bit-reversal and
butterfly \cite{Book,the:Noxim-simulator-User}. For the hotspot traffic, two central
nodes in the mesh network have $10\%$ higher probability to receive
packets from other nodes. Figures \ref{fig:Latency-simulation-results}
summarizes the comparisons on the latency under the XY routing algorithm.
For all the cases, the proposed FSNoC out-performs other three schemes.
As observed from the figure, the improvement highly depends on the
traffic patterns. For example, comparing the results in random traffic
with transpose or bitreversal traffic, we can find the uneven traffics
achieve more latency improvement as these non-uniform patterns provide
more chances for the BiNoC and FSNoC architecture to use the $2X$
bandwidth during the transmission. Specifically, we can compare the
maximum packet injection rate sustainable by the network of each architectures
\cite{application_specific_plaesi}. The maximum injection rate ($pir$) is calculated
as the rate (packets/cycle/node) at which the corresponding throughput
reaches $95\%$ of the saturation. For instance, for the random traffic
pattern, the $pir$ of FSNoC, BiNoC, $2X$ speedup NoC and Typical
NoC are $0.020$, $0.021$,$0.022$ and $0.0255$ (\textit{packets/cycle}),
respectively. While for transpose traffic, the $pir$ of four architectures
are $0.008$, $0.008$, $0.014$, $0.016$ (\textit{packets/cycle}).
In summary, the BiNoC architecture improves $pir$ by $10\%-72\%$
over the typical and $2X$ input speedup schemes. In addition to that,
the FSNoC further increases the saturation injection point by $12\%-22\%$.

We then evaluate the NoC performance regarding the influence of different
packet length ($4-16$ flits) and buffer depth ($8-20$ flits per
VC). Figure \ref{fig:Latency-simulation-results-1} shows the maximum
throughput that four architectures can obtain for the random traffic
pattern on an $8\times8$ mesh. All the throughput values are normalized
to that of a typical NoC design with constant packet length of $4$
flits and buffer depth of $20$ flits. As shown in the figure, the
throughput improvement is more obvious when a longer packet length
is adopted as the single packet transfer time can be more significantly
reduced by allowing sending two flits at one time. Overall, the proposed
FSNoC design achieves the highest throughput over a wide variety of
packet length and buffer depth combinations. 
\begin{figure}
\includegraphics[width=0.95\columnwidth]{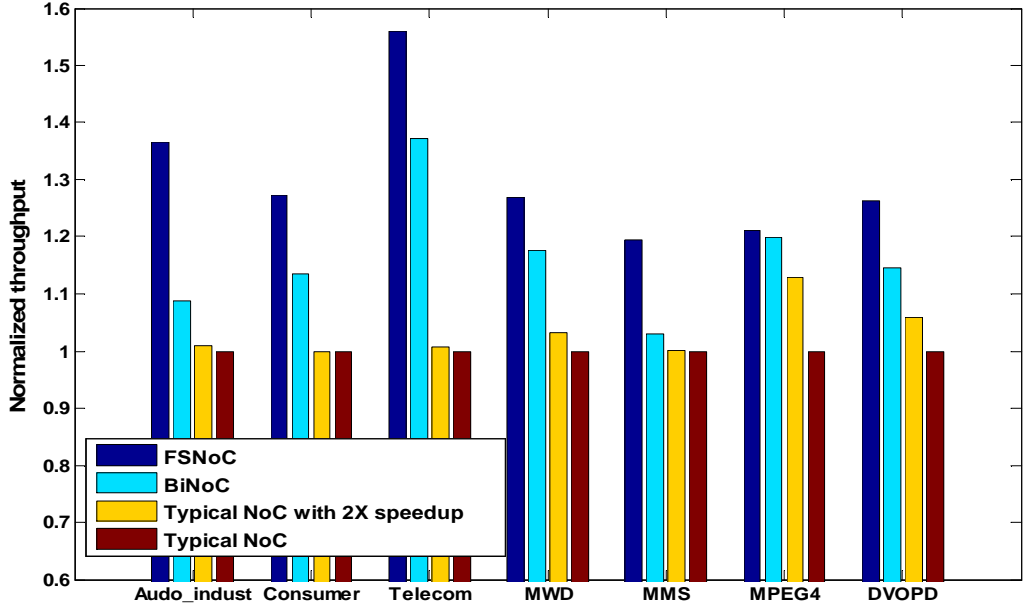}
\caption{\label{fig:Benchmark-throughput-simulation}Benchmark throughput comparison}
\end{figure}

In Fig. \ref{fig:Histogram-of-the-4}-a, we show the latency of four
architectures running oddeven routing algorithm under random and transpose
traffic patterns. For the uneven traffic patterns such as transpose,
the oddeven routing algorithm evenly distribute the load during the
routing, therefore, the improvement of FSNoC over BiNoC is smaller
than the case of XY routing. However, as can be seen from the figure,
the FSNoC still achieves $6.5\%$ improvement in the saturation
injection point for the random and transpose traffic. In Fig. \ref{fig:Histogram-of-the-4}-b,
we evaluate the influence of number of virtual channels on the latency
performance. For all architectures, the buffer size per input port
is fixed as in Table \ref{tab:NoC-architectures-used}. As shown in Fig. \ref{fig:Histogram-of-the-4}-b,
FSNoC consistently achieves the highest performance for both the wormhole
(WH) architecture and two virtual channel architectures. The idea
of flit-level speedup can also be applied to NoCs with four uni-directional
links between router pairs (\textit{i.e.,} $4L$ link architecture
in \cite{5715603,BiNoC1}). Because the $4L$ link architecture
always provides $2X$ bandwidth for each direction, it gives the upper-bound
performance that BiNoC can achieve. Conventional $4L$ link design
only allows two different packets to use the $2X$ channel. Therefore,
we can improve the performance by allowing two flits from the same
packet to use $2X$ links (\textit{i.e.,} flit-level speedup). In
Fig. \ref{fig:Histogram-of-the-4}, we compare the latency performance
of four architectures, namely BiNoC, FSNoC, Typical NoC with $4L$
link and FSNoC with $4L$ link, using random and transpose traffic
patterns. As can be seen from the figure, for uniform traffic patterns
such as random, the $4L$ link architectures (\textit{e.g.,} Typical
$4L$ and FSNoC $4L$) achieve higher performance than the FSNoC and
BiNoC because the static bandwidth assignment avoids the direction
switching overhead existed in bi-directional designs. For uneven traffic
patterns such as transpose, the Typical $4L$ and FSNoC $4L$ architectures
have similar performance as BiNoC and FSNoC, respectively. This is
because the direction switching is not so frequent in these patterns.
Hence, the FSNoC and BiNoC architectures work just like the $4L$
architecture under these cases. \vspace{-3mm}

\subsection{\label{sub:Simulation-results-on-1}Simulation results on real benchmarks\vspace{-0.9mm}
}

\begin{figure}
\includegraphics[width=0.95\columnwidth]{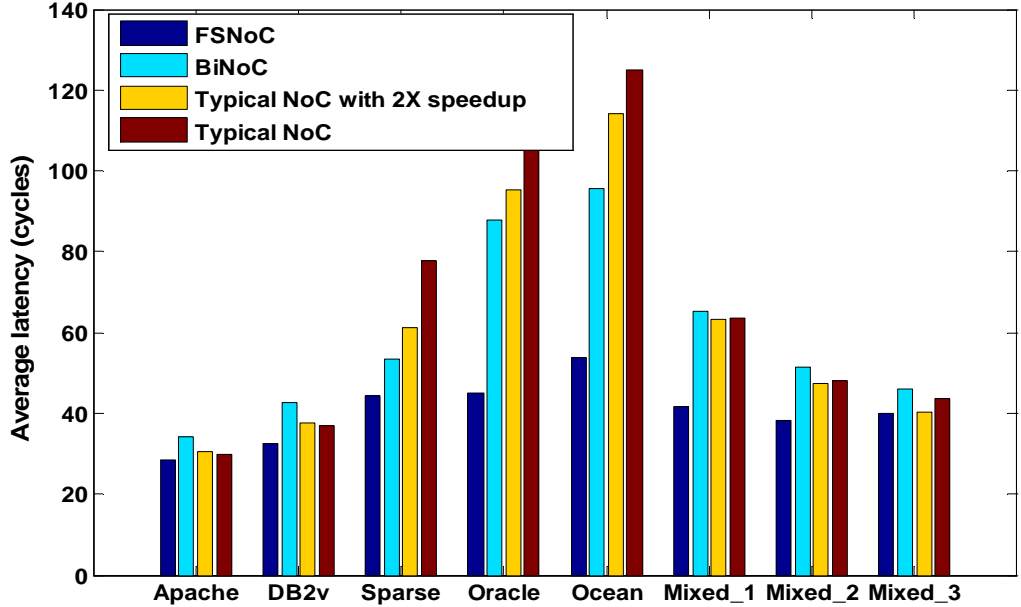}
\caption{\label{fig:Histogram-of-the-1}Latency comparisons for the SPEC-web
benchmarks }
\end{figure}


We also evaluate the FSNoC performance using several real benchmarks.
In Fig. \ref{fig:Histogram-of-the-1}, we compare the saturation throughput
for seven MPSoC applications. Of note, for these MPSoC benchmarks
such as MMS, MWD\textit{,} we assume the packets are injected according
to a Poisson process whose mean rate is scaled by a parameter $c$
according to the communication data volume on the edges of the task
graphs. We then tune the scaling factor $c$ to find out the maximum
throughput that various architectures can support for the comparison
in Fig. \ref{fig:Histogram-of-the-1}. All the saturation throughputs
are normalized to that of the typical NoC design. As shown in the
figure, while the bi-directional switching scheme out-performs the
two unidirectional schemes, our proposed flit-level speedup scheme
consistently further improves the performance. The relative improvement
of FSNoC over BiNoC ranges from 2\% to 17\% depending on the applications.
In Fig. \ref{fig:Histogram-of-the-1}, we use the collected traces from SPEC-web benchmarks to
compare the average communication latency in the CMP platforms. As
shown in the figures, for all the SPEC applications, the proposed FSNoC design achieves fairly good reduction in latency. Especially
for those applications with medium or high workload (\textit{e.g.,}
\textit{Oracle} and \textit{Ocean} in Fig. \ref{fig:Histogram-of-the-1}),
the improvement is more significant as the FSNoC significantly reduce
the network congestion under these workloads by providing more flexibility
to use the inter-router bandwidth.

\begin{figure}
\includegraphics[width=0.95\columnwidth]{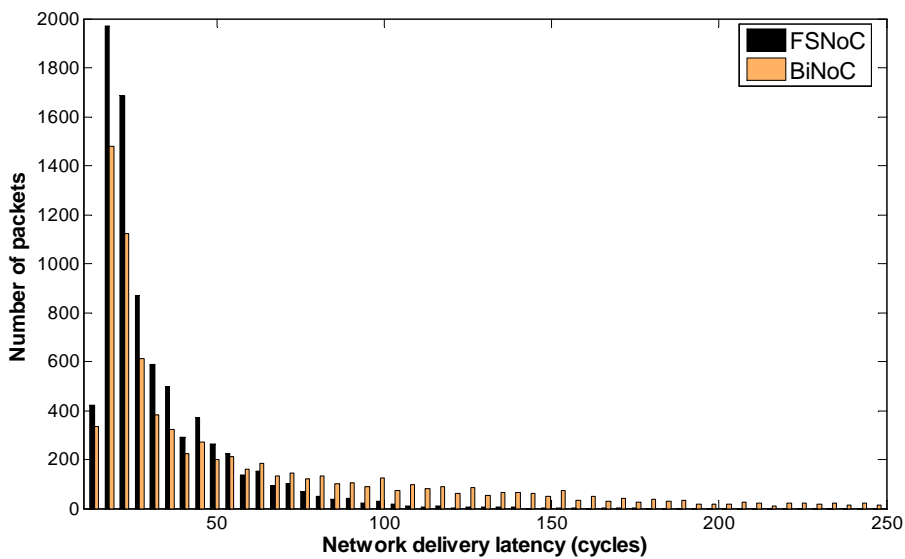}
\caption{\label{fig:Histogram-of-the}Histogram of the delivery time for Telecom }
\end{figure}

In Figure \ref{fig:Histogram-of-the}, we also show an example histogram
of the packet delivery time under 80\% saturation injection factor
for the telecom application in E3S application. As shown in the figure,
not only the average latency but also the maximum delay are significantly
reduced. As the worst-case delay is more important for the applications
with real-time deadline constraints, the FSNoC also provides a good
prospective to satisfy the QoS requirements for these applications.

\subsection{Simulation results for FSNoC with link pipelines }

\begin{figure}
\includegraphics[width=0.99\columnwidth]{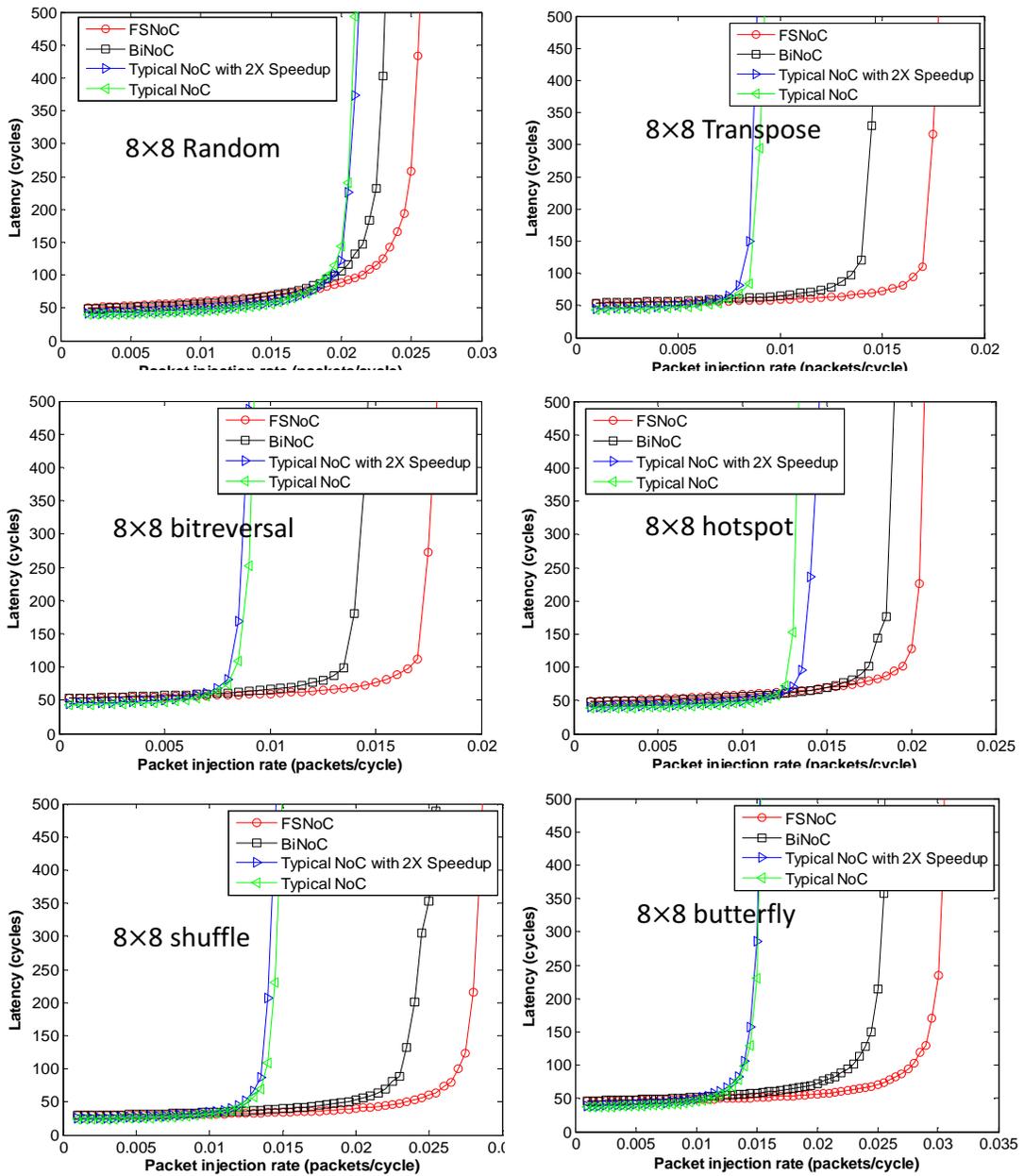}
\caption{\label{fig:Histogram-of-the-3}Latency comparisons for NoCs with 2-stage
pipelined links }
\end{figure}

Regarding the influences of long wire links, we next evaluate the
latency performance of four architectures with $2$-stage pipeline
registers inserted in all links. In Fig. \ref{fig:Histogram-of-the-3},
the latency simulation results for four synthetic traffic patterns
on an $8\times8$ mesh are shown. As can be seen from the figure,
for these two traffic patterns, the network latency has increased
at light loads. For example, for the random traffic, comparing Fig.
\ref{fig:Latency-simulation-results} with Fig. \ref{fig:Histogram-of-the-3},
it can be observed the zero load latency has increased from $28$
cycles to $40$ cycles because it takes $3$ cycles in total for a
flit to reach the neighboring router. Moreover, the saturation injection
point (\textit{i.e.,} $pir$) has also decreased slightly for all
the architectures. For random traffic, the $pir$ values of four architectures
have reduced to $0.018$, $0.018$, $0.021$ and $0.024$\textit{
(packet/cycle)}, respectively. This is because in order to avoid both
wire and register conflicts described in section \ref{sub:Long-wire-links-with},
the OWC module should determine the direction control ($DS$) signals
by considering not only the current cycle pressure information but
also that of previous two cycles, which reduces the chances to use
the $2X$ channel bandwidth. However, as can be seen from the figure,
the pipelined FSNoC still outperforms the other three architectures
for both random and transpose traffics.\vspace{-3mm}

\subsection{\label{sub:Implementation-overhead}Implementation overhead}

As discussed in \cite{5715603}, although the $4L$ link architectures
have almost the same overhead as BiNoC in the router design, the more
inter-router wiring resources required is not cheap as technologies
keep scaling down. Rather than increasing physical bandwidth by adding
more additional uni-directional links, NoCs based on bi-directional
channels are advocated to reduce the routing congestion and spaces.
Therefore, in the hardware overhead evaluation, we majorly implement
and compare the four router architectures described in Table \ref{tab:NoC-architectures-used}. We
implement each scheme in Verilog and synthesized the design using
Synopsys Design Vision based on TSMC 65nm library. According to our synthesis results, for all the four architectures, the critical paths are located in the ST stage. For both the BiNoC and FSNoC architectures, the maximum path delay $0.96ns$  is dictated by the $10\times10$  crossbar and hence they run at the same maximum clock frequency. Table \ref{tab:NoC-architectures-used-1} and \ref{tab:NoC-architectures-used-1-1} summarize the area and power breakdown of different NoC architectures. For FSNoC, the direction control module (i.e., Dir. ctrl. in Table \ref{tab:NoC-architectures-used-1} and \ref{tab:NoC-architectures-used-1-1}) includes the Request extractor (RE) and Output width controller (OWC). As shown in the tables, the input buffers consume most of the router area and power for all the four architectures. From the synthesis results, the area overhead of 2X  input speedup, BiNoC and the proposed FSNoC over the conventional NoC are $6\%$ , $15\%$  and $20\%$ , respectively. For the power comparison, we assume a $50\%$  switching activity factor of a random payload as in \cite{5715603}. As can be seen from the table, the power overhead of 2X  input speedup, BiNoC and FSNoC over conventional NoC are $5\%$ , $13\%$  and $14\%$ , respectively.
\begin{table}
\caption{\label{tab:NoC-architectures-used-1}Area breakdown of different NoC
architectures}
\begin{tabular}{|p{0.20\columnwidth}|p{0.08\columnwidth}|p{0.06\columnwidth}|p{0.08\columnwidth}|p{0.06\columnwidth}|p{0.08\columnwidth}|p{0.06\columnwidth}|p{0.08\columnwidth}|p{0.06\columnwidth}|}
\hline 
\multirow{2}{0.18\columnwidth} & \multicolumn{8}{c|}{Area($\mu m^{2}$)($\%$)}\tabularnewline
\cline{2-9} 
 & \multicolumn{2}{c|}{{ Typical NoC}} & \multicolumn{2}{c|}{{ 2X speedup}} & \multicolumn{2}{c|}{{ BiNoC}} & \multicolumn{2}{c|}{{ FSNoC}}\tabularnewline
\hline 
\hline 
\textit{ Input buf. } & $109774$  & $89.7$ & $110515$ & $88.5$ & $120545$ & $85.2$ & $125268$ & $85.2$ \tabularnewline
\hline 
\textit{ RC} & $1806$ & $1.5$ & $1806$ & $1.4$ & $1806$ & $1.3$ & $1806$ & $1.2$ \tabularnewline
\hline 
\textit{ VC allocator} & $8806$  & $8.0$ & $8806$ & $7.1$ & $8806$ & $6.2$ & $8806$ & $6.0$ \tabularnewline
\hline 
\textit{ SA allocator} & $583$ & $0.5$ & $1138$ & $0.9$ & $1365$ & $1.0$ & $1671$ & $1.1$ \tabularnewline
\hline 
\textit{ Crossbar} & $1429$ & $1.2$ & $2558$ & $2.0$ & $6371$ & $5.3$ & $6371$ & $4.3$ \tabularnewline
\hline 
\textit{ Dir. control }%
\footnote{\textit{ Including RE and OWC module}%
} & $N/A$  & $N/A$ & $N/A$ & $N/A$  & $2567$ &$1.8$  & $3065$ & $2.1$\tabularnewline
\hline 
\textit{ Total} & $122398$  & $100$ & $124823$ & $100$ & $141460$ & $100$ & $146989$ &  $100$\tabularnewline
\hline 
\textit{ Norm. area} & \multicolumn{2}{c|}{$1.0$}  & \multicolumn{2}{c|}{$1.06$} & \multicolumn{2}{c|}{$1.15$} & \multicolumn{2}{c|}{$1.20$}\tabularnewline
\hline 
\end{tabular}
\end{table}

\begin{table}
\caption{\label{tab:NoC-architectures-used-1-1}Power breakdown of different
NoC architectures}
\begin{tabular}{|p{0.20\columnwidth}|p{0.08\columnwidth}|p{0.06\columnwidth}|p{0.08\columnwidth}|p{0.06\columnwidth}|p{0.08\columnwidth}|p{0.06\columnwidth}|p{0.08\columnwidth}|p{0.06\columnwidth}|}
\hline 
\multirow{2}{0.18\columnwidth} & \multicolumn{8}{c|}{Power($mw$)($\%$)}\tabularnewline
\cline{2-9} 
 & \multicolumn{2}{c|}{{ Typical NoC}} & \multicolumn{2}{c|}{{ 2X speedup}} & \multicolumn{2}{c|}{{ BiNoC}} & \multicolumn{2}{c|}{{ FSNoC}}\tabularnewline
\hline 
\hline 
\textit{ Input buf. } & $10.30$  & $89.4$ & $10.7$ & $88.4$ & $11.3$ & $87.1$ & $11.5$ & $87.2$ \tabularnewline
\hline 
\textit{ RC} & $0.10$ & $0.9$ & $0.10$ & $0.8$ & $0.10$ & $0.8$ & $0.10$ & $0.8$ \tabularnewline
\hline 
\textit{ VC allocator} & $0.94$ & $8.2$ & $0.94$ & $7.8$ & $0.94$ & $7.2$ & $0.94$ & $7.1$ \tabularnewline
\hline 
\textit{ SA allocator} & $0.04$  & $0.3$ & $0.06$  & $0.5$  & $0.06$ & $0.5$ & $0.06$ & $0.5$ \tabularnewline
\hline 
\textit{ Crossbar} & $0.18$ & $1.6$ & $0.30$  & $2.5$  &  $0.52$ & $4.0$ & $0.52$ & $3.9$ \tabularnewline
\hline 
\textit{ Dir. control } & $N/A$ & $N/A$  & $N/A$  & $N/A$ & $0.05$ & $0.4$ & $0.06$ & $0.5$ \tabularnewline
\hline 
\textit{ Total} & $11.52$ & $100$ & $12.10$ & $100$  & $12.97$ & $100$ & $13.18$ & $100$ \tabularnewline
\hline 
\textit{ Norm. area} & \multicolumn{2}{c|}{$1.00$} &  \multicolumn{2}{c|}{$1.05$} &  \multicolumn{2}{c|}{$1.13$}&  \multicolumn{2}{c|}{$1.14$}\tabularnewline
\hline 
\end{tabular}
\end{table}

\section{Conclusion}

In this chapter, we have proposed a flit-level speedup scheme for improving
the NoC performance using self-reconfigurable bi-directional links.
In order to support transmitting two flits within the same packet
at the same cycle, a novel channel direction control protocol is proposed
to dynamically configure the link directions. The corresponding design
of the input buffer organization and the switch allocator are also
proposed. Also, we have extended the channel direction control scheme
to work under the existence of long wire links. From the simulation
results, significant improvement in latency and throughput are achieved
for both synthetic traffic and the real benchmarks.

\chapter{A Traffic-aware Adaptive Routing Algorithm on a Highly Reconfigurable Network-on-Chip Architecture}

\textit{In this chapter, we propose a flexible NoC architecture and a dynamic distributed routing algorithm which can enhance the NoC communication performance with minimal energy overhead. In particular, our proposed NoC architecture exploits the following two features: i) self-reconfigurable bidirectional channels to increase the effective bandwidth and ii) express virtual paths, as well as localized hub routers, to bypass some intermediate nodes at run time in the network. A deadlock-free and traffic-aware dynamic routing algorithm is further developed for the proposed architecture, which can take advantage of the increased flexibility in the proposed architecture. Both the channels self-reconfiguration and routing decisions are made in a distributed fashion, based on a function of the localized traffic conditions, in order to maximize the performance and minimize the energy costs at the macroscopic level. Our simulation results show that the proposed approach can reduce the network latency by $30\% -80\%$ in most cases compared to a conventional unidirectional mesh topology, while incurring less than 15\% power overhead.
}

\section{Introduction}

The NoC architecture and the corresponding routing strategy play an important role in optimizing the system performance in terms of throughput and latency. Indeed, enhancing the NoC architecture with additional flexibility is appealing since it can offer more opportunities to minimize the impact of traffic congestion by bypassing the intermediate router pipeline stages for some packets at run time \cite{Kumar}. At the same time, a routing strategy, which fully utilizes the characteristics of both the underlying communication infrastructure and the NoC traffic, is equally important since the routing decision made at the run time is critical in optimizing the NoC performance and resources utilization \cite{NoP,fractal_traffic}.

For NoC architecture, the bandwidth and the average distance between the nodes have a significant impact on the overall network performance. Most of the existing NoC architectures have employed a two-dimensional mesh topology with two unidirectional links connecting the neighboring tiles. Although the mesh topology is well suited for silicon implementation \cite{RCA}, the channel bandwidth is not optimally used \cite{BiNoC1} and it may suffer from long packet latencies due to the lack of short paths between remotely located nodes \cite{long-range-link}. 

For routing strategy design, several adaptive routing algorithms have been proposed for conventional unidirectional mesh NoCs \cite{NoP,RCA,DBAR,ANCS}. However, dynamic routing has been far less explored for performance improvement.

In this chapter, we propose a reconfigurable NoC architecture which combines the advantages of providing higher effective bandwidth between the neighboring nodes and shorter paths between remotely located nodes \cite{traffic-aware-noc}. Our proposed NoC architecture exploits the following two capabilities, namely: i) self reconfigurable bidirectional channels (BiNoC) \cite{BiNoC1,5090667} and ii) static express virtual channels (EVCs)\cite{Kumar} with regional hub routers, to dynamically create short-cut paths for transmitting packets. In order to exploit the added flexibility provided by the proposed architecture, we also propose a new fitness-based adaptive routing algorithm for optimizing the network performance.

%


Experimental results obtained for synthetic traffic and real benchmarks show that our approach can improve the latency by as much as 80\% , while involving less than 15\% overhead in power dissipation.

The remaining of this chapter is organized as follows: In section 6.2, we review related work on NoC architecture and adaptive routing algorithm design. In section 6.3, we describe the newly proposed NoC architecture. Section 6.4 presents the adaptive routing algorithm for the proposed NoC. The practical considerations and simulation results are discussed in section 6.5. Finally, section 6.6 concludes this work.
\begin{figure}
\centering
\epsfig{file=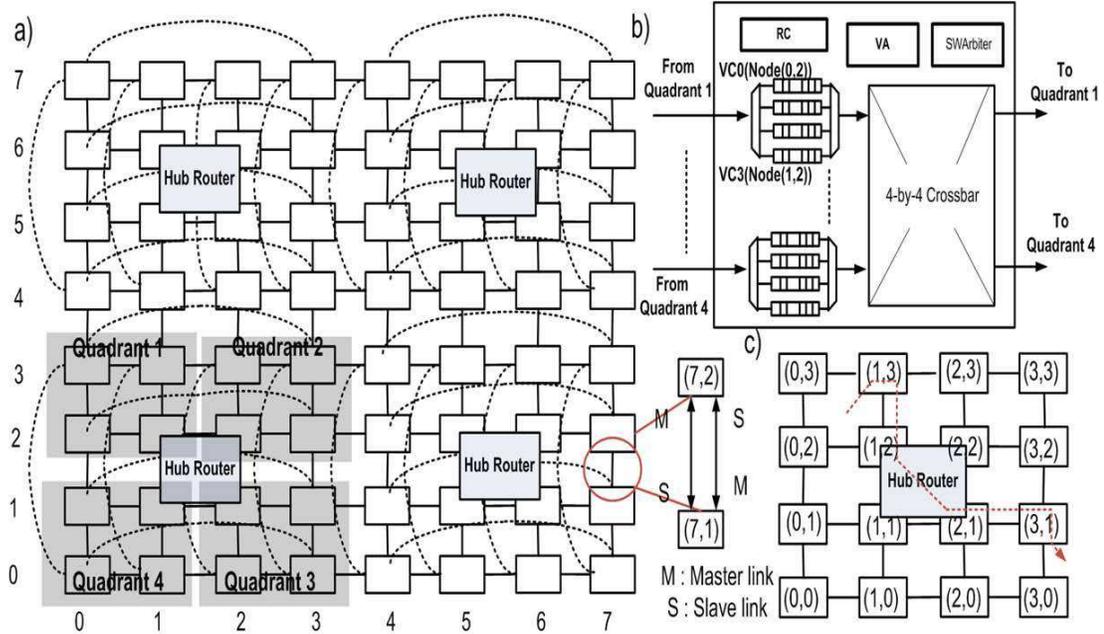,width=1.00\linewidth}
\caption{\label{figure_1_proposed_noc} a) Proposed NoC architecture for an $8\times8$ mesh with bi-directional links, EVCs and hub routers  b) The hub router architecture  with each VC corresponding to a node in the region c) An example of routing from (1,3) to (3,1) using hub routers }
\end{figure}
\section{Background}
In NoC architecture design, many efforts are focused on increasing the total bandwidth and reducing the average node distance by proposing new topologies \cite{Chifeng,Grot}. In \cite{Chifeng}, a Diagonally-linked Mesh topology is proposed which employs physical diagonal express links between routers to reduce the distances between nodes. In \cite{Grot}, a new topology called Multidrop Express Channels (MECS) is proposed; this uses a one-to-many communication model to enable a high degree of connectivity. Compared to the mesh based NoC, these new topologies can improve the network bandwidth and connectivity at the cost of adding more physical resources such as the link wires and increased the router complexity. 

To improve the trade-off between the design complexity and the performance, there are also approaches exploiting the mesh topology due to its modularity and scalability. In \cite{Kumar}, the NoC architecture with static and dynamic express channels (EVC) is proposed. Instead of adding physical links, a flow control mechanism allows packets to virtually bypass intermediate routers along their path. Static EVC approach uses the express paths of uniform lengths and distinguishes the nodes as either EVC source (sink) nodes or bypass nodes according to whether the EVC originates (terminates) at those nodes. Dynamic EVCs make every node in the network a source (sink) and allow EVCs of various lengths to originate from a node\cite{Kumar}.  

Besides the idea of bypassing intermediate routers, another important direction in optimizing the mesh NoC architecture is to enable the network to increase the effective bandwidth at run time. Indeed, very often, there is much traffic in one direction, while the channel for the opposite direction is idle. NoCs with run-time reconfigurable bi-directional channels have been recently proposed to fully utilize the bandwidth according to the run-time traffic conditions (BiNoC) \cite{BiNoC1,5090667,5715603,zhiliang}. However, adaptive routing is less exploited in these works for simplicity reasons.

Previous adaptive routing algorithms are mainly designed for unidirectional NoCs. In \cite{NoP}, a neighbors-on-path (NoP) selection strategy is proposed to make each node routing selection based on the condition of the nodes adjacent to the neighbors. In \cite{RCA}, a regional contention awareness scheme (RCA) was first proposed to utilize both the local and non-local information to improve the load balancing in NoCs. In \cite{DBAR}, the RCA scheme was improved by using the adaptive routing scheme DBAR to leverage the local and non-local network information. In \cite{ANCS}, a destination-based adaptive routing algorithm (DAR) is proposed where every node estimates the delay to every other node in the network. However, for a more flexible NoC platform such as the ones employing bi-directional links and EVC channels, these schemes need to be modified with a congestion fitness function to reflect the run time traffic status.

Starting from these overarching ideas, in this work, we aim at improving the NoC performance with respect to both architecture design and routing algorithm development. Towards this end, we propose a flexible NoC architecture which combines the advantages of higher effective bandwidth and router bypass capability and develop an adaptive routing algorithm to optimize the network run-time performance.
\section{New NoC architecture }
In this section, we first present the motivation for adding regional hub routers into the NoC and then elaborate on the new reconfigurable NoC architecture.
\subsection{Motivation for adding hub routers }
Because of the complexity of the dynamic EVC approach \cite{Kumar}, in this work, we adopt a 3-hop static EVC network; this provides a better trade off between performance and design complexity. We also define a region as a $4\times4$ mesh whose boundary is made up of the EVC source/sink nodes. A regional hub router is added into each region based on the following key observations:

1)  The regional traffic represents a significant portion of the total traffic distribution. Here, we define the regional traffic as the communication flow whose source and destination reside within the same region. This is especially true for application-specific NoCs where highly communicating nodes are usually mapped close to each other. For example, in \cite{5715603}, the regional traffic with more than $90\%$ of nodes communicating with other nodes residing within a region of three hops away is used in the evaluation. 

2) While the static EVCs provide express paths crossing over NoC regions, they typically lack sufficient paths needed to handle the traffic within the NoC regions. If both the source and destination nodes are bypass nodes in a static EVC, then the packets cannot use the EVCs to bypass the intermediate nodes.

3) The static EVCs enable packets to bypass the intermediate nodes following a fixed express path. Also, the EVCs are restricted to a single dimension and cannot turn directly. Consequently, it will be beneficial if we can dynamically build shortcut paths for regional traffic rather than being limited by the single dimension requirement.

Therefore, in this work, we propose to add a light weight hub router to minimize the burden of the regional traffic in NoCs. More precisely, we additionally allocate a dedicated (\textit{i.e.,} static) express path for each node to route to the hub router and utilize the hub router to connect the nodes within the same region; this way, when the hub router is used, the intermediate pipeline stages of the normal routers can be bypassed altogether.
\subsection{Overview of proposed NoC architecture }
Figure \ref{figure_1_proposed_noc}-a shows the proposed flexible NoC architecture for $8\times8$ meshes. In this example, the static EVC hop length is equal to 3. Each region is a small $4\times4$ mesh. Other static EVC hop length and region size can be used for different sizes of the meshes. The solid line in Figure \ref{figure_1_proposed_noc}-a represents the physical channels between the tiles which are made up of two bi-directional links. Both bidirectional links can be reconfigured as sending or receiving. For each router, we define one link as a high priority link (master link) and the other as a low priority link (slave link) as in \cite{5715603}. By default, the master link is the sending link and the slave link is the receiving link when there is traffic in both directions. As shown in Figure \ref{figure_1_proposed_noc}-a, the master link of one router corresponds to the slave link of its neighbor. The dash line in the figure represents the 3-hop EVCs.

We further divide each $4\times4$ region into four quadrants (see Figure \ref{figure_1_proposed_noc}-a). In order to reduce the additional complexity added to the NoC backbone, the regional hub router is a simple and generic four ports router with four virtual channels at each input port as shown in Figure \ref{figure_1_proposed_noc}-b. Every input port corresponds to a quadrant in the region and each node in the quadrant is assigned a virtual channel at the input port of the hub router. For example, VC0 of the first input port (i.e. the quadrant 1 port) of the hub router is assigned to node (0,2) in Figure \ref{figure_1_proposed_noc}-a.

At run time, the hub router plays the role of setting a connection between the source and destination pairs within different quadrants of the same region. For example, in Figure \ref{figure_1_proposed_noc}-c, if node (1,3) in quadrant one needs to communicate with node (3,1) in the quadrant three, it cannot use the static EVCs. Instead, it can communicate via the hub router. The packet will first follow the dedicated express hub path bypassing router (1,2) pipeline stages to reach the input port corresponding to quadrant one of the hub router. After the hub router grants its connection from quadrant one input port to the quadrant three output port in the switch allocation stage, the packet will traverse the crossbar in the hub router and bypass router (2,1) pipeline stages to arrive at the destination node (3,1) directly. Compared with using normal paths to send packets from (1,3) to (3,1), which requires five router pipeline stages, only one router pipeline in the hub router is needed.  

In the proposed architecture, we utilize the virtual channel allocator and switch allocator in the hub router to arbitrate the contentions. For instance, if both nodes (1,3) and (1,0) want to send packets to (3,1) using the hub router, after arriving at the hub router, these two packets will request the same virtual channel which corresponds to node (3,1) in the quadrant three output port. The virtual channel allocator will grant one of the requests first, while the other packet needs to wait in the input channel buffer until the virtual channel becomes available again. Similarly, when there are multiple requests from different input quadrant ports to the same output quadrant, for instance, if both the packets from node (1,3) to (3,1) and node (1,1) to (3,0) want to use the quadrant four output port in the hub router, the switch allocator will only serve one request each time.
\begin{figure}
\centering
\epsfig{file=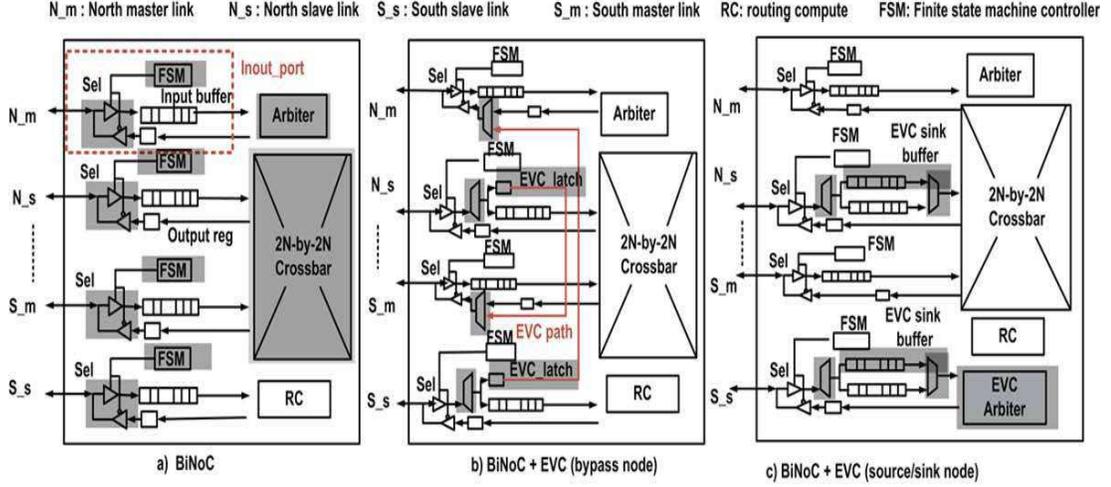,width=1.00\linewidth}
\caption{\label{figure_2_router_architecture} Router architecture of  a) BiNoC; b) BiNoC + EVC bypass node; c) BiNoC+EVC source/sink node}
\end{figure}
\subsection{Router implementation }
In this section, we present the router design that supports the BiNoC with EVC and region hub. First, we will discuss the additional hardware required to support the BiNoC approach. Then the micro-architecture that supports static EVCs will be presented. Finally, we present the modifications required to support the region-hub routing.

Figure \ref{figure_2_router_architecture}-a shows the router design for BiNoC with the extra logic required (see the shaded area) to support the bi-directional switching. Compared to a unidirectional wormhole router, in BiNoC each input and output port need to be modified to inout ports controlled by the corresponding finite state machines (FSMs) \cite{5715603}. Also, the arbiter and the crossbar are modified to handle the requests received from the master and slave ports. At run time, the FSM associated to every inout port configures the port mode (\textit{i.e.,} input or output) according to the requests existing in current router and the backpressure feedback from the neighboring routers as discussed in \cite{5715603}.

\begin{figure}
\epsfig{file=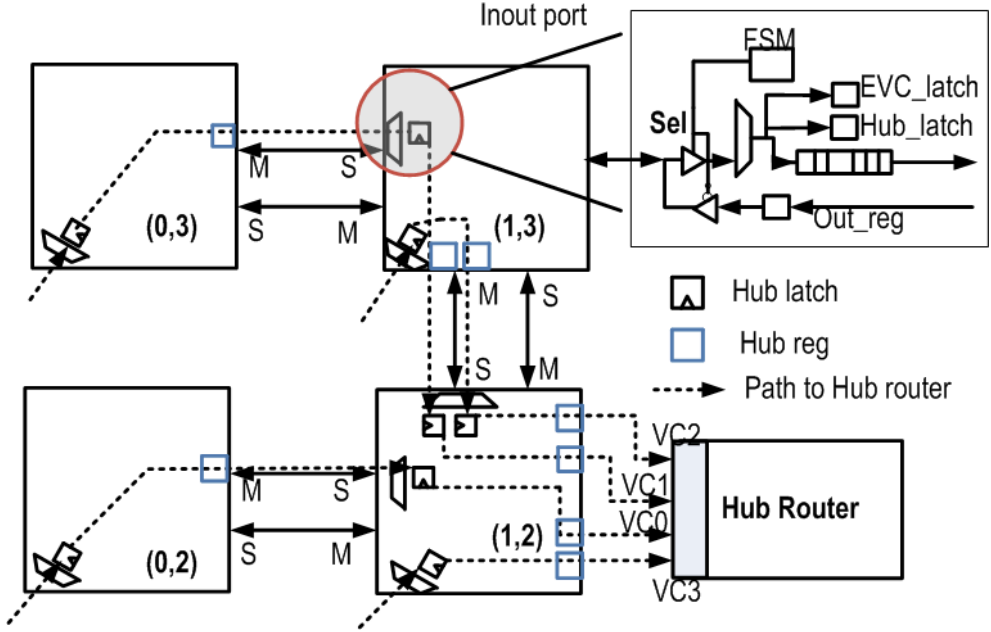,width=0.90\columnwidth}
\caption{\label{figure_3_express_path} The express paths to the hub router}
\end{figure} 

Figure \ref{figure_2_router_architecture}-b shows the router micro-architecture that supports static EVCs for BiNoC. In this work, we build the EVC path using the output ports corresponding to the master links, because the master link has a higher priority in transmitting flits according to the channel direction controlled by the FSM. Compared to the BiNoC router in Figure \ref{figure_2_router_architecture}-a, the extra logic need to support the EVC architecture is shaded in Figure \ref{figure_2_router_architecture}-b and Figure \ref{figure_2_router_architecture}-c. In static EVCs, there are two types of nodes, namely the bypass nodes and the EVC source/sink nodes. In the bypass nodes (as shown in Figure \ref{figure_2_router_architecture}-b), an EVC latch is added to hold the packets using the EVC. Once a packet arrives at the EVC latch, the packet will bypass all the intermediate pipeline stages of the current router (such as switch allocation, crossbar traversal) and is directly sent to the output port in the same dimension. In Figure \ref{figure_2_router_architecture}-b, the EVC paths along the north/south direction that a packet will take to bypass the router pipeline is highlighted. The architecture of the EVC source/sink routers is shown in Figure \ref{figure_2_router_architecture}-c. Here, we add an EVC sink buffer at the slave link port to hold the packets from the upstream nodes.  In addition, an EVC arbiter is added to handle the arbitration requests from all the input channels (both the normal input and EVC sink buffers). Once a packet successfully wins the allocation, an EVC flag in the flit will be asserted. Based on this flag, the downstream router will decide whether to put the flits into the EVC latch or into the normal input buffer. 

In the proposed regional hub routing, for every node, there is one static express path assigned to connect to the hub router. Figure \ref{figure_3_express_path} shows the paths assigned to the four nodes in the quadrant one of Figure \ref{figure_1_proposed_noc}. We add hub latches to multiplex with the normal buffers at the input port and hub registers to multiplex with the output registers at the output port (shown in Figure \ref{figure_3_express_path}). The hub latch and register pair forms an express path for a dedicated node to bypass the pipeline stage within the router.

\begin{figure}
\centering
\epsfig{file=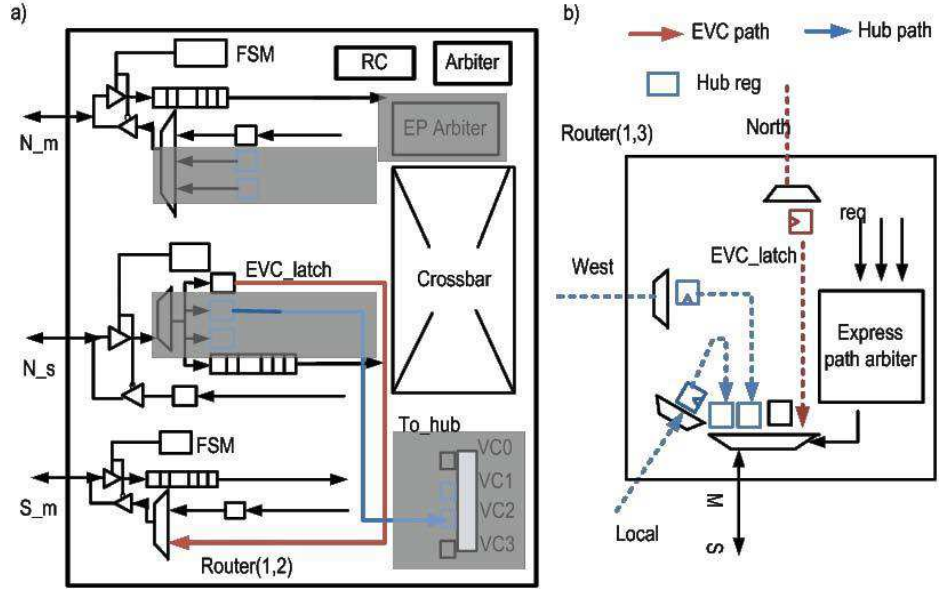,width=0.85\columnwidth}
\caption{\label{figure_4_router_arch} a)Router architecture of node (1,2) b) Contention between EVC and Hub paths}
\end{figure}

Figure \ref{figure_4_router_arch}-a shows the router design for node (1,2) in Figure \ref{figure_3_express_path} which connects to the hub router using an additional port. The extra logic needed compared to Figure \ref{figure_2_router_architecture}-c is shaded. As shown in Figure \ref{figure_4_router_arch}-a, the EVC paths are highlighted in red lines while the express paths formed by hub latch/register pairs are highlighted in blue lines.

At run time, there may be contentions between the hub latches or between the EVC and hub latches. For example, for router (1,3) in Figure \ref{figure_4_router_arch}-b, the hub latch in the local port and the hub latch in the west port may contend with the EVC latch in the north port to use the south output link. In order to resolve such contentions, the express path arbiter in Figure \ref{figure_4_router_arch}-b is needed in every router to handle the hub and EVC latch requests to make sure that, at any one time, only one packet can bypass the pipeline stages and use the physical channel. In this work, we allocate a higher priority to the EVC paths, because the hub paths only serve the regional traffic, while the EVC paths usually work for the traffic meant for longer distances. For example, the express path arbiter will first grant the north EVC latch to use the south output link. The packets from the west and local hub latches, as well as the input buffers will wait in the output port. Then, the express path arbiter will grant the requests from hub latches to use the output link with a higher priority than the normal paths. In order to avoid the starvation of flits in normal buffers and registers, the arbiter will grant the requests from normal paths after serving the EVC and hub latches for $n$ consecutive cycles as in \cite{Kumar}. In this work, $n$ is set to be 30 according to our simulation results.
\section{Routing algorithm design}
In this section, we present the details of our adaptive routing algorithm for the proposed flexible NoC. We first describe the scheme to choose among the static EVCs, express hub paths and BiNoC for sending a packet. Next, we present the adaptive routing algorithm when using the bi-directional channels to route packets. We develop a fitness function as a metric to evaluate the neighboring node availability based on the channels occupancy. The fitness function not only captures the buffer status but also takes the traffic dynamics, as well as the characteristics of the bi-directional channel into account. Then, a K-step speculative routing selection strategy is proposed to evaluate each candidate direction, which dynamically selects the nodes with the best fitness values that reside on a path towards the destination.
\subsection{Choosing the express hub and EVC paths }
In the proposed NoC architecture, we can use EVCs, hub routers, as well as the bi-directional channels to send packets around. The network interface (NI) in the source tile determines whether to use the express hub paths for the packet first. More specifically, the NI first compares the source and destination address of the packet. If the source and destination nodes are located in different quadrants of the same region, and if the destination node cannot be reached by EVCs from the source directly, then the NI will check the availability of the hub latch in the local port of the neighboring router. If the hub latch is free, the packet will be sent to it and follow the express hub path to route towards the destination. Otherwise, the packet will be sent to the normal buffer in the local port of the neighboring router. 

In the EVC source/sink nodes, for each input channel, the routing computation module decides whether to use the EVCs based on the EVC latch availability and the average waiting time of the sink buffer in the EVC sink node. If the waiting time is larger than a threshold T (which means that there exists heavy congestion at the EVC sink node), the EVC will not be chosen for routing. 

At last, for the packets which cannot use EVCs or the express hub paths, the adaptive routing algorithm proposed for bi-directional NoC is used to find an output direction based on the run time traffic status.
\subsection{Adaptive routing for BiNoC}
\subsubsection{Channel and direction fitness}
In NoCs utilizing bidirectional channels, for each direction, there are two input/output channels corresponding to the master and slave links, respectively. We use two metrics, namely channel fitness and direction fitness, to reflect the suitability of an input channel to receive packets. 

In the channel fitness function, in order to predict the dynamics of the network traffic which may exhibit a non-Markovian behavior \cite{fractal_traffic}, we include the channel average waiting time to reflect the channels "memory". The fitness value of an input channel is represented as:
\begin{equation}
fitness_{in}=\alpha\times{m_{in}}\times{w_{in}^{-1}}
\end{equation}
where $\alpha$ is a scaling parameter, $m_{in}$ is the average number of free slots in the channel and $w_{in}$ is the average waiting time that a packet spends in the channel. In this work, according to the simulation results with different $\alpha$ values ranging from $0.5$ to $1.5$, we observe $\alpha$ equals to 0.8 offers the highest performance.

\begin{table}[t]
\centering
\caption{\label{notations_routing} Notations in the routing for proposed NoC}
\begin{tabular}{|c|p{0.70\columnwidth}|} \hline
Parameter & Description\\ \hline\hline
$N$ & Number of nodes in the mesh network\\ \hline
$P$ & The set of directions $P={N,E,S,W,L}$\\ \hline
$f[i][j][k]$ & Fitness of the input channel where $i\in{P}$, $j\in{J}$, $k\in{N}$\\ \hline
$P_{block}[i][j]$ & Blocking probability of the input channel in master inout port in direction $i$ of node $k$ \\ \hline
$F[i][k]$ & Fitness value of direction $i$ of node $k$ \\ \hline
$S_{l}[i][k]$ & Total score from current node $l$ to node $k$ through direction $i$\\ \hline
$CH[i][k]$ & Set of (node, direction) pairs chosen by K-step speculative algorithm from current node to the destination node $k$ along direction $i$\\
\hline\end{tabular}
\end{table}
During the routing stage, since we have not yet determined whether the master or the slave output port will be used, we need to evaluate the input channels at both ports of the downstream router to get a more appropriate metric for choosing the output direction. The direction fitness function is used to indicate the overall availability of one direction in the downstream router. One issue when designing the direction fitness function is to consider the characteristics of the bi-directional channels since the effective bandwidth between two neighboring routers varies with the actual traffic over time. One simple way is to add together the two channels fitness values regardless of them being of master or slave type. However, because the master port has a higher priority in sending and its input channel can only receive flits when there is no sending request at the same port. Then, if we only consider the channel fitness without regarding to the probability of receiving flits using this channel or the actual channel utilization, it may not lead to an optimal selection.
\begin{algorithm}
\caption{\label{fitness_cal} Channel direction fitness calculation}
\textbf{Input:} Average waiting time: $w[i][j][k]$; Average available slots $m[i][j][k]$; master port input channel block probability $P_{block}[i][j]$; scaling parameter $\alpha$ \\
\textbf{Output:} Channel fitness: $f[i][j][k]$ and the direction fitness function $F[i][k]$
\begin{algorithmic}[1]
\FORALL{node $k \in N$ }
\FORALL{$i \in P$}
\STATE $F[i][k]=0$
\FORALL{$j \in P$}
\item{$f[i][j][k]= \alpha \times{m[i][j][k]}\times{w[i][j][k]^{-1}}$}
\IF {$j==0$}
\STATE $F[i][k] += (1-P_{block}[i][j])\times{f[i][j][k]}$
\ELSE
\STATE $F[i][k] += f[i][j][k]$
\ENDIF
\ENDFOR \COMMENT {for all input direction $i$}
\ENDFOR \COMMENT {for all output direction $j$}
\ENDFOR \COMMENT{for all the nodes in the NoC}
\end{algorithmic}
\end{algorithm}

In order to deal with this issue, we calculate the fitness of a particular direction by distinguishing the input channels at the master and slave ports. For the input channel in the slave port, because the channel can always be used whenever there is a request from a neighbor, its fitness directly contributes to the fitness function calculation. For the input channels at the master inout port, we calculate the average block probability $P_{block}$ due to the write conflict during a time window. This is the probability when the FSM controller configures the port into the output mode and lets the output register use the link for sending packets instead of receiving the packets from the neighboring nodes. The fitness of direction i under consideration is then:
\begin{equation}
F_{i}=fitness_{slave\_in}+(1-P_{block})\times{fitness_{master\_in}}
\end{equation}
In Algorithm \ref{fitness_cal}, we present the idea of calculating the fitness function periodically for any given direction. The notations used for the algorithm are listed in Table \ref{notations_routing}. Two counters are needed in each input channel to calculate the average waiting time of packets and the number of available buffer slots during any time window. According to these two metrics, the input channel fitness at the master and slave inout ports can be calculated. An extra counter at the master port FSM is used to record the average blocking probability $P_{block}$. The fitness of the direction under evaluation is then the weighted sum of the two input channels fitness values. 
\subsubsection{Fitness based adaptive routing algorithm}
In order to avoid deadlocks at run time, the odd-even turn rules \cite{OE} are used in the minimal path based routing computation. Under this assumption, for any packet in the current router, there are at most two feasible output directions.

As discussed in \cite{RCA,NoP,ANCS}, if the adaptive routing algorithm only considers the local congestion metric, it will be too slow to re-act to congestion in the more distant parts of the network as it relies on network backpressure to propagate the congestion state.  In this work, we propose a scheme that combines the local and remote channel fitness status in order to make a better routing decision (see Algorithm 2).

\begin{figure}
\centering
\epsfig{file=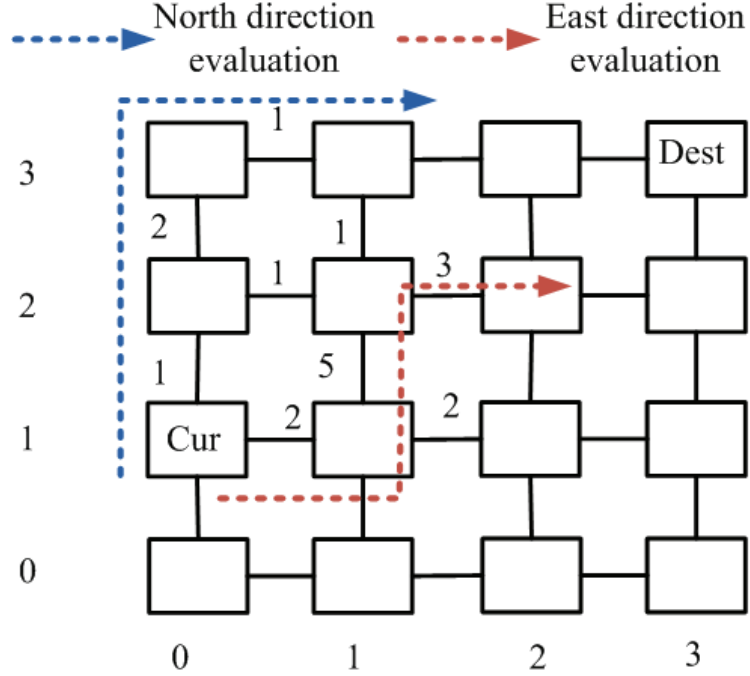,width=0.68\columnwidth}
\caption{\label{figure_5_K_step} K-step (K=3) speculative routing}
\end{figure}

Different from previous approaches in \cite{NoP,RCA,DBAR}, which utilize fixed local and non-local neighbors to evaluate each candidate direction in the router, our proposed adaptive routing algorithm uses a K-step speculative routing to dynamically select a set of nodes on the routing path which have the best fitness value towards the destination currently. Based on the fitness value of these nodes, the routing algorithm decides the optimal output port for the current router.
 
  Figure \ref{figure_5_K_step} shows a routing scenario for the proposed K-step speculative routing with $K=3$. The values on the link indicate the direction fitness level measured at each router. When evaluating the east output direction for the current node, node (1,1) is chosen first. Then, the router compares the two candidate directions from node (1,1) to the destination. Since the north output direction at node (1,1) has a larger direction fitness value (5) than the east output direction. The north neighboring node (1,2) is chosen second. Similarly, for node (1,2),  its east neighboring node (2,2) will be selected. Therefore, nodes (1,1), (1,2) and (2,2) form a 3-step speculative path towards the destination using the east output direction of the current router. Similarly, nodes (0,2), (0,3) and (1,3) form a speculative path from the north output direction towards the destination. 
 \begin{algorithm}
\caption{\label{K-step-lookahead} K-step speculative selection algorithm}
\textbf{Input:} Current node $\textbf{cur}$: Destination node $\textbf{dst}$; Set of candidate output directions $P_{c}$ \\
\textbf{Output:} Output direction $P_{o}\in{P_{c}}$\\
\textbf{Function:}$Candidate\_dir(a,b)$: A set of candidate directions for node $a$
\begin{algorithmic}[1]
\FORALL{direction $i\in{P_{c}}$}
\STATE $S_{cur}[i][\textbf{dst}]=0$;
\STATE $CH[i][\textbf{dst}]\leftarrow(cur,i)$;
\STATE $check\_length=0$;
\WHILE {$check\_length<K \&\& !CH[i][\textbf{dst}].IsEmpty()$}
\STATE $(check\_node,check\_dir)=CH[i][\textbf{dst}].Pop()$;
\STATE $next\_node = check\_node.get\_neighbor(check\_dir)$;
\IF {$next\_node!=dst$}
\STATE $S_{cur}[i][dst]+=F[reverse(check_{dir})][next\_node]$;
\STATE $C\_set = Candidate\_dir(next\_node, dst)$;
\IF {$F[C\_set[1]][next\_node]>F[C\_set[0]][next\_node]$}
\STATE $CH[i][dst]\leftarrow(next\_node,C\_set[1])$;
\ELSE 
\STATE $CH[i][dst]\leftarrow(next\_node,C\_set[0])$;
\ENDIF
\ENDIF
\STATE $check\_length++$;
\ENDWHILE
\ENDFOR
\STATE $P_{0}=Select\_highest\_score\_dir(S_{cur},P_{c})$;
\end{algorithmic}
\end{algorithm}

By adding together the direction fitness values of each path, the scores of using the north and the east output port can be calculated. The router then chooses the direction with the best score for the current router. If the direction fitness value changes at run time, the nodes selected for evaluating each candidate port will also change accordingly to capture the network dynamics.

Algorithm \ref{K-step-lookahead} presents the details of the proposed $K$-step speculative routing algorithm. At run time, each router periodically collects the channel and direction fitness values of the neighbors located in a region less than $K$-hops away. Then, for each output candidate, we perform a $K$-step look ahead routing towards the destinations and select a set of $K$ consecutive nodes and directions to form a speculative routing path towards the destinations using current candidate direction. By comparing the total fitness values along the paths between the two candidate directions, the direction with the best path fitness value is chosen as the output port for the current router. 
\begin{figure}[h]
\centering
\epsfig{file=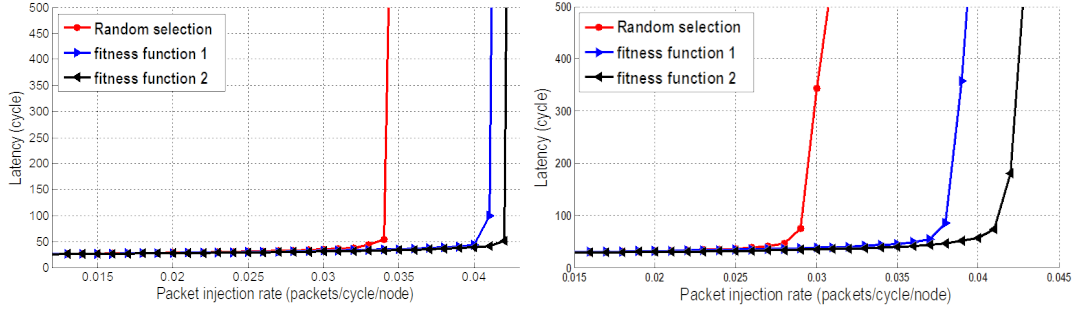,width=0.99\linewidth}
\caption{\label{figure_6_comp} Comparison of the fitness functions for $8\times8$ meshes using bi-directional channels}
\end{figure}
\begin{figure}[h]
\centering
\epsfig{file=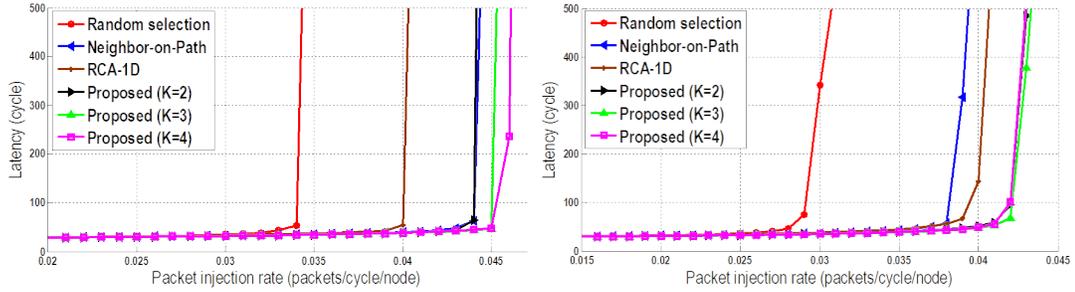,width=0.99\linewidth}
\caption{\label{figure_7_comp} Comparison of different adaptive routing schemes for $8\times8$ meshes}
\end{figure}

In this work, we adopt a light weight traffic status propagation network similar to \cite{ANCS} which is used to propagate the direction fitness information to the nodes within $K$-hops distance in the network. More specifically, every $T$ cycles, the direction fitness value updated by the current router is propagated $K$ hops away throughout the network in a hop-by-hop manner. After collecting all the direction fitness values from the neighboring routers within K-hops distance, these values are stored in the router and used for the proposed K-step speculative routing algorithm. For the case $K\leq2$, we can further reduce the network complexity by using a specific control channel between the neighboring nodes as in \cite{NoP} instead of a status propagation network to exchange the traffic information. The signal sent through the control channel contains the fitness of the direction $F[i][k]$ associated to the current router and its 1-hop neighbors. According to the simulation results shown in section 5, there is not too much performance improvement for $K\geq2$  compared to $K=2$. Hence we propose to set $K=2$ (i.e. look only two hops ahead) to achieve the best trade-off between the performance and additional complexity.
\section{Simulation results}
\subsection{Simulation setup}
We evaluate the proposed NoC architecture, as well as the adaptive routing algorithm, using a cycle accurate NoC simulator implemented in C++. Simulations were done for different mesh network sizes: $4\times4$ and $8\times8$. To investigate the benefits of our NoC architecture and the performance improvements of our adaptive traffic aware routing algorithm, we compare our approach with the following NoC architectures: 1) classical unidirectional NoC;
2) bi-directional NoC \cite{5715603}; 3) bi-directional NoC with static EVCs \cite{Kumar}.
 
Throughout the experiments, we assume that each input channel has a buffer depth of 6 flits for the NoC using bi-directional channels and a buffer depth of 12 flits for the uni-directional NoC. Both synthetic traffic (\textit{i.e.,} uniform, transpose) and real benchmark traces were used in the simulations. For uniform traffic, each node sends packets to other nodes in the network with an equal probability. For transpose traffic, each node at the $(i,j)$ mesh location sends packets to the node at the $(j,i)$ location on the mesh. 

For the real benchmark workloads, we consider several 16-node multithreaded commercial and scientific workloads from DBmbench \cite{DBmbench} and SPECweb99 \cite{Spec}. This type of traffic includes the memory request/response coherence traffic between PEs and caches. 
\begin{figure}
\epsfig{file=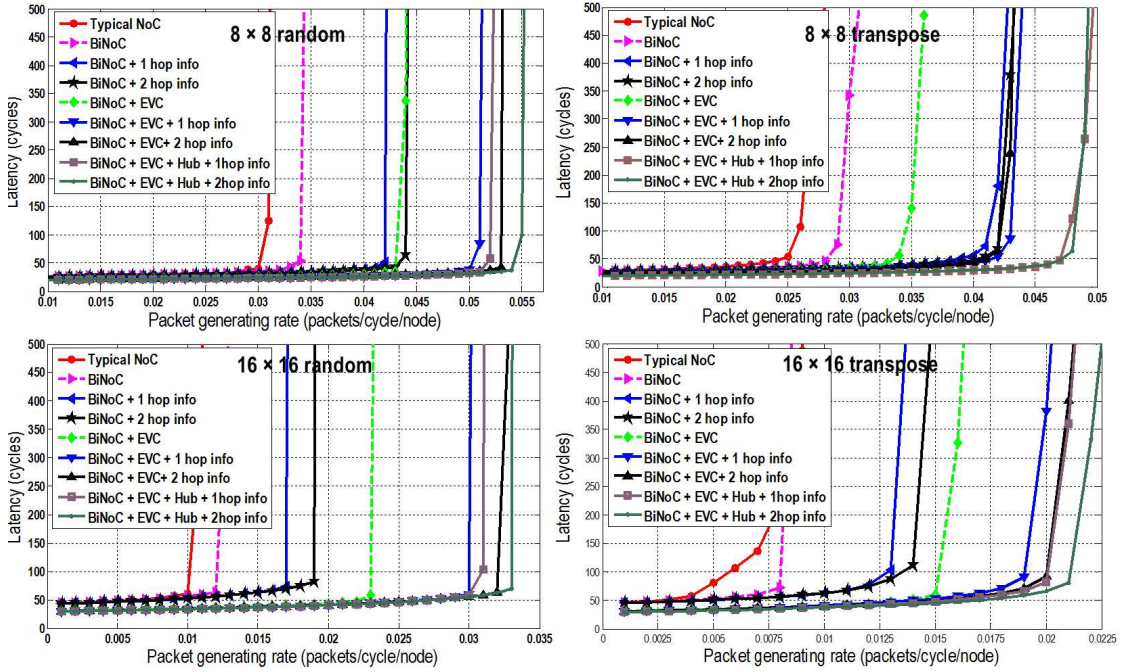,width=1.02\linewidth,clip=}
\caption{\label{figure_8_comp} Comparison of different NoC architectures under various mesh size and traffic patterns. }
\end{figure}
\subsection{Comparison of adaptive routing for BiNoC}
We compare our proposed adaptive routing algorithm with some widely adopted routing algorithms to evaluate the performance under bi-directional channels. More precisely, we compare the benefits of using a fitness function and a K-step speculative selection algorithm to capture the network traffic status. First, we compare different fitness functions used to evaluate the router channels. In this comparison, we only gather the traffic status from the immediate neighbors and use different fitness functions to evaluate the channels status. The baseline for this comparison is the random selection strategy of the Odd-even routing algorithm \cite{OE}. Two direction fitness functions are used in the simulation:

1) Fitness function 1 is the direct sum of the input channel fitness without distinguishing the master and slave channels 

2) Fitness function 2 uses the link blocking probability to weight the channels in the combination of the master and slave channels fitness values. 

Figure \ref{figure_6_comp} shows the comparison results obtained for an $8\times8$ mesh under random and transpose traffic patterns. As shown in this figure, using a fitness function to represent the channel traffic status significantly improves the network critical load (by more than 22.5\%). Furthermore, if we consider the characteristic of the bidirectional links by distinguishing the master and slave channels, an extra 5\% - 8\% performance gain can be achieved.

Next, using the proposed fitness function for the bidirectional NoC, we make a comparison of different adaptive routing schemes which consider both the immediate neighbors and non-local nodes to evaluate the directions in the routing selection phase. We evaluate our proposed K-step speculative selection scheme with different K values (K=2, 3, 4). At the same time we compare our scheme with two widely adopted schemes, namely neighbor-on-path \cite{NoP} and regional-contention-awareness (RCA-1D) \cite{RCA} under random, and transpose traffic patterns on an $8\times8$ mesh.  In order to minimize the intra-region interference in RCA-1D \cite{DBAR,ANCS}, we modify the RCA-1D scheme to integrate the destination into the selection function as in the DBAR approach \cite{DBAR}.

As shown from Figure \ref{figure_7_comp}, the proposed K-step speculative scheme achieves the best performance results by selectively choosing the intermediate nodes towards destination at run time. Also, it can be seen that there is no significant improvement in the performance for a K value larger than 2. Since the complexity of the fitness value propagation increases dramatically when $K\geq2$, we use $K=2$ throughout the remaining simulations. 

\subsection{Effect of adding EVC and hub routers}
In this section, we evaluate the performance improvement resulted by adding EVC and regional hub routers into the BiNoC. Both random selection and K-step speculative selection strategies are used in these simulations. The following architectures were evaluated and compared: the typical NoC with unidirectional channels, the BiNoC proposed in \cite{5715603}, the BiNoC + EVC which combines the features of Bi-directional switching in \cite{5715603} and express virtual channels in \cite{Kumar}, and the proposed NoC with regional hubs.

Figure \ref{figure_8_comp} summarizes the simulation results. We compare the average latency performance under various traffic injection rate for $4\times4$ and $8\times8$ meshes. Several observations can be made based on these results:
 
1) BiNoC outperforms the typical NoC design. Adding EVC into the network can further improve the network criticality by 9.8\%-80\% for different traffic patterns and mesh sizes. Furthermore, if the regional hub router is added, the throughput can be further improved by 9.3\%-15\%.

2) For all the NoC architectures using the adaptive routing algorithm, the fitness function employed to evaluate the traffic status can significantly improve the network performance. As shown in Figure \ref{figure_8_comp}, using the 1 hop away neighboring info can improve the network criticality by 13\% - 30\%, while using 2 hop neighbor info can further improve the performance by 3\% - 11\% in most cases.

\begin{figure}
\centering
\epsfig{file=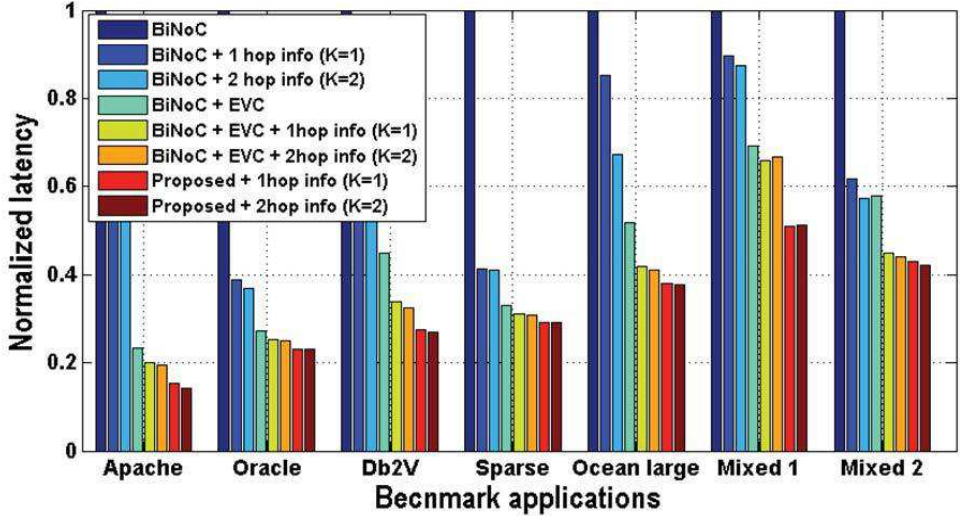,width=0.90\columnwidth}
\caption{\label{figure_9_latency} Latency comparison for benchmarks}
\end{figure}

\subsection{Results using real world workloads}
In this section, we show the latency comparison for various real world benchmarks with an $8\times8$ mesh size. We report seven experiments. The first five experiments collect the traces from the same applications, \textit{i.e.,} Apache, Oracle, Db2V, sparse and Ocean which were mapped onto an $8\times8$ mesh. The last two experiments named Mixed-1 and Mixed-2 combine the traces from different applications and map them to the same NoC platform. The simulation time of each benchmark is set to $5,000,000$ clock circles with a $2,000$ cycles warmup period.

\begin{figure}
\centering
\epsfig{file=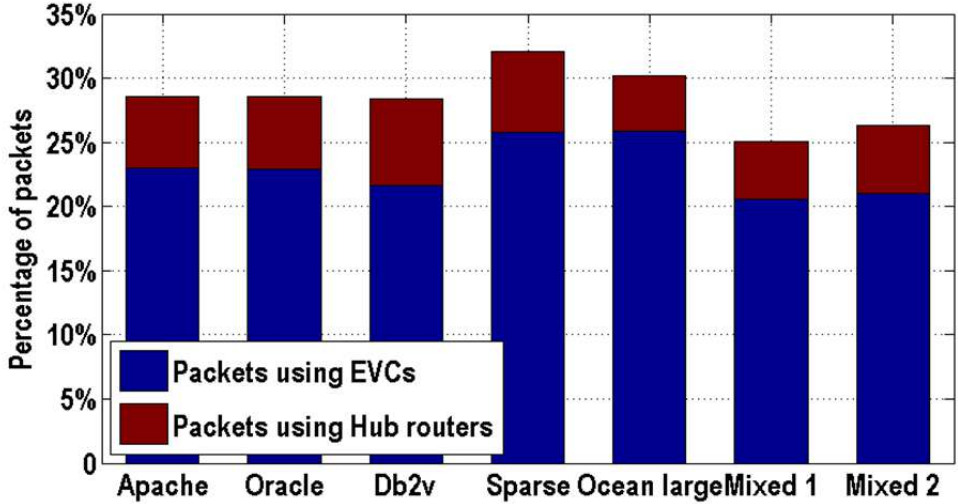,width=0.90\columnwidth}
\caption{\label{figure_10_percentage} Percentage of packets using EVCs and hub routers}
\end{figure}

As shown in Figure \ref{figure_9_latency}, compared with the baseline BiNoC design with random selection strategy, which is normalized to 1, by adopting a fitness function to evaluate the channels status can reduce the average latency by $15\%-60\%$.  For example, after applying the fitness functions, the latency of the sparse benchmark is reduced to $42\%$ compared with BiNoC routing. The static EVC approach further improves the performance by $6\%-28\%$ compared with its BiNoC counterparts for both random selection or fitness function based selection strategies. As in the sparse application, the EVC+BiNoC+2 hop info routing improves the latency by $25.2\%$ over the BiNoC + 2 hop info routing. If a central hub router is added into each region in our proposed NoC architecture, the performance can be further improved by $5\%-10\%$. As shown in Figure \ref{figure_9_latency}, for the sparse application, the proposed NoC architecture with $K=2$ speculative routing improves the latency by $6.5\%$. 
\begin{figure}
\centering
\epsfig{file=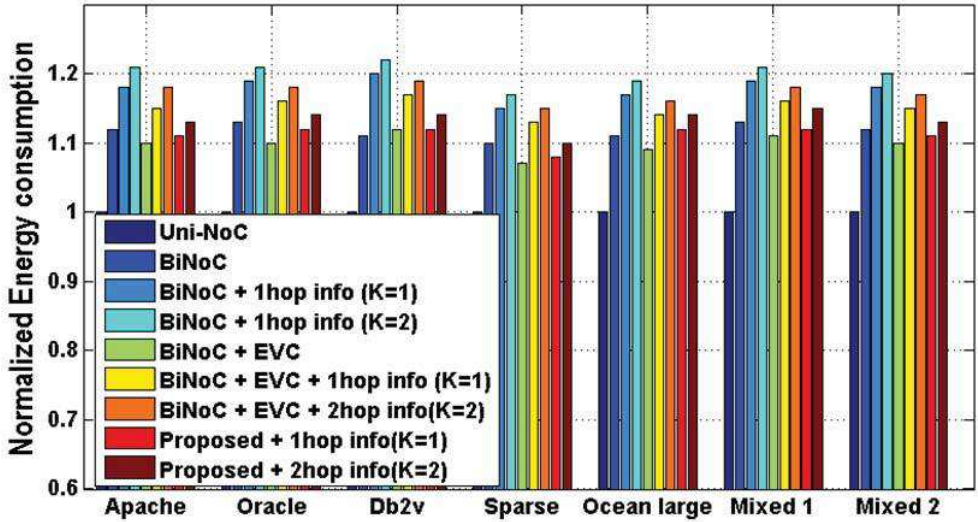,width=0.90\columnwidth,clip=}
\caption{\label{figure_11_energy} Energy comparison for benchmarks}
\end{figure}

Figure \ref{figure_10_percentage} shows the percentage of packets that utilize the EVCs and hub routers when running different benchmarks. As shown, on average, $5.5\%$ of packets utilize the express hub paths to send packets to the destination directly, while $22.9\%$ packets utilize EVCs at least once at run time.

Next, we evaluate the energy overhead of different NoC architectures while running these applications. For this purpose, we have modified the Orion power model \cite{orion2.0} to consider the extra energy consumption over the unidirectional NoC. For the BiNoC, the additional energy consumption is mainly due to the larger crossbar size ($2N\times{2N}$), and the channel direction control logic \cite{5715603}. We modified the energy consumption of the crossbar in the Orion power model and added the energy overhead for the link direction switching according to \cite{5715603}. For BiNoC + EVC architecture, the additional energy consumption for the EVC latch and the EVC arbiter is discussed in \cite{Kumar} and we add this to the Orion model. For the energy consumption of the proposed architecture, the hub latches in the normal router are modeled similarly to the EVC, while the energy consumption of the hub router is modeled as a four-port, four-VC normal router as in the Orion power model. For the calculation of the fitness values for each channel, we need additional adders and multipliers to calculate the fitness values every 100 cycles. This energy overhead is modeled in a similar way as in \cite{NoP}.

In Figure \ref{figure_11_energy}, the energy consumption of the unidirectional NoC architecture is normalized to one. The BiNoC approach has about 10\%-20\% energy overhead due to a large crossbar size and the link direction switching overhead \cite{5715603}. The EVC and regional hub approach can effectively reduce the energy by allowing some packets use the express paths and bypassing the router pipeline stages. In Figure \ref{figure_11_energy}, the energy overhead for the proposed NoC architecture with 2-step selection strategy is 13\% on average. 


\subsection{Overhead evaluation}
\begin{figure}
\centering
\epsfig{file=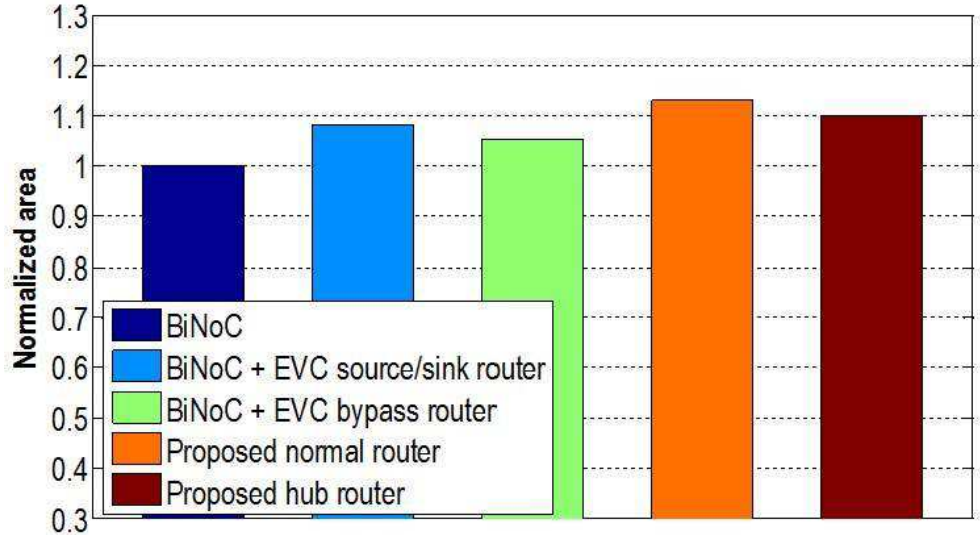,width=0.90\columnwidth,clip=}
\caption{\label{figure_13_area} Area overhead comparison}
\end{figure}

We have modified the Orion 2.0 area model \cite{orion2.0} to evaluate the area overhead. In this evaluation, we target a 1.0GHz operating frequency under the 65nm technology. When evaluating the BiNoC, the bi-directional link direction control logic is considered as in \cite{5715603}. As shown in Figure \ref{figure_13_area}, the area numbers of source/sink nodes and bypass nodes in the BiNoC+EVC architecture, as well as the normal router and hub router in the proposed architecture are compared against the baseline BiNoC design which is normalized to one. Router (1,2) in Figure \ref{figure_3_express_path} is chosen when evaluating the normal routers in the proposed architecture, since this router is the most complicated one with an extra port connected to the hub router. In Figure \ref{figure_13_area}, compared to BiNoC baseline router, the normal router of the proposed architecture has an overhead of about 13\%. Together with the additional hub router in each $4\times4$ region, the proposed NoC architecture has an area overhead of 19\%. 

\section{Conclusion}
In this chapter, we have proposed a flexible NoC architecture which utilizes bidirectional channels, EVC channels and regional hub routers to improve the NoC performance. We have also proposed a traffic-aware, adaptive routing algorithm that considers the characteristic of the bidirectional links. Simulation results on synthetic traffic, as well as real world benchmarks, show that the proposed NoC architecture, together with the routing algorithm, can significantly reduce the latency by as much as $80\%$ with a small energy and area overhead.  

\chapter{Conclusions and future work}
\section{Research summary}
The work discussed in this thesis focus on the issues of achieving a high performance NoC design for future multicore systems. In particular, starting from the offline design space exploration, we have proposed a more accurate NoC model to predict the latency performance under various routing algorithm and placement choices. We first identified the limitations of conventional simulations and queuing-theory-based performance evaluation methods. Motivated by combining the high accuracy in simulations and fast speed in analytical models, we have applied machine learning techniques to combine the advantages of simulations and mathematical formalism. In Chapter 2, we have presented a support vector regression based latency model for evaluating NoC performance in the synthesis inner loop. Through learning from the typical training data, the SVR-NoC achieves better accuracy and similar speedup performance compared to the queuing models, which will benefit the exploration of the design space offline. 
\\Then, in Chapter 3 and Chapter 4,  we have explored the NoC routing algorithm designs for different objectives and constraints. Specifically, we have discussed the routing algorithms for the thermal-awareness and fault-tolerance purposes. In Chapter 3, in order to reduce the hotspot temperature of the whole system while maintaining the latency and throughput performance, we have proposed an application-specific path set finding algorithm which has a high adaptivity to distribute traffic and ensures deadlock-free property. Then, a linear programming problem is formulated to compute the optimal ratio of using the paths offline. The router architecture which supports the proposed ratio-based adaptive routing is also discussed. In Chapter 4, we have proposed an adaptive routing algorithm to tackle run time router faults. In this chapter, we proposed a dynamic buffer swapping algorithm and a dynamic crossbar Mux swapping algorithm to maximally maintain the network connectivity under the buffer and crossbar faults, respectively. Higher packet acceptance rate and lower latency can be achieved compared to previous re-routing based methods.
\\In Chapter 5 and Chapter 6, we have proceeded to explore the architectures for optimizing the system performance. We have proposed our new NoC design based on the usage of the emerging bi-directional channels. More precisely, we have designed two new flexible NoC architectures using bi-directional links to further improve the throughput and latency performance. In Chapter 5, the overall NoC topology is unchanged. We focused on the router design to improve the usage of bi-directional channel bandwidth. Towards this end, we have proposed a flit-level speedup scheme (FSNoC) to allow two flits from the same packet to be transmitted simultaneously. In this way, higher bandwidth utilization and throughput can be achieved compared to the conventional NoCs equipped with bi-directional channels. In Chapter 6, we have proposed a new NoC topology which takes the the advantage of several express paths provided in the system: 1) bi-directional channels, 2) express paths and 3) regional hub router based shortcut paths. An adaptive routing algorithm is designed to dynamically choose among these paths. We have demonstrated the effectiveness of the proposed new architecture using both synthetic traffic and real multicore workloads. 
\section{Future research directions}
Additional work is needed to extend this thesis work. In future, we expect to conduct the researches in the following directions:
\\ \textbf{1) Exploring the current learning based NoC models:} We will explore and compare several other learning methods besides support vector regression (\textit{e.g.,} neural networks, Gaussian process). We will compare the model accuracy with the SVR-based model. Also, we will explore the inclusion of different features to improve the learning accuracy and reduce the training data size as well as learning time. We plan to study the importance of each features in the feature set and develop the criteria of how to choose features in the training stage. Moreover, we will develop a more flexible platform to extract arbitrary training data from the simulation.
\\ \textbf{2) Applying learning techniques to model the worst-case delay:} Besides the average latency modeling in this work, we will also apply the learning techniques to predict the worst-case delays for the NoCs with real-time deadline requirement. We will explore the effects of different learning methods on the worst case delay bound tightness. 
\\ \textbf{3) Hardware implementation for the flexible NoC architecture:} For our proposed NoC architecture, we plan to implement a FPGA-based prototype and carry out a thorough comparison and analysis on the power, area overhead.
\\ \textbf{4) More applications based on the proposed NoC platform:} In future, we also expect to see more applications which is implemented based on the proposed NoC architectures. Specifically, we have started to apply the proposed NoC platform for a specific biological application, \textit{i.e.,} the protein folding calculation and prediction.
\\ \textbf{5) Hardware implementations and detail evaluations of FSNoC under the existence of long wire links:} As the bi-directional repeaters introduced in Chapter 5 may increase the critical path length and power overhead, a more accurate hardware overhead evaluation of FSNoC should also consider the physical VLSI implementation and optimization of the bi-repeaters.
\\ \textbf{6) Run-time thermal aware routing for reducing the hotspot temperature:} The proposed thermal aware routing is based on application task graph, which only characterizes the average-case communication bandwidth requirement. Therefore, the routing is done in the offline phase and cannot capture the burst traffic arrivals at run time. In order to consider the fluctuation of power/thermal profile at run time, a fine-grained on-line routing algorithm is needed which determines the routing path dynamically based on the current temperature of the routers and PEs.

\AtEndEnvironment{thebibliography}{
 
}

\bibliographystyle{IEEEbib}
\bibliography{FSNoC,Learning,CODES_ISSS,Thermal_Springer,VLSI_SOC,Introduction,NoC_review,my_paper}

\begin{thebibliography}{100}

\bibitem{transistor_count}
``Transistor count statistics,''
  \url{http://www.wikipedia.org/wiki/Transistor_count}, 2013.

\bibitem{Computer_architecture_book}
John~L. Hennessy and David~A. Patterson,
\newblock {\em Computer Architecture, Fourth Edition: A Quantitative Approach},
\newblock Morgan Kaufmann Publishers Inc., San Francisco, CA, USA, 2012.

\bibitem{digital_system_design}
Jan~M. Rabaey,
\newblock {\em Digital Integrated Circuits: A Design Perspective},
\newblock Prentice-Hall, Inc., Upper Saddle River, NJ, USA, 2008.

\bibitem{isscc_noc_tutorial}
``Design of energy-efficient on-chip networks,''
  \url{http://www.rle.mit.edu/isg/documents/2010_NOC_tutorial_vladimir.pdf},
  2010.

\bibitem{Routing_algorithm_springer}
Maurizio Palesi and Masoud Daneshtalab~(Eds.),
\newblock {\em Routing Algorithms in Networks-on-Chip},
\newblock Springer, 2014.

\bibitem{NoCbook_Peh}
N.~E. Jerger and L.~S. Peh,
\newblock {\em On-chip Networks},
\newblock Morgan, New York, 2009.

\bibitem{Ge-type}
Y.~Wu, G.~Min, M.~Ould-Khaoua, H.~Yin, and L.~Wang,
\newblock ``Analytical modelling of networks in multicomputer systems under
  bursty and batch arrival traffic,''
\newblock {\em The Journal of Supercomputing}, vol. 51, no. 2, pp. 115--130,
  2010.

\bibitem{ML}
C.M. Bishop,
\newblock {\em Pattern recognition and machine learning},
\newblock Springer, 2006.

\bibitem{5158098}
Shu-Yen Lin, Wen-Chung Shen, Chan-Cheng Hsu, Chih-Hao Chao, and An-Yeu Wu,
\newblock ``Fault-tolerant router with built-in self-test/self-diagnosis and
  fault-isolation circuits for 2d-mesh based chip multiprocessor systems,''
\newblock in {\em VLSI Design, Automation and Test, 2009. VLSI-DAT '09.
  International Symposium on}, 2009, pp. 72--75.

\bibitem{5467330}
A.~Kohler, G.~Schley, and M.~Radetzki,
\newblock ``Fault tolerant network on chip switching with graceful performance
  degradation,''
\newblock {\em Computer-Aided Design of Integrated Circuits and Systems, IEEE
  Transactions on}, vol. 29, no. 6, pp. 883--896, 2010.

\bibitem{Moore_law}
Robert~R. Schaller,
\newblock ``Moore's law: Past, present, and future,''
\newblock {\em IEEE Spectr.}, vol. 34, no. 6, pp. 52--59, June 1997.

\bibitem{Xeon}
``Intel xeon processors,'' \url{http://www.wikipedia.org/wiki/Xeon}, 2013.

\bibitem{Nvidia_gpu}
``Nvidia geforce gtx780,''
  \url{http://www.techpowerup.com/reviews/NVIDIA/GeForce_GTX_780_Ti/1.html},
  2013.

\bibitem{jounarth_thesis}
Lap~Fai Leung,
\newblock {\em Designing high-performance and low-energy real-time embedded
  systems based on single-core and multi-cores structures (Ph.D. Thesis)},
\newblock HKUST, Hong Kong, 2007.

\bibitem{CMP_ISSCC}
F.~Clermidy, C.~Bernard, R.~Lemaire, J.~Martin, I.~Miro-Panades, Y.~Thonnart,
  P.~Vivet, and N.~Wehn,
\newblock ``A 477mw noc-based digital baseband for mimo 4g sdr,''
\newblock in {\em Solid-State Circuits Conference Digest of Technical Papers
  (ISSCC), 2010 IEEE International}, 2010, pp. 278--279.

\bibitem{Intel_teraflop}
S.~R. Vangal, J.~Howard, G.~Ruhl, S.~Dighe, H.~Wilson, J.~Tschanz, D.~Finan,
  A.~Singh, T.~Jacob, S.~Jain, V.~Erraguntla, C.~Roberts, Y.~Hoskote,
  N.~Borkar, and S.~Borkar,
\newblock ``An 80-tile sub-100-w teraflops processor in 65-nm cmos,''
\newblock {\em Solid-State Circuits, IEEE Journal of}, vol. 43, no. 1, pp.
  29--41, 2008.

\bibitem{optical_noc}
Yaoyao Ye, Jiang Xu, Xiaowen Wu, Wei Zhang, Xuan Wang, M.~Nikdast, Zhehui Wang,
  and Weichen Liu,
\newblock ``System-level modeling and analysis of thermal effects in optical
  networks-on-chip,''
\newblock {\em Very Large Scale Integration (VLSI) Systems, IEEE Transactions
  on}, vol. 21, no. 2, pp. 292--305, Feb 2013.

\bibitem{optical_noc_2}
S.~Poddar, P.~Ghosal, P.~Mukherjee, S.~Samui, and H.~Rahaman,
\newblock ``Design of an noc with on-chip photonic interconnects using adaptive
  cdma links,''
\newblock in {\em SOC Conference (SOCC), 2012 IEEE International}, Sept 2012,
  pp. 352--357.

\bibitem{vicis}
D.~Fick, A.~DeOrio, Jin Hu, V.~Bertacco, D.~Blaauw, and D~Sylvester,
\newblock ``Vicis: A reliable network for unreliable silicon,''
\newblock in {\em Design Automation Conference, 2009. DAC '09. 46th ACM/IEEE},
  2009, pp. 812--817.

\bibitem{Book}
William Dally and Brian Towles,
\newblock {\em Principles and Practices of Interconnection Networks},
\newblock Morgan Kaufmann, 2003.

\bibitem{GALS}
Y.~Thonnart, P.~Vivet, and F.~Clermidy,
\newblock ``A fully-asynchronous low-power framework for gals noc
  integration,''
\newblock in {\em Design, Automation Test in Europe Conference Exhibition
  (DATE), 2010}, 2010, pp. 33--38.

\bibitem{Comparison_NoC}
``A comparison of network-on-chip and busses,''
  \url{http://www.ict.kth.se/courses/IL2207/0810/docs/noc_whitepaper.pdf},
  2008.

\bibitem{application_specific_plaesi}
M.~Palesi, R.~Holsmark, S.~Kumar, and V.~Catania,
\newblock ``Application specific routing algorithms for networks on chip,''
\newblock {\em Parallel and Distributed Systems, IEEE Transactions on}, vol.
  20, no. 3, pp. 316--330, 2009.

\bibitem{TVLSI12}
A.~E. Kiasari, Z.~Lu, and A.~Jantsch,
\newblock ``An analytical latency model for networks-on-chip,''
\newblock {\em Very Large Scale Integration (VLSI) Systems, IEEE Trans. on},
  vol. PP, no. 99, pp. 1 --11, 2012.

\bibitem{mesh_topo}
M.~Mirza-Aghatabar, S.~Koohi, S.~Hessabi, and M.~Pedram,
\newblock ``An empirical investigation of mesh and torus noc topologies under
  different routing algorithms and traffic models,''
\newblock in {\em Digital System Design Architectures, Methods and Tools, 2007.
  DSD 2007. 10th Euromicro Conference on}, 2007, pp. 19--26.

\bibitem{flattern_butterfly}
J.~Kim, J.~Balfour, and W.J. Dally,
\newblock ``Flattened butterfly topology for on-chip networks,''
\newblock in {\em Microarchitecture, 2007. MICRO 2007. 40th Annual IEEE/ACM
  International Symposium on}, 2007, pp. 172--182.

\bibitem{3D-mesh}
M.~Ahmed and R.~Kumar,
\newblock ``Parameterized path-based, randomized, oblivious, minimal routing in
  3d mesh noc,''
\newblock in {\em TENCON 2012 - 2012 IEEE Region 10 Conference}, 2012, pp.
  1--6.

\bibitem{fat_tree}
V.~Dvorak and J.~Jaros,
\newblock ``Optimizing collective communications on 2d-mesh and fat tree noc,''
\newblock in {\em Networks (ICN), 2010 Ninth International Conference on},
  2010, pp. 22--27.

\bibitem{raw_chip}
M.B. Taylor, J.~Kim, J.~Miller, D.~Wentzlaff, F.~Ghodrat, B.~Greenwald,
  H.~Hoffman, P.~Johnson, Jae-Wook Lee, W.~Lee, A.~Ma, A.~Saraf, M.~Seneski,
  N.~Shnidman, V.~Strumpen, M.~Frank, S.~Amarasinghe, and A.~Agarwal,
\newblock ``The raw microprocessor: a computational fabric for software
  circuits and general-purpose programs,''
\newblock {\em Micro, IEEE}, vol. 22, no. 2, pp. 25--35, 2002.

\bibitem{interconnection_network}
Jose Duato, Sudhakar Yalamanchili, and Ni~Lionel,
\newblock {\em Interconnection Networks: An Engineering Approach},
\newblock Morgan Kaufmann Publishers Inc., San Francisco, CA, USA, 2002.

\bibitem{virtual-cut-through}
Parviz Kermani and Leonard Kleinrock,
\newblock ``Virtual cut-through: a new computer communication switching
  technique,''
\newblock {\em Computer Networks}, vol. 3, pp. 267--286, 1979.

\bibitem{look_ahead_router}
Ling Xin and Chiu-Sing Choy,
\newblock ``A low-latency noc router with lookahead bypass,''
\newblock in {\em Circuits and Systems (ISCAS), Proceedings of 2010 IEEE
  International Symposium on}, May 2010, pp. 3981--3984.

\bibitem{ICCAD_latency}
Mingche Lai, Lei Gao, Nong Xiao, and Zhiying Wang,
\newblock ``An accurate and efficient performance analysis approach based on
  queuing model for network on chip,''
\newblock in {\em proc. ICCAD 2009.}, nov. 2009, pp. 563 --570.

\bibitem{Guz07networkdelays}
Z.~Guz, I.~Walter, E.~Bolotin, I.~Cidon, R.~Ginosar, and A.~Kolodny,
\newblock ``Network delays and link capacities in application-specific wormhole
  nocs,'' 2007.

\bibitem{umit}
U.Y. Ogras, P.~Bogdan, and R.~Marculescu,
\newblock ``An analytical approach for network-on-chip performance analysis,''
\newblock {\em Computer-Aided Design of Integrated Circuits and Systems, IEEE
  Trans. on}, vol. 29, no. 12, pp. 2001 --2013, dec. 2010.

\bibitem{Delay_hetergeneous_noc}
Y.~Ben-Itzhak, I.~Cidon, and A.~Kolodny,
\newblock ``Delay analysis of wormhole based heterogeneous noc,''
\newblock in {\em Networks on Chip (NoCS), 2011 Fifth IEEE/ACM International
  Symposium on}, 2011, pp. 161--168.

\bibitem{fractal_traffic}
P.~Bogdan and R.~Marculescu,
\newblock ``Non-stationary traffic analysis and its implications on multicore
  platform design,''
\newblock {\em Computer-Aided Design of Integrated Circuits and Systems, IEEE
  Transactions on}, vol. 30, no. 4, pp. 508--519, 2011.

\bibitem{Vapnik}
V.~Vapnik,
\newblock {\em Statistical Learning theory},
\newblock John Wiley and Sons, 1998.

\bibitem{prob_model}
O.~Lysne,
\newblock ``Towards a generic analytical model of wormhole routing networks,''
\newblock {\em microprocessors and microsystems}, vol. 21, no. 7-8, pp.
  491--498, 1998.

\bibitem{network_calculus_1}
M.~Bakhouya, S.~Suboh, J.~Gaber, and T.~El-Ghazawi,
\newblock ``Analytical modeling and evaluation of on-chip interconnects using
  network calculus,''
\newblock in {\em proc. NoCS 2009.}, may 2009, pp. 74 --79.

\bibitem{ICCAD}
N.~Nikitin and J.~Cortadella,
\newblock ``A performance analytical model for network-on-chip with constant
  service time routers,''
\newblock in {\em proc. ICCAD 2009.}, nov. 2009, pp. 571 --578.

\bibitem{Hu_ananalytical}
Po-Chi Hu and Leonard Kleinrock,
\newblock ``An analytical model for wormhole routing with finite size input
  buffers,'' 1997.

\bibitem{learning-aspdac-area-power}
A.B. Kahng, Bill Lin, and K.~Samadi,
\newblock ``Improved on-chip router analytical power and area modeling,''
\newblock in {\em Design Automation Conference (ASP-DAC), 2010 15th Asia and
  South Pacific}, 2010, pp. 241--246.

\bibitem{orion2.0}
A.B. Kahng, Bin Li, Li-Shiuan Peh, and K.~Samadi,
\newblock ``Orion 2.0: A power-area simulator for interconnection networks,''
\newblock {\em Very Large Scale Integration (VLSI) Systems, IEEE Transactions
  on}, vol. 20, no. 1, pp. 191--196, 2012.

\bibitem{area-power-model-dac}
Andrew~B. Kahng, Bill Lin, and Siddhartha Nath,
\newblock ``Explicit modeling of control and data for improved noc router
  estimation,''
\newblock in {\em Proceedings of the 49th Annual Design Automation Conference},
  New York, NY, USA, 2012, DAC '12, pp. 392--397, ACM.

\bibitem{da-cheng-islped}
Da-Cheng Juan and Diana Marculescu,
\newblock ``Power-aware performance increase via core/uncore reinforcement
  control for chip-multiprocessors,''
\newblock in {\em Proceedings of the 2012 ACM/IEEE International Symposium on
  Low Power Electronics and Design}, New York, NY, USA, 2012, ISLPED '12, pp.
  97--102, ACM.

\bibitem{learning-routing}
M.~Ebrahimi, M.~Daneshtalab, F.~Farahnakian, J.~Plosila, P.~Liljeberg,
  M.~Palesi, and H.~Tenhunen,
\newblock ``Haraq: Congestion-aware learning model for highly adaptive routing
  algorithm in on-chip networks,''
\newblock in {\em Networks on Chip (NoCS), 2012 Sixth IEEE/ACM International
  Symposium on}, 2012, pp. 19--26.

\bibitem{neural-network-routing}
Terrence Mak, Peter~Y.K. Cheung, Wayne Luk, and Kai~Pui Lam,
\newblock ``A dp-network for optimal dynamic routing in network-on-chip,''
\newblock in {\em Proceedings of the 7th IEEE/ACM International Conference on
  Hardware/Software Codesign and System Synthesis}, New York, NY, USA, 2009,
  CODES+ISSS '09, pp. 119--128, ACM.

\bibitem{Ge-model-two}
D.~D. Kouvatsos, A.~SALAM, and M.~Ould-Khaoua,
\newblock ``Performance modeling of wormhole-routed hypercubes with bursty
  traffic and finite buffers,''
\newblock {\em Int J Simul Pract Syst Sci Technol}, vol. 6, pp. 69--81, 2005.

\bibitem{SC_model}
M.~Arjomand and H.~Sarbazi-Azad,
\newblock ``A comprehensive power-performance model for nocs with multi-flit
  channel buffers,''
\newblock in {\em Proceedings of the 23rd international conference on
  Supercomputing}, New York, NY, USA, 2009, ICS '09, pp. 470--478, ACM.

\bibitem{Geyong-min}
G.~Min and M.~Ould-Khaoua,
\newblock ``A performance model for wormhole-switched interconnection networks
  under self-similar traffic,''
\newblock {\em Computers, IEEE Transactions on}, vol. 53, no. 5, pp. 601--613,
  2004.

\bibitem{NoC13-latency-model}
E.~Fischer and G.P. Fettweis,
\newblock ``An accurate and scalable analytic model for round-robin arbitration
  in network-on-chip,''
\newblock in {\em Networks on Chip (NoCS), 2013 Seventh IEEE/ACM International
  Symposium on}, 2013, pp. 1--8.

\bibitem{KPC-toolbox}
G.~Casale, E.~Z. Zhang, and E.~Smirni,
\newblock ``Kpc-toolbox: Simple yet effective trace fitting using markovian
  arrival processes,''
\newblock in {\em Quantitative Evaluation of Systems, 2008. QEST '08. Fifth
  International Conference on}, 2008, pp. 83--92.

\bibitem{VC_multiplexing}
M.~Ould-Khaoua,
\newblock ``A performance model for duato's fully adaptive routing algorithm in
  k-ary n-cubes,''
\newblock {\em Computers, IEEE Transactions on}, vol. 48, no. 12, pp.
  1297--1304, 1999.

\bibitem{VC_multiplexing_dally}
W.~Dally,
\newblock ``Virtual-channel flow control,''
\newblock {\em Parallel and Distributed Systems, IEEE Transactions on}, vol. 3,
  no. 2, pp. 194--205, 1992.

\bibitem{Europar-vc}
A.~E. Kiasari, D.~Rahmati, H.~Sarbazi-Azad, and S.~Hessabi,
\newblock ``A markovian performance model for networks-on-chip,''
\newblock in {\em Parallel, Distributed and Network-Based Processing, 2008. PDP
  2008. 16th Euromicro Conference on}, 2008, pp. 157--164.

\bibitem{Maximum_entropy}
M.~A. El-Affendi and D.~D. Kouvatsos,
\newblock ``A maximum entropy analysis of the m/g/1 and g/m/1 queueing systems
  at equilibrium,''
\newblock {\em Acta Informatica}, vol. 19, no. 4, pp. 339--355, 1983.

\bibitem{svr-noc}
Zhiliang Qian, Da-Cheng Juan, Paul Bogdan, Chi-Ying Tsui, Diana Marculescu, and
  Radu Marculescu,
\newblock ``Svr-noc: A performance analysis tool for network-on-chips using
  learning-based support vector regression model,''
\newblock in {\em Design, Automation Test in Europe Conference Exhibition
  (DATE), 2013}, March 2013, pp. 354--357.

\bibitem{libsvm}
{\em libsvm,http://www.csie.ntu.edu.tw/~cjlin/libsvm},
\newblock 2012.

\bibitem{lssvm}
{\em LS-SVMlab,http://www.esat.kuleuven.be/sista/lssvmlab},
\newblock 2012.

\bibitem{booksim}
{\em Booksim 2.0, http://nocs.stanford.edu/booksim.html},
\newblock 2012.

\bibitem{noc_design_65nm}
A.~Pullini, F.~Angiolini, P.~Meloni, D.~Atienza, S.~Murali, L.~Raffo,
  G.~De~Micheli, and L.~Benini,
\newblock ``Noc design and implementation in 65nm technology,''
\newblock in {\em Networks-on-Chip, 2007. NOCS 2007. First International
  Symposium on}, 2007, pp. 273--282.

\bibitem{thermal_noc}
L.~Shang, Li-Shiuan Peh, A~Kumar, and N.K. Jha,
\newblock ``Temperature-aware on-chip networks,''
\newblock {\em Micro, IEEE}, vol. 26, no. 1, pp. 130--139, 2006.

\bibitem{hotspot_elimination}
B.C. Schafer and Taewhan Kim,
\newblock ``Hotspots elimination and temperature flattening in vlsi circuits,''
\newblock {\em Very Large Scale Integration (VLSI) Systems, IEEE Transactions
  on}, vol. 16, no. 11, pp. 1475--1487, 2008.

\bibitem{thermal_aware_IP}
W.~Hung, C.~Addo-Quaye, T.~Theocharides, Y.~Xie, N.~Vijakrishnan, and M.J.
  Irwin,
\newblock ``Thermal-aware ip virtualization and placement for networks-on-chip
  architecture,''
\newblock in {\em Computer Design: VLSI in Computers and Processors, 2004. ICCD
  2004. Proceedings. IEEE International Conference on}, 2004, pp. 430--437.

\bibitem{router_example}
S.~Vangal, A.~Singh, J.~Howard, S.~Dighe, N.~Borkar, and A.~Alvandpour,
\newblock ``A 5.1ghz 0.34mm2 router for network-on-chip applications,''
\newblock in {\em VLSI Circuits, 2007 IEEE Symposium on}, 2007, pp. 42--43.

\bibitem{1411933}
Jingcao Hu and R.~Marculescu,
\newblock ``Energy- and performance-aware mapping for regular noc
  architectures,''
\newblock {\em Computer-Aided Design of Integrated Circuits and Systems, IEEE
  Transactions on}, vol. 24, no. 4, pp. 551 -- 562, april 2005.

\bibitem{Noc_synthesis}
D.~Bertozzi, A.~Jalabert, S.~Murali, R.~Tamhankar, S.~Stergiou, L.~Benini, and
  G.~De~Micheli,
\newblock ``Noc synthesis flow for customized domain specific multiprocessor
  systems-on-chip,''
\newblock {\em Parallel and Distributed Systems, IEEE Transactions on}, vol.
  16, no. 2, pp. 113--129, 2005.

\bibitem{antnet_routing}
M.~Daneshtalab, A.~Sobhani, A.~Afzali-Kusha, O.~Fatemi, and Z.~Navabi,
\newblock ``Noc hot spot minimization using antnet dynamic routing algorithm,''
\newblock in {\em Application-specific Systems, Architectures and Processors,
  2006. ASAP '06. International Conference on}, 2006, pp. 33--38.

\bibitem{hotspot_prevention}
G.M. Link and N.~Vijaykrishnan,
\newblock ``Hotspot prevention through runtime reconfiguration in
  network-on-chip,''
\newblock in {\em Design, Automation and Test in Europe, 2005. Proceedings},
  2005, pp. 648--649 Vol. 1.

\bibitem{inducing_thermal_wareness}
D.~Atienza and E.~Martinez,
\newblock ``Inducing thermal-awareness in multicore systems using
  networks-on-chip,''
\newblock in {\em VLSI, 2009. ISVLSI '09. IEEE Computer Society Annual
  Symposium on}, 2009, pp. 187--192.

\bibitem{fluidity_concept_noc}
Ying-Cherng Lan, M.C. Chen, A.P. Su, Yu-Hen Hu, and Sao-Jie Chen,
\newblock ``Fluidity concept for noc: A congestion avoidance and relief routing
  scheme,''
\newblock in {\em SOC Conference, 2008 IEEE International}, 2008, pp. 65--70.

\bibitem{routing_traffic_migration}
Chih-Hao Chao, Kun-Chih Chen, and An-Yeu Wu,
\newblock ``Routing-based traffic migration and buffer allocation schemes for
  3-d network-on-chip systems with thermal limit,''
\newblock {\em Very Large Scale Integration (VLSI) Systems, IEEE Transactions
  on}, vol. 21, no. 11, pp. 2118--2131, 2013.

\bibitem{my_asp_dac_2011}
Zhiliang Qian and Chi-Ying Tsui,
\newblock ``A thermal-aware application specific routing algorithm for
  network-on-chip design,''
\newblock in {\em Design Automation Conference (ASP-DAC), 2011 16th Asia and
  South Pacific}, Jan 2011, pp. 449--454.

\bibitem{deadlock_duato}
J.~Duato,
\newblock ``A necessary and sufficient condition for deadlock-free adaptive
  routing in wormhole networks,''
\newblock {\em Parallel and Distributed Systems, IEEE Transactions on}, vol. 6,
  no. 10, pp. 1055--1067, 1995.

\bibitem{tarjan_algorithm}
Robert Tarjan,
\newblock ``Depth first search and linear graph algorithms,''
\newblock {\em SIAM Journal on Computing}, 1972.

\bibitem{hotspot}
``Hotspot 5.0 temperature modeling tool,''
  \url{http://lava.cs.virginia.edu/HotSpot}, 2008.

\bibitem{matlab_cvx}
``Cvx: Matlab software for disciplined convex programming,''
  \url{http://cvxr.com/cvx}, 2010.

\bibitem{the:Noxim-simulator-User}
``Noxim simulator,'' \url{http://www.noxim.org}, 2010.

\bibitem{nigram}
``Nigram simulator,'' \url{http://nigram.ecs.soton.ac.uk}.

\bibitem{xpipe_luca_benini}
D.~Bertozzi and L.~Benini,
\newblock ``Xpipes: a network-on-chip architecture for gigascale
  systems-on-chip,''
\newblock {\em Circuits and Systems Magazine, IEEE}, vol. 4, no. 2, pp. 18--31,
  2004.

\bibitem{bandwidth_aware_routing}
M.~Palesi, S.~Kumar, and V.~Catania,
\newblock ``Bandwidth-aware routing algorithms for networks-on-chip
  platforms,''
\newblock {\em Computers Digital Techniques, IET}, vol. 3, no. 5, pp. 413--429,
  2009.

\bibitem{turn_model}
Christopher~J. Glass and Lionel~M. Ni,
\newblock ``The turn model for adaptive routing,''
\newblock {\em SIGARCH Comput. Archit. News}, vol. 20, no. 2, pp. 278--287,
  Apr. 1992.

\bibitem{isca_routing}
Rajendra~V. Boppana and Suresh Chalasani,
\newblock ``A comparison of adaptive wormhole routing algorithms,''
\newblock in {\em Proceedings of the 20th Annual International Symposium on
  Computer Architecture}, New York, NY, USA, 1993, ISCA '93, pp. 351--360, ACM.

\bibitem{4555858}
Zhen Zhang, A.~Greiner, and S.~Taktak,
\newblock ``A reconfigurable routing algorithm for a fault-tolerant 2d-mesh
  network-on-chip,''
\newblock in {\em Design Automation Conference, 2008. DAC 2008. 45th ACM/IEEE},
  2008, pp. 441--446.

\bibitem{resilient_routing}
D.~Fick, A.~DeOrio, G.~Chen, V.~Bertacco, D~Sylvester, and D.~Blaauw,
\newblock ``A highly resilient routing algorithm for fault-tolerant nocs,''
\newblock in {\em Design, Automation Test in Europe Conference Exhibition,
  2009. DATE '09.}, 2009, pp. 21--26.

\bibitem{4542002}
A.~Hosseini, T.~Ragheb, and Y.~Massoud,
\newblock ``A fault-aware dynamic routing algorithm for on-chip networks,''
\newblock in {\em Circuits and Systems, 2008. ISCAS 2008. IEEE International
  Symposium on}, 2008, pp. 2653--2656.

\bibitem{1633499}
Dongkook Park, C.~Nicopoulos, Jongman Kim, N.~Vijaykrishnan, and C.R. Das,
\newblock ``Exploring fault-tolerant network-on-chip architectures,''
\newblock in {\em Dependable Systems and Networks, 2006. DSN 2006.
  International Conference on}, 2006, pp. 93--104.

\bibitem{my_vlsi_soc}
Zhiliang Qian, Ying-Fei Teh, and Chi-Ying Tsui,
\newblock ``A fault-tolerant network-on-chip design using dynamic
  reconfiguration of partial-faulty routing resources,''
\newblock in {\em VLSI and System-on-Chip (VLSI-SoC), 2011 IEEE/IFIP 19th
  International Conference on}, Oct 2011, pp. 192--195.

\bibitem{4669214}
D.~Frazzetta, G.~Dimartino, M.~Palesi, S.~Kumar, and V.~Catania,
\newblock ``Efficient application specific routing algorithms for noc systems
  utilizing partially faulty links,''
\newblock in {\em Digital System Design Architectures, Methods and Tools, 2008.
  DSD '08. 11th EUROMICRO Conference on}, 2008, pp. 18--25.

\bibitem{OE}
Ge-Ming Chiu,
\newblock ``The odd-even turn model for adaptive routing,''
\newblock {\em Parallel and Distributed Systems, IEEE Transactions on}, vol.
  11, no. 7, pp. 729--738, 2000.

\bibitem{250114}
J.~Duato,
\newblock ``A new theory of deadlock-free adaptive routing in wormhole
  networks,''
\newblock {\em Parallel and Distributed Systems, IEEE Transactions on}, vol. 4,
  no. 12, pp. 1320--1331, 1993.

\bibitem{524561}
K.V. Anjan and T.M. Pinkston,
\newblock ``An efficient, fully adaptive deadlock recovery scheme: Disha,''
\newblock in {\em Computer Architecture, 1995. Proceedings., 22nd Annual
  International Symposium on}, 1995, pp. 201--210.

\bibitem{4492729}
M.~Palesi, G.~Longo, S.~Signorino, R.~Holsmark, S.~Kumar, and V.~Catania,
\newblock ``Design of bandwidth aware and congestion avoiding efficient routing
  algorithms for networks-on-chip platforms,''
\newblock in {\em Networks-on-Chip, 2008. NoCS 2008. Second ACM/IEEE
  International Symposium on}, 2008, pp. 97--106.

\bibitem{976921}
L.~Benini and G.~De~Micheli,
\newblock ``Networks on chips: a new soc paradigm,''
\newblock {\em Computer}, vol. 35, no. 1, pp. 70 --78, jan 2002.

\bibitem{Vichar}
C.A. Nicopoulos, Dongkook Park, Jongman Kim, N.~Vijaykrishnan, M.S. Yousif, and
  C.R. Das,
\newblock ``Vichar: A dynamic virtual channel regulator for network-on-chip
  routers,''
\newblock in {\em Microarchitecture, 2006. MICRO-39. 39th Annual IEEE/ACM
  International Symposium on}, 2006, pp. 333--346.

\bibitem{reconfigurable-links-GLVLSI}
Jie Meng, Chao Chen, Ayse~Kivilcim Coskun, and Ajay Joshi,
\newblock ``Run-time energy management of manycore systems through
  reconfigurable interconnects,''
\newblock in {\em Proceedings of the 21st edition of the great lakes symposium
  on Great lakes symposium on VLSI}, New York, NY, USA, 2011, GLSVLSI '11, pp.
  43--48, ACM.

\bibitem{PACT}
Lei Wang, Poornachandran Kumar, Ki~Hwan Yum, and Eun~Jung Kim,
\newblock ``Apcr: an adaptive physical channel regulator for on-chip
  interconnects,''
\newblock in {\em Proceedings of the 21st international conference on Parallel
  architectures and compilation techniques}. 2012, PACT '12, pp. 87--96, ACM.

\bibitem{ISCA}
Asit~K. Mishra, N.~Vijaykrishnan, and Chita~R. Das,
\newblock ``A case for heterogeneous on-chip interconnects for cmps,''
\newblock in {\em Proceedings of the 38th annual international symposium on
  Computer architecture}, New York, NY, USA, 2011, ISCA '11, pp. 389--400, ACM.

\bibitem{5090667}
M.A. Al~Faruque, T.~Ebi, and J.~Henkel,
\newblock ``Configurable links for runtime adaptive on-chip communication,''
\newblock in {\em Design, Automation Test in Europe Conference Exhibition,
  2009. DATE '09.}, april 2009, pp. 256 --261.

\bibitem{5715603}
Ying-Cherng Lan, Hsiao-An Lin, Shih-Hsin Lo, Yu~Hen Hu, and Sao-Jie Chen,
\newblock ``A bidirectional noc (binoc) architecture with dynamic
  self-reconfigurable channel,''
\newblock {\em Computer-Aided Design of Integrated Circuits and Systems, IEEE
  Transactions on}, vol. 30, no. 3, pp. 427 --440, march 2011.

\bibitem{BiNoC1}
Myong~Hyon Cho, M.~Lis, Keun~Sup Shim, M.~Kinsy, T.~Wen, and S~Devadas,
\newblock ``Oblivious routing in on-chip bandwidth-adaptive networks,''
\newblock in {\em Parallel Architectures and Compilation Techniques, 2009. PACT
  '09. 18th International Conference on}, 2009, pp. 181--190.

\bibitem{5982012}
Wen-Chung Tsai, Deng-Yuan Zheng, Sao-Jie Chen, and Yu-Hen Hu,
\newblock ``A fault-tolerant noc scheme using bidirectional channel,''
\newblock in {\em Design Automation Conference (DAC), 2011 48th ACM/EDAC/IEEE},
  june 2011, pp. 918 --923.

\bibitem{5470359}
Shih-Hsin Lo, Ying-Cherng Lan, Hsin-Hsien Yeh, Wen-Chung Tsai, Yu-Hen Hu, and
  Sao-Jie Chen,
\newblock ``Qos aware binoc architecture,''
\newblock in {\em Parallel Distributed Processing (IPDPS), 2010 IEEE
  International Symposium on}, april 2010, pp. 1 --10.

\bibitem{phit-noc}
R.~Hesse, J.~Nicholls, and N.E. Jerger,
\newblock ``Fine-grained bandwidth adaptivity in networks-on-chip using
  bidirectional channels,''
\newblock in {\em Networks on Chip (NoCS), 2012 Sixth IEEE/ACM International
  Symposium on}, 2012, pp. 132--141.

\bibitem{1310774}
R.~Mullins, A.~West, and S.~Moore,
\newblock ``Low-latency virtual-channel routers for on-chip networks,''
\newblock in {\em Computer Architecture, 2004. Proceedings. 31st Annual
  International Symposium on}, june 2004, pp. 188 -- 197.

\bibitem{long-wire-link}
A.K. Lusala, P.~Manet, B.~Rousseau, and J.~Legat,
\newblock ``Noc implementation in fpga using torus topology,''
\newblock in {\em Field Programmable Logic and Applications, 2007. FPL 2007.
  International Conference on}, 2007, pp. 778--781.

\bibitem{long-range-link}
U.Y. Ogras and R.~Marculescu,
\newblock ``"it's a small world after all": Noc performance optimization via
  long-range link insertion,''
\newblock {\em Very Large Scale Integration (VLSI) Systems, IEEE Transactions
  on}, vol. 14, no. 7, pp. 693--706, 2006.

\bibitem{repeaters-noc}
A.~Morgenshtein, I.~Cidon, A.~Kolodny, and R.~Ginosar,
\newblock ``Low-leakage repeaters for noc interconnects,''
\newblock in {\em Circuits and Systems, 2005. ISCAS 2005. IEEE International
  Symposium on}, 2005, pp. 600--603 Vol. 1.

\bibitem{E3S}
{\em E3S benchmarks, http://ziyang.eecs.umich.edu/dickrp/e3s/}.

\bibitem{Spec}
{\em Spec benchmarks, http://www.spec.org/web99},
\newblock 1999.

\bibitem{DBAR}
Sheng Ma, N.E. Jerger, and Zhiying Wang,
\newblock ``Dbar: An efficient routing algorithm to support multiple concurrent
  applications in networks-on-chip,''
\newblock in {\em Computer Architecture (ISCA), 2011 38th Annual International
  Symposium on}, 2011, pp. 413--424.

\bibitem{Kumar}
A~Kumar, Li-Shiuan Peh, P.~Kundu, and N.K. Jha,
\newblock ``Toward ideal on-chip communication using express virtual
  channels,''
\newblock {\em Micro, IEEE}, vol. 28, no. 1, pp. 80--90, 2008.

\bibitem{NoP}
G.~Ascia, V.~Catania, M.~Palesi, and D.~Patti,
\newblock ``Neighbors-on-path: A new selection strategy for on-chip networks,''
\newblock in {\em Embedded Systems for Real Time Multimedia, Proceedings of the
  2006 IEEE/ACM/IFIP Workshop on}, 2006, pp. 79--84.

\bibitem{RCA}
P.~Gratz, B.~Grot, and S.W. Keckler,
\newblock ``Regional congestion awareness for load balance in
  networks-on-chip,''
\newblock in {\em High Performance Computer Architecture, 2008. HPCA 2008. IEEE
  14th International Symposium on}, 2008, pp. 203--214.

\bibitem{ANCS}
R.S. Ramanujam and Bill Lin,
\newblock ``Destination-based adaptive routing on 2d mesh networks,''
\newblock in {\em Architectures for Networking and Communications Systems
  (ANCS), 2010 ACM/IEEE Symposium on}, 2010, pp. 1--12.

\bibitem{traffic-aware-noc}
Zhiliang Qian, Paul Bogdan, Guopeng Wei, Chi-Ying Tsui, and Radu Marculescu,
\newblock ``A traffic-aware adaptive routing algorithm on a highly
  reconfigurable network-on-chip architecture,''
\newblock in {\em Proceedings of the Eighth IEEE/ACM/IFIP International
  Conference on Hardware/Software Codesign and System Synthesis}, New York, NY,
  USA, 2012, CODES+ISSS '12, pp. 161--170, ACM.

\bibitem{Chifeng}
Chifeng Wang, Wen-Hsiang Hu, and N.~Bagherzadeh,
\newblock ``Congestion-aware network-on-chip router architecture,''
\newblock in {\em Computer Architecture and Digital Systems (CADS), 2010 15th
  CSI International Symposium on}, 2010, pp. 137--144.

\bibitem{Grot}
B.~Grot, J.~Hestness, S.W. Keckler, and O.~Mutlu,
\newblock ``Express cube topologies for on-chip interconnects,''
\newblock in {\em High Performance Computer Architecture, 2009. HPCA 2009. IEEE
  15th International Symposium on}, 2009, pp. 163--174.

\bibitem{zhiliang}
Zhiliang Qian, Ying-Fei Teh, and Chi-Ying Tsui,
\newblock ``A flit-level speedup scheme for network-on-chips using
  self-reconfigurable bi-directional channels,''
\newblock in {\em Design, Automation Test in Europe Conference Exhibition
  (DATE), 2012}, 2012, pp. 1295--1300.

\bibitem{DBmbench}
Minglong Shao, Anastassia Ailamaki, and Babak Falsafi,
\newblock ``Dbmbench: Fast and accurate database workload representation on
  modern microarchitecture,''
\newblock in {\em Proceedings of the 2005 Conference of the Centre for Advanced
  Studies on Collaborative Research}. 2005, CASCON '05, pp. 254--267, IBM
  Press.

\end{thebibliography}

\end{document}